%% file: fst24-fusion.tex
\documentclass{sty/nseJournal}

\input tex/header

\input tex/pf

\usepackage[parfill]{parskip}
\usepackage[german]{varioref}
\usepackage{url}
\usepackage{amsmath} 
\usepackage{dcolumn}
\usepackage{tikz}
\usetikzlibrary{shapes,arrows}
\usetikzlibrary{intersections}

\usepackage{pgfplots}
\usepackage{pgfplotstable}
\usepgfplotslibrary{groupplots}
\pgfplotsset{compat=1.5}
\pgfplotsset{compat = newest}  
\usepgfplotslibrary{groupplots}
\usetikzlibrary{matrix}

\begin{document}

\title{Spectral Element Simulation of Liquid Metal Magnetohydrodynamics}


\addAuthor{Yichen Guo}{d}   
\addAuthor{Paul Fischer}{a,b,c} 
\addAuthor{\correspondingAuthor{Misun Min}}{a}

\correspondingEmail{mmin@mcs.anl.gov}  

\addAffiliation{a}{Mathematics and Computer Science Division, Argonne National Laboratory\\ 
                   9700 S. Cass Avenue, Lemont, IL 60439}
\addAffiliation{b}{University of Illinois Urbana-Champaign, Department of Computer Science\\ 
                   201 N Goodwin Avenue, Urbana, IL 61801}
\addAffiliation{c}{University of Illinois Urbana-Champaign, Mechanical Science \& Engineering\\
                   1206 W. Green Steet, Urbana, IL 61801}
\addAffiliation{d}{Virginia Tech, Department of Mathematics\\ 
                   225 Stranger Street, Blacksburg, VA 24060}                       

\addKeyword{MHD}
\addKeyword{Fusion blanket}
\addKeyword{Liquid metal} 
\addKeyword{Spectral element method}
\addKeyword{Nek5000}
\addKeyword{NekRS}

\titlePage

 \input tex/abstract
 \input tex/introduction

 \input tex/governing

\input tex/paul      
 \input tex/yichen
 \input tex/conclusion

 \input tex/acknowledgments


\bibliographystyle{sty/ans_js}                                            
\bibliography{bibs/emmd}

\end{document}

%% file: tex/header.tex
\newcommand{\dt}{\dftl{t}}

\newcommand{\cH}{\mathcal{H}}

\newcommand{\be}{\begin{equation}}
\newcommand{\ee}{\end{equation}}
\newcommand{\bes}{\begin{equation*}}
\newcommand{\ees}{\end{equation*}}
\newcommand{\bse}{\begin{subequations}}
\newcommand{\ese}{\end{subequations}}

\newcommand{\bB}{\mathbf{B}}
\newcommand{\uu}{\mathbf{u}}

%% file: tex/pf.tex
\def\d{\partial}
\def\dO{\partial \Omega}

\def\ee{{\hat {\underline e}}}

\def\dt{ \Delta t }

\def\cH{{\cal H}}

\def\scriptO{{{\it O}\kern -.42em {\it `}\kern + .20em}}
\def\VV{{{\rm V}\kern - .62em {\rm V} }}
\def\RR{{{\rm l}\kern - .15em {\rm R} }}
\def\PP{{{\rm l}\kern - .15em {\rm P} }}
\def\L2{{{\sf L}^2}}
\def\H1{{{\sf H}^1}}
\def\PN2{{\PP_{N}-\PP_{N-2}}}

\def\complex{{{\rm C} \kern - .53em {\rm l} \kern + .38em}}

\def\a1{{ | \lambda_{\min} |}}

\def\l1{{   \lambda_{\min}  }}
\def\bhu{{\hat   {\bf u}}}
\def\bhB{{\hat   {\bf B}}}

\def\bz{{\bf z}}

\def\bg{{\bf g}}

\def\bhu{{\hat   {\bf u}}}

\def\bnh{{\hat   {\bf n}}}

\def\bu0{{\underline {\bf 0}}}

\def\br{{\bf r}}
\def\bu{{\bf u}}

\def\bx{{\bf x}}

\def\bB{{\bf B}} 

\def\Oh{{\hat \Omega}}

\def\be{{\bf e}}

\def\bh{{\bf h}}

\def\uf{{\underline f}}

\def\uq{{\underline q}}

\def\uu{{\underline u}}

\def\u0{{\underline 0}}
\def\1u{{\underline 1}}

\newcommand{\pp}[2]{\frac{\partial #1}{\partial #2} }

\def\tb{{\tilde b}}

\def\N2{{\frac{N}{2}}}

\newlength{\dhatheight}

\def\ha{\mathit{Ha}}

\def\C{\mathbb{C}}
\def\N{\mathbb{N}}

\def\CNN{\C^{N\times N}\kern-2pt}

  \def\BA{{\bf A}}

  \def\BP{{\bf P}} \def\CP{{\cal P}}

  \def\BV{{\bf V}}

\def\BAs{\BA{\kern-1pt}}      
\def\BPs{\BP{\kern-1.2pt}}    
\def\BVs{\BV{\kern-1.2pt}}    
\def\CPs{\CP{\kern-1.2pt}}    

%% file: tex/abstract.tex
\abstract{
A spectral-element-based formulation of incompressible MHD is presented in the
context of the open-source fluid-thermal code, Nek5000/RS.  The formulation supports
magnetic fields in a solid domain that surrounds the fluid domain.  Several 
steady-state and time-transient model problems are presented as part of the
code verification process.   Nek5000/RS is designed for large-scale turbulence
simulations, which will be the next step with this new MHD capability.
}

%% file: tex/introduction.tex
\section{Introduction}

Incompressible magnetohydrodynamics (MHD) describes the motion of electrically
conducting fluids in a magnetic field. This coupled system for the velocity,
$\bu$, and magnetic field, $\bB$, has been widely applied in various fields,
including astrophysics and fusion.  For fusion, there is significant interest
in developing liquid metal breeder blankets, which involves transport under
strong magnetic fields and high thermal loads.  Here, we describe an extension
of the spectral element thermal-fluid code, Nek5000/RS, to support
incompressible MHD.  Nek5000/RS is based on rapidly convergent matrix-free
spectral element methods (SEM) that yield significant savings in computational
resources for turbulent flow simulations.  Nek5000 has scaled to $P > 10^6$
CPU cores \cite{fischer15} and NekRS has scaled to $P = 72,000$ GCDs
on OLCF's AMD-based Frontier supercomputer \cite{gb23}.

Due to the strong coupling between a magnetic field and a velocity field
resulting from the induction effect, the MHD system is more complicated than
the Navier-Stokes equations that govern incompressible fluid motion alone.  The
MHD system involves two vector fields and, invariably, two Poisson equations,
one for pressure and a magnetic pseudo-pressure, that act as Lagrange
multipliers to keep each of the vector fields divergence-free.  Additionally,
the $\bB$-field can take on nontrivial values in the solid vessel that
surrounds the fluid domain, such that two domains are required to specify the
governing equations.


%% file: tex/governing.tex
\section{Governing Equations}  

We consider incompressible MHD defined on $\Omega_M=\Omega_F \bigcup \Omega_S$,
where $\Omega_F$ is the fluid domain and $\Omega_S$ is the solid domain.   The
governing equations are
\begin{eqnarray}
 \label{mhd1} \pp{\bu}{t}-\nu \nabla^{2} \bu +\nabla p  
               &=& \bB \cdot \nabla \bB-\bu \cdot \nabla \bu, \\ 
 \label{mhd3} \nabla \cdot \mathbf{u}  &=&0, \\ 
 \label{mhd2} \pp{\bB}{t}-\eta \nabla^{2} \bB
               +\nabla q &=& \bB \cdot \nabla \bu-\bu \cdot \nabla \bB, \\ 
 \label{mhd4} \nabla \cdot \bB &=& 0,
 \end{eqnarray}
with $\bu$, $\bB$, $p$, and $q$ the respective velocity, magnetic field,
hydrodynamic pressure (divided by the density), and magnetic pressure.
The kinematic viscosity and magnetic diffusivity are $\nu$ and $\eta$,
respectively.
We define initial conditions $\bu(\bx,t=0)=\bu^0$ and $\bB(\bx,t=0)=\bB^0$
and boundary conditions $\bu(\bx,t)=\bu_b$ where 
Dirichlet conditions are applied on $\dO_{FD}\subset\dO_F$
and $\bB(\bx,t)=\bB_b$ on $\dO_{D}\subset\dO_M$.
Homogeneous Neumann conditions are defined on $\dO_{FN}\subset\dO_F$ 
and $\dO_{MN}\subset\dO_M$.

For the quadratic interaction terms on the right-hand side of (\ref{mhd1}) and
(\ref{mhd2}) it is convenient to introduce Els\"asser variables defined by
$\bz^{\pm}=\bu \pm \bB$, which give 
 \begin{eqnarray}
 \bu &=&  (\bz^+ + \bz^-)/2,\\
 \bB &=&  (\bz^+ - \bz^-)/2.
 \end{eqnarray}
Defining $\nu^\pm=(\nu\pm\eta)/2$, 
(equivalently $\nu= \nu^+ + \nu^-$ and $\eta= \nu^+ - \nu^-$), 
we can express the left-hand sides of 
Eqs.~(\ref{mhd1}) and (\ref{mhd2}) as the following:
  \begin{eqnarray}
  \label{elsass1}
  lhs^+ &=& \frac{\d \bz^+}{\d t} - \nu^+\nabla^2\bz^+ - \nu^-\nabla^2 \bz^- +\nabla (p+q),\\
  \label{elsass2}
  lhs^- &=& \frac{\d \bz^-}{\d t} - \nu^+\nabla^2\bz^- - \nu^-\nabla^2 \bz^+ +\nabla (p-q).
  \end{eqnarray}
Then the nonlinear terms on the right-hand sides of Eqs.~(\ref{mhd1}) and 
(\ref{mhd2}) can be expressed by $rhs^+$ and $rhs^-$, respectively, 
as the following:
  \begin{eqnarray}
  \label{elsass3}
  rhs^+ &=& -\bz^- \cdot \nabla \bz^+,\\
  \label{elsass4}
  rhs^- &=& -\bz^+ \cdot \nabla \bz^-.   
  \end{eqnarray}
Thus we can write (\ref{elsass1})--(\ref{elsass4}) in the following compact form:
  \begin{eqnarray} 
  \label{elsass0}
  \frac{\d \bz^\pm}{\d t} - \nu^+\nabla^2\bz^\pm - \nu^-\nabla^2 \bz^\mp +\nabla (p\pm q)=
  -\bz^\mp \cdot \nabla \bz^\pm.
  \end{eqnarray}

The attraction of (\ref{elsass0}) is its simplicity and symmetry, particularly
regarding the $rhs^\pm$, which requires evaluation of only two convective substeps
rather than the four that are required in the primitive-variable formulation
(\ref{mhd1})--(\ref{mhd2}).  Unfortunately, if $\Omega_F \neq \Omega_M$, the full
Els\"asser formulation presents several challenge because $\bu$ and $\bB$ have
boundary conditions in differing locations.  Our approach therefore is to use
(\ref{elsass0}) to generate the convection terms and to then convert
these back to primitive variables using
  \begin{eqnarray}
   rhs^{\bu} &=& (rhs^+ \, + \; rhs^-)/2, \\[1.3ex]
   rhs^{\bB} &=& (rhs^+ \, -  \;rhs^-)/2 .
  \end{eqnarray}
In this way, the $rhs^\pm$ terms require only two straightforward advection
updates, (which can be done using either BDF$k$/EXT$k$ or
characterisics~\cite{patel18}).   Note that $z^{\pm}$ is defined everywhere by
simply setting the fluid velocity $\bu\equiv 0$ in $\Omega_S$.

Finally, we have the following equations in nondimensional form
for solving Eqs.~(\ref{mhd1})--({\ref{mhd4}):
  \begin{eqnarray}
  \label{solve-mhd1}
  \pp {\bu}{t}-\frac{1}{Re} \nabla^{2} \bu +\nabla p
  &=& rhs^\bu,\\
  \label{solve-mhd2}
  \pp {\bB}{t}- \frac{1}{Rm} \nabla \cdot r_w \nabla \bB 
  +\nabla q &=& rhs^\bB,\\
  \label{solve-mhd3}
  \nabla \cdot \mathbf{u}  &=&0, \\ 
  \label{solve-mhd4}
  \nabla \cdot \bB &=& 0, 
\end{eqnarray}
where $Re=UL/\nu$ and $Rm=UL/\eta$ are the Reynolds 
number and the magnetic Reynolds number, respectively, and 
 $r_w$ reflects the change in $\eta$ as one moves from
the liquid to the supporting wall, $\Omega_S$,
\begin{eqnarray}
   r_w    &=& 1 \;\;\; \mbox{in } \Omega_F \\[1ex]
   r_w    &=& \frac{\eta_S}{\eta} \;\;\; \mbox{in } \Omega_S,
\end{eqnarray}
where $\eta_S$ is the magnetic diffusivity in the solid,
$\Omega_S := \Omega_M \backslash \Omega_F.$

\section{Numerical Formulation}    
In this section we describe our temporal and spatial discretizations including
the weak formulation.  Following the formulation used in Nek5000/RS
\cite{fischer17}, the temporal discretizations are based on $k$th-order
backward difference formulas and extrapolation (BDF$k$/EXT$k$, $k=1,$ 2, or 3).
Spatial discretizations are based on the spectral element method \cite{pat84,dfm02}.

\subsection{Temporal Discretization}
The terms on the left of (\ref{solve-mhd1})--(\ref{solve-mhd2}) are treated
implicitly, with the temporal derivatve computed at time $t^n$ using a
$k$th-order backward difference formulation (BDFk),
 \begin{eqnarray} 
  \label{eq:bdfk1}
  \left.  \pp{\bu}{t} \right|^{}_{t^n} & \equiv &
  \frac{1}{\dt} \sum_{j=0}^k \beta_j \bu^{n-j} \;+\; O(\dt^k),\\
  \label{eq:bdfk2}
  \left.  \pp{\bB}{t} \right|^{}_{t^n} & \equiv &
  \frac{1}{\dt} \sum_{j=0}^k \beta_j \bB^{n-j} \;+\; O(\dt^k),
 \end{eqnarray}
where $\bu^{n-j}$ and $\bB^{n-j}$  denote the velocity and the magnetic field 
at time $t^{n-j}$, respectively.  For the case
of uniform timestep size $\dt$, $t^{n}=n \dt$, and the BDF$k$ coefficients are
$[\beta_0, \dots, \beta_k]$=[-1,1] for $k=1$,
[3/2, -4/2, 1/2] for $k=2$, and
[11/6, -18/6, 9/6, -2/6] for $k=3$.
The terms on the right of (\ref{solve-mhd1})--(\ref{solve-mhd2}), which include nonlinear 
advection and any forcing terms are treated with $k$th-order extrapolation.  
If $\bg := rhs^\bu$ and $\bh := rhs^\bB$,
then the right-hand side of (\ref{solve-mhd1})--(\ref{solve-mhd2}) are
  \begin{eqnarray}
  \label{eq:extk1}
  \left.  \bg \right|^{}_{t^n} & \equiv &
  \sum_{j=1}^k \alpha_j \bg^{n-j} \;+\; O(\dt^k),\\
  \left.  \bh \right|^{}_{t^n} & \equiv &
  \sum_{j=1}^k \alpha_j \bh^{n-j} \;+\; O(\dt^k).
  \label{eq:extk2}
  \end{eqnarray}
For uniform $\dt$, we have $[\alpha_1, \dots, \alpha_k]$=[1] for $k=1$, [2, -1]
for $k=2$, and [3, -3, 1] for $k=3$.
For the semidiscretization in time, we drop the $O(\dt^k)$ residual terms
and use (\ref{eq:bdfk1})--(\ref{eq:extk2}) in (\ref{solve-mhd1})--(\ref{solve-mhd2}).  
All other terms are evaluated implicitly at time $t^n$.

The first step in the splitting scheme is simply to sum all quantities known
from previous timesteps, which amounts to constructing tentative velocity 
and magnetic fields,
  \begin{eqnarray} 
  \label{eq:adv1}
  \bhu & := &
   \sum_{j=1}^k \left( -\beta_j \bu^{n-j} \;+\; \dt \alpha_j \bg^{n-j} \right),\\
  \bhB & := &
  \label{eq:adv2}
   \sum_{j=1}^k \left( -\beta_j \bB^{n-j} \;+\; \dt \alpha_j \bh^{n-j} \right).  
  \end{eqnarray}
As this step is just data collection, no boundary conditions are imposed
in (\ref{eq:adv1})--(\ref{eq:adv2}).
The remaining implict equations are linear, with the velocity and magnetic
field solutions completely decoupled.  Within each field, the 
individual {\em components} of each are still coupled because of the
the divergence-free constraints, (\ref{solve-mhd3}) and (\ref{solve-mhd4}).  
Formally taking the divergence of (\ref{mhd1}) and (\ref{mhd2}) 
and requiring $\nabla \cdot \bu^n=0$ and $\nabla \cdot \bB^n=0$ lead to 
independent Poisson equations for the pressure and the magnetic pressure, 
  \begin{eqnarray} 
   \label{eq:p1}
   -\nabla^2 p^n &=& -\nabla \cdot \bhu \;\; \mbox{in} \, \Omega_F,\\
   \label{eq:p2}
   -\nabla^2 q^n &=& -\nabla \cdot \bhB  \;\; \mbox{in} \, \Omega_M.
\end{eqnarray}
Boundary conditions for (\ref{eq:p1}) and (\ref{eq:p2}) are found by taking 
the dot product of (\ref{solve-mhd1})--(\ref{solve-mhd2}) with the 
outward-pointing unit normal on $\Omega_F$ and $\Omega_M$, respectively,
which yields
  \begin{eqnarray} 
   \label{eq:pbc1}
   \biggl.
   \nabla p^n \cdot \bnh  \biggr|^{}_{\dO_F}
    &=&
   \left.
   \left(\frac{\bhu-\beta_0 \bu^n}{\dt} + \frac{1}{Re} \nabla^2 \bu^n \right)
        \cdot \bnh \right|^{}_{\dO_F},\\
   \label{eq:pbc2}
   \biggl.
   \nabla q^m \cdot \bnh  \biggr|^{}_{\dO_M}
    &=&
   \left.
   \left(\frac{\bhB-\beta_0 \bB^n}{\dt} + \frac{1}{Rm} \nabla \cdot \frac{1}{r_w} \nabla \bB^n \right)
        \cdot \bnh \right|^{}_{\dO_M}.  
   \end{eqnarray}
We remark that $\bu^n$ is known on $\dO_F$ while $\nabla^2 \bu^n$ is not.
Similarly, $\bB^n$ is known on $\dO_M$ while $\nabla^2 \bB^n$ is not.
As with the nonlinear term, the unknown viscous and diffusivity terms can be 
treated explicitly after applying the vector identity
   \begin{eqnarray} 
   \label{eq:pcc1}
   \nabla^2 \bu^n  &=& \nabla(\nabla \cdot \bu^n) \,-\, \nabla \times (\nabla \times \bu^n),\\
   \label{eq:pcc2}
   \nabla^2 \bB^n  &=& \nabla(\nabla \cdot \bB^n) \,-\, \nabla \times (\nabla \times \bB^n).
\end{eqnarray}
The first terms on the right of (\ref{eq:pcc1})--(\ref{eq:pcc2}) 
are set to zero because of (\ref{solve-mhd3})--(\ref{solve-mhd4}), 
so one only needs to extrapolate the curl of the vorticity,
$\nabla \times (\nabla \times \bu^n)$ on $\dO_{FD}$ and
$\nabla \times (\nabla \times \bB^n)$ on $\dO_{MD}$~\cite{tomboulides89}.
Once the fluid pressure and magnetic pressure are known from (\ref{eq:p1})--(\ref{eq:pbc2}),
we can compute each component of $\bu^n = [u^n_1 \; u^n_2 \; u^n_3]$
and $\bB^n = [B^n_1 \; B^n_2 \; B^n_3]$ by solving a
sequence of Helmholtz problems for $\bu$ and $\bB$,
  \begin{eqnarray} 
  \label{eq:visc1}
  - \frac{\dt}{Re} \nabla^2 u^n_i
  \;+\;
  \beta_0 u^n_i
  &=& \dt \left( {\hat u}_i - \pp{p^m}{x_i} \right),
  \;\; i=1,\dots,3, \\
    \label{eq:visc2}
  - \frac{\dt}{Rm} \nabla^2 B^n_i
  \;+\;
  \beta_0 B^n_i
  &=& \dt \left( {\hat B}_i - \pp{q^m}{x_i} \right),
  \;\; i=1,\dots,3,    
  \end{eqnarray}
subject to appropriate boundary conditions on each component of $\bu^n$ and $\bB^n$.  In
the case of mixed boundary conditions or variable diffusivities, 
we replace (\ref{eq:visc1})--(\ref{eq:visc2}) by a pair of independent systems
in which the individual components of velocity and magnetic field components
are coupled.   In either case, the viscous and diffusivity system will be
diagonally dominant for modest to high Reynolds number applications given that 
$\dt/Re \ll 1$ and $\dt/Rm \ll 1$. 
For these cases,
Jacobi preconditioned conjugate gradient iteration is an efficient solution
strategy for the fully discretized form of (\ref{eq:visc1})--(\ref{eq:visc2}).

\subsection{Spatial Discretization}
We use the spectral element method (SEM) \cite{pat84}, which is a
high-order $p$-type finite element method (FEM) restricted to elements that
are isoparametric images of the unit cube, $\Oh := [-1,1]^3$.
On each element, $\Omega^e$, $e=1,\dots,E$,
the solution and geometry are represented by tensor-product sums of
$N$th-order Lagrange polynomials in $\Oh$,
 \begin{eqnarray} 
 \label{eq:uijk1}
 \biggl.  u(\br) \biggr|^{}_{\Omega^e}
   &=& \sum_{k=0}^N \sum_{j=0}^N \sum_{i=0}^N u_{ijk}^e h_i(r) h_j(s) h_k(t), 
 \end{eqnarray}
and 
 \begin{eqnarray} 
 \label{eq:uijk2}
 \biggl.  \bx(\br) \biggr|^{}_{\Omega^e}
   &=& \sum_{k=0}^N \sum_{j=0}^N \sum_{i=0}^N \bx_{ijk}^e h_i(r) h_j(s) h_k(t).
 \end{eqnarray}
With nodal points $\xi_j$ taken to be the Gauss-Lobatto-Legendre (GLL)
quadrature points, this basis is {\em stable} and does not exhibit the Runge
phenomenon often associated with high-order polynomials.  Production values of
$N$ for NekRS are typically in the range $5$--$10$, but we have run single
element cases with Nek5000 up to $N=100$.  
As noted in the early work of Orszag
\cite{sao80}, restriction to tensor-product forms permits the use of fast {\em
matrix-free} formulations in which operator evaluation is effected in $O(EN^4)$
operations and $O(EN^3)$ memory references for an $E$-element mesh, even in the
case of deformed geometries.  The number of grid points is $n \approx EN^3$, so
the storage is only $O(n)$ and the work only $O(Nn)$, which is in sharp
contrast to the standard FEM, where $O(nN^3)$ storage and work complexities
typically compel a restriction to $N<4$.

The elliptic subproblems (\ref{eq:p1})--(\ref{eq:p2}) and
(\ref{eq:visc1})--(\ref{eq:visc2}) are cast in terms of variational projection operators.  
For example, for $u:=u_i^n$ (the $i$th component of the velocity), the discrete
form of (\ref{eq:visc1}) reads, {\em Find $u \in X^N_0 \subset \cH_0^1$ such that}
  \begin{eqnarray} 
  \label{eq:var}
  \frac{\dt}{Re} a(v,u) \;+\; \beta_0 \,(v,u) &=& (v,f)
  \;\;\; \forall \, v \, \in X_0^N.
  \end{eqnarray}
Here,
$(p,q):=\int_{\Omega} p q \, d\bx$ is the standard $L^2$ inner product on $\Omega$;
$a(p,q):=\int_{\Omega} \nabla p \cdot \nabla q \, d\bx$ is the $L^2$ inner product
of the gradients; $X_0^N$ is the trial/test space comprising the spectral element
basis functions; and $\cH_0^1$ is the space of square-integrable functions on
$\Omega$ that vanish on $\dO_D$ whose gradient is also square-integrable.
Formally expanding $u$ in terms of trial functions $\phi_j(\bx)$
that span $X_0^N$, $u(\bx) := \sum_{j=1}^n \phi_j(\bx) u_j$, and
setting $v=\phi_i$, $i=1,\dots,n$, leads to the matrix form of
(\ref{eq:var}),
  \begin{eqnarray} 
  \label{eq:mat}
  \frac{\dt}{Re} A \uu \;+\; \beta_0 \,B \uu &=& B\uf,
  \end{eqnarray}
where
$\uu=[u_1 \; \dots u_n]^T$ is the vector of unknown basis coefficients
and $\uf$ is the vector of all known data, including the extrapolated
nonlinear terms, boundary terms, and BDF$k$ terms.

Here, we introduce respective stiffness, mass, and advection matrices
having entries
\begin{eqnarray} \label{eq:abc}
   A_{ij}&=&\int \nabla \phi_i \cdot \nabla \phi_j \, d\bx \\[1.2ex]
   B_{ij}&=&\delta_{ij} \int \phi_i \, d\bx \\[1.2ex]
   C(\bz)_{ij}&=&\int \phi_i \, \bz \cdot \nabla \phi_j \, d\bx,
\end{eqnarray}
where $\bz$ is a given advecting field (e.g., evaluation of 
(\ref{elsass3}) reads $-C(\bz^-) {\underline \bz^+}$).
For the stiffness matrix, $A$, and mass matrix, $B$, it is standard practice in
the SEM to replace integration by quadrature on the nodal points, which is
effective because the rapid convergence of GLL quadrature for smooth functions.  
By contrast, it is important to use proper integration for the advection
operator in order to ensure that $C$ is skew-symmetric \cite{johan13}.
Over-integration with a 3/2s-rule increases operator complexity by about
$4\times$, which is one of the reasons why the parsimonious Elsasser
formulation of the nonlinear terms, given by (\ref{elsass3})--(\ref{elsass4})
is attractive.  The Elsasser formulation also makes clear that the physics
of the nonlinear terms are effectively advection operators, with implied
Courant-Friedrichs-Lewy (CFL) stability constraints that scale with 
characteristic velocity ($U_0$) and applied magnetic field strength ($B_0$).
Stability regions for BDF$k$/EXT$k$ can be found in \cite{fischer17}.

%% file: tex/paul.tex
\subsection{Transient Hartmann Flow}


In this subsection, we examine time-transient laminar Hartmann flow in a square
duct.  Since the flow is primarily in the $x$-direction and at low to moderate
Reynolds numbers, as considered here, the solution remains invariant in $x$ and
varies only in the $y$ and $z$ directions.  Under these conditions, we can
derive a simplified set of equations suitable for analysis and compare the
analysis with the full MHD solution generated by NekRS.  This allows us to
assess both steady-state and asymptotic time-transient behavior as part of the
code verification process.

In the test cases, we assume that we have conducting sidewalls of finite
thickness $\delta$.  We define the fluid domain as $\Omega_F = [0,
L]\times[-1,1]^2$ with Dirichlet boundary conditions $\bu|_{\dO_{FD}} =
(0,0,0)$ and the magnetic field domain as $\Omega_M = [0,
L]\times[-1-\delta,1+\delta]^2$ with Dirichlet boundary condition
$\bB|_{\dO_{MD}} \equiv \bB_0 = (0,B_0,0)$.  The initial conditions of velocity
and magnetic field are zero.  With an applied pressure gradient,
$\frac{\partial p}{\partial x} = -\frac{1}{Re}$, the flow evolves to a steady
state.  We define $r_w=1$ in the fluid domain, $\Omega_F$, and $r_w$ to be the
ratio of the conductivity of the fluid domain to the electric conductivity of
the sidewall, $\Omega_S := \Omega_M \backslash \Omega_F$.  When $r_w \gg 1$, it
is electrically insulating wall, which is equivalent to a Dirichlet condition
for $\bB$ on $\Omega_M$.  When $r_w \ll 1$, it is a conducting wall, which
corresponds to a Neumann condition for $\bB$ on $\Omega_M$.  The Hartmann
number $\ha$ is defined as $\ha := B_0 \sqrt{ReRm}$, which is a measure of the
applied field strength.  

Our solutions are marched to steady state from initial conditions $\bu = (0,0,0)$
and $\bB = \bB_0$.  With the chosen nondimensionalization, the steady state
solution is independent of $Re$ and $Rm$, so we take these to be unity in the
fluid domain, unless otherwise indicated.  The magnetic diffusivity in the
solid domain is consequently $r_w$, which is set to $r_w=10^{2}$, 1, and
$10^{-5}$.  The first case corresponds to (nearly) electrically insulating
walls while the last case corresponds to highly conducting walls. In the first
case, the magnetic field is approximately equal to $\bB_0$ everywhere in the
wall and the same computational results can be attained by setting $\bB :=
\bB_0$ at the fluid boundary $\d\Omega_F$.

Figure~\ref{vcontour} shows the steady state axial velocity fields for the
three boundary condition cases at $\ha =$ 10, 50, and 100. Similar plots for
$\ha=$ 10 and 50 can be found in \cite{muller2001magnetofluiddynamics}.  It can
be seen that thin bounday layers, known as Hartmann layers, form at the
Hartmann walls for large $\ha$, while somewhat thicker layers form at the
sidewalls. Figure~\ref{vmesh} shows the same data replotted as a surface mesh.
For the insulating case in particular the flattening of the velocity profiles
at high Ha is evident---the effect of the Lorentz force coupled with the
induction equation is to suppress gradients of u in the channel center.
Figures~\ref{bcontour} and ~\ref{bmesh} show the corresponding distributions
steady state axial magnetic fields for the cases of Figs.~\ref{vcontour} and
\ref{vmesh}.  Note that the magnetic field profile for large $\ha$ is essentially
linear in $y$ when the walls are highly conducting and does not form boundary
layers in $\Omega_M$. The case of perfectly conducting walls corresponds to
Neumann conditions for $\bB$  on $\Omega_F$ and results in a singular condition
for the steady-state solution because $\bB$ may be displaced by an arbitrary
constant. This singularity motivated the consideration of walls having large,
but finite, conductivity.

\input tex/paul_fig2  
\input tex/paul_fig4  
\input tex/paul_fig1  
\input tex/paul_fig3  

\newpage

\begin{figure}[t]
  \begin{tikzpicture}
  \begin{groupplot}[group style={
                        group name=myplot,
                        group size= 3 by 1,
                         horizontal sep = 50pt,
                          vertical sep= 2cm
  },height=6cm,width=5.cm]
   \nextgroupplot[  
      xlabel={time},
      ylabel={$u_z(0,0,t)$},
      xmax = 2,
      y tick label style={/pgf/number format/fixed, /pgf/number format/precision=2},
      scaled y ticks=false,
      ]
   \input{figs/data/ha010.tex}
  \nextgroupplot[ 
      xlabel={time},
      xmax = 1,
      y tick label style={/pgf/number format/fixed, /pgf/number format/precision=2},
      scaled y ticks=false,
  ]
  \input{figs/data/ha050.tex}
  \nextgroupplot[ 
    xlabel={time},
    xmax = 1,
    y tick label style={/pgf/number format/fixed, /pgf/number format/precision=3},
    scaled y ticks=false,
]
\input{figs/data/ha100.tex}
      \end{groupplot}
  \path (myplot c1r1.south west|-current bounding box.south)--
        coordinate(legendpos)
        (myplot c3r1.south east|-current bounding box.south);
  \matrix[
      matrix of nodes,
      anchor=south,
      draw,
      inner sep=0.2em,
      draw
    ]at([yshift=-6ex]legendpos)
    {
      \ref{1e2}&  $r_w = 10^{2}$ &[5pt]
      \ref{1e0}& $r_w = 1$ &[5pt]
      \ref{1e-5}& $r_w = 10^{-5}$ & [5pt] \\}; 
  \end{tikzpicture}
  \caption{Time histories of center-point velocity for $\ha=10,50,100$ (left,
middle, and right). $Re= Rm = 1$.}     \label{fig:time_his1}
  \end{figure}

\noindent {\bf Analysis of Transient Response.}

Because we are interested in simulation of turbulent flows, Nek5000/RS
is intrinsically designed to solve the time-dependent MHD equations.
All of the solutions of the preceding sections were thus computed in
a time-stepping fashion and the transient responses in fact reveal 
interesting physics.  Examples are shown in Fig. \ref{fig:time_his1},
where decaying and, potentially, oscillatory responses are evident,
depending on $\ha$ and $r_w$.

We analyze this behavior by deriving a simplified set of equations
that model the transient response.
We seek solutions of the form
$\bu=(u,0,0)$, $\bB=(\tb,B_0,0)$,
where $u=u(y,z,t)$ is the $x$-component of the velocity and $\tb=\tb(y,z,t)$ is
the $x$-component of the magnetic field.  We note of course that this is not
the only set of solutions---less trivial solutions will arise for sufficiently
high $(Re,Rm)$.  For sufficiently low Reynolds numbers, however, the flow is
laminar and axially-independent, so that (\ref{mhd1})--(\ref{mhd4}) reduce to 
\begin{eqnarray} 
\label{eq:hart11}
\pp{u}{t} \,-\; \frac{1}{Re}
\nabla^{2} {u}   &=& B_0 \pp{\tb}{y} \,+\, \frac{1}{Re} 
                             \hspace{.2500in} \mbox{in } \Omega_F, \\[1ex]
\label{eq:hart22}
\pp{\tb}{t} \,-\, \frac{r_w}{Rm}
\nabla^{2} \tb   &=& B_0 \pp{u}{y} 
                             \hspace{.7000in} \mbox{in } \Omega_M .
\end{eqnarray}
We apply homogeneous conditions for $u$ on $\dO_F$.
The boundary condition for magnetic field is defined later.
Note that $r_w \equiv 1$ in $\Omega_F$.  

To simplify the description of the steady-state solution, we presently
take $r_w=1$ and multiply
\eqref{eq:hart11} by $Re$ and \eqref{eq:hart22} by $Rm$ to yield
\begin{equation} \label{sh3}
  \begin{aligned}
  Re   \pp{u}{t} \, - \, \nabla^2 u   \,-\, B_0 Re \pp{\tb}{y} &=& 1, \\[1ex]
  Rm \pp{\tb}{t} \, - \, \nabla^2 \tb \,-\, B_0 Rm \pp{u}{y}   &=& 0.
  \end{aligned}
\end{equation}
Recall that the Hartmann number is defined as $\ha := B_0 \sqrt{ReRm}$.
Let $\tb := \sqrt{\frac{Rm}{Re}} b$.  Then
\begin{equation} 
  \label{eq:hart1dd}
  \begin{aligned}
  Re \pp{u}{t} \, - \, \nabla^2 u \,-\, \ha \pp{b}{y} &= 1, \\      
  Rm \pp{b}{t} \, - \, \nabla^2 b \,-\, \ha \pp{u}{y} &= 0.
  \end{aligned}
\end{equation} 

For the case of insulating walls, we have $u=0$ on $\d\Omega_F$ and $b=0$ on $\d\Omega_M$,
and we can conveniently analyze (\ref{eq:hart1dd}).
  If $Re=Rm$, we rewrite the equation \eqref{eq:hart1dd} 
in Elsasser variables $z^+ = u+b$ and $z^-=u-b$, 
\begin{equation}
\begin{aligned}
\label{hart1elss}
{Re}\pp{z^+}{t}- \nabla^2 z^+ - {\ha}\pp{z^+}{y} & = 1,\\
{Re}\pp{z^-}{t}- \nabla^2 z^- + {\ha}\pp{z^-}{y} & = 1.
\end{aligned}
\end{equation}
Coupled with homogeneous Dirichlet boundary conditions, (\ref{hart1elss})
represents a pair of independent convection-diffusion equations. Consistent
with the behavior in Fig.~\ref{fig:time_his1}, these parabolic equations do not
produce oscillatory solutions unless the initial condition is wavy. The
boundary layer character of these equations as $t\rightarrow \infty$ is also
well-understood in the high-$\ha$ limit.

We remark that the conducting sidewall case $r_w = 10^{-5}$ is not simplified
by the Elsasser formulation because the boundary conditions differ between $u$
and $b$; the velocity $u$ satisifes homogeneous Dirichlet conditions while the
magnetic field $b$ satisfies homogeneous Neumann conditions.  

Let $(u_0,b_0)$ be the steady-state solution to (\ref{eq:hart1dd}) satisfying,
\begin{equation}
\begin{aligned}
\label{eq:hart1ss}
   - \, \nabla^2 u_0 \,-\, \ha \pp{b_0}{y} &=& 1, \\      
   - \, \nabla^2 b_0 \,-\, \ha \pp{u_0}{y} &=& 0,
\end{aligned}
\end{equation}
which is governed by the single independent parameter, $\ha$, and the
yet to be prescribed boundary conditions on $b$. 
For the fully conducting sidewall, Fig.
\ref{bmesh} shows that the magnetic field is almost constant along x-direction.
Thus, we assume that $b_0$ only depends on $y$.
The weak form for \eqref{eq:hart1ss} is
\begin{equation}
\begin{aligned}
  \label{eq:u0b0}
\left(\nabla u_0, \nabla \phi \right) - \ha \left(\frac{\partial b_0}{\partial y}, \phi \right) & =  \left(1, \phi \right), \quad \forall \phi\in H_0^1(\Omega_F)\\
\left(\frac{\partial b_0}{\partial y}, \frac{\partial \psi}{\partial y} \right) & = \ha \left(\frac{\partial u_0}{\partial y} , \psi \right), \quad \forall \psi\in H^1(\Omega_M),
\end{aligned}
\end{equation}
Using integration-by-parts and the homogeneous Dirichlet conditions on $u_0$,
we obtain 
\[
  \left(\frac{\partial b_0}{\partial y}, \frac{\partial \psi}{\partial y} \right) = - \ha \left(u_0, \frac{\partial \psi}{\partial y} \right).
\]
The last equality implies 
\begin{equation}
  \left(\frac{\partial b_0}{\partial y}, \phi \right) =  - \ha \left(u_0, \phi \right).
  \label{eq:b0}
\end{equation}
Inserting 
(\ref{eq:b0}) into (\ref{eq:u0b0}), we obtain a reaction diffusion equation for $u_0$,
\begin{equation}
  \left(\nabla u_0, \nabla \phi \right) + \ha^2 \left( u_0, \phi \right)  =  \left(1, \phi \right)
\label{eq:u0}
\end{equation}
Equation (\ref*{eq:u0}) is a singular perturbation problem with homogeneous
Dirichlet boundary condition.  When $\ha$ is large, the center-point value,
$u_0(0,0)$, is close to $1/\ha^2$ \cite[Part III, section 1]{roos2008robust}.

To analyze the transient response, define $\bar{u} = u -u_0$ and $\bar{b} = b
-b_0$.  The functions $\bar{u}$ and $\bar{b}$ satisfy the homogeneous
equivalent of (\ref{eq:hart1dd}),
\begin{equation}
\begin{aligned}
\label{eq:hart1dzero}
   Re \pp{\bar u}{t} \,-\, \nabla^2 {\bar u} \, - \, \ha \pp{\bar b}{y} & = 0, \\     
   Rm \pp{\bar b}{t} \,-\, \nabla^2 {\bar b} \, - \, \ha \pp{\bar u}{y} & = 0,
\end{aligned}
\end{equation}

We proceed with the ansatz that $u$ has a modal expansion of the form
\begin{eqnarray}
\label{expand1}
    u( y, z,t)=\sum_{k=0}^{\infty} 
               \sum_{l=0}^{\infty} 
               \sum_{k=0}^{\infty} \sum_{l=0}^{\infty} \hat{u}_{k l}(t) \cos \left(\frac{ \pi}{2}  (2k+1) y \right) \cos \left(\frac{ \pi}{2} (2l+1)z \right),
\end{eqnarray}
which satisfies the homogeneous Dirichlet boundary conditions and matches the even symmetry 
of the data and initial conditions. Given that (\ref{eq:hart1dd}) contains 
a first derivative of $u$ with respect to $y$, we must assume that $b$ has the form
\begin{eqnarray}
\label{expand2}
    b( y, z,t)=\sum_{k=0}^{\infty} \sum_{l=0}^{\infty} \hat{b}_{k l}(t) \sin \left( \frac{ \pi}{2}  (2k+1) y \right) \cos \left(\frac{ \pi}{2} (2l+1)z \right),
\end{eqnarray}
which is consistent with (\ref{eq:hart1dd}) and (\ref{expand1}). 
We are interested in the time-transient response of the most slowly decaying mode. 
Inserting (\ref{expand1}) and (\ref{expand2}) into (\ref{eq:hart1dd}) and 
retaining only the lowest ($k = l = 0$) modes leads to
\begin{equation}
  \begin{aligned}
    & R e \hat{u}_t+\frac{\pi^2}{2} \hat{u}-\ha \frac{\pi}{2} \hat{b}=0 \\
    & R m \hat{b}_t+\frac{\pi^2}{2} \hat{b}+\ha \frac{\pi}{2} \hat{u}=0
  \end{aligned}
  \label{eq:mode}
\end{equation}
where $\hat u:=\hat u_{00}$ and $\hat b:=\hat b_{00}$. 


 Let $q:=(\hat u,\hat b)^T$. Then (\ref{eq:mode}) has the form $\uq_t + A\uq = 0$,with
 \begin{eqnarray} \label{eq:eig1}
 A:=\left[\begin{array}{cc}
       \frac{\pi^2}{2Re}  & - \frac{\pi}{2} \frac{\ha}{Re} \\
                          &                                \\
       \frac{\pi}{2} \frac{\ha}{Rm} & \frac{\pi^2}{2Rm}  \\
      \end{array}\right].
 \end{eqnarray}
 The eigenvalues of $A$ are of the form $s \pm iw$, where $s$ and $w$ are real with values
 \begin{equation}
  \label{eig1a}
  \begin{aligned}
  &s &=& \frac{\pi^2}{4}\left(\frac{1}{R e}+\frac{1}{R m}\right),\\ 
  &w &=& \frac{\pi}{2} \frac{\ha}{\sqrt{R e R m}}\left(1-\frac{\pi^2}{4 \ha^2}
                      \frac{(R m-R e)^2}{R m R e}\right)^{\frac{1}{2}}.
  \end{aligned}
\end{equation}


Combining the first-mode transient with the steady state solution to (\ref{eq:hart1dd})
leads to the asymptotic transient response,
\begin{equation} \label{eq:soln}
u \; \sim \; \hat{u}_0 e^{-s t} \sin (\omega t+\phi)+\frac{1}{\ha^2},
\end{equation}
To leading order in $\ha$ (exact for the case $Re = Rm$), (\ref{eig1a}) implies
a period $\tau = 4/\ha$ with a decaying envelope of the form $e^{st}$, which is in excellent 
agreement with the behavior observed in the simultations. 


Figure~\ref{fig:time_his2} shows the decay rate of the centerline velocity for
the case $\ha = 10, 50, 100$ with $Re = Rm = 1$ compared with the model
\begin{eqnarray}
 u \, \sim \, \hat u_0 e^{-st} \sin(\omega t + \phi) + s_0,
\end{eqnarray}
where $\hat u_0, s, \omega, \phi$ and $s_0$ are determined by a nonlinear fit
using Levenberg-Marquardt.

\input tex/fig_his2

\input tex/fig_his3

Table~\ref{table:compare} shows the approximate values of the decay rate and
frequency via the nonlinear fit and the analytic values from \eqref{eig1a} for
various $\ha=10,50,100$.  The results indicate that the approximate decay rate is
slightly higher than the analytic value, likely due to the analysis considering
only the dominant term, while higher-order terms exhibit faster decay rates.
In contrast, the approximate frequency and steady-state value show good
agreement with the analytic values.  Figure~\ref{fig:time_his3} illustrates the
time history of the center-point velocity, using the analytic decay rate and
frequency from equation \eqref{eig1a}, while employing a nonlinear fit to
determine only  $\hat{u}_0$ and $\phi$.  The asymptotic results are in good
agreement with the numerical results.

\input tex/table1
\input tex/table2

%% file: tex/paul_fig2.tex
\begin{figure}
  \centering
  \begin{subfigure}[t]{0.37\textwidth} 
      \makebox[20pt]{\raisebox{40pt}{\rotatebox[origin=c]{0}{
      $\begin{matrix}
          r_w \\ 10^{2}
      \end{matrix}$
      }}}%
      \includegraphics[width=5.5cm]{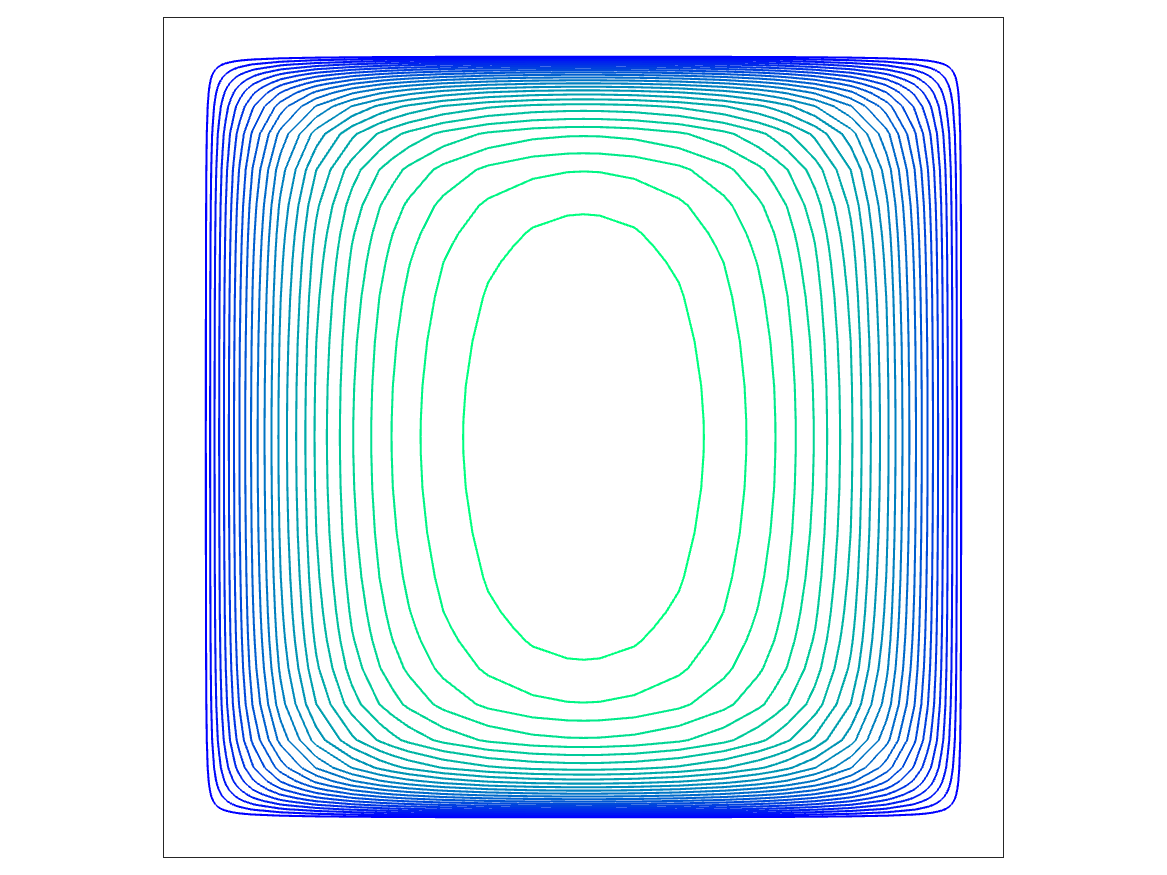}\\[5pt]
      \makebox[20pt]{\raisebox{40pt}{\rotatebox[origin=c]{0}{
      $\begin{matrix}
          r_w \\ 1
      \end{matrix}$
      }}}%
      \includegraphics[width=5.5cm]{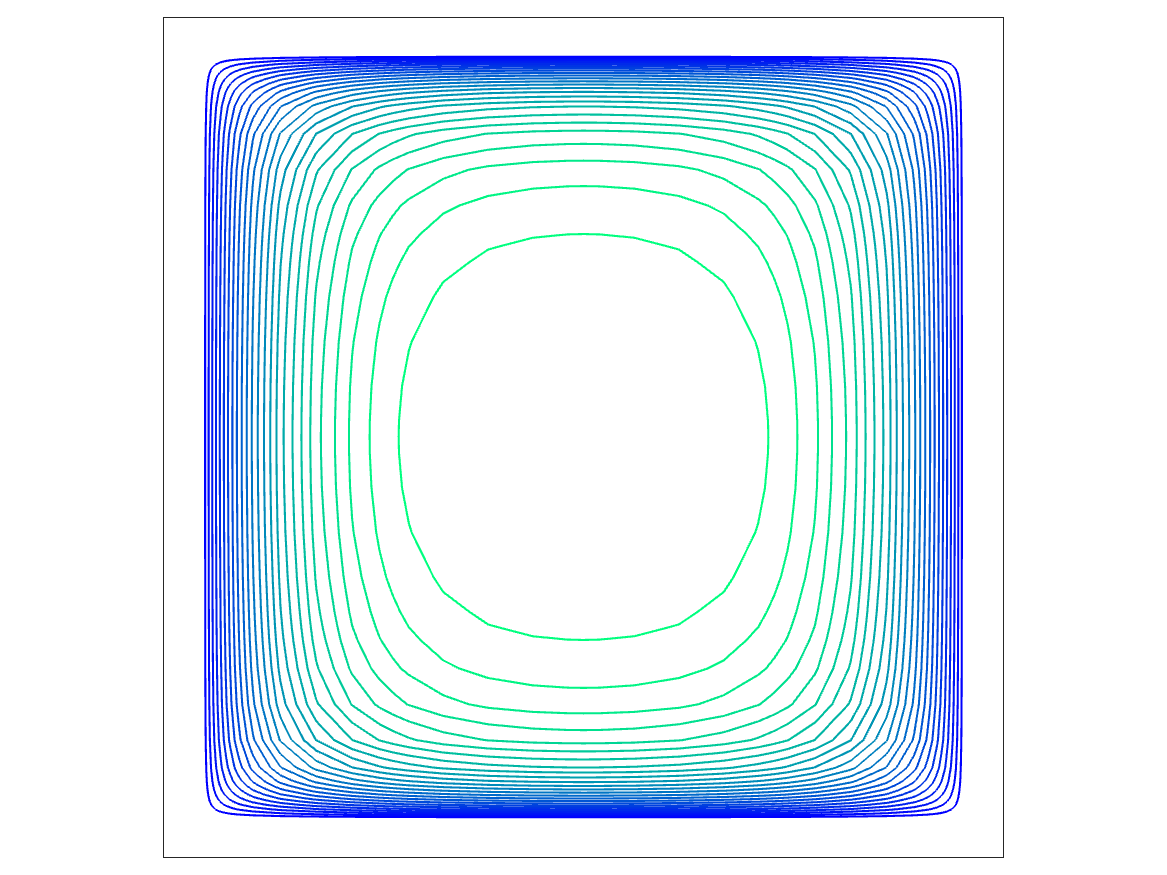}\\[5pt]
      \makebox[20pt]{\raisebox{40pt}{\rotatebox[origin=c]{0}{
      $\begin{matrix}
          r_w \\ 10^{-5}
      \end{matrix}$
      }}}%
      \includegraphics[width=5.5cm]{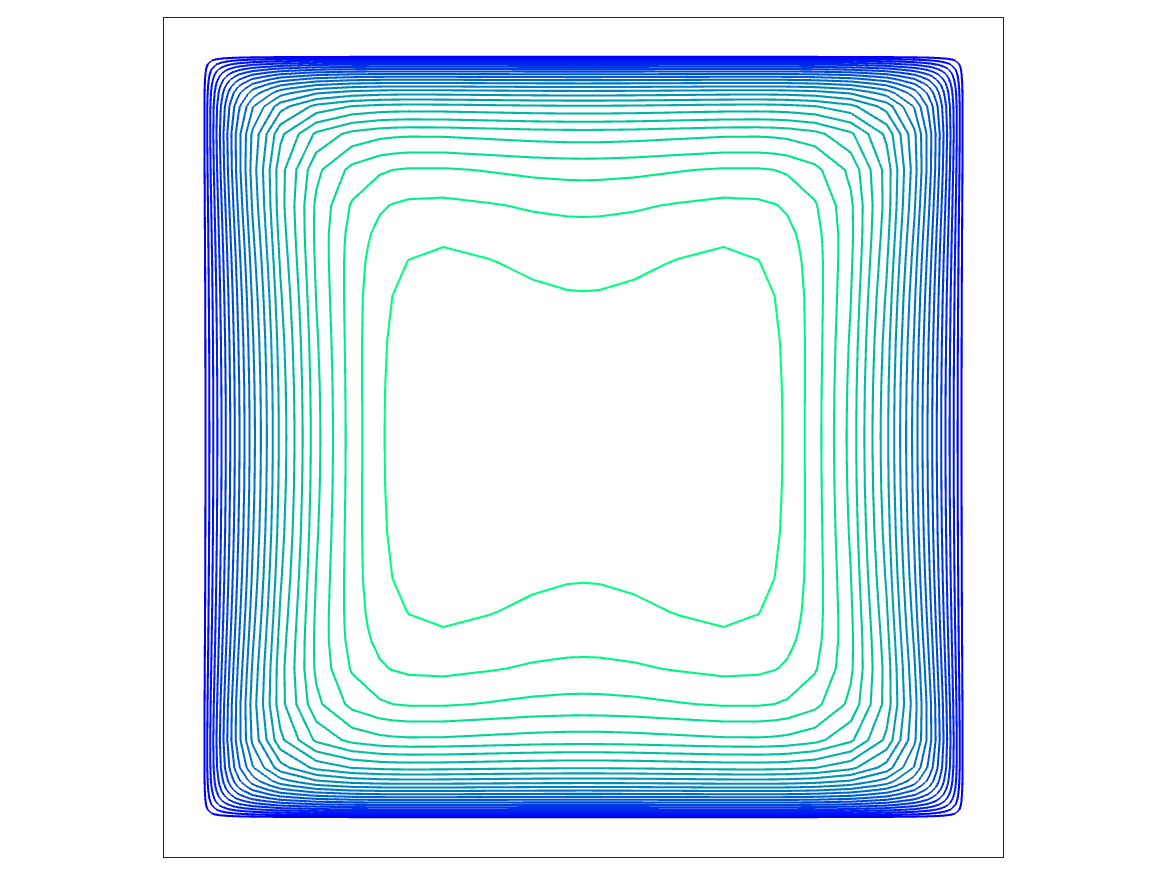}
      \caption{$\ha = 10$} 
  \end{subfigure}
  \hfill
  \begin{subfigure}[t]{0.31\textwidth} 
      \includegraphics[width=5.5cm]{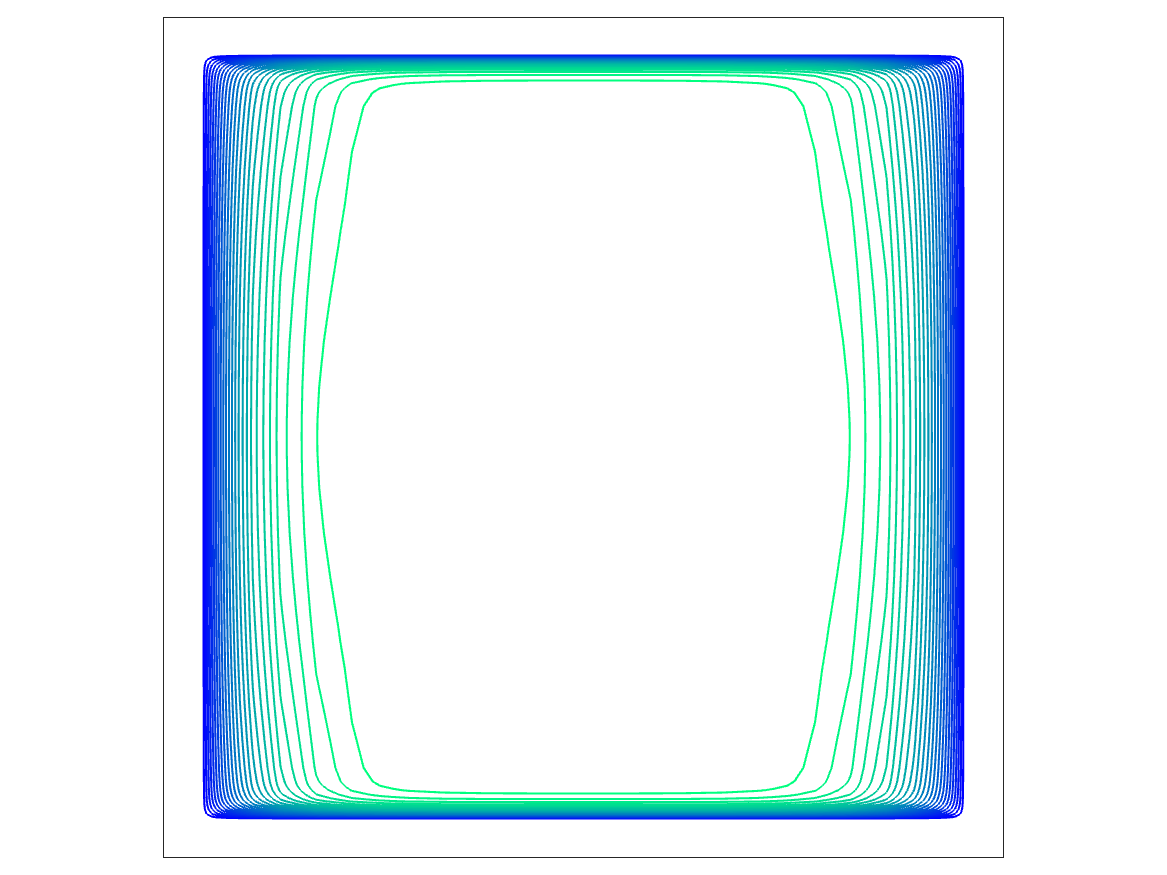}\\[5pt]
      \includegraphics[width=5.5cm]{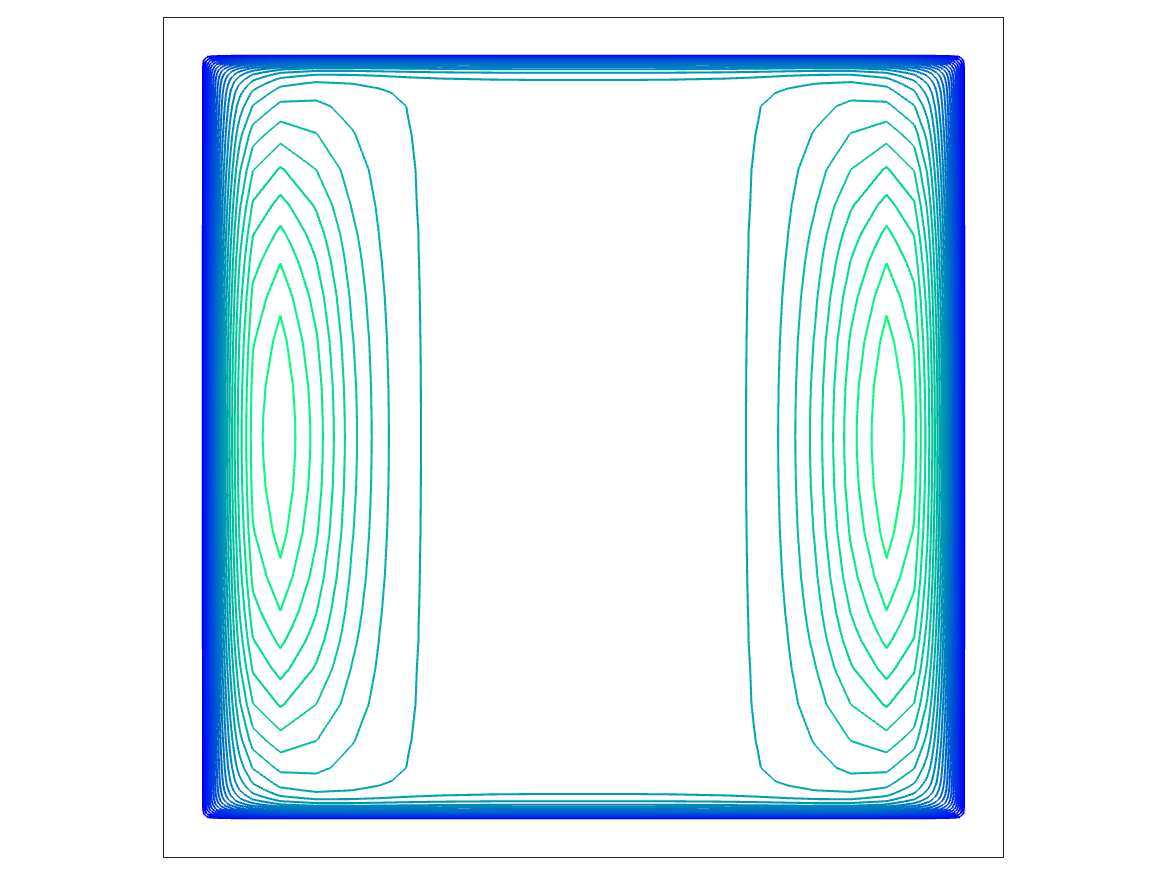}\\[5pt]
      \includegraphics[width=5.5cm]{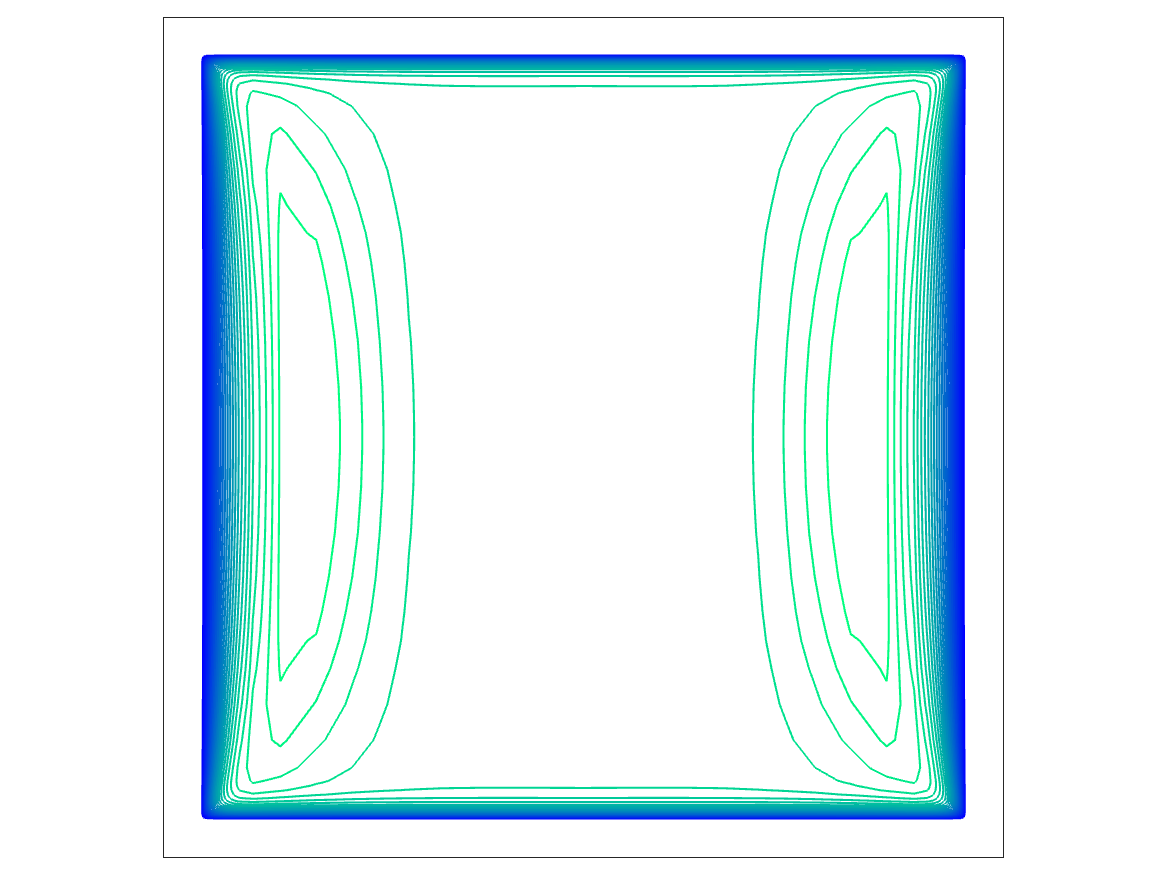}
      \caption{$\ha = 50$}
  \end{subfigure}
  \hfill
  \begin{subfigure}[t]{0.3\textwidth} 
      \includegraphics[width=5.5cm]{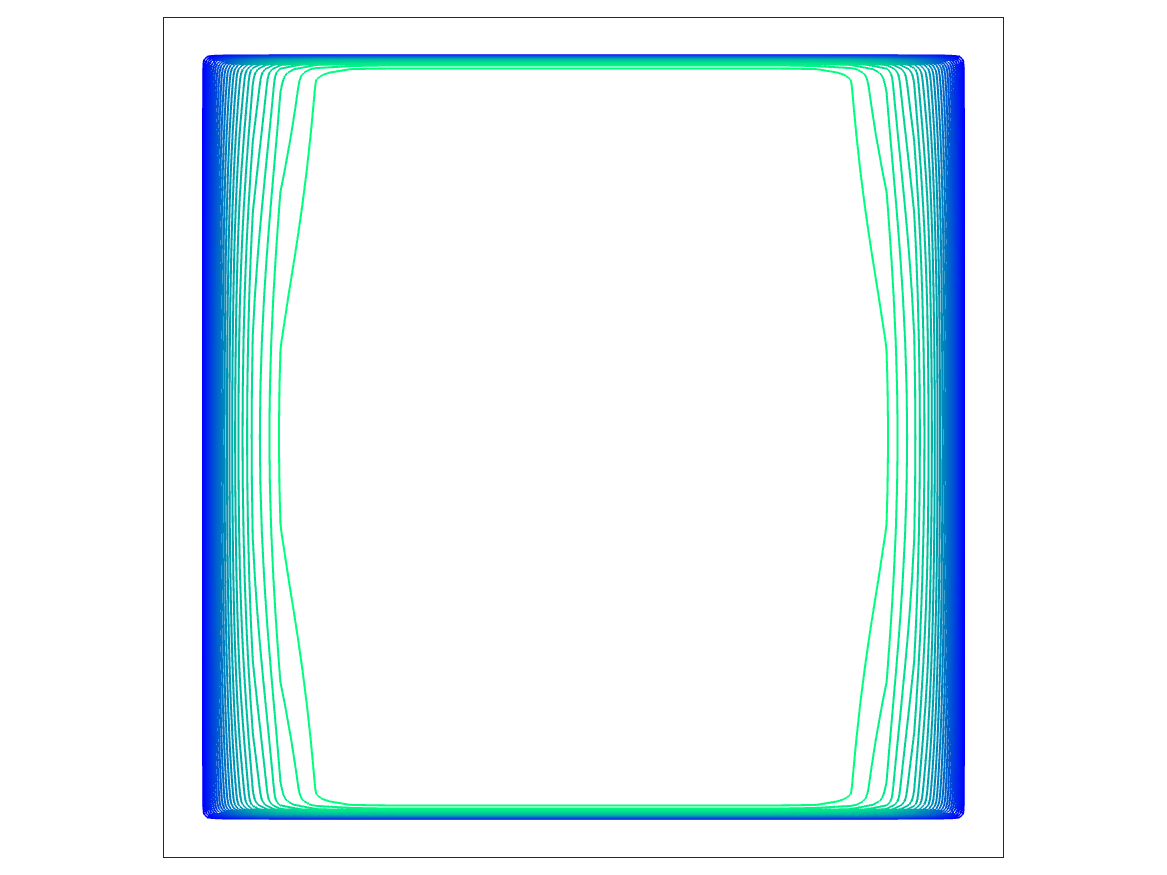}\\[5pt]
      \includegraphics[width=5.5cm]{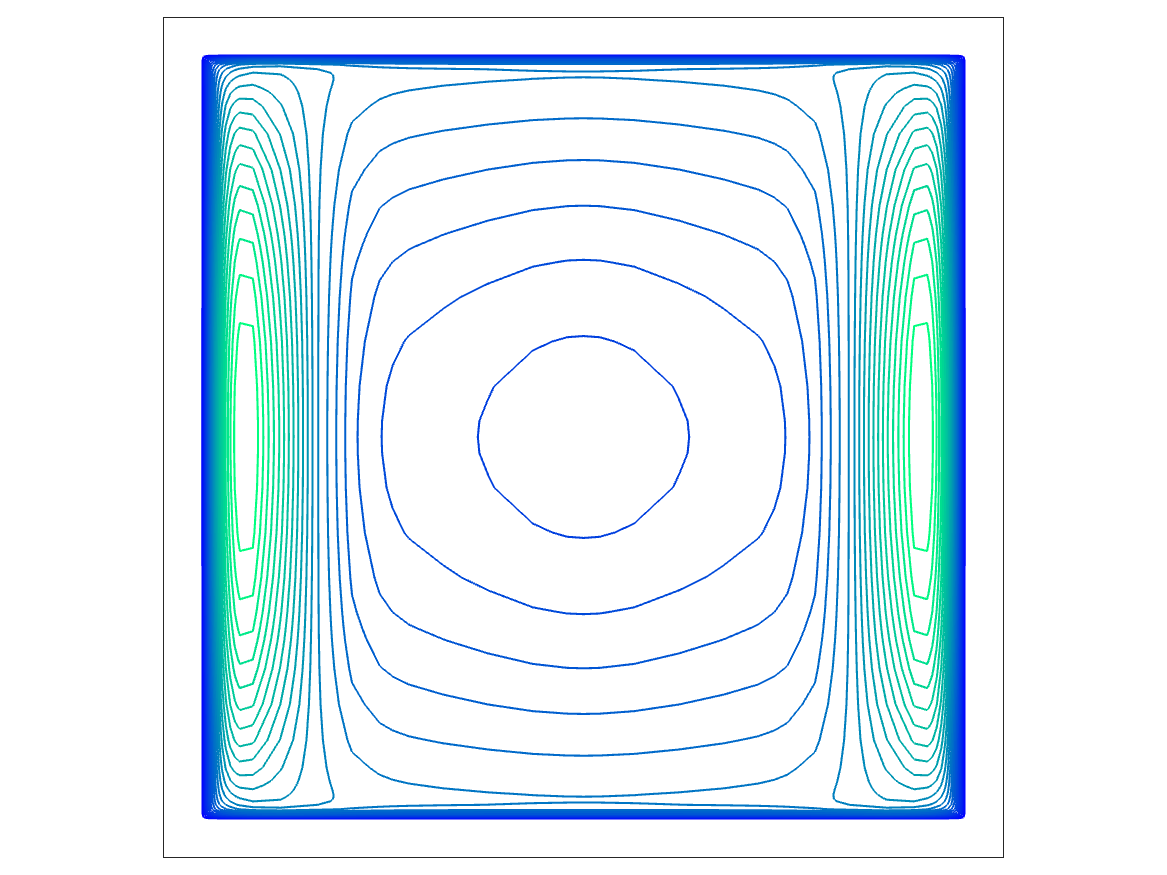}\\[5pt]
      \includegraphics[width=5.5cm]{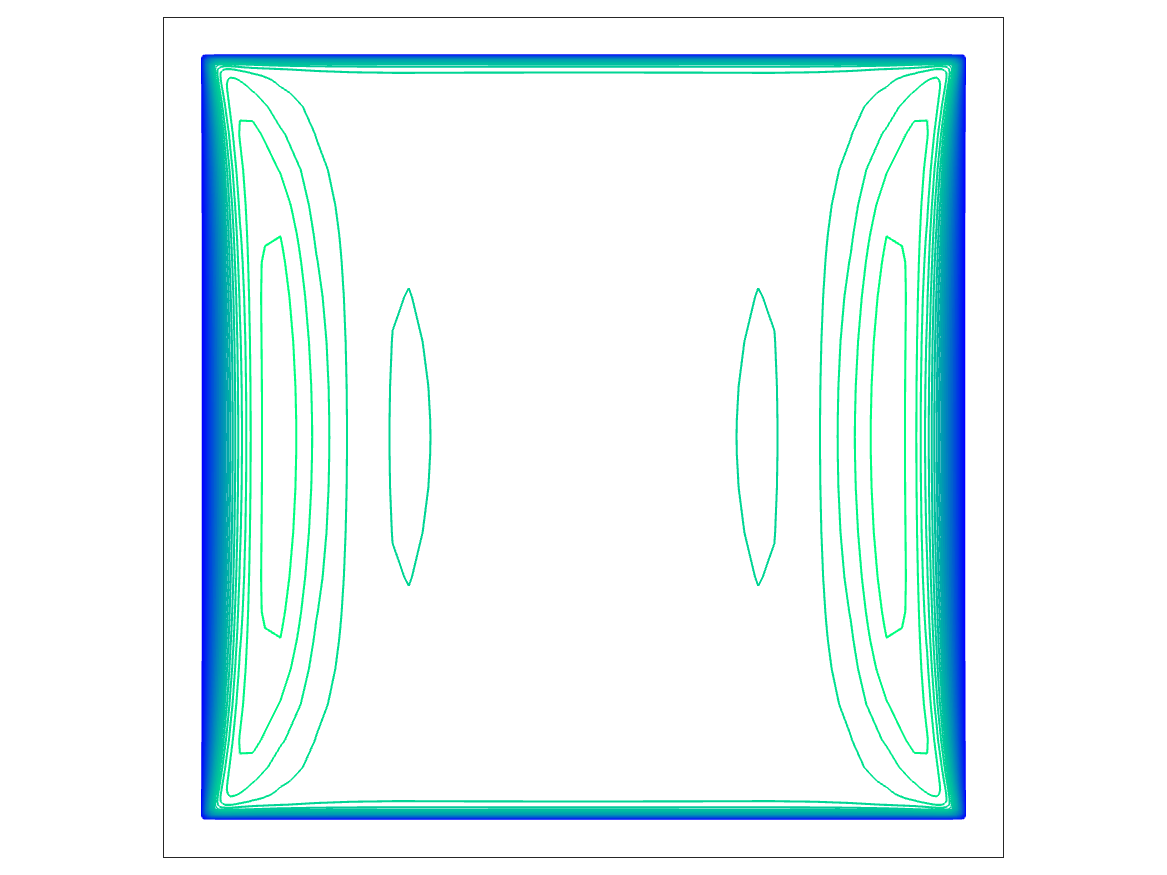}
      \caption{$\ha = 100$} 
  \end{subfigure}
  \caption{Velocity distribution contour for 
   $r_w= 10^{2}$,  $r_w= 1$, and $r_w= 10^{-5}$ (bottom). 
  $B_0=\ha/\sqrt{Re  Rm}$= 10, 50, and 100 are used from left to right.} 
  \label{vcontour}
\end{figure}

%% file: tex/paul_fig4.tex
\begin{figure}
\centering
\begin{subfigure}[t]{0.37\textwidth} 
    \makebox[20pt]{\raisebox{40pt}{\rotatebox[origin=c]{0}{
    $\begin{matrix}
        r_w \\ 10^{2}
    \end{matrix}$
    }}}%
    \includegraphics[width=5.5cm]{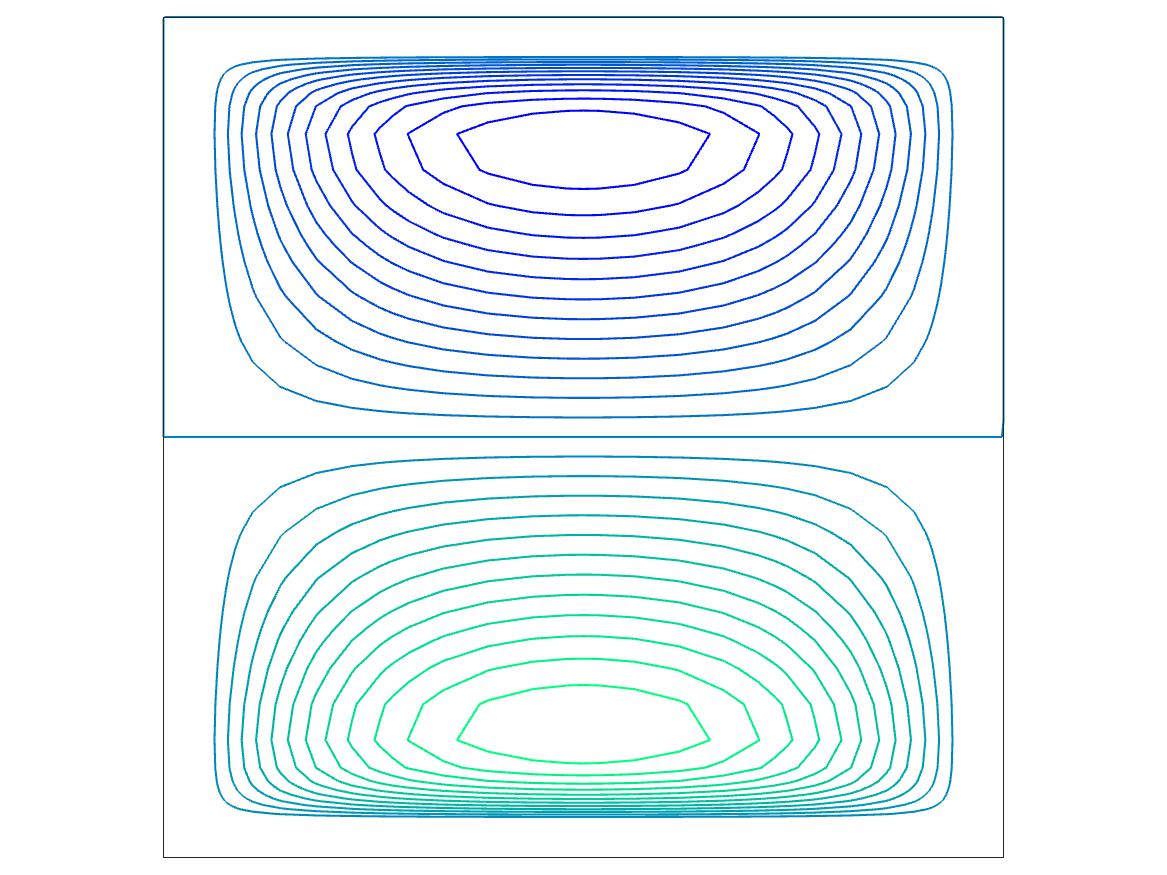}\\[5pt]
    \makebox[20pt]{\raisebox{40pt}{\rotatebox[origin=c]{0}{
    $\begin{matrix}
        r_w \\ 1
    \end{matrix}$
    }}}%
    \includegraphics[width=5.5cm]{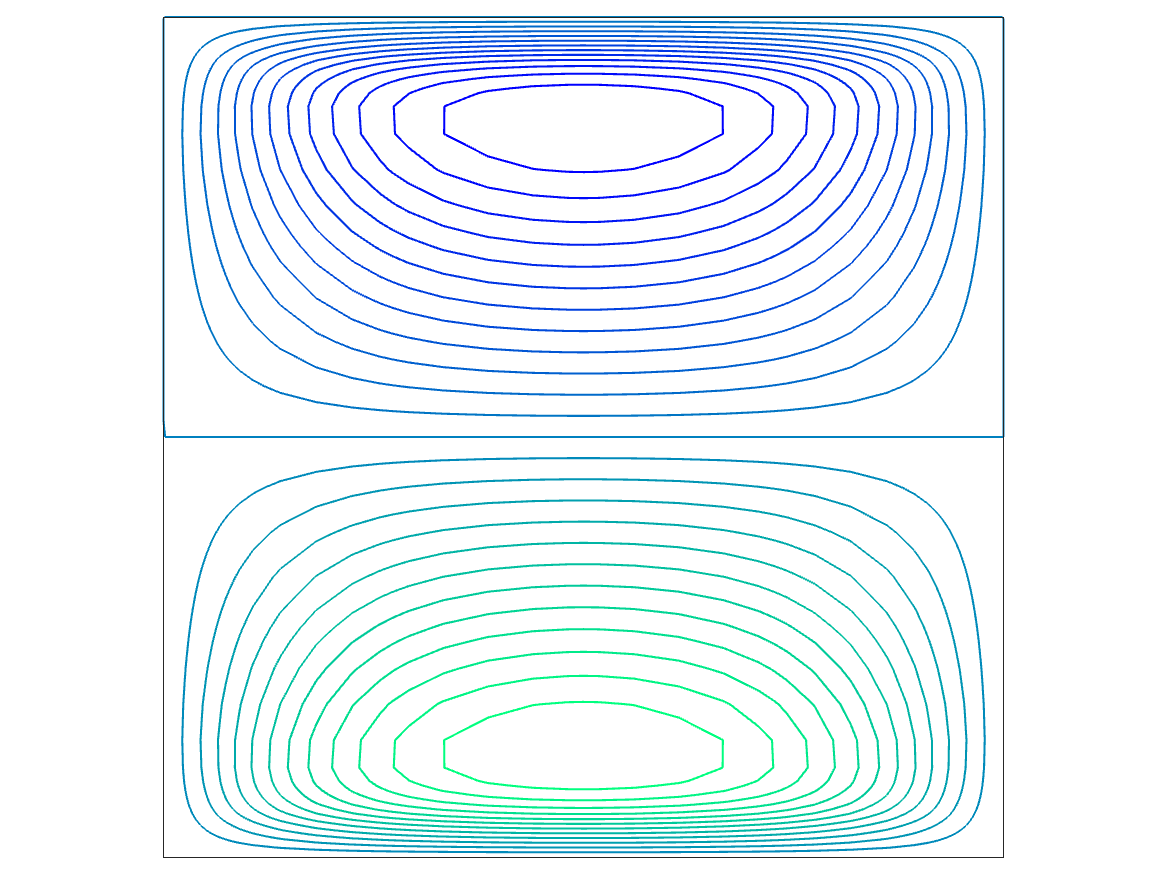}\\[5pt]
    \makebox[20pt]{\raisebox{40pt}{\rotatebox[origin=c]{0}{
    $\begin{matrix}
        r_w \\ 10^{-5}
    \end{matrix}$
    }}}%
    \includegraphics[width=5.5cm]{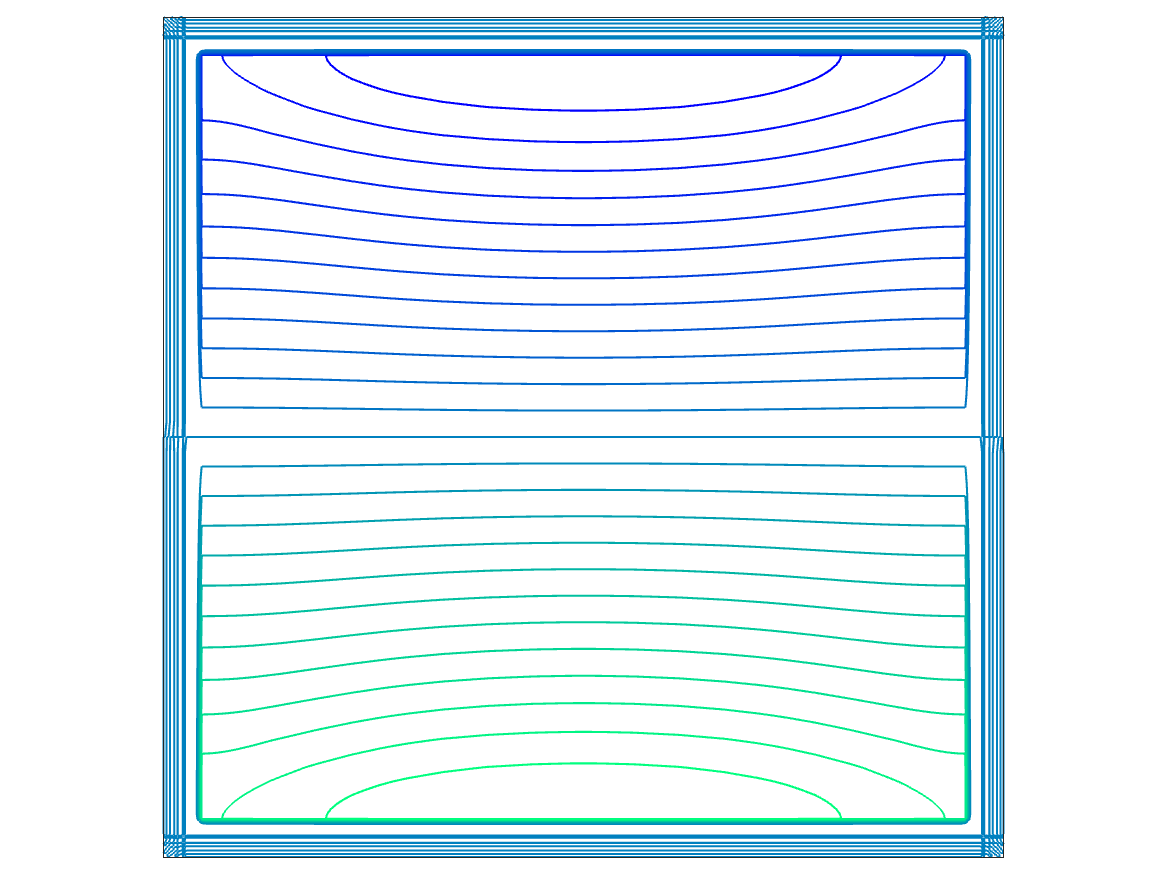}
    \caption{$\ha = 10$} 
\end{subfigure}
\hfill
\begin{subfigure}[t]{0.31\textwidth} 
    \includegraphics[width=5.5cm]{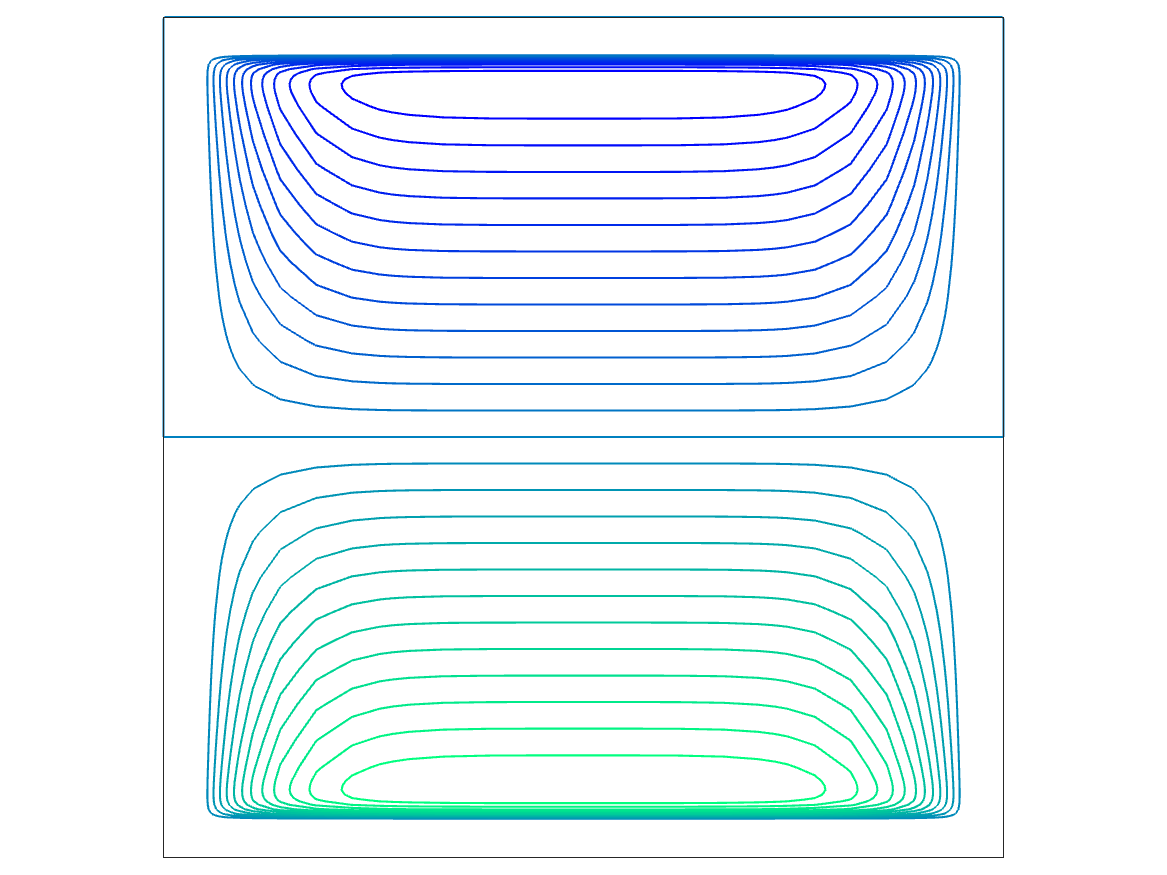}\\[5pt]
    \includegraphics[width=5.5cm]{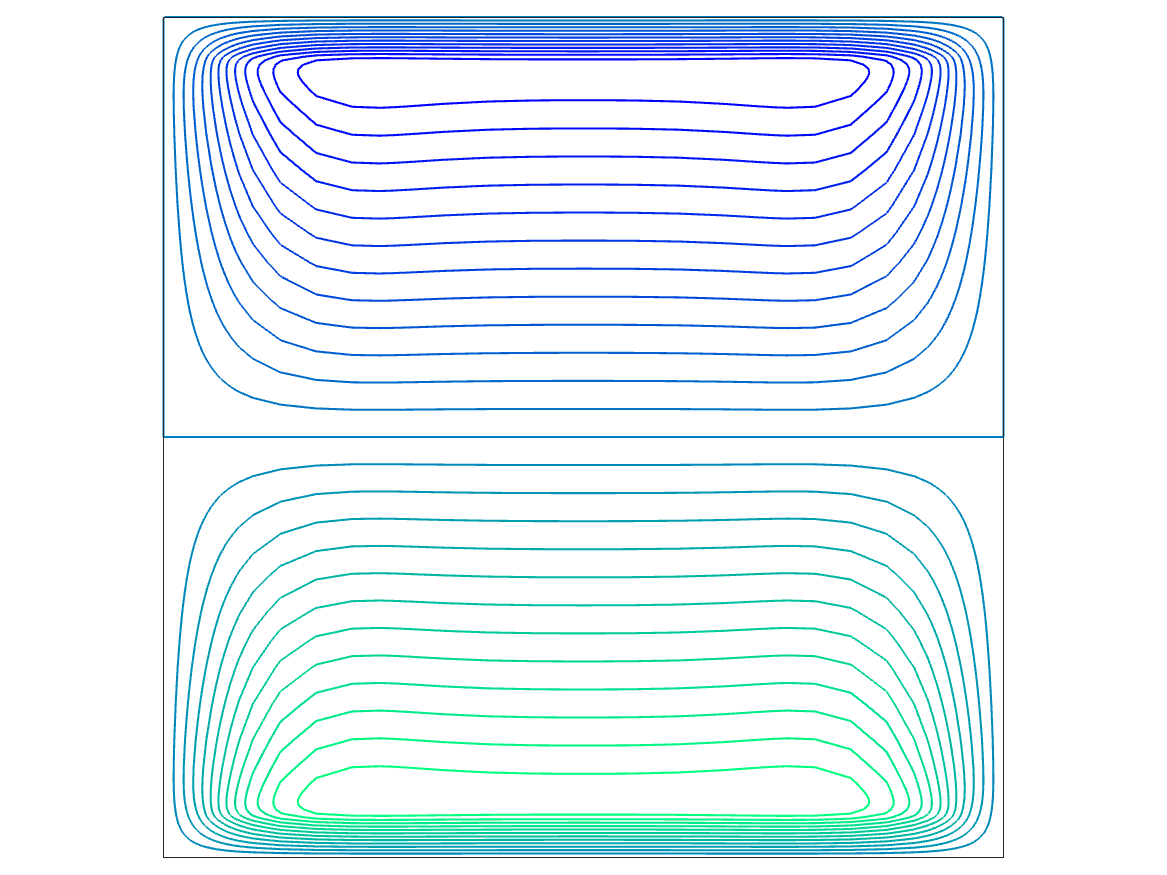}\\[5pt]
    \includegraphics[width=5.5cm]{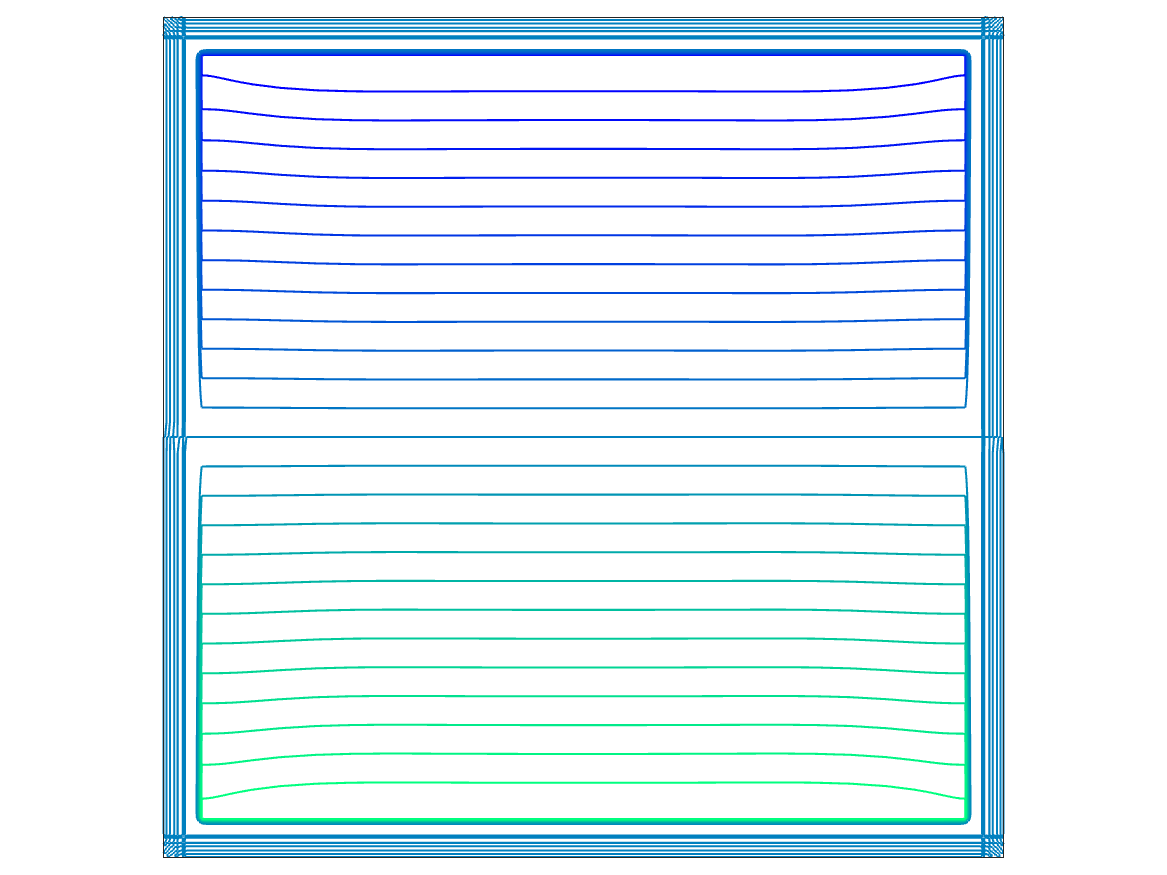}
    \caption{$\ha = 50$}
\end{subfigure}
\hfill
\begin{subfigure}[t]{0.3\textwidth} 
    \includegraphics[width=5.5cm]{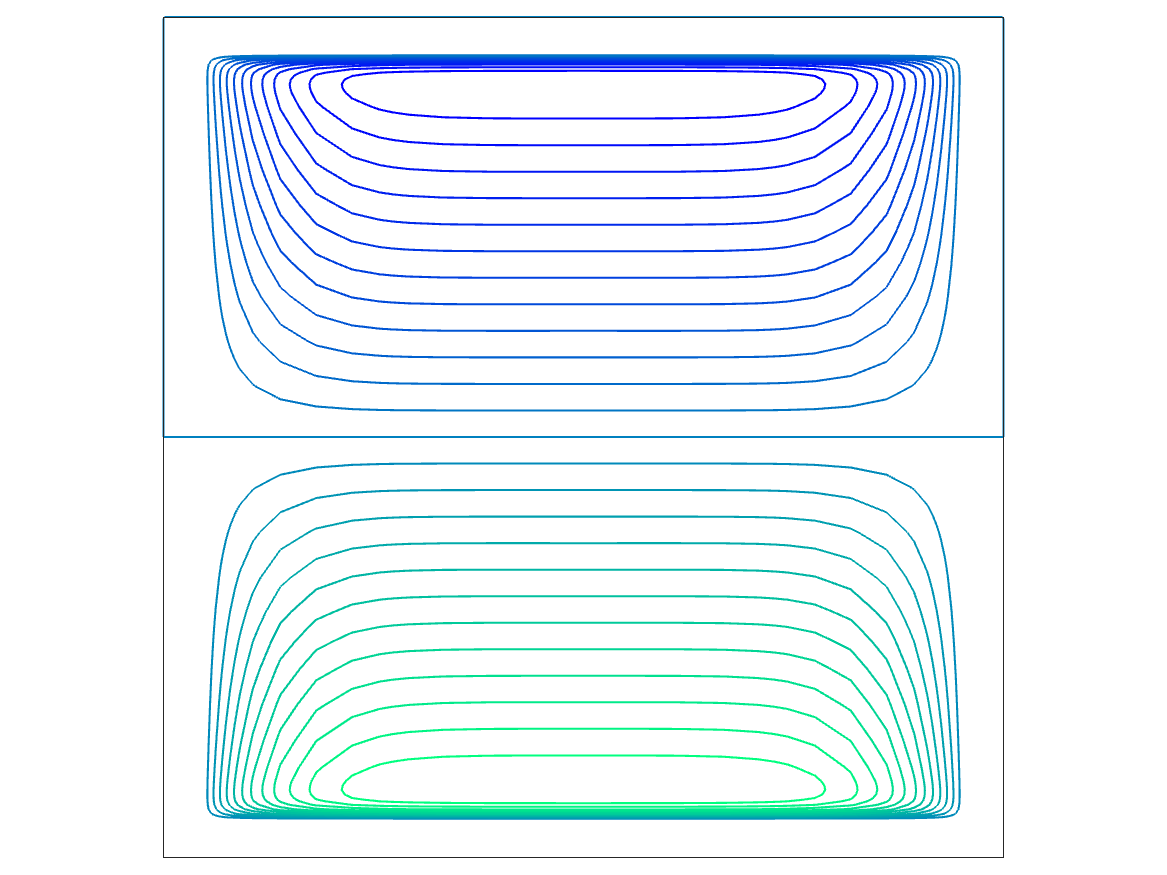}\\[5pt]
    \includegraphics[width=5.5cm]{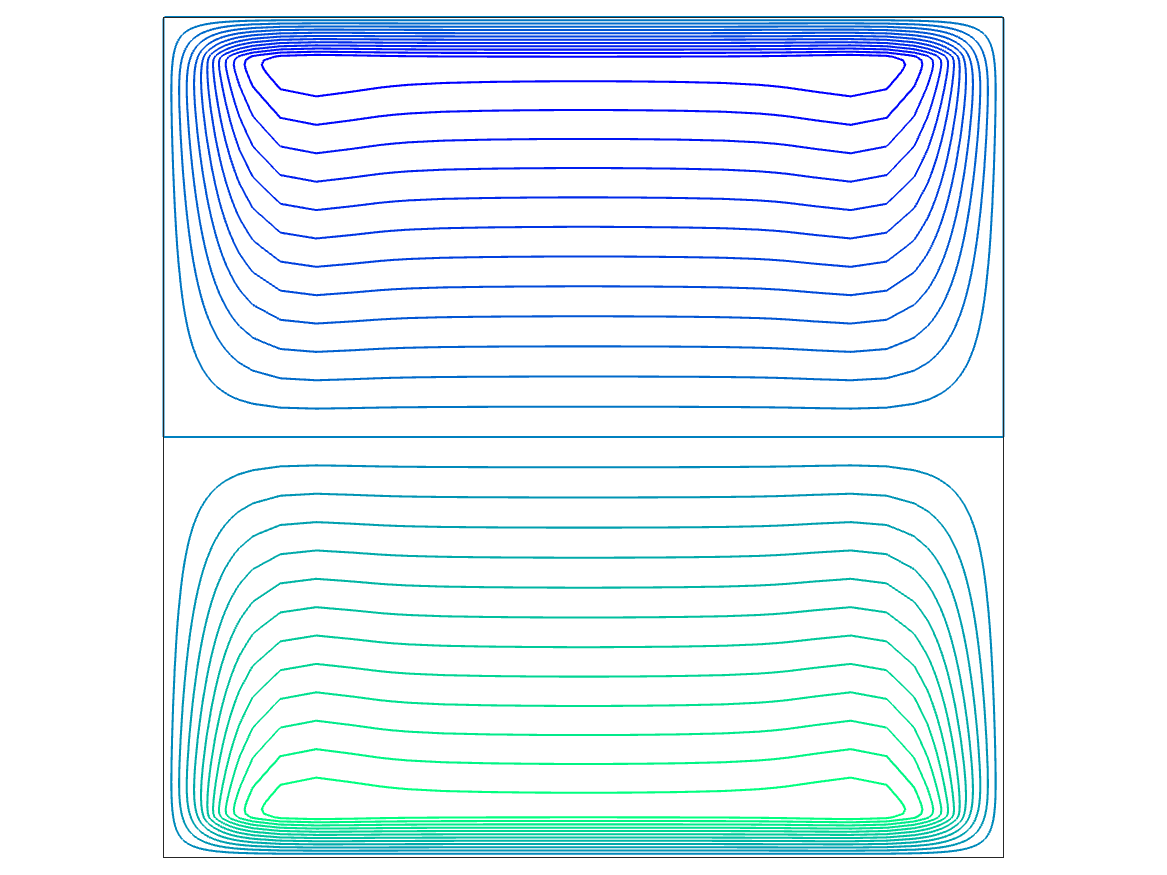}\\[5pt]
    \includegraphics[width=5.5cm]{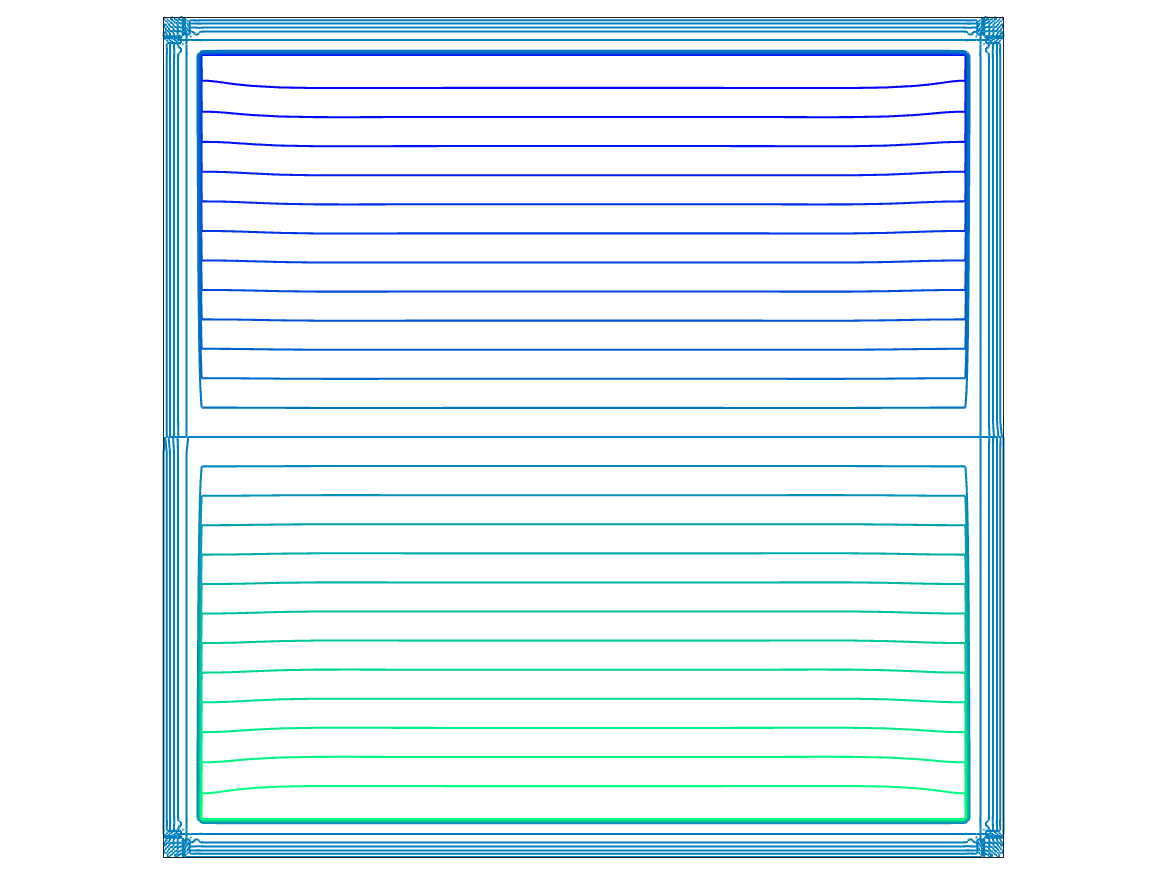}
    \caption{$\ha = 100$} 
\end{subfigure}
\caption{Magnetic distribution contour for
$r_w= 10^{2}$, $r_w= 1$, and $r_w= 10^{-5}$ (bottom).
$B_0=\ha/\sqrt{Re Rm}$= 10, 50, and 100 are used from left to right.}
\label{bcontour}
\end{figure}

%% file: tex/paul_fig1.tex
\begin{figure}
    \centering
    \begin{subfigure}[t]{0.37\textwidth} 
        \makebox[20pt]{\raisebox{40pt}{\rotatebox[origin=c]{0}{
        $\begin{matrix}
            r_w \\ 10^{2}
        \end{matrix}$
        }}}%
        \includegraphics[width=5.7cm]{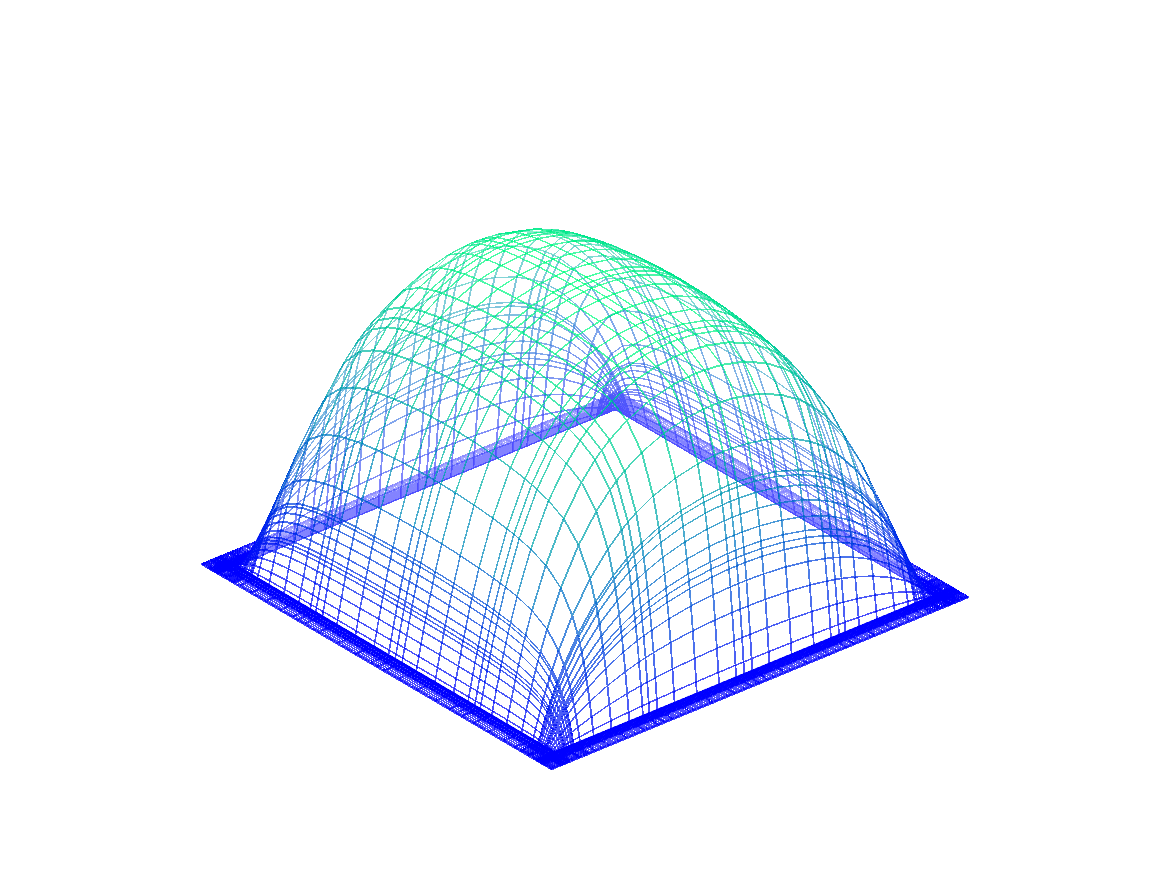}\\[5pt]
        \makebox[20pt]{\raisebox{40pt}{\rotatebox[origin=c]{0}{
        $\begin{matrix}
            r_w \\ 1
        \end{matrix}$
        }}}%
        \includegraphics[width=5.7cm]{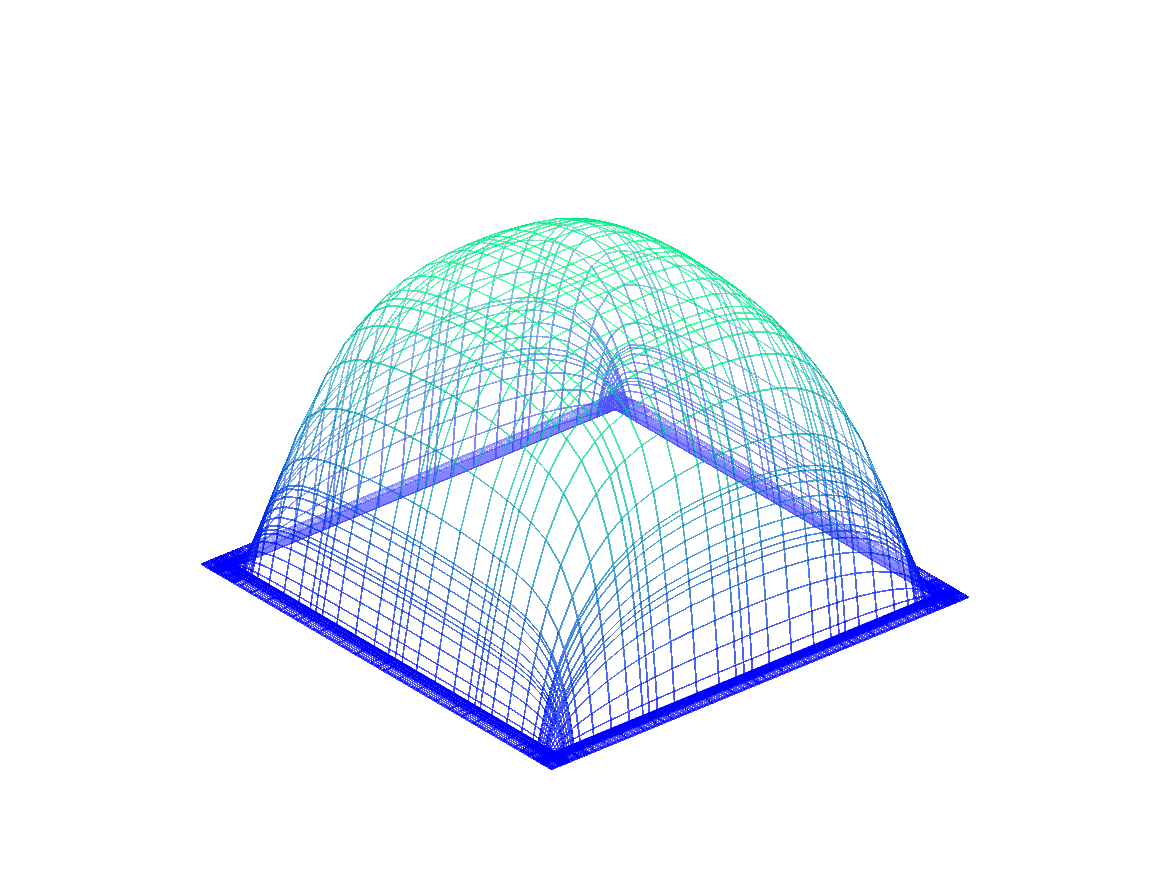}\\[5pt]
        \makebox[20pt]{\raisebox{40pt}{\rotatebox[origin=c]{0}{
        $\begin{matrix}
            r_w \\ 10^{-5}
        \end{matrix}$
        }}}%
        \includegraphics[width=5.7cm]{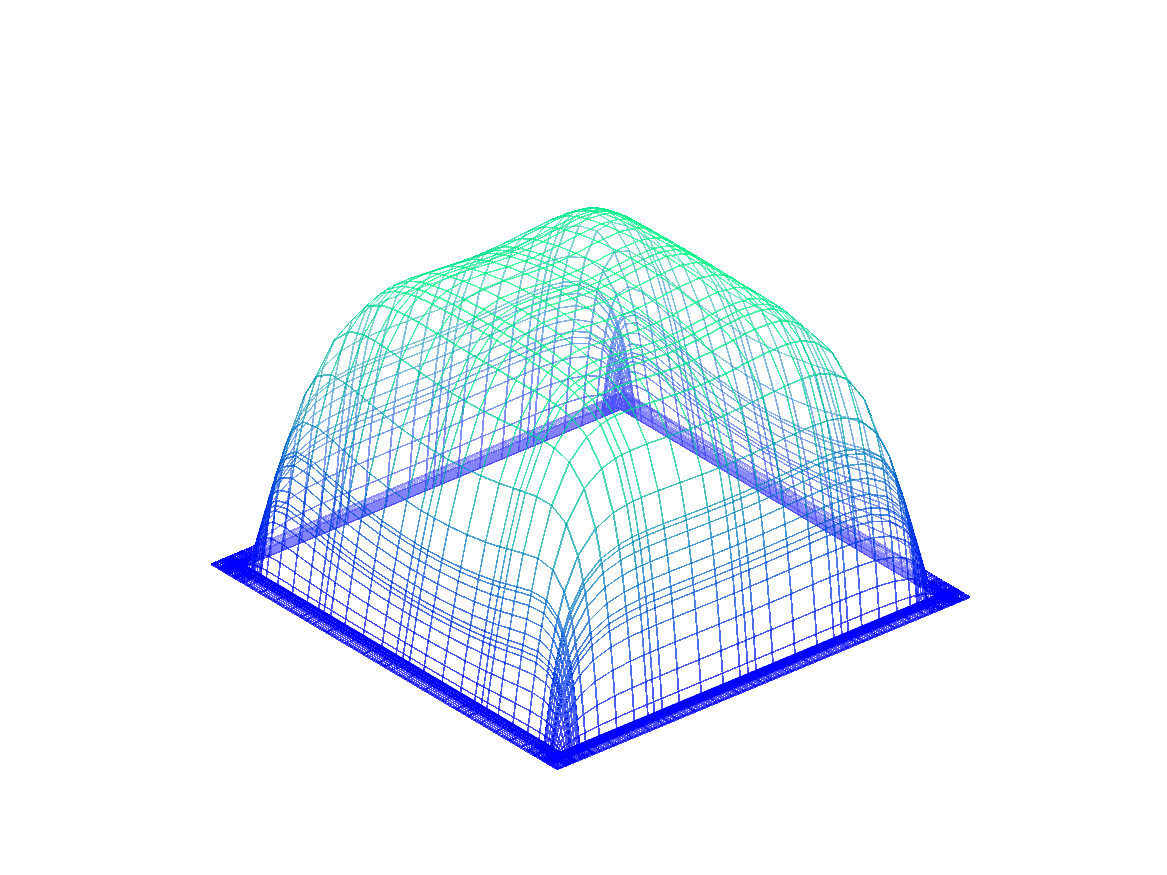}
        \caption{$\ha = 10$} 
    \end{subfigure}
    \hfill
    \begin{subfigure}[t]{0.31\textwidth} 
        \includegraphics[width=5.7cm]{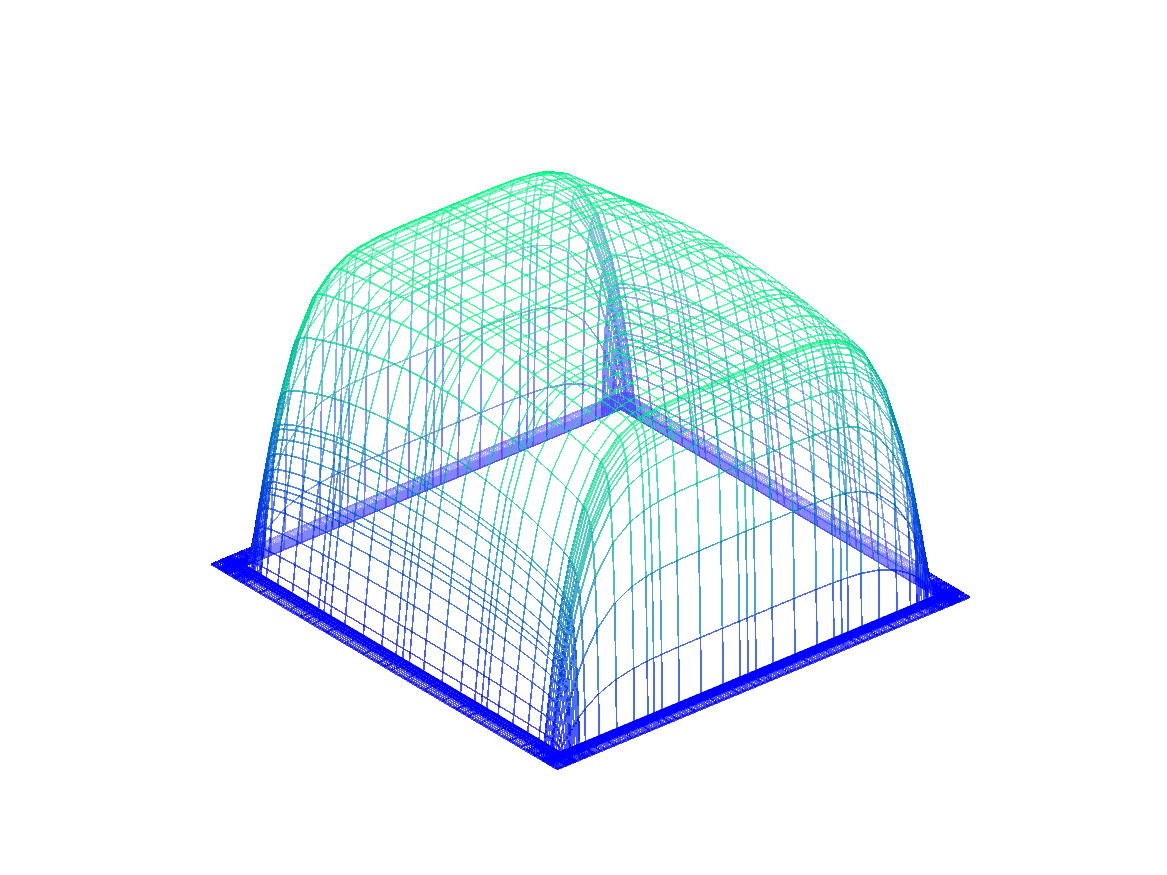}\\[5pt]
        \includegraphics[width=5.7cm]{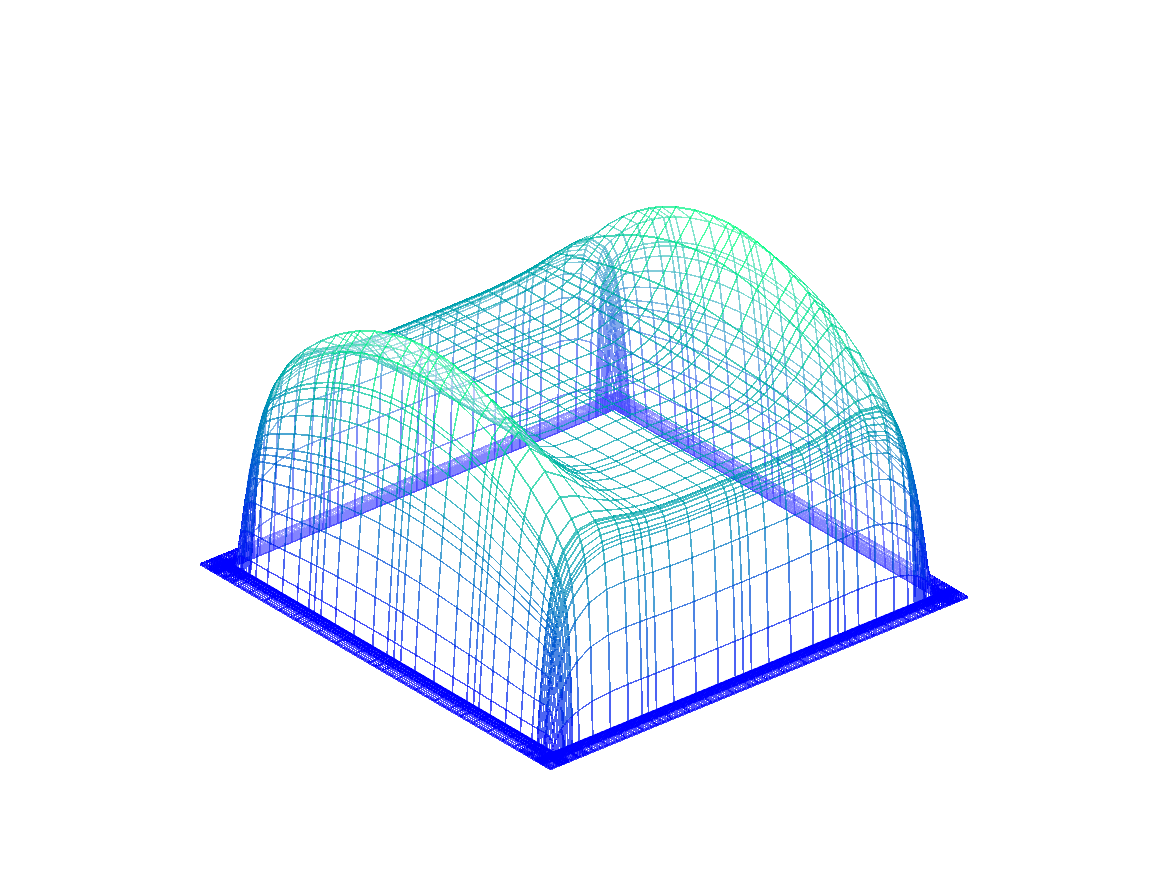}\\[5pt]
        \includegraphics[width=5.7cm]{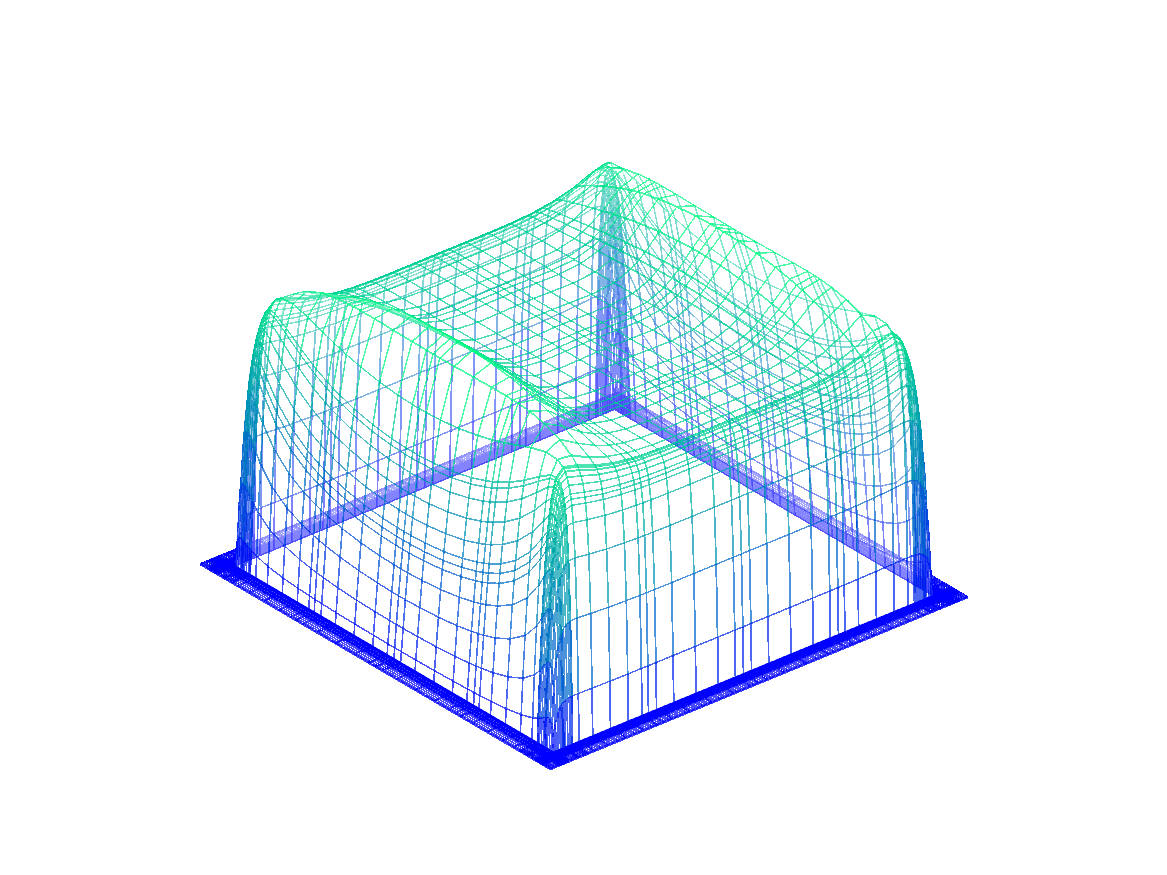}
        \caption{$\ha = 50$}
    \end{subfigure}
    \hfill
    \begin{subfigure}[t]{0.3\textwidth} 
        \includegraphics[width=5.7cm]{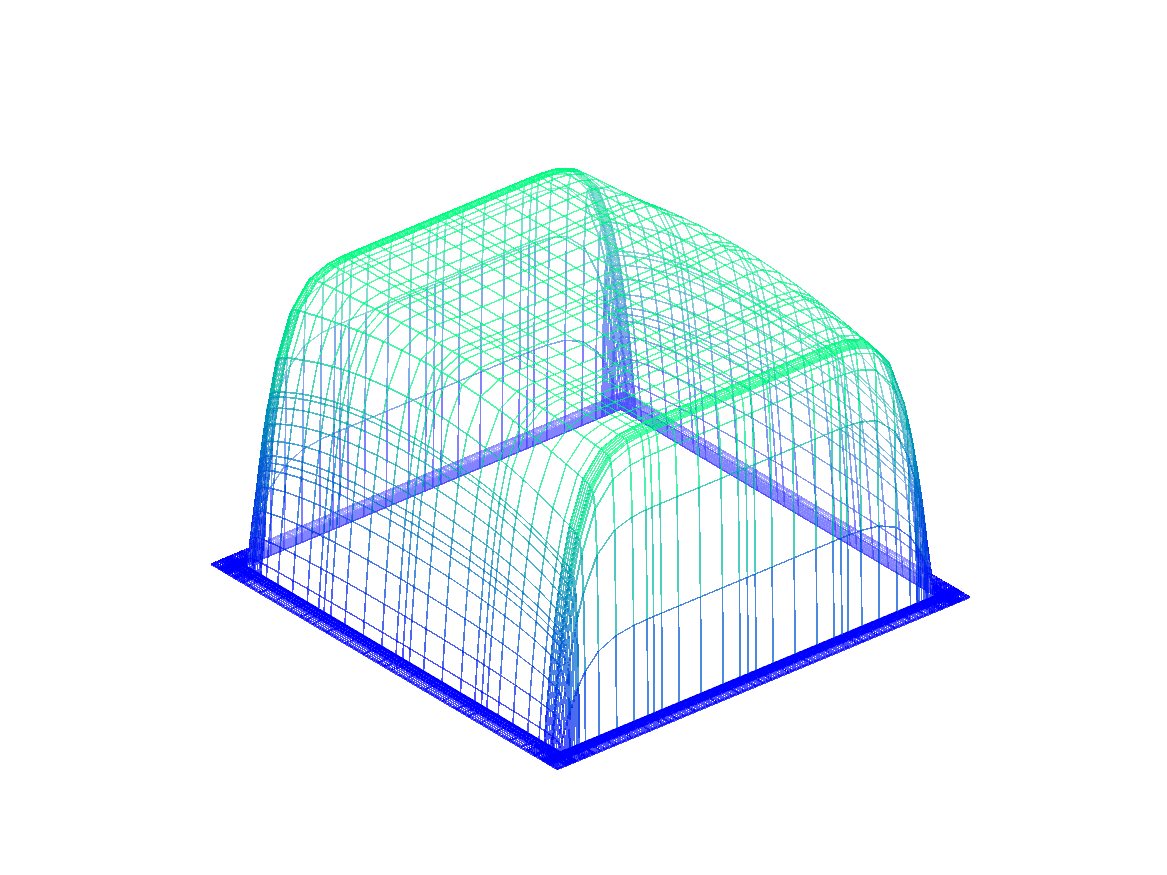}\\[5pt]
        \includegraphics[width=5.7cm]{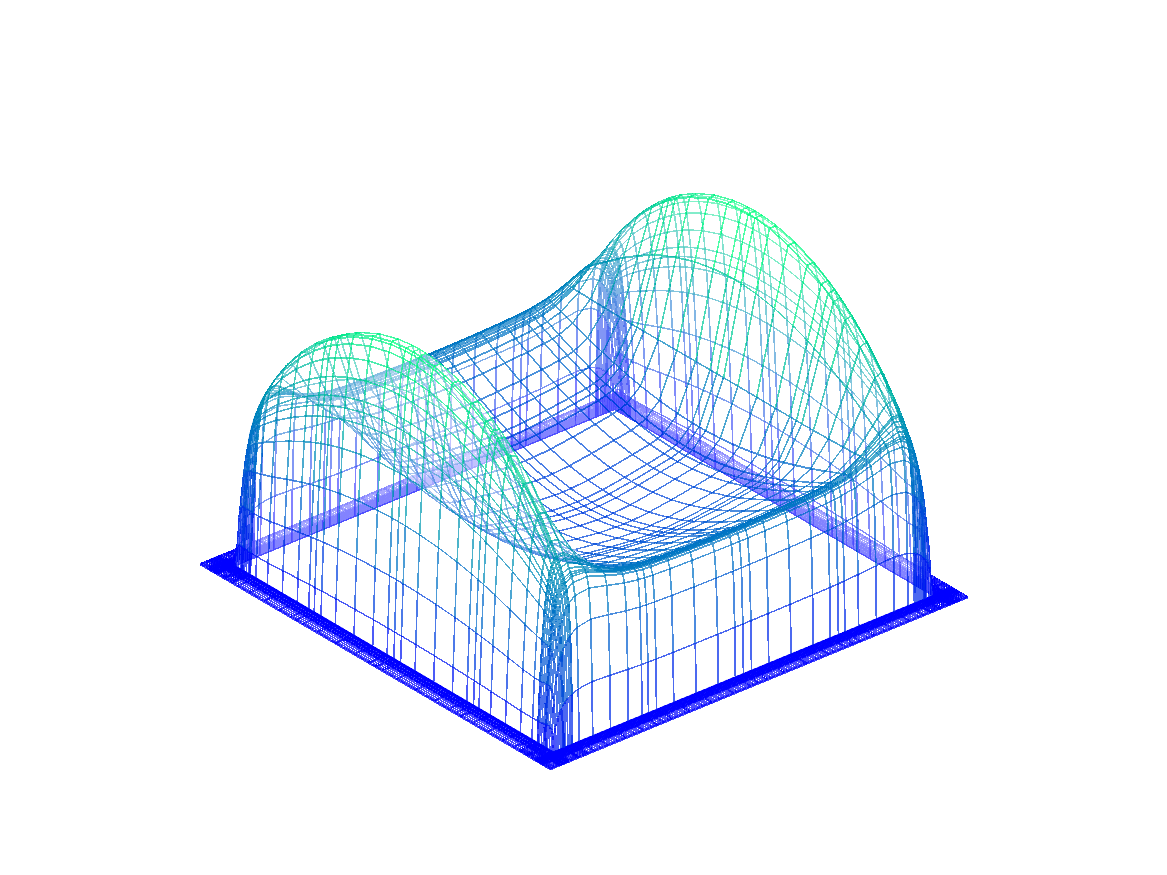}\\[5pt]
        \includegraphics[width=5.7cm]{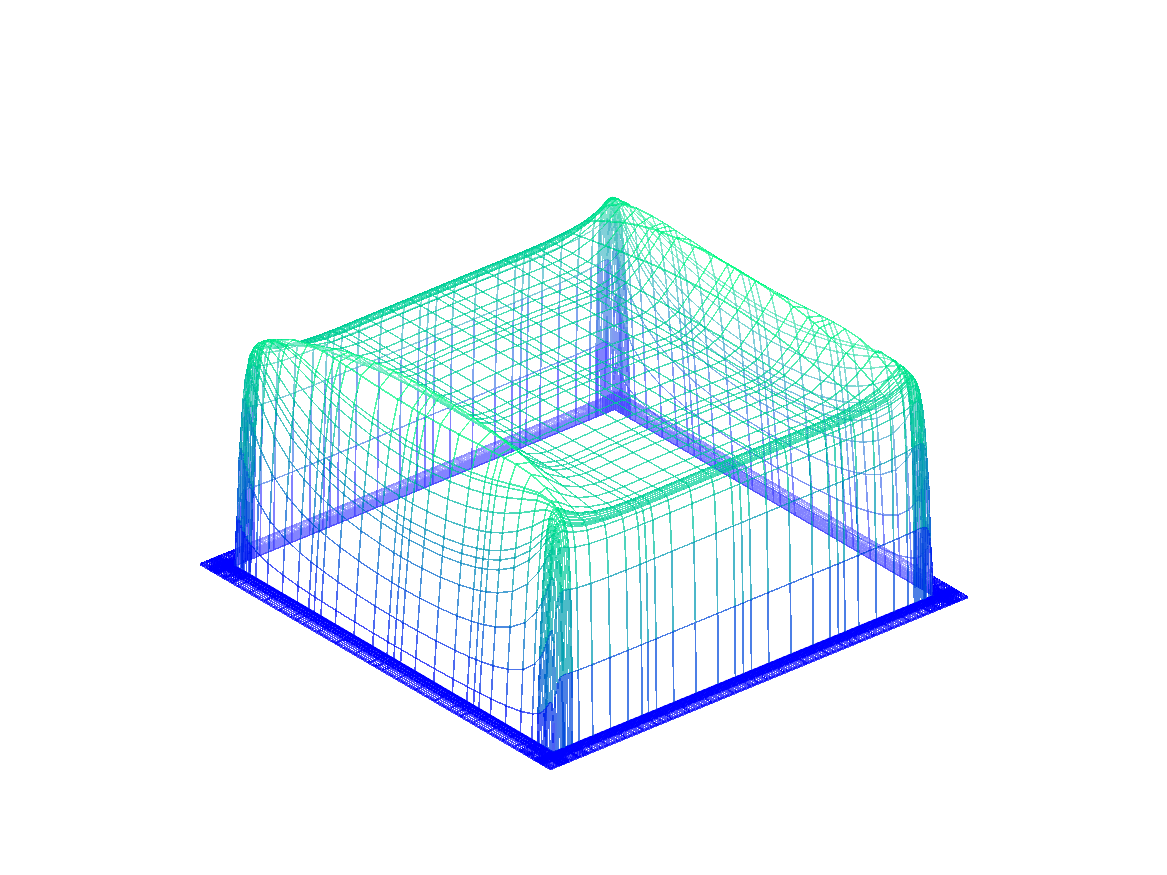}
        \caption{$\ha = 100$} 
    \end{subfigure}
    \caption{Velocity distribution profile for
    $r_w= 10^{2}$, $r_w= 1$, and $r_w= 10^{-5}$ (bottom).
             $B_0=\ha/\sqrt{Re Rm}$= 10, 50, and 100 are used from left to right.}
    \label{vmesh}
\end{figure}

%% file: tex/paul_fig3.tex
\begin{figure}
  \centering
  \begin{subfigure}[t]{0.37\textwidth} 
      \makebox[20pt]{\raisebox{40pt}{\rotatebox[origin=c]{0}{
      $\begin{matrix}
          r_w \\ 10^{2}
      \end{matrix}$
      }}}%
      \includegraphics[width=5.5cm]{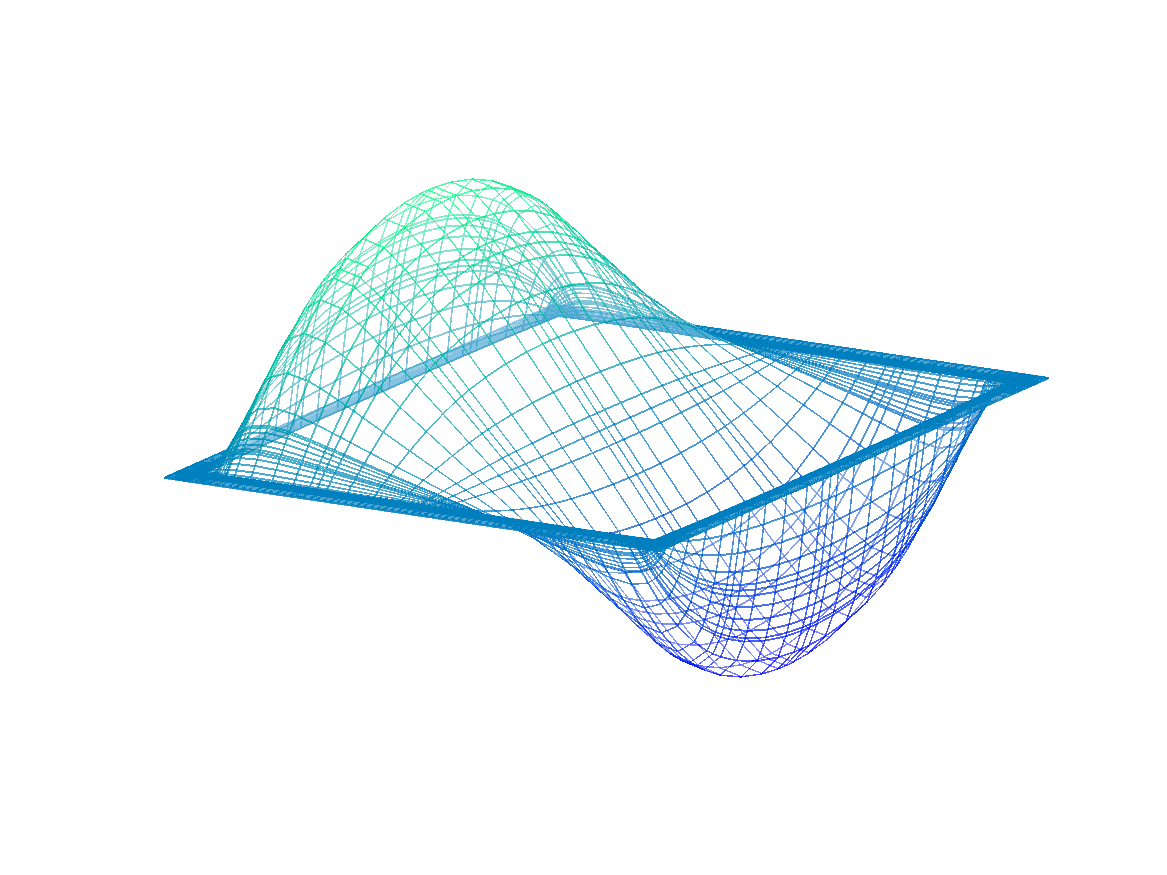}\\[5pt]
      \makebox[20pt]{\raisebox{40pt}{\rotatebox[origin=c]{0}{
      $\begin{matrix}
          r_w \\ 1
      \end{matrix}$
      }}}%
      \includegraphics[width=5.5cm]{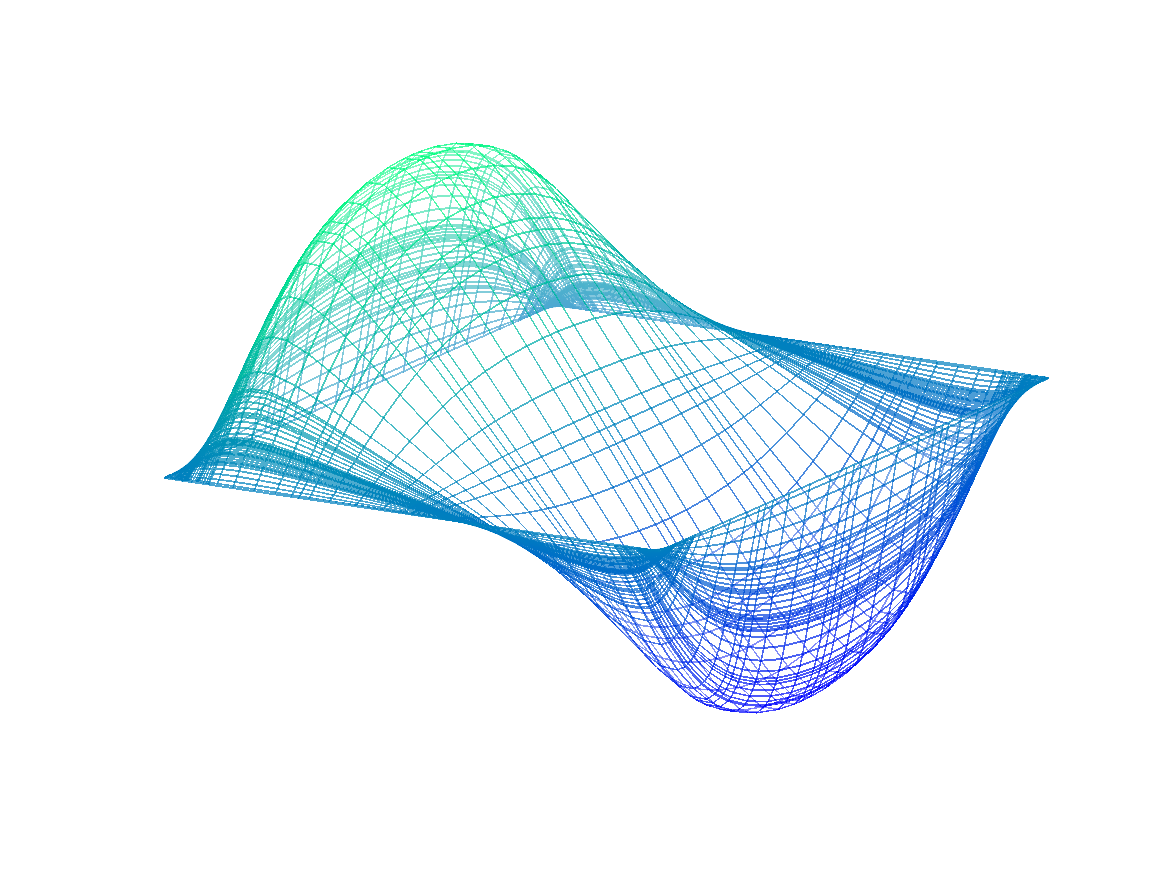}\\[5pt]
      \makebox[20pt]{\raisebox{40pt}{\rotatebox[origin=c]{0}{
      $\begin{matrix}
          r_w \\ 10^{-5}
      \end{matrix}$
      }}}%
      \includegraphics[width=5.5cm]{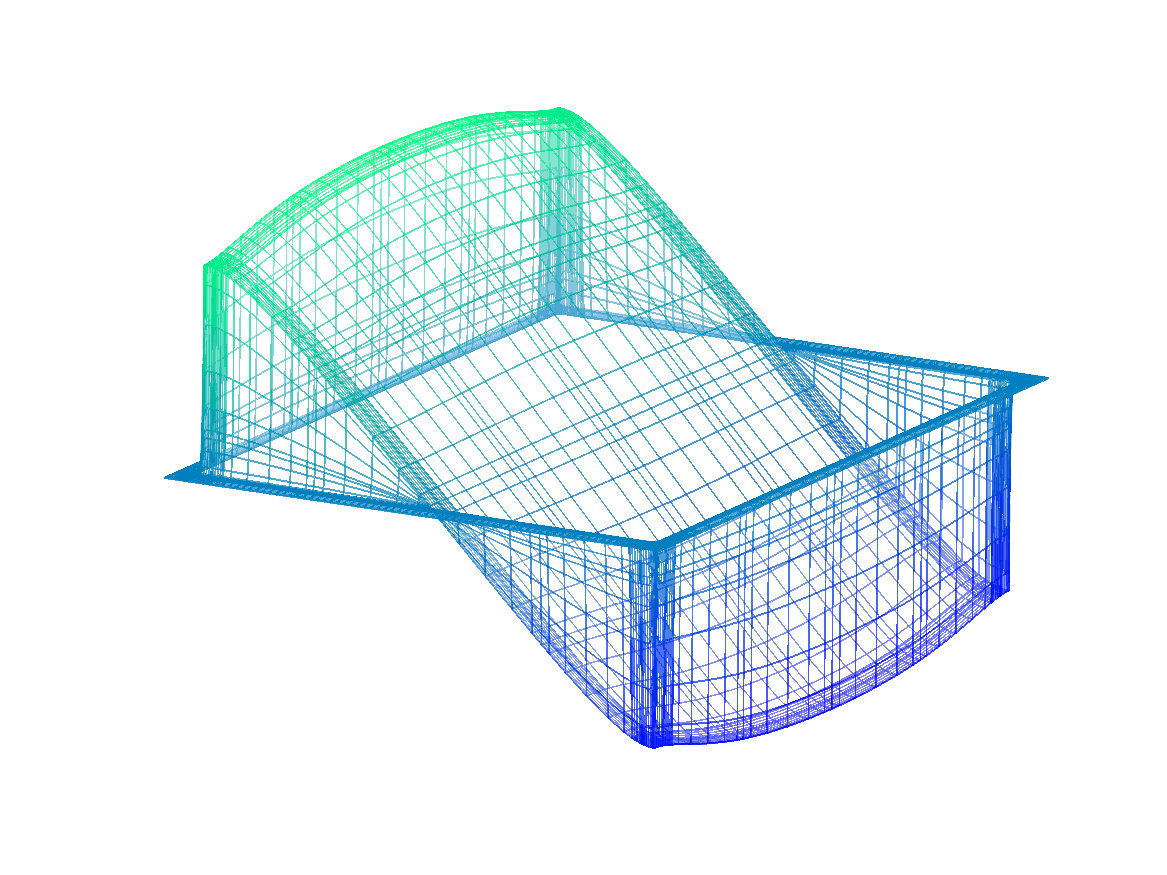}
      \caption{$\ha = 10$} 
  \end{subfigure}
  \hfill
  \begin{subfigure}[t]{0.31\textwidth} 
      \includegraphics[width=5.5cm]{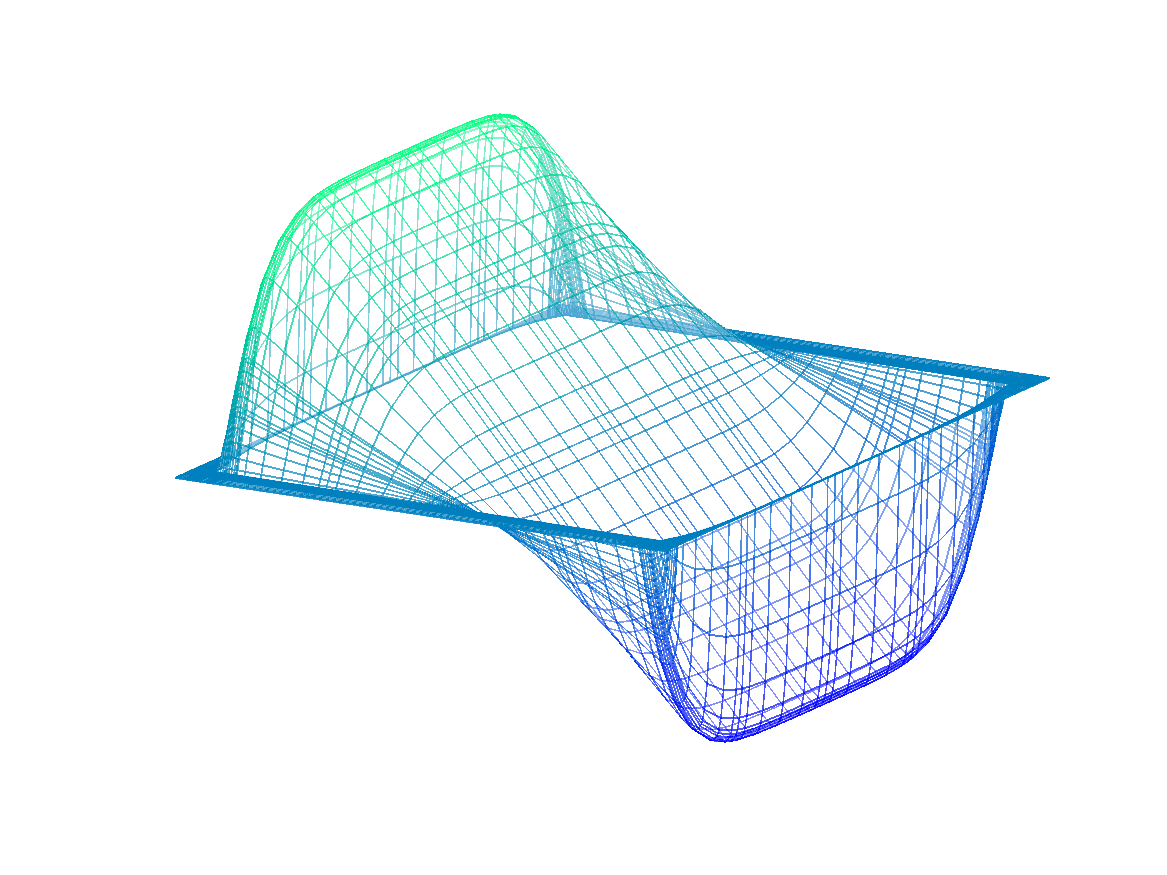}\\[5pt]
      \includegraphics[width=5.5cm]{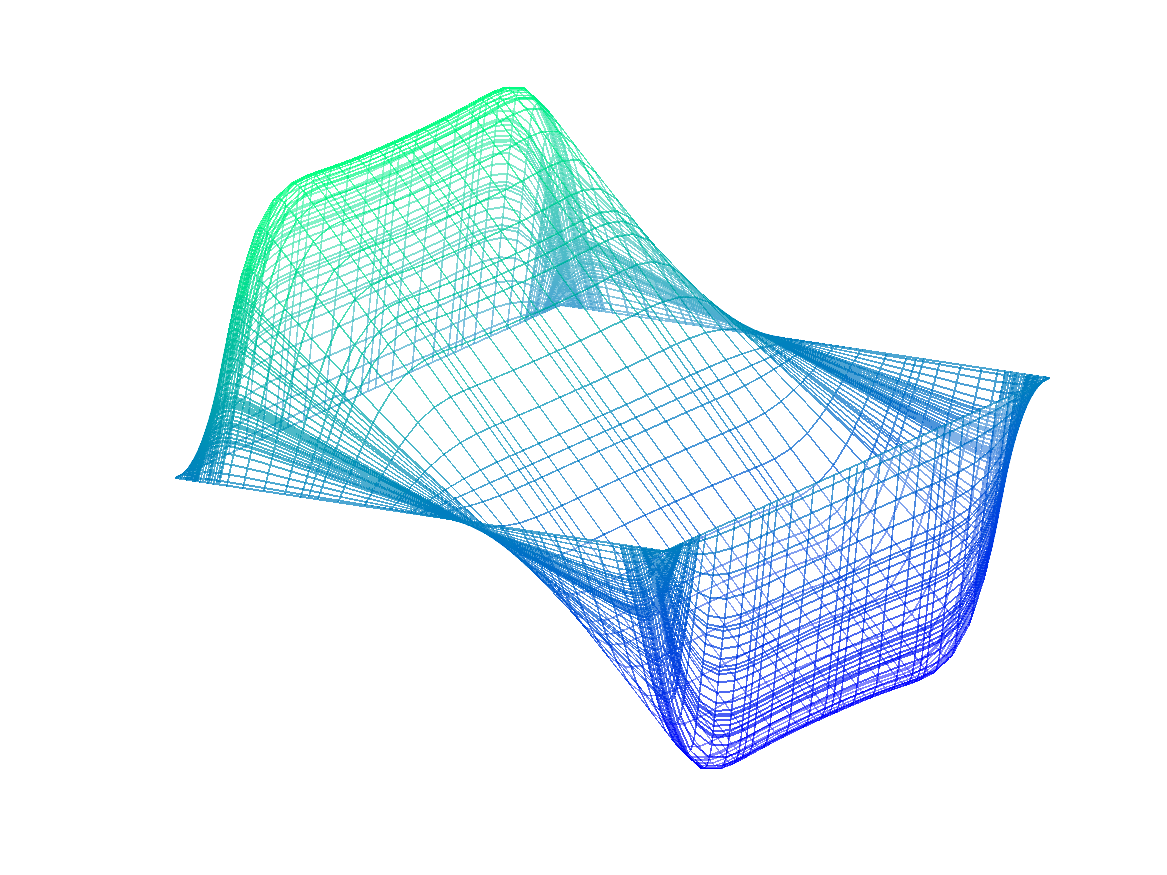}\\[5pt]
      \includegraphics[width=5.5cm]{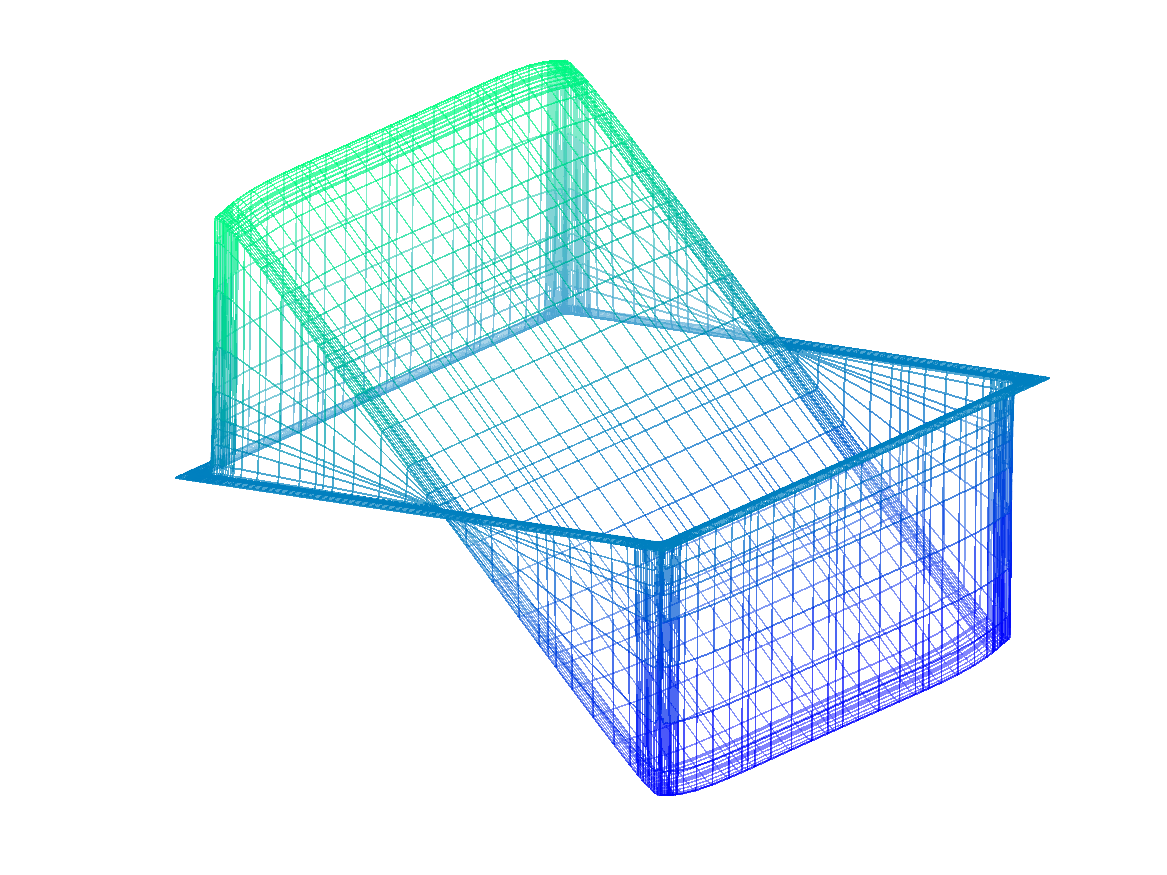}
      \caption{$\ha = 50$}
  \end{subfigure}
  \hfill
  \begin{subfigure}[t]{0.3\textwidth} 
      \includegraphics[width=5.5cm]{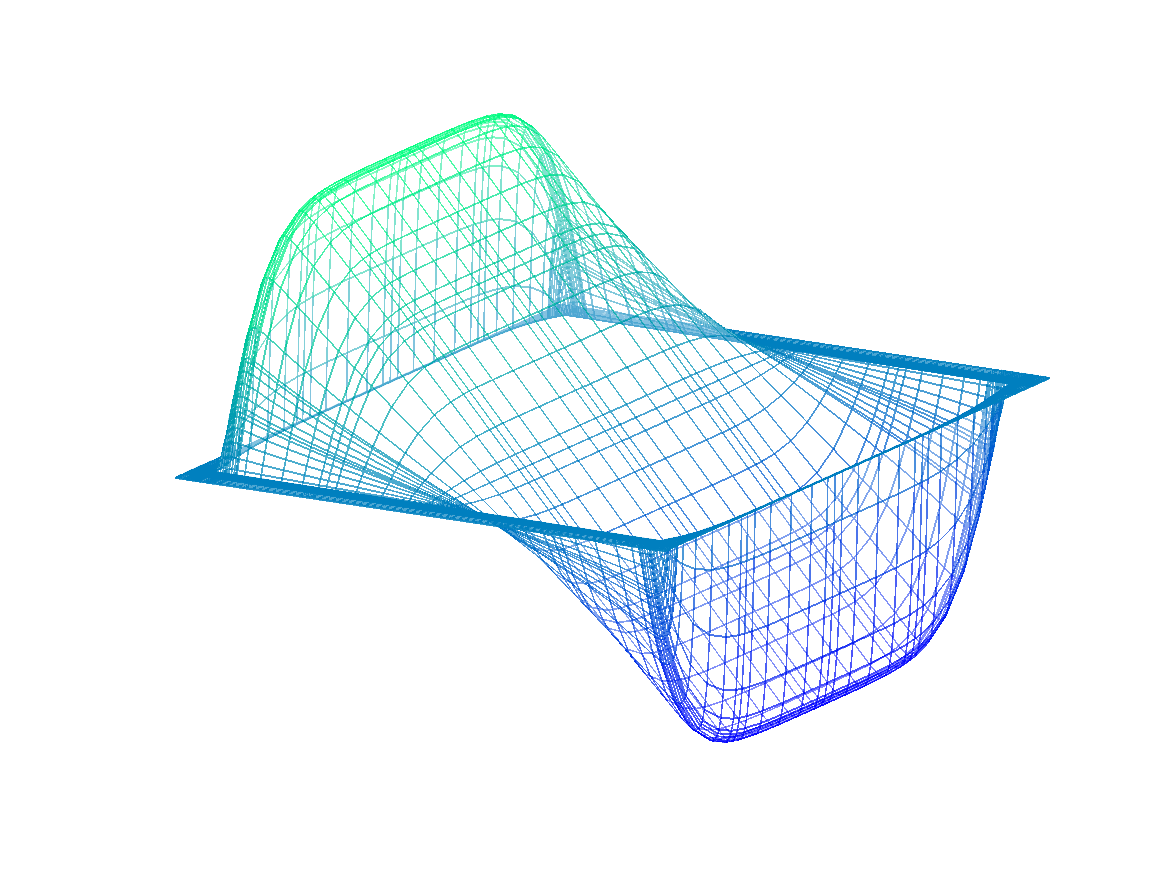}\\[5pt]
      \includegraphics[width=5.5cm]{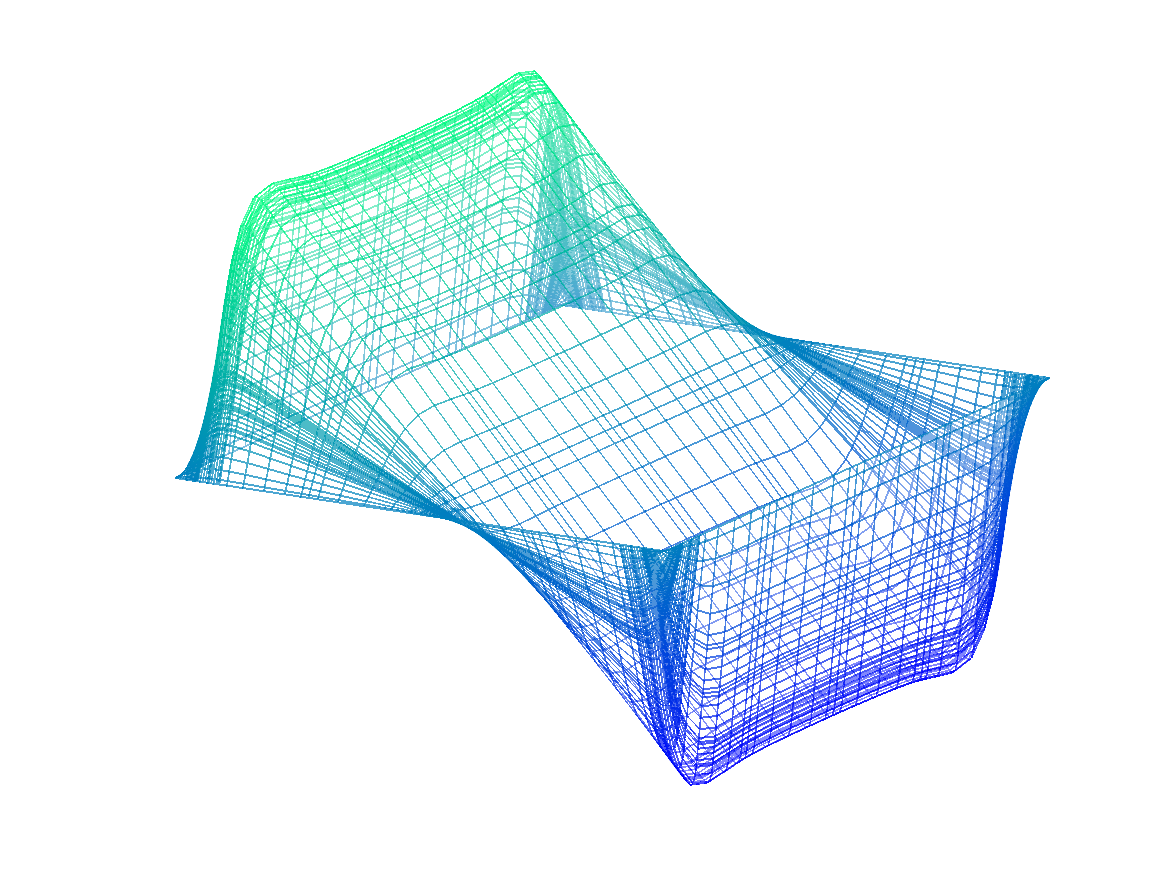}\\[5pt]
      \includegraphics[width=5.5cm]{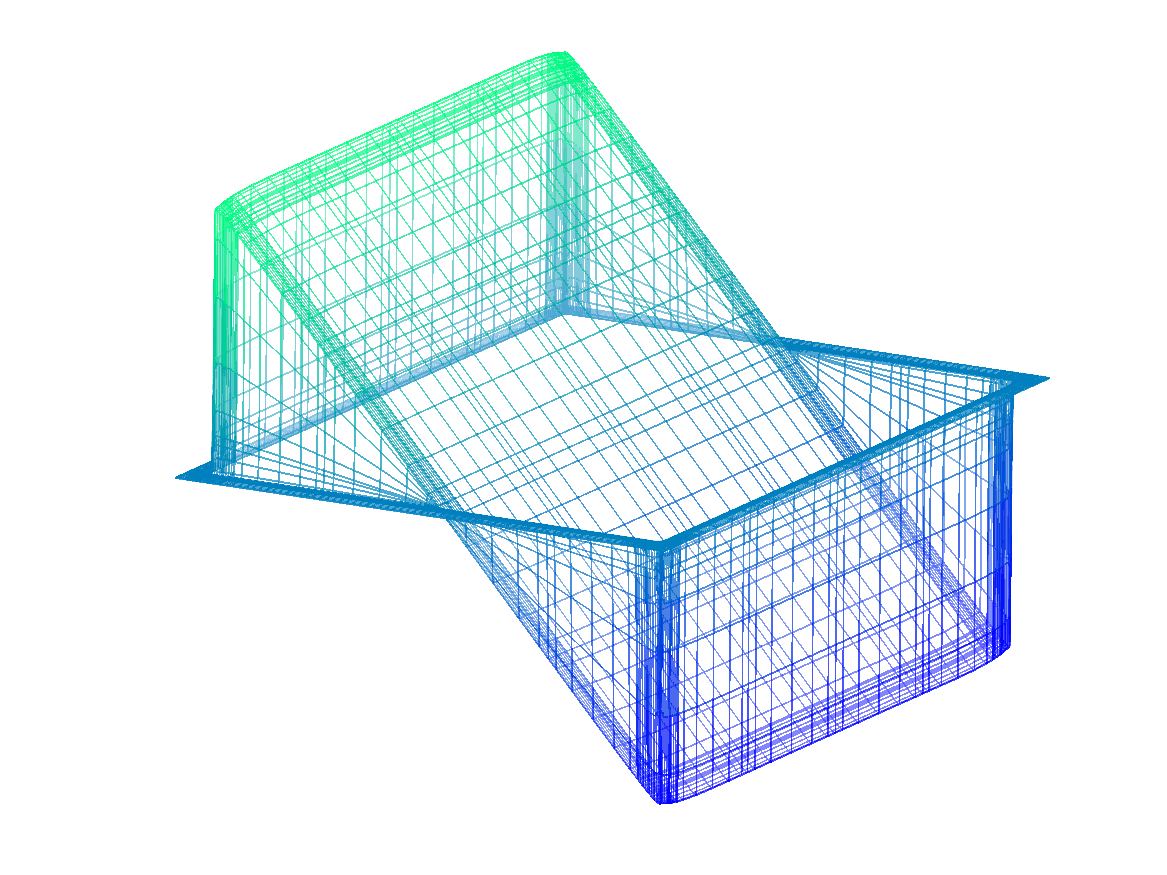}
      \caption{$\ha = 100$} 
  \end{subfigure}
  \caption{Magnetic distribution profile for
  $r_w= 10^{2}$,  $r_w= 1$, and  $r_w= 10^{-5}$ (bottom).
  $B_0=\ha/\sqrt{Re  Rm}$= 10, 50, and 100 are used from left to right.}
  \label{bmesh}
\end{figure}

%% file: figs/data/ha010.tex
\addplot[
line width=1.0pt, color=gray, solid]
coordinates {
(0.000010, 0.000010)
(0.003010, 0.003010)
(0.006010, 0.006010)
(0.009010, 0.009010)
(0.012010, 0.012010)
(0.015010, 0.015010)
(0.018010, 0.018010)
(0.021010, 0.021010)
(0.024010, 0.024009)
(0.027010, 0.027007)
(0.030010, 0.030000)
(0.033010, 0.032984)
(0.036010, 0.035954)
(0.039010, 0.038901)
(0.042010, 0.041816)
(0.045010, 0.044690)
(0.048010, 0.047512)
(0.051010, 0.050273)
(0.054010, 0.052962)
(0.057010, 0.055572)
(0.060010, 0.058096)
(0.063010, 0.060526)
(0.066010, 0.062859)
(0.069010, 0.065091)
(0.072010, 0.067220)
(0.075010, 0.069244)
(0.078010, 0.071163)
(0.081010, 0.072978)
(0.084010, 0.074691)
(0.087010, 0.076303)
(0.090010, 0.077818)
(0.093010, 0.079239)
(0.096010, 0.080568)
(0.099010, 0.081811)
(0.102010, 0.082970)
(0.105010, 0.084049)
(0.108010, 0.085054)
(0.111010, 0.085987)
(0.114010, 0.086853)
(0.117010, 0.087655)
(0.120010, 0.088398)
(0.123010, 0.089086)
(0.126010, 0.089721)
(0.129010, 0.090307)
(0.132010, 0.090848)
(0.135010, 0.091347)
(0.138010, 0.091806)
(0.141010, 0.092228)
(0.144010, 0.092617)
(0.147010, 0.092974)
(0.150010, 0.093302)
(0.153010, 0.093602)
(0.156010, 0.093878)
(0.159010, 0.094131)
(0.162010, 0.094362)
(0.165010, 0.094574)
(0.168010, 0.094768)
(0.171010, 0.094945)
(0.174010, 0.095107)
(0.177010, 0.095255)
(0.180010, 0.095389)
(0.183010, 0.095512)
(0.186010, 0.095624)
(0.189010, 0.095726)
(0.192010, 0.095819)
(0.195010, 0.095903)
(0.198010, 0.095979)
(0.201010, 0.096049)
(0.204010, 0.096112)
(0.207010, 0.096169)
(0.210010, 0.096221)
(0.213010, 0.096267)
(0.216010, 0.096310)
(0.219010, 0.096348)
(0.222010, 0.096382)
(0.225010, 0.096413)
(0.228010, 0.096441)
(0.231010, 0.096466)
(0.234010, 0.096489)
(0.237010, 0.096509)
(0.240010, 0.096527)
(0.243010, 0.096543)
(0.246010, 0.096558)
(0.249010, 0.096570)
(0.252010, 0.096582)
(0.255010, 0.096592)
(0.258010, 0.096601)
(0.261010, 0.096609)
(0.264010, 0.096616)
(0.267010, 0.096622)
(0.270010, 0.096628)
(0.273010, 0.096632)
(0.276010, 0.096636)
(0.279010, 0.096640)
(0.282010, 0.096643)
(0.285010, 0.096646)
(0.288010, 0.096648)
(0.291010, 0.096650)
(0.294010, 0.096651)
(0.297010, 0.096653)
(0.300010, 0.096654)
(0.303010, 0.096655)
(0.306010, 0.096655)
(0.309010, 0.096656)
(0.312010, 0.096656)
(0.315010, 0.096657)
(0.318010, 0.096657)
(0.321010, 0.096657)
(0.324010, 0.096657)
(0.327010, 0.096657)
(0.330010, 0.096657)
(0.333010, 0.096656)
(0.336010, 0.096656)
(0.339010, 0.096656)
(0.342010, 0.096656)
(0.345010, 0.096655)
(0.348010, 0.096655)
(0.351010, 0.096655)
(0.354010, 0.096654)
(0.357010, 0.096654)
(0.360010, 0.096654)
(0.363010, 0.096653)
(0.366010, 0.096653)
(0.369010, 0.096653)
(0.372010, 0.096652)
(0.375010, 0.096652)
(0.378010, 0.096652)
(0.381010, 0.096651)
(0.384010, 0.096651)
(0.387010, 0.096651)
(0.390010, 0.096651)
(0.393010, 0.096650)
(0.396010, 0.096650)
(0.399010, 0.096650)
(0.402010, 0.096650)
(0.405010, 0.096650)
(0.408010, 0.096649)
(0.411010, 0.096649)
(0.414010, 0.096649)
(0.417010, 0.096649)
(0.420010, 0.096649)
(0.423010, 0.096648)
(0.426010, 0.096648)
(0.429010, 0.096648)
(0.432010, 0.096648)
(0.435010, 0.096648)
(0.438010, 0.096648)
(0.441010, 0.096648)
(0.444010, 0.096648)
(0.447010, 0.096648)
(0.450010, 0.096647)
(0.453010, 0.096647)
(0.456010, 0.096647)
(0.459010, 0.096647)
(0.462010, 0.096647)
(0.465010, 0.096647)
(0.468010, 0.096647)
(0.471010, 0.096647)
(0.474010, 0.096647)
(0.477010, 0.096647)
(0.480010, 0.096647)
(0.483010, 0.096647)
(0.486010, 0.096647)
(0.489010, 0.096647)
(0.492010, 0.096647)
(0.495010, 0.096647)
(0.498010, 0.096647)
(0.501010, 0.096647)
(0.504010, 0.096647)
(0.507010, 0.096647)
(0.510010, 0.096647)
(0.513010, 0.096647)
(0.516010, 0.096647)
(0.519010, 0.096647)
(0.522010, 0.096647)
(0.525010, 0.096647)
(0.528010, 0.096646)
(0.531010, 0.096646)
(0.534010, 0.096646)
(0.537010, 0.096646)
(0.540010, 0.096646)
(0.543010, 0.096646)
(0.546010, 0.096646)
(0.549010, 0.096646)
(0.552010, 0.096646)
(0.555010, 0.096646)
(0.558010, 0.096646)
(0.561010, 0.096646)
(0.564010, 0.096646)
(0.567010, 0.096646)
(0.570010, 0.096646)
(0.573010, 0.096646)
(0.576010, 0.096646)
(0.579010, 0.096646)
(0.582010, 0.096646)
(0.585010, 0.096646)
(0.588010, 0.096646)
(0.591010, 0.096646)
(0.594010, 0.096646)
(0.597010, 0.096646)
(0.600010, 0.096646)
(0.603010, 0.096646)
(0.606010, 0.096646)
(0.609010, 0.096646)
(0.612010, 0.096646)
(0.615010, 0.096646)
(0.618010, 0.096646)
(0.621010, 0.096646)
(0.624010, 0.096646)
(0.627010, 0.096646)
(0.630010, 0.096646)
(0.633010, 0.096646)
(0.636010, 0.096646)
(0.639010, 0.096646)
(0.642010, 0.096646)
(0.645010, 0.096646)
(0.648010, 0.096646)
(0.651010, 0.096646)
(0.654010, 0.096646)
(0.657010, 0.096646)
(0.660010, 0.096646)
(0.663010, 0.096646)
(0.666010, 0.096646)
(0.669010, 0.096646)
(0.672010, 0.096646)
(0.675010, 0.096646)
(0.678010, 0.096646)
(0.681010, 0.096646)
(0.684010, 0.096646)
(0.687010, 0.096646)
(0.690010, 0.096646)
(0.693010, 0.096646)
(0.696010, 0.096646)
(0.699010, 0.096646)
(0.702010, 0.096646)
(0.705010, 0.096646)
(0.708010, 0.096646)
(0.711010, 0.096646)
(0.714010, 0.096646)
(0.717010, 0.096646)
(0.720010, 0.096646)
(0.723010, 0.096646)
(0.726010, 0.096646)
(0.729010, 0.096646)
(0.732010, 0.096646)
(0.735010, 0.096646)
(0.738010, 0.096646)
(0.741010, 0.096646)
(0.744010, 0.096646)
(0.747010, 0.096646)
(0.750010, 0.096646)
(0.753010, 0.096646)
(0.756010, 0.096646)
(0.759010, 0.096646)
(0.762010, 0.096646)
(0.765010, 0.096646)
(0.768010, 0.096646)
(0.771010, 0.096646)
(0.774010, 0.096646)
(0.777010, 0.096646)
(0.780010, 0.096646)
(0.783010, 0.096646)
(0.786010, 0.096646)
(0.789010, 0.096646)
(0.792010, 0.096646)
(0.795010, 0.096646)
(0.798010, 0.096646)
(0.801010, 0.096646)
(0.804010, 0.096646)
(0.807010, 0.096646)
(0.810010, 0.096646)
(0.813010, 0.096646)
(0.816010, 0.096646)
(0.819010, 0.096646)
(0.822010, 0.096646)
(0.825010, 0.096646)
(0.828010, 0.096646)
(0.831010, 0.096646)
(0.834010, 0.096646)
(0.837010, 0.096646)
(0.840010, 0.096646)
(0.843010, 0.096646)
(0.846010, 0.096646)
(0.849010, 0.096646)
(0.852010, 0.096646)
(0.855010, 0.096646)
(0.858010, 0.096646)
(0.861010, 0.096646)
(0.864010, 0.096646)
(0.867010, 0.096646)
(0.870010, 0.096646)
(0.873010, 0.096646)
(0.876010, 0.096646)
(0.879010, 0.096646)
(0.882010, 0.096646)
(0.885010, 0.096646)
(0.888010, 0.096646)
(0.891010, 0.096646)
(0.894010, 0.096646)
(0.897010, 0.096646)
(0.900010, 0.096646)
(0.903010, 0.096646)
(0.906010, 0.096646)
(0.909010, 0.096646)
(0.912010, 0.096646)
(0.915010, 0.096646)
(0.918010, 0.096646)
(0.921010, 0.096646)
(0.924010, 0.096646)
(0.927010, 0.096646)
(0.930010, 0.096646)
(0.933010, 0.096646)
(0.936010, 0.096646)
(0.939010, 0.096646)
(0.942010, 0.096646)
(0.945010, 0.096646)
(0.948010, 0.096646)
(0.951010, 0.096646)
(0.954010, 0.096646)
(0.957010, 0.096646)
(0.960010, 0.096646)
(0.963010, 0.096646)
(0.966010, 0.096646)
(0.969010, 0.096646)
(0.972010, 0.096646)
(0.975010, 0.096646)
(0.978010, 0.096646)
(0.981010, 0.096646)
(0.984010, 0.096646)
(0.987010, 0.096646)
(0.990010, 0.096646)
(0.993010, 0.096646)
(0.996010, 0.096646)
(0.999010, 0.096646)
(1.002010, 0.096646)
(1.005010, 0.096646)
(1.008010, 0.096646)
(1.011010, 0.096646)
(1.014010, 0.096646)
(1.017010, 0.096646)
(1.020010, 0.096646)
(1.023010, 0.096646)
(1.026010, 0.096646)
(1.029010, 0.096646)
(1.032010, 0.096646)
(1.035010, 0.096646)
(1.038010, 0.096646)
(1.041010, 0.096646)
(1.044010, 0.096646)
(1.047010, 0.096646)
(1.050010, 0.096646)
(1.053010, 0.096646)
(1.056010, 0.096646)
(1.059010, 0.096646)
(1.062010, 0.096646)
(1.065010, 0.096646)
(1.068010, 0.096646)
(1.071010, 0.096646)
(1.074010, 0.096646)
(1.077010, 0.096646)
(1.080010, 0.096646)
(1.083010, 0.096646)
(1.086010, 0.096646)
(1.089010, 0.096646)
(1.092010, 0.096646)
(1.095010, 0.096646)
(1.098010, 0.096646)
(1.101010, 0.096646)
(1.104010, 0.096646)
(1.107010, 0.096646)
(1.110010, 0.096646)
(1.113010, 0.096646)
(1.116010, 0.096646)
(1.119010, 0.096646)
(1.122010, 0.096646)
(1.125010, 0.096646)
(1.128010, 0.096646)
(1.131010, 0.096646)
(1.134010, 0.096646)
(1.137010, 0.096646)
(1.140010, 0.096646)
(1.143010, 0.096646)
(1.146010, 0.096646)
(1.149010, 0.096646)
(1.152010, 0.096646)
(1.155010, 0.096646)
(1.158010, 0.096646)
(1.161010, 0.096646)
(1.164010, 0.096646)
(1.167010, 0.096646)
(1.170010, 0.096646)
(1.173010, 0.096646)
(1.176010, 0.096646)
(1.179010, 0.096646)
(1.182010, 0.096646)
(1.185010, 0.096646)
(1.188010, 0.096646)
(1.191010, 0.096646)
(1.194010, 0.096646)
(1.197010, 0.096646)
(1.200010, 0.096646)
(1.203010, 0.096646)
(1.206010, 0.096646)
(1.209010, 0.096646)
(1.212010, 0.096646)
(1.215010, 0.096646)
(1.218010, 0.096646)
(1.221010, 0.096646)
(1.224010, 0.096646)
(1.227010, 0.096646)
(1.230010, 0.096646)
(1.233010, 0.096646)
(1.236010, 0.096646)
(1.239010, 0.096646)
(1.242010, 0.096646)
(1.245010, 0.096646)
(1.248010, 0.096646)
(1.251010, 0.096646)
(1.254010, 0.096646)
(1.257010, 0.096646)
(1.260010, 0.096646)
(1.263010, 0.096646)
(1.266010, 0.096646)
(1.269010, 0.096646)
(1.272010, 0.096646)
(1.275010, 0.096646)
(1.278010, 0.096646)
(1.281010, 0.096646)
(1.284010, 0.096646)
(1.287010, 0.096646)
(1.290010, 0.096646)
(1.293010, 0.096646)
(1.296010, 0.096646)
(1.299010, 0.096646)
(1.302010, 0.096646)
(1.305010, 0.096646)
(1.308010, 0.096646)
(1.311010, 0.096646)
(1.314010, 0.096646)
(1.317010, 0.096646)
(1.320010, 0.096646)
(1.323010, 0.096646)
(1.326010, 0.096646)
(1.329010, 0.096646)
(1.332010, 0.096646)
(1.335010, 0.096646)
(1.338010, 0.096646)
(1.341010, 0.096646)
(1.344010, 0.096646)
(1.347010, 0.096646)
(1.350010, 0.096646)
(1.353010, 0.096646)
(1.356010, 0.096646)
(1.359010, 0.096646)
(1.362010, 0.096646)
(1.365010, 0.096646)
(1.368010, 0.096646)
(1.371010, 0.096646)
(1.374010, 0.096646)
(1.377010, 0.096646)
(1.380010, 0.096646)
(1.383010, 0.096646)
(1.386010, 0.096646)
(1.389010, 0.096646)
(1.392010, 0.096646)
(1.395010, 0.096646)
(1.398010, 0.096646)
(1.401010, 0.096646)
(1.404010, 0.096646)
(1.407010, 0.096646)
(1.410010, 0.096646)
(1.413010, 0.096646)
(1.416010, 0.096646)
(1.419010, 0.096646)
(1.422010, 0.096646)
(1.425010, 0.096646)
(1.428010, 0.096646)
(1.431010, 0.096646)
(1.434010, 0.096646)
(1.437010, 0.096646)
(1.440010, 0.096646)
(1.443010, 0.096646)
(1.446010, 0.096646)
(1.449010, 0.096646)
(1.452010, 0.096646)
(1.455010, 0.096646)
(1.458010, 0.096646)
(1.461010, 0.096646)
(1.464010, 0.096646)
(1.467010, 0.096646)
(1.470010, 0.096646)
(1.473010, 0.096646)
(1.476010, 0.096646)
(1.479010, 0.096646)
(1.482010, 0.096646)
(1.485010, 0.096646)
(1.488010, 0.096646)
(1.491010, 0.096646)
(1.494010, 0.096646)
(1.497010, 0.096646)
(1.500010, 0.096646)
(1.503010, 0.096646)
(1.506010, 0.096646)
(1.509010, 0.096646)
(1.512010, 0.096646)
(1.515010, 0.096646)
(1.518010, 0.096646)
(1.521010, 0.096646)
(1.524010, 0.096646)
(1.527010, 0.096646)
(1.530010, 0.096646)
(1.533010, 0.096646)
(1.536010, 0.096646)
(1.539010, 0.096646)
(1.542010, 0.096646)
(1.545010, 0.096646)
(1.548010, 0.096646)
(1.551010, 0.096646)
(1.554010, 0.096646)
(1.557010, 0.096646)
(1.560010, 0.096646)
(1.563010, 0.096646)
(1.566010, 0.096646)
(1.569010, 0.096646)
(1.572010, 0.096646)
(1.575010, 0.096646)
(1.578010, 0.096646)
(1.581010, 0.096646)
(1.584010, 0.096646)
(1.587010, 0.096646)
(1.590010, 0.096646)
(1.593010, 0.096646)
(1.596010, 0.096646)
(1.599010, 0.096646)
(1.602010, 0.096646)
(1.605010, 0.096646)
(1.608010, 0.096646)
(1.611010, 0.096646)
(1.614010, 0.096646)
(1.617010, 0.096646)
(1.620010, 0.096646)
(1.623010, 0.096646)
(1.626010, 0.096646)
(1.629010, 0.096646)
(1.632010, 0.096646)
(1.635010, 0.096646)
(1.638010, 0.096646)
(1.641010, 0.096646)
(1.644010, 0.096646)
(1.647010, 0.096646)
(1.650010, 0.096646)
(1.653010, 0.096646)
(1.656010, 0.096646)
(1.659010, 0.096646)
(1.662010, 0.096646)
(1.665010, 0.096646)
(1.668010, 0.096646)
(1.671010, 0.096646)
(1.674010, 0.096646)
(1.677010, 0.096646)
(1.680010, 0.096646)
(1.683010, 0.096646)
(1.686010, 0.096646)
(1.689010, 0.096646)
(1.692010, 0.096646)
(1.695010, 0.096646)
(1.698010, 0.096646)
(1.701010, 0.096646)
(1.704010, 0.096646)
(1.707010, 0.096646)
(1.710010, 0.096646)
(1.713010, 0.096646)
(1.716010, 0.096646)
(1.719010, 0.096646)
(1.722010, 0.096646)
(1.725010, 0.096646)
(1.728010, 0.096646)
(1.731010, 0.096646)
(1.734010, 0.096646)
(1.737010, 0.096646)
(1.740010, 0.096646)
(1.743010, 0.096646)
(1.746010, 0.096646)
(1.749010, 0.096646)
(1.752010, 0.096646)
(1.755010, 0.096646)
(1.758010, 0.096646)
(1.761010, 0.096646)
(1.764010, 0.096646)
(1.767010, 0.096646)
(1.770010, 0.096646)
(1.773010, 0.096646)
(1.776010, 0.096646)
(1.779010, 0.096646)
(1.782010, 0.096646)
(1.785010, 0.096646)
(1.788010, 0.096646)
(1.791010, 0.096646)
(1.794010, 0.096646)
(1.797010, 0.096646)
(1.800010, 0.096646)
(1.803010, 0.096646)
(1.806010, 0.096646)
(1.809010, 0.096646)
(1.812010, 0.096646)
(1.815010, 0.096646)
(1.818010, 0.096646)
(1.821010, 0.096646)
(1.824010, 0.096646)
(1.827010, 0.096646)
(1.830010, 0.096646)
(1.833010, 0.096646)
(1.836010, 0.096646)
(1.839010, 0.096646)
(1.842010, 0.096646)
(1.845010, 0.096646)
(1.848010, 0.096646)
(1.851010, 0.096646)
(1.854010, 0.096646)
(1.857010, 0.096646)
(1.860010, 0.096646)
(1.863010, 0.096646)
(1.866010, 0.096646)
(1.869010, 0.096646)
(1.872010, 0.096646)
(1.875010, 0.096646)
(1.878010, 0.096646)
(1.881010, 0.096646)
(1.884010, 0.096646)
(1.887010, 0.096646)
(1.890010, 0.096646)
(1.893010, 0.096646)
(1.896010, 0.096646)
(1.899010, 0.096646)
(1.902010, 0.096646)
(1.905010, 0.096646)
(1.908010, 0.096646)
(1.911010, 0.096646)
(1.914010, 0.096646)
(1.917010, 0.096646)
(1.920010, 0.096646)
(1.923010, 0.096646)
(1.926010, 0.096646)
(1.929010, 0.096646)
(1.932010, 0.096646)
(1.935010, 0.096646)
(1.938010, 0.096646)
(1.941010, 0.096646)
(1.944010, 0.096646)
(1.947010, 0.096646)
(1.950010, 0.096646)
(1.953010, 0.096646)
(1.956010, 0.096646)
(1.959010, 0.096646)
(1.962010, 0.096646)
(1.965010, 0.096646)
(1.968010, 0.096646)
(1.971010, 0.096646)
(1.974010, 0.096646)
(1.977010, 0.096646)
(1.980010, 0.096646)
(1.983010, 0.096646)
(1.986010, 0.096646)
(1.989010, 0.096646)
(1.992010, 0.096646)
(1.995010, 0.096646)
(1.998010, 0.096646)
(2.001010, 0.096646)
(2.004010, 0.096646)
(2.007010, 0.096646)
(2.010010, 0.096646)
(2.013010, 0.096646)
(2.016010, 0.096646)
(2.019010, 0.096646)
(2.022010, 0.096646)
(2.025010, 0.096646)
(2.028010, 0.096646)
(2.031010, 0.096646)
(2.034010, 0.096646)
(2.037010, 0.096646)
(2.040010, 0.096646)
(2.043010, 0.096646)
(2.046010, 0.096646)
(2.049010, 0.096646)
(2.052010, 0.096646)
(2.055010, 0.096646)
(2.058010, 0.096646)
(2.061010, 0.096646)
(2.064010, 0.096646)
(2.067010, 0.096646)
(2.070010, 0.096646)
(2.073010, 0.096646)
(2.076010, 0.096646)
(2.079010, 0.096646)
(2.082010, 0.096646)
(2.085010, 0.096646)
(2.088010, 0.096646)
(2.091010, 0.096646)
(2.094010, 0.096646)
(2.097010, 0.096646)
(2.100010, 0.096646)
(2.103010, 0.096646)
(2.106010, 0.096646)
(2.109010, 0.096646)
(2.112010, 0.096646)
(2.115010, 0.096646)
(2.118010, 0.096646)
(2.121010, 0.096646)
(2.124010, 0.096646)
(2.127010, 0.096646)
(2.130010, 0.096646)
(2.133010, 0.096646)
(2.136010, 0.096646)
(2.139010, 0.096646)
(2.142010, 0.096646)
(2.145010, 0.096646)
(2.148010, 0.096646)
(2.151010, 0.096646)
(2.154010, 0.096646)
(2.157010, 0.096646)
(2.160010, 0.096646)
(2.163010, 0.096646)
(2.166010, 0.096646)
(2.169010, 0.096646)
(2.172010, 0.096646)
(2.175010, 0.096646)
(2.178010, 0.096646)
(2.181010, 0.096646)
(2.184010, 0.096646)
(2.187010, 0.096646)
(2.190010, 0.096646)
(2.193010, 0.096646)
(2.196010, 0.096646)
(2.199010, 0.096646)
(2.202010, 0.096646)
(2.205010, 0.096646)
(2.208010, 0.096646)
(2.211010, 0.096646)
(2.214010, 0.096646)
(2.217010, 0.096646)
(2.220010, 0.096646)
(2.223010, 0.096646)
(2.226010, 0.096646)
(2.229010, 0.096646)
(2.232010, 0.096646)
(2.235010, 0.096646)
(2.238010, 0.096646)
(2.241010, 0.096646)
(2.244010, 0.096646)
(2.247010, 0.096646)
(2.250010, 0.096646)
(2.253010, 0.096646)
(2.256010, 0.096646)
(2.259010, 0.096646)
(2.262010, 0.096646)
(2.265010, 0.096646)
(2.268010, 0.096646)
(2.271010, 0.096646)
(2.274010, 0.096646)
(2.277010, 0.096646)
(2.280010, 0.096646)
(2.283010, 0.096646)
(2.286010, 0.096646)
(2.289010, 0.096646)
(2.292010, 0.096646)
(2.295010, 0.096646)
(2.298010, 0.096646)
(2.301010, 0.096646)
(2.304010, 0.096646)
(2.307010, 0.096646)
(2.310010, 0.096646)
(2.313010, 0.096646)
(2.316010, 0.096646)
(2.319010, 0.096646)
(2.322010, 0.096646)
(2.325010, 0.096646)
(2.328010, 0.096646)
(2.331010, 0.096646)
(2.334010, 0.096646)
(2.337010, 0.096646)
(2.340010, 0.096646)
(2.343010, 0.096646)
(2.346010, 0.096646)
(2.349010, 0.096646)
(2.352010, 0.096646)
(2.355010, 0.096646)
(2.358010, 0.096646)
(2.361010, 0.096646)
(2.364010, 0.096646)
(2.367010, 0.096646)
(2.370010, 0.096646)
(2.373010, 0.096646)
(2.376010, 0.096646)
(2.379010, 0.096646)
(2.382010, 0.096646)
(2.385010, 0.096646)
(2.388010, 0.096646)
(2.391010, 0.096646)
(2.394010, 0.096646)
(2.397010, 0.096646)
(2.400010, 0.096646)
(2.403010, 0.096646)
(2.406010, 0.096646)
(2.409010, 0.096646)
(2.412010, 0.096646)
(2.415010, 0.096646)
(2.418010, 0.096646)
(2.421010, 0.096646)
(2.424010, 0.096646)
(2.427010, 0.096646)
(2.430010, 0.096646)
(2.433010, 0.096646)
(2.436010, 0.096646)
(2.439010, 0.096646)
(2.442010, 0.096646)
(2.445010, 0.096646)
(2.448010, 0.096646)
(2.451010, 0.096646)
(2.454010, 0.096646)
(2.457010, 0.096646)
(2.460010, 0.096646)
(2.463010, 0.096646)
(2.466010, 0.096646)
(2.469010, 0.096646)
(2.472010, 0.096646)
(2.475010, 0.096646)
(2.478010, 0.096646)
(2.481010, 0.096646)
(2.484010, 0.096646)
(2.487010, 0.096646)
(2.490010, 0.096646)
(2.493010, 0.096646)
(2.496010, 0.096646)
(2.499010, 0.096646)
(2.502010, 0.096646)
(2.505010, 0.096646)
(2.508010, 0.096646)
(2.511010, 0.096646)
(2.514010, 0.096646)
(2.517010, 0.096646)
(2.520010, 0.096646)
(2.523010, 0.096646)
(2.526010, 0.096646)
(2.529010, 0.096646)
(2.532010, 0.096646)
(2.535010, 0.096646)
(2.538010, 0.096646)
(2.541010, 0.096646)
(2.544010, 0.096646)
(2.547010, 0.096646)
(2.550010, 0.096646)
(2.553010, 0.096646)
(2.556010, 0.096646)
(2.559010, 0.096646)
(2.562010, 0.096646)
(2.565010, 0.096646)
(2.568010, 0.096646)
(2.571010, 0.096646)
(2.574010, 0.096646)
(2.577010, 0.096646)
(2.580010, 0.096646)
(2.583010, 0.096646)
(2.586010, 0.096646)
(2.589010, 0.096646)
(2.592010, 0.096646)
(2.595010, 0.096646)
(2.598010, 0.096646)
(2.601010, 0.096646)
(2.604010, 0.096646)
(2.607010, 0.096646)
(2.610010, 0.096646)
(2.613010, 0.096646)
(2.616010, 0.096646)
(2.619010, 0.096646)
(2.622010, 0.096646)
(2.625010, 0.096646)
(2.628010, 0.096646)
(2.631010, 0.096646)
(2.634010, 0.096646)
(2.637010, 0.096646)
(2.640010, 0.096646)
(2.643010, 0.096646)
(2.646010, 0.096646)
(2.649010, 0.096646)
(2.652010, 0.096646)
(2.655010, 0.096646)
(2.658010, 0.096646)
(2.661010, 0.096646)
(2.664010, 0.096646)
(2.667010, 0.096646)
(2.670010, 0.096646)
(2.673010, 0.096646)
(2.676010, 0.096646)
(2.679010, 0.096646)
(2.682010, 0.096646)
(2.685010, 0.096646)
(2.688010, 0.096646)
(2.691010, 0.096646)
(2.694010, 0.096646)
(2.697010, 0.096646)
(2.700010, 0.096646)
(2.703010, 0.096646)
(2.706010, 0.096646)
(2.709010, 0.096646)
(2.712010, 0.096646)
(2.715010, 0.096646)
(2.718010, 0.096646)
(2.721010, 0.096646)
(2.724010, 0.096646)
(2.727010, 0.096646)
(2.730010, 0.096646)
(2.733010, 0.096646)
(2.736010, 0.096646)
(2.739010, 0.096646)
(2.742010, 0.096646)
(2.745010, 0.096646)
(2.748010, 0.096646)
(2.751010, 0.096646)
(2.754010, 0.096646)
(2.757010, 0.096646)
(2.760010, 0.096646)
(2.763010, 0.096646)
(2.766010, 0.096646)
(2.769010, 0.096646)
(2.772010, 0.096646)
(2.775010, 0.096646)
(2.778010, 0.096646)
(2.781010, 0.096646)
(2.784010, 0.096646)
(2.787010, 0.096646)
(2.790010, 0.096646)
(2.793010, 0.096646)
(2.796010, 0.096646)
(2.799010, 0.096646)
(2.802010, 0.096646)
(2.805010, 0.096646)
(2.808010, 0.096646)
(2.811010, 0.096646)
(2.814010, 0.096646)
(2.817010, 0.096646)
(2.820010, 0.096646)
(2.823010, 0.096646)
(2.826010, 0.096646)
(2.829010, 0.096646)
(2.832010, 0.096646)
(2.835010, 0.096646)
(2.838010, 0.096646)
(2.841010, 0.096646)
(2.844010, 0.096646)
(2.847010, 0.096646)
(2.850010, 0.096646)
(2.853010, 0.096646)
(2.856010, 0.096646)
(2.859010, 0.096646)
(2.862010, 0.096646)
(2.865010, 0.096646)
(2.868010, 0.096646)
(2.871010, 0.096646)
(2.874010, 0.096646)
(2.877010, 0.096646)
(2.880010, 0.096646)
(2.883010, 0.096646)
(2.886010, 0.096646)
(2.889010, 0.096646)
(2.892010, 0.096646)
(2.895010, 0.096646)
(2.898010, 0.096646)
(2.901010, 0.096646)
(2.904010, 0.096646)
(2.907010, 0.096646)
(2.910010, 0.096646)
(2.913010, 0.096646)
(2.916010, 0.096646)
(2.919010, 0.096646)
(2.922010, 0.096646)
(2.925010, 0.096646)
(2.928010, 0.096646)
(2.931010, 0.096646)
(2.934010, 0.096646)
(2.937010, 0.096646)
(2.940010, 0.096646)
(2.943010, 0.096646)
(2.946010, 0.096646)
(2.949010, 0.096646)
(2.952010, 0.096646)
(2.955010, 0.096646)
(2.958010, 0.096646)
(2.961010, 0.096646)
(2.964010, 0.096646)
(2.967010, 0.096646)
(2.970010, 0.096646)
(2.973010, 0.096646)
(2.976010, 0.096646)
(2.979010, 0.096646)
(2.982010, 0.096646)
(2.985010, 0.096646)
(2.988010, 0.096646)
(2.991010, 0.096646)
(2.994010, 0.096646)
(2.997010, 0.096646)
};
\label{1e2} 
\addplot[
line width=1.0pt, color=blue, solid ]
coordinates {
(0.000050, 0.000050)
(0.015050, 0.015050)
(0.030050, 0.030038)
(0.045050, 0.044677)
(0.060050, 0.057765)
(0.075050, 0.068044)
(0.090050, 0.075045)
(0.105050, 0.079016)
(0.120050, 0.080537)
(0.135050, 0.080244)
(0.150050, 0.078697)
(0.165050, 0.076350)
(0.180050, 0.073560)
(0.195050, 0.070606)
(0.210050, 0.067706)
(0.225050, 0.065022)
(0.240050, 0.062665)
(0.255050, 0.060699)
(0.270050, 0.059147)
(0.285050, 0.058000)
(0.300050, 0.057223)
(0.315050, 0.056768)
(0.330050, 0.056579)
(0.345050, 0.056595)
(0.360050, 0.056763)
(0.375050, 0.057030)
(0.390050, 0.057354)
(0.405050, 0.057699)
(0.420050, 0.058038)
(0.435050, 0.058351)
(0.450050, 0.058625)
(0.465050, 0.058852)
(0.480050, 0.059032)
(0.495050, 0.059165)
(0.510050, 0.059255)
(0.525050, 0.059308)
(0.540050, 0.059331)
(0.555050, 0.059329)
(0.570050, 0.059310)
(0.585050, 0.059280)
(0.600050, 0.059243)
(0.615050, 0.059203)
(0.630050, 0.059164)
(0.645050, 0.059128)
(0.660050, 0.059096)
(0.675050, 0.059070)
(0.690050, 0.059049)
(0.705050, 0.059034)
(0.720050, 0.059023)
(0.735050, 0.059017)
(0.750050, 0.059014)
(0.765050, 0.059014)
(0.780050, 0.059017)
(0.795050, 0.059020)
(0.810050, 0.059024)
(0.825050, 0.059029)
(0.840050, 0.059033)
(0.855050, 0.059038)
(0.870050, 0.059041)
(0.885050, 0.059044)
(0.900050, 0.059047)
(0.915050, 0.059048)
(0.930050, 0.059050)
(0.945050, 0.059050)
(0.960050, 0.059051)
(0.975050, 0.059051)
(0.990050, 0.059050)
(1.005050, 0.059050)
(1.020050, 0.059050)
(1.035050, 0.059049)
(1.050050, 0.059049)
(1.065050, 0.059048)
(1.080050, 0.059048)
(1.095050, 0.059047)
(1.110050, 0.059047)
(1.125050, 0.059047)
(1.140050, 0.059047)
(1.155050, 0.059047)
(1.170050, 0.059047)
(1.185050, 0.059047)
(1.200050, 0.059047)
(1.215050, 0.059047)
(1.230050, 0.059047)
(1.245050, 0.059047)
(1.260050, 0.059047)
(1.275050, 0.059047)
(1.290050, 0.059047)
(1.305050, 0.059047)
(1.320050, 0.059047)
(1.335050, 0.059047)
(1.350050, 0.059047)
(1.365050, 0.059047)
(1.380050, 0.059047)
(1.395050, 0.059047)
(1.410050, 0.059047)
(1.425050, 0.059047)
(1.440050, 0.059047)
(1.455050, 0.059047)
(1.470050, 0.059047)
(1.485050, 0.059047)
(1.500050, 0.059047)
(1.515050, 0.059047)
(1.530050, 0.059047)
(1.545050, 0.059047)
(1.560050, 0.059047)
(1.575050, 0.059047)
(1.590050, 0.059047)
(1.605050, 0.059047)
(1.620050, 0.059047)
(1.635050, 0.059047)
(1.650050, 0.059047)
(1.665050, 0.059047)
(1.680050, 0.059047)
(1.695050, 0.059047)
(1.710050, 0.059047)
(1.725050, 0.059047)
(1.740050, 0.059047)
(1.755050, 0.059047)
(1.770050, 0.059047)
(1.785050, 0.059047)
(1.800050, 0.059047)
(1.815050, 0.059047)
(1.830050, 0.059047)
(1.845050, 0.059047)
(1.860050, 0.059047)
(1.875050, 0.059047)
(1.890050, 0.059047)
(1.905050, 0.059047)
(1.920050, 0.059047)
(1.935050, 0.059047)
(1.950050, 0.059047)
(1.965050, 0.059047)
(1.980050, 0.059047)
(1.995050, 0.059047)
};
\label{1e0}
\addplot[
line width=1.0pt, color=red, solid ]
coordinates {
(0.000010, 0.000010)
(0.010010, 0.010010)
(0.020010, 0.020010)
(0.030010, 0.029997)
(0.040010, 0.039836)
(0.050010, 0.049128)
(0.060010, 0.057315)
(0.070010, 0.063909)
(0.080010, 0.068614)
(0.090010, 0.071344)
(0.100010, 0.072176)
(0.110010, 0.071284)
(0.120010, 0.068898)
(0.130010, 0.065260)
(0.140010, 0.060606)
(0.150010, 0.055158)
(0.160010, 0.049121)
(0.170010, 0.042684)
(0.180010, 0.036024)
(0.190010, 0.029309)
(0.200010, 0.022696)
(0.210010, 0.016333)
(0.220010, 0.010350)
(0.230010, 0.004865)
(0.240010, -0.000026)
(0.250010, -0.004249)
(0.260010, -0.007753)
(0.270010, -0.010507)
(0.280010, -0.012506)
(0.290010, -0.013760)
(0.300010, -0.014302)
(0.310010, -0.014177)
(0.320010, -0.013443)
(0.330010, -0.012169)
(0.340010, -0.010430)
(0.350010, -0.008307)
(0.360010, -0.005883)
(0.370010, -0.003241)
(0.380010, -0.000462)
(0.390010, 0.002375)
(0.400010, 0.005197)
(0.410010, 0.007938)
(0.420010, 0.010538)
(0.430010, 0.012947)
(0.440010, 0.015122)
(0.450010, 0.017030)
(0.460010, 0.018649)
(0.470010, 0.019962)
(0.480010, 0.020963)
(0.490010, 0.021655)
(0.500010, 0.022047)
(0.510010, 0.022154)
(0.520010, 0.021997)
(0.530010, 0.021602)
(0.540010, 0.020999)
(0.550010, 0.020219)
(0.560010, 0.019295)
(0.570010, 0.018263)
(0.580010, 0.017156)
(0.590010, 0.016006)
(0.600010, 0.014846)
(0.610010, 0.013704)
(0.620010, 0.012605)
(0.630010, 0.011573)
(0.640010, 0.010628)
(0.650010, 0.009784)
(0.660010, 0.009053)
(0.670010, 0.008444)
(0.680010, 0.007961)
(0.690010, 0.007605)
(0.700010, 0.007375)
(0.710010, 0.007265)
(0.720010, 0.007268)
(0.730010, 0.007374)
(0.740010, 0.007572)
(0.750010, 0.007851)
(0.760010, 0.008195)
(0.770010, 0.008592)
(0.780010, 0.009028)
(0.790010, 0.009489)
(0.800010, 0.009961)
(0.810010, 0.010433)
(0.820010, 0.010893)
(0.830010, 0.011330)
(0.840010, 0.011737)
(0.850010, 0.012106)
(0.860010, 0.012432)
(0.870010, 0.012709)
(0.880010, 0.012936)
(0.890010, 0.013111)
(0.900010, 0.013234)
(0.910010, 0.013307)
(0.920010, 0.013331)
(0.930010, 0.013312)
(0.940010, 0.013252)
(0.950010, 0.013156)
(0.960010, 0.013030)
(0.970010, 0.012879)
(0.980010, 0.012710)
(0.990010, 0.012527)
(1.000010, 0.012336)
(1.010010, 0.012142)
(1.020010, 0.011951)
(1.030010, 0.011766)
(1.040010, 0.011592)
(1.050010, 0.011432)
(1.060010, 0.011289)
(1.070010, 0.011164)
(1.080010, 0.011059)
(1.090010, 0.010976)
(1.100010, 0.010913)
(1.110010, 0.010871)
(1.120010, 0.010850)
(1.130010, 0.010848)
(1.140010, 0.010863)
(1.150010, 0.010894)
(1.160010, 0.010938)
(1.170010, 0.010994)
(1.180010, 0.011059)
(1.190010, 0.011131)
(1.200010, 0.011207)
(1.210010, 0.011285)
(1.220010, 0.011364)
(1.230010, 0.011441)
(1.240010, 0.011515)
(1.250010, 0.011583)
(1.260010, 0.011646)
(1.270010, 0.011701)
(1.280010, 0.011748)
(1.290010, 0.011788)
(1.300010, 0.011818)
(1.310010, 0.011840)
(1.320010, 0.011853)
(1.330010, 0.011858)
(1.340010, 0.011856)
(1.350010, 0.011847)
(1.360010, 0.011832)
(1.370010, 0.011811)
(1.380010, 0.011787)
(1.390010, 0.011759)
(1.400010, 0.011729)
(1.410010, 0.011697)
(1.420010, 0.011665)
(1.430010, 0.011633)
(1.440010, 0.011602)
(1.450010, 0.011572)
(1.460010, 0.011545)
(1.470010, 0.011521)
(1.480010, 0.011500)
(1.490010, 0.011481)
(1.500010, 0.011467)
(1.510010, 0.011456)
(1.520010, 0.011448)
(1.530010, 0.011444)
(1.540010, 0.011443)
(1.550010, 0.011445)
(1.560010, 0.011450)
(1.570010, 0.011457)
(1.580010, 0.011466)
(1.590010, 0.011477)
(1.600010, 0.011489)
(1.610010, 0.011501)
(1.620010, 0.011514)
(1.630010, 0.011527)
(1.640010, 0.011540)
(1.650010, 0.011552)
(1.660010, 0.011564)
(1.670010, 0.011575)
(1.680010, 0.011584)
(1.690010, 0.011592)
(1.700010, 0.011599)
(1.710010, 0.011604)
(1.720010, 0.011608)
(1.730010, 0.011610)
(1.740010, 0.011611)
(1.750010, 0.011611)
(1.760010, 0.011610)
(1.770010, 0.011607)
(1.780010, 0.011604)
(1.790010, 0.011600)
(1.800010, 0.011595)
(1.810010, 0.011590)
(1.820010, 0.011585)
(1.830010, 0.011580)
(1.840010, 0.011574)
(1.850010, 0.011569)
(1.860010, 0.011564)
(1.870010, 0.011559)
(1.880010, 0.011555)
(1.890010, 0.011551)
(1.900010, 0.011548)
(1.910010, 0.011546)
(1.920010, 0.011544)
(1.930010, 0.011543)
(1.940010, 0.011542)
(1.950010, 0.011541)
(1.960010, 0.011542)
(1.970010, 0.011542)
(1.980010, 0.011543)
(1.990010, 0.011545)
(2.000010, 0.011547)
(2.010010, 0.011549)
(2.020010, 0.011551)
(2.030010, 0.011553)
(2.040010, 0.011555)
(2.050010, 0.011557)
(2.060010, 0.011559)
(2.070010, 0.011561)
(2.080010, 0.011563)
(2.090010, 0.011564)
(2.100010, 0.011566)
(2.110010, 0.011567)
(2.120010, 0.011568)
(2.130010, 0.011568)
(2.140010, 0.011569)
(2.150010, 0.011569)
(2.160010, 0.011569)
(2.170010, 0.011569)
(2.180010, 0.011568)
(2.190010, 0.011568)
(2.200010, 0.011567)
(2.210010, 0.011566)
(2.220010, 0.011565)
(2.230010, 0.011564)
(2.240010, 0.011564)
(2.250010, 0.011563)
(2.260010, 0.011562)
(2.270010, 0.011561)
(2.280010, 0.011560)
(2.290010, 0.011559)
(2.300010, 0.011559)
(2.310010, 0.011558)
(2.320010, 0.011558)
(2.330010, 0.011557)
(2.340010, 0.011557)
(2.350010, 0.011557)
(2.360010, 0.011557)
(2.370010, 0.011557)
(2.380010, 0.011557)
(2.390010, 0.011557)
(2.400010, 0.011557)
(2.410010, 0.011557)
(2.420010, 0.011558)
(2.430010, 0.011558)
(2.440010, 0.011558)
(2.450010, 0.011559)
(2.460010, 0.011559)
(2.470010, 0.011559)
(2.480010, 0.011560)
(2.490010, 0.011560)
(2.500010, 0.011560)
(2.510010, 0.011560)
(2.520010, 0.011561)
(2.530010, 0.011561)
(2.540010, 0.011561)
(2.550010, 0.011561)
(2.560010, 0.011561)
(2.570010, 0.011561)
(2.580010, 0.011561)
(2.590010, 0.011561)
(2.600010, 0.011561)
(2.610010, 0.011560)
(2.620010, 0.011560)
(2.630010, 0.011560)
(2.640010, 0.011560)
(2.650010, 0.011560)
(2.660010, 0.011560)
(2.670010, 0.011560)
(2.680010, 0.011559)
(2.690010, 0.011559)
(2.700010, 0.011559)
(2.710010, 0.011559)
(2.720010, 0.011559)
(2.730010, 0.011559)
(2.740010, 0.011559)
(2.750010, 0.011559)
(2.760010, 0.011559)
(2.770010, 0.011558)
(2.780010, 0.011558)
(2.790010, 0.011558)
(2.800010, 0.011558)
(2.810010, 0.011558)
(2.820010, 0.011559)
(2.830010, 0.011559)
(2.840010, 0.011559)
(2.850010, 0.011559)
(2.860010, 0.011559)
(2.870010, 0.011559)
(2.880010, 0.011559)
(2.890010, 0.011559)
(2.900010, 0.011559)
(2.910010, 0.011559)
(2.920010, 0.011559)
(2.930010, 0.011559)
(2.940010, 0.011559)
(2.950010, 0.011559)
(2.960010, 0.011559)
(2.970010, 0.011559)
(2.980010, 0.011559)
(2.990010, 0.011559)
(3.000010, 0.011559)
(3.010010, 0.011559)
(3.020010, 0.011559)
(3.030010, 0.011559)
(3.040010, 0.011559)
(3.050010, 0.011559)
(3.060010, 0.011559)
(3.070010, 0.011558)
(3.080010, 0.011558)
(3.090010, 0.011558)
(3.100010, 0.011558)
(3.110010, 0.011558)
(3.120010, 0.011558)
(3.130010, 0.011558)
(3.140010, 0.011558)
(3.150010, 0.011558)
(3.160010, 0.011558)
(3.170010, 0.011558)
(3.180010, 0.011558)
(3.190010, 0.011558)
(3.200010, 0.011558)
(3.210010, 0.011558)
(3.220010, 0.011558)
(3.230010, 0.011558)
(3.240010, 0.011558)
(3.250010, 0.011558)
(3.260010, 0.011558)
(3.270010, 0.011558)
(3.280010, 0.011558)
(3.290010, 0.011558)
(3.300010, 0.011558)
(3.310010, 0.011558)
(3.320010, 0.011558)
(3.330010, 0.011558)
(3.340010, 0.011558)
(3.350010, 0.011558)
(3.360010, 0.011558)
(3.370010, 0.011558)
(3.380010, 0.011558)
(3.390010, 0.011558)
(3.400010, 0.011558)
(3.410010, 0.011558)
(3.420010, 0.011558)
(3.430010, 0.011558)
(3.440010, 0.011558)
(3.450010, 0.011558)
(3.460010, 0.011558)
(3.470010, 0.011558)
(3.480010, 0.011558)
(3.490010, 0.011558)
(3.500010, 0.011558)
(3.510010, 0.011558)
(3.520010, 0.011558)
(3.530010, 0.011558)
(3.540010, 0.011558)
(3.550010, 0.011558)
(3.560010, 0.011558)
(3.570010, 0.011558)
(3.580010, 0.011558)
(3.590010, 0.011558)
(3.600010, 0.011557)
(3.610010, 0.011557)
(3.620010, 0.011557)
(3.630010, 0.011557)
(3.640010, 0.011557)
(3.650010, 0.011557)
(3.660010, 0.011557)
(3.670010, 0.011557)
(3.680010, 0.011557)
(3.690010, 0.011557)
(3.700010, 0.011557)
(3.710010, 0.011557)
(3.720010, 0.011557)
(3.730010, 0.011557)
(3.740010, 0.011557)
(3.750010, 0.011557)
(3.760010, 0.011557)
(3.770010, 0.011557)
(3.780010, 0.011557)
(3.790010, 0.011557)
(3.800010, 0.011557)
(3.810010, 0.011557)
(3.820010, 0.011557)
(3.830010, 0.011557)
(3.840010, 0.011557)
(3.850010, 0.011557)
(3.860010, 0.011557)
(3.870010, 0.011557)
(3.880010, 0.011557)
(3.890010, 0.011557)
(3.900010, 0.011557)
(3.910010, 0.011557)
(3.920010, 0.011557)
(3.930010, 0.011557)
(3.940010, 0.011557)
(3.950010, 0.011557)
(3.960010, 0.011557)
(3.970010, 0.011557)
(3.980010, 0.011557)
(3.990010, 0.011557)
(4.000010, 0.011557)
(4.010010, 0.011557)
(4.020010, 0.011557)
(4.030010, 0.011557)
(4.040010, 0.011557)
(4.050010, 0.011557)
(4.060010, 0.011557)
(4.070010, 0.011557)
(4.080010, 0.011557)
(4.090010, 0.011557)
(4.100010, 0.011557)
(4.110010, 0.011557)
(4.120010, 0.011557)
(4.130010, 0.011557)
(4.140010, 0.011557)
(4.150010, 0.011557)
(4.160010, 0.011557)
(4.170010, 0.011557)
(4.180010, 0.011557)
(4.190010, 0.011557)
(4.200010, 0.011557)
(4.210010, 0.011557)
(4.220010, 0.011557)
(4.230010, 0.011557)
(4.240010, 0.011557)
(4.250010, 0.011557)
(4.260010, 0.011557)
(4.270010, 0.011557)
(4.280010, 0.011557)
(4.290010, 0.011557)
(4.300010, 0.011557)
(4.310010, 0.011557)
(4.320010, 0.011557)
(4.330010, 0.011557)
(4.340010, 0.011557)
(4.350010, 0.011557)
(4.360010, 0.011557)
(4.370010, 0.011557)
(4.380010, 0.011556)
(4.390010, 0.011556)
(4.400010, 0.011556)
(4.410010, 0.011556)
(4.420010, 0.011556)
(4.430010, 0.011556)
(4.440010, 0.011556)
(4.450010, 0.011556)
(4.460010, 0.011556)
(4.470010, 0.011556)
(4.480010, 0.011556)
(4.490010, 0.011556)
(4.500010, 0.011556)
(4.510010, 0.011556)
(4.520010, 0.011556)
(4.530010, 0.011556)
(4.540010, 0.011556)
(4.550010, 0.011556)
(4.560010, 0.011556)
(4.570010, 0.011556)
(4.580010, 0.011556)
(4.590010, 0.011556)
(4.600010, 0.011556)
(4.610010, 0.011556)
(4.620010, 0.011556)
(4.630010, 0.011556)
(4.640010, 0.011556)
(4.650010, 0.011556)
(4.660010, 0.011556)
(4.670010, 0.011556)
(4.680010, 0.011556)
(4.690010, 0.011556)
(4.700010, 0.011556)
(4.710010, 0.011556)
(4.720010, 0.011556)
(4.730010, 0.011556)
(4.740010, 0.011556)
(4.750010, 0.011556)
(4.760010, 0.011556)
(4.770010, 0.011556)
(4.780010, 0.011556)
(4.790010, 0.011556)
(4.800010, 0.011556)
(4.810010, 0.011556)
(4.820010, 0.011556)
(4.830010, 0.011556)
(4.840010, 0.011556)
(4.850010, 0.011556)
(4.860010, 0.011556)
(4.870010, 0.011556)
(4.880010, 0.011556)
(4.890010, 0.011556)
(4.900010, 0.011556)
(4.910010, 0.011556)
(4.920010, 0.011556)
(4.930010, 0.011556)
(4.940010, 0.011556)
(4.950010, 0.011556)
(4.960010, 0.011556)
(4.970010, 0.011556)
(4.980010, 0.011556)
(4.990010, 0.011556)
};
\label{1e-5}

%% file: figs/data/ha050.tex
\addplot[
line width=1.5pt, color=gray, solid ]
coordinates {
(0.000050, 0.000050)
(0.005050, 0.005050)
(0.010050, 0.010050)
(0.015050, 0.014932)
(0.020050, 0.018404)
(0.025050, 0.019590)
(0.030050, 0.019721)
(0.035050, 0.019631)
(0.040050, 0.019512)
(0.045050, 0.019391)
(0.050050, 0.019274)
(0.055050, 0.019174)
(0.060050, 0.019105)
(0.065050, 0.019071)
(0.070050, 0.019059)
(0.075050, 0.019057)
(0.080050, 0.019059)
(0.085050, 0.019061)
(0.090050, 0.019064)
(0.095050, 0.019066)
(0.100050, 0.019067)
(0.105050, 0.019068)
(0.110050, 0.019069)
(0.115050, 0.019069)
(0.120050, 0.019069)
(0.125050, 0.019069)
(0.130050, 0.019069)
(0.135050, 0.019069)
(0.140050, 0.019069)
(0.145050, 0.019069)
(0.150050, 0.019069)
(0.155050, 0.019069)
(0.160050, 0.019069)
(0.165050, 0.019069)
(0.170050, 0.019069)
(0.175050, 0.019069)
(0.180050, 0.019069)
(0.185050, 0.019069)
(0.190050, 0.019069)
(0.195050, 0.019069)
(0.200050, 0.019069)
(0.205050, 0.019069)
(0.210050, 0.019069)
(0.215050, 0.019069)
(0.220050, 0.019069)
(0.225050, 0.019069)
(0.230050, 0.019069)
(0.235050, 0.019069)
(0.240050, 0.019069)
(0.245050, 0.019069)
(0.250050, 0.019069)
(0.255050, 0.019069)
(0.260050, 0.019069)
(0.265050, 0.019069)
(0.270050, 0.019069)
(0.275050, 0.019069)
(0.280050, 0.019069)
(0.285050, 0.019069)
(0.290050, 0.019069)
(0.295050, 0.019069)
(0.300050, 0.019069)
(0.305050, 0.019069)
(0.310050, 0.019069)
(0.315050, 0.019069)
(0.320050, 0.019069)
(0.325050, 0.019069)
(0.330050, 0.019069)
(0.335050, 0.019069)
(0.340050, 0.019069)
(0.345050, 0.019069)
(0.350050, 0.019069)
(0.355050, 0.019069)
(0.360050, 0.019069)
(0.365050, 0.019069)
(0.370050, 0.019069)
(0.375050, 0.019069)
(0.380050, 0.019069)
(0.385050, 0.019069)
(0.390050, 0.019069)
(0.395050, 0.019069)
(0.400050, 0.019069)
(0.405050, 0.019069)
(0.410050, 0.019069)
(0.415050, 0.019069)
(0.420050, 0.019069)
(0.425050, 0.019069)
(0.430050, 0.019069)
(0.435050, 0.019069)
(0.440050, 0.019069)
(0.445050, 0.019069)
(0.450050, 0.019069)
(0.455050, 0.019069)
(0.460050, 0.019069)
(0.465050, 0.019069)
(0.470050, 0.019069)
(0.475050, 0.019069)
(0.480050, 0.019069)
(0.485050, 0.019069)
(0.490050, 0.019069)
(0.495050, 0.019069)
(0.500050, 0.019069)
(0.505050, 0.019069)
(0.510050, 0.019069)
(0.515050, 0.019069)
(0.520050, 0.019069)
(0.525050, 0.019069)
(0.530050, 0.019069)
(0.535050, 0.019069)
(0.540050, 0.019069)
(0.545050, 0.019069)
(0.550050, 0.019069)
(0.555050, 0.019069)
(0.560050, 0.019069)
(0.565050, 0.019069)
(0.570050, 0.019069)
(0.575050, 0.019069)
(0.580050, 0.019069)
(0.585050, 0.019069)
(0.590050, 0.019069)
(0.595050, 0.019069)
(0.600050, 0.019069)
(0.605050, 0.019069)
(0.610050, 0.019069)
(0.615050, 0.019069)
(0.620050, 0.019069)
(0.625050, 0.019069)
(0.630050, 0.019069)
(0.635050, 0.019069)
(0.640050, 0.019069)
(0.645050, 0.019069)
(0.650050, 0.019069)
(0.655050, 0.019069)
(0.660050, 0.019069)
(0.665050, 0.019069)
(0.670050, 0.019069)
(0.675050, 0.019069)
(0.680050, 0.019069)
(0.685050, 0.019069)
(0.690050, 0.019069)
(0.695050, 0.019069)
(0.700050, 0.019069)
(0.705050, 0.019069)
(0.710050, 0.019069)
(0.715050, 0.019069)
(0.720050, 0.019069)
(0.725050, 0.019069)
(0.730050, 0.019069)
(0.735050, 0.019069)
(0.740050, 0.019069)
(0.745050, 0.019069)
(0.750050, 0.019069)
(0.755050, 0.019069)
(0.760050, 0.019069)
(0.765050, 0.019069)
(0.770050, 0.019069)
(0.775050, 0.019069)
(0.780050, 0.019069)
(0.785050, 0.019069)
(0.790050, 0.019069)
(0.795050, 0.019069)
(0.800050, 0.019069)
(0.805050, 0.019069)
(0.810050, 0.019069)
(0.815050, 0.019069)
(0.820050, 0.019069)
(0.825050, 0.019069)
(0.830050, 0.019069)
(0.835050, 0.019069)
(0.840050, 0.019069)
(0.845050, 0.019069)
(0.850050, 0.019069)
(0.855050, 0.019069)
(0.860050, 0.019069)
(0.865050, 0.019069)
(0.870050, 0.019069)
(0.875050, 0.019069)
(0.880050, 0.019069)
(0.885050, 0.019069)
(0.890050, 0.019069)
(0.895050, 0.019069)
(0.900050, 0.019069)
(0.905050, 0.019069)
(0.910050, 0.019069)
(0.915050, 0.019069)
(0.920050, 0.019069)
(0.925050, 0.019069)
(0.930050, 0.019069)
(0.935050, 0.019069)
(0.940050, 0.019069)
(0.945050, 0.019069)
(0.950050, 0.019069)
(0.955050, 0.019069)
(0.960050, 0.019069)
(0.965050, 0.019069)
(0.970050, 0.019069)
(0.975050, 0.019069)
(0.980050, 0.019069)
(0.985050, 0.019069)
(0.990050, 0.019069)
(0.995050, 0.019069)
(1.000050, 0.019069)
(1.005050, 0.019069)
(1.010050, 0.019069)
(1.015050, 0.019069)
(1.020050, 0.019069)
(1.025050, 0.019069)
(1.030050, 0.019069)
(1.035050, 0.019069)
(1.040050, 0.019069)
(1.045050, 0.019069)
(1.050050, 0.019069)
(1.055050, 0.019069)
(1.060050, 0.019069)
(1.065050, 0.019069)
(1.070050, 0.019069)
(1.075050, 0.019069)
(1.080050, 0.019069)
(1.085050, 0.019069)
(1.090050, 0.019069)
(1.095050, 0.019069)
(1.100050, 0.019069)
(1.105050, 0.019069)
(1.110050, 0.019069)
(1.115050, 0.019069)
(1.120050, 0.019069)
(1.125050, 0.019069)
(1.130050, 0.019069)
(1.135050, 0.019069)
(1.140050, 0.019069)
(1.145050, 0.019069)
(1.150050, 0.019069)
(1.155050, 0.019069)
(1.160050, 0.019069)
(1.165050, 0.019069)
(1.170050, 0.019069)
(1.175050, 0.019069)
(1.180050, 0.019069)
(1.185050, 0.019069)
(1.190050, 0.019069)
(1.195050, 0.019069)
(1.200050, 0.019069)
(1.205050, 0.019069)
(1.210050, 0.019069)
(1.215050, 0.019069)
(1.220050, 0.019069)
(1.225050, 0.019069)
(1.230050, 0.019069)
(1.235050, 0.019069)
(1.240050, 0.019069)
(1.245050, 0.019069)
(1.250050, 0.019069)
(1.255050, 0.019069)
(1.260050, 0.019069)
(1.265050, 0.019069)
(1.270050, 0.019069)
(1.275050, 0.019069)
(1.280050, 0.019069)
(1.285050, 0.019069)
(1.290050, 0.019069)
(1.295050, 0.019069)
(1.300050, 0.019069)
(1.305050, 0.019069)
(1.310050, 0.019069)
(1.315050, 0.019069)
(1.320050, 0.019069)
(1.325050, 0.019069)
(1.330050, 0.019069)
(1.335050, 0.019069)
(1.340050, 0.019069)
(1.345050, 0.019069)
(1.350050, 0.019069)
(1.355050, 0.019069)
(1.360050, 0.019069)
(1.365050, 0.019069)
(1.370050, 0.019069)
(1.375050, 0.019069)
(1.380050, 0.019069)
(1.385050, 0.019069)
(1.390050, 0.019069)
(1.395050, 0.019069)
(1.400050, 0.019069)
(1.405050, 0.019069)
(1.410050, 0.019069)
(1.415050, 0.019069)
(1.420050, 0.019069)
(1.425050, 0.019069)
(1.430050, 0.019069)
(1.435050, 0.019069)
(1.440050, 0.019069)
(1.445050, 0.019069)
(1.450050, 0.019069)
(1.455050, 0.019069)
(1.460050, 0.019069)
(1.465050, 0.019069)
(1.470050, 0.019069)
(1.475050, 0.019069)
(1.480050, 0.019069)
(1.485050, 0.019069)
(1.490050, 0.019069)
(1.495050, 0.019069)
(1.500050, 0.019069)
(1.505050, 0.019069)
(1.510050, 0.019069)
(1.515050, 0.019069)
(1.520050, 0.019069)
(1.525050, 0.019069)
(1.530050, 0.019069)
(1.535050, 0.019069)
(1.540050, 0.019069)
(1.545050, 0.019069)
(1.550050, 0.019069)
(1.555050, 0.019069)
(1.560050, 0.019069)
(1.565050, 0.019069)
(1.570050, 0.019069)
(1.575050, 0.019069)
(1.580050, 0.019069)
(1.585050, 0.019069)
(1.590050, 0.019069)
(1.595050, 0.019069)
(1.600050, 0.019069)
(1.605050, 0.019069)
(1.610050, 0.019069)
(1.615050, 0.019069)
(1.620050, 0.019069)
(1.625050, 0.019069)
(1.630050, 0.019069)
(1.635050, 0.019069)
(1.640050, 0.019069)
(1.645050, 0.019069)
(1.650050, 0.019069)
(1.655050, 0.019069)
(1.660050, 0.019069)
(1.665050, 0.019069)
(1.670050, 0.019069)
(1.675050, 0.019069)
(1.680050, 0.019069)
(1.685050, 0.019069)
(1.690050, 0.019069)
(1.695050, 0.019069)
(1.700050, 0.019069)
(1.705050, 0.019069)
(1.710050, 0.019069)
(1.715050, 0.019069)
(1.720050, 0.019069)
(1.725050, 0.019069)
(1.730050, 0.019069)
(1.735050, 0.019069)
(1.740050, 0.019069)
(1.745050, 0.019069)
(1.750050, 0.019069)
(1.755050, 0.019069)
(1.760050, 0.019069)
(1.765050, 0.019069)
(1.770050, 0.019069)
(1.775050, 0.019069)
(1.780050, 0.019069)
(1.785050, 0.019069)
(1.790050, 0.019069)
(1.795050, 0.019069)
(1.800050, 0.019069)
(1.805050, 0.019069)
(1.810050, 0.019069)
(1.815050, 0.019069)
(1.820050, 0.019069)
(1.825050, 0.019069)
(1.830050, 0.019069)
(1.835050, 0.019069)
(1.840050, 0.019069)
(1.845050, 0.019069)
(1.850050, 0.019069)
(1.855050, 0.019069)
(1.860050, 0.019069)
(1.865050, 0.019069)
(1.870050, 0.019069)
(1.875050, 0.019069)
(1.880050, 0.019069)
(1.885050, 0.019069)
(1.890050, 0.019069)
(1.895050, 0.019069)
(1.900050, 0.019069)
(1.905050, 0.019069)
(1.910050, 0.019069)
(1.915050, 0.019069)
(1.920050, 0.019069)
(1.925050, 0.019069)
(1.930050, 0.019069)
(1.935050, 0.019069)
(1.940050, 0.019069)
(1.945050, 0.019069)
(1.950050, 0.019069)
(1.955050, 0.019069)
(1.960050, 0.019069)
(1.965050, 0.019069)
(1.970050, 0.019069)
(1.975050, 0.019069)
(1.980050, 0.019069)
(1.985050, 0.019069)
(1.990050, 0.019069)
(1.995050, 0.019069)
(2.000050, 0.019069)
(2.005050, 0.019069)
(2.010050, 0.019069)
(2.015050, 0.019069)
(2.020050, 0.019069)
(2.025050, 0.019069)
(2.030050, 0.019069)
(2.035050, 0.019069)
(2.040050, 0.019069)
(2.045050, 0.019069)
(2.050050, 0.019069)
(2.055050, 0.019069)
(2.060050, 0.019069)
(2.065050, 0.019069)
(2.070050, 0.019069)
(2.075050, 0.019069)
(2.080050, 0.019069)
(2.085050, 0.019069)
(2.090050, 0.019069)
(2.095050, 0.019069)
(2.100050, 0.019069)
(2.105050, 0.019069)
(2.110050, 0.019069)
(2.115050, 0.019069)
(2.120050, 0.019069)
(2.125050, 0.019069)
(2.130050, 0.019069)
(2.135050, 0.019069)
(2.140050, 0.019069)
(2.145050, 0.019069)
(2.150050, 0.019069)
(2.155050, 0.019069)
(2.160050, 0.019069)
(2.165050, 0.019069)
(2.170050, 0.019069)
(2.175050, 0.019069)
(2.180050, 0.019069)
(2.185050, 0.019069)
(2.190050, 0.019069)
(2.195050, 0.019069)
(2.200050, 0.019069)
(2.205050, 0.019069)
(2.210050, 0.019069)
(2.215050, 0.019069)
(2.220050, 0.019069)
(2.225050, 0.019069)
(2.230050, 0.019069)
(2.235050, 0.019069)
(2.240050, 0.019069)
(2.245050, 0.019069)
(2.250050, 0.019069)
(2.255050, 0.019069)
(2.260050, 0.019069)
(2.265050, 0.019069)
(2.270050, 0.019069)
(2.275050, 0.019069)
(2.280050, 0.019069)
(2.285050, 0.019069)
(2.290050, 0.019069)
(2.295050, 0.019069)
(2.300050, 0.019069)
(2.305050, 0.019069)
(2.310050, 0.019069)
(2.315050, 0.019069)
(2.320050, 0.019069)
(2.325050, 0.019069)
(2.330050, 0.019069)
(2.335050, 0.019069)
(2.340050, 0.019069)
(2.345050, 0.019069)
(2.350050, 0.019069)
(2.355050, 0.019069)
(2.360050, 0.019069)
(2.365050, 0.019069)
(2.370050, 0.019069)
(2.375050, 0.019069)
(2.380050, 0.019069)
(2.385050, 0.019069)
(2.390050, 0.019069)
(2.395050, 0.019069)
(2.400050, 0.019069)
(2.405050, 0.019069)
(2.410050, 0.019069)
(2.415050, 0.019069)
(2.420050, 0.019069)
(2.425050, 0.019069)
(2.430050, 0.019069)
(2.435050, 0.019069)
(2.440050, 0.019069)
(2.445050, 0.019069)
(2.450050, 0.019069)
(2.455050, 0.019069)
(2.460050, 0.019069)
(2.465050, 0.019069)
(2.470050, 0.019069)
(2.475050, 0.019069)
(2.480050, 0.019069)
(2.485050, 0.019069)
(2.490050, 0.019069)
(2.495050, 0.019069)
(2.500050, 0.019069)
(2.505050, 0.019069)
(2.510050, 0.019069)
(2.515050, 0.019069)
(2.520050, 0.019069)
(2.525050, 0.019069)
(2.530050, 0.019069)
(2.535050, 0.019069)
(2.540050, 0.019069)
(2.545050, 0.019069)
(2.550050, 0.019069)
(2.555050, 0.019069)
(2.560050, 0.019069)
(2.565050, 0.019069)
(2.570050, 0.019069)
(2.575050, 0.019069)
(2.580050, 0.019069)
(2.585050, 0.019069)
(2.590050, 0.019069)
(2.595050, 0.019069)
(2.600050, 0.019069)
(2.605050, 0.019069)
(2.610050, 0.019069)
(2.615050, 0.019069)
(2.620050, 0.019069)
(2.625050, 0.019069)
(2.630050, 0.019069)
(2.635050, 0.019069)
(2.640050, 0.019069)
(2.645050, 0.019069)
(2.650050, 0.019069)
(2.655050, 0.019069)
(2.660050, 0.019069)
(2.665050, 0.019069)
(2.670050, 0.019069)
(2.675050, 0.019069)
(2.680050, 0.019069)
(2.685050, 0.019069)
(2.690050, 0.019069)
(2.695050, 0.019069)
(2.700050, 0.019069)
(2.705050, 0.019069)
(2.710050, 0.019069)
(2.715050, 0.019069)
(2.720050, 0.019069)
(2.725050, 0.019069)
(2.730050, 0.019069)
(2.735050, 0.019069)
(2.740050, 0.019069)
(2.745050, 0.019069)
(2.750050, 0.019069)
(2.755050, 0.019069)
(2.760050, 0.019069)
(2.765050, 0.019069)
(2.770050, 0.019069)
(2.775050, 0.019069)
(2.780050, 0.019069)
(2.785050, 0.019069)
(2.790050, 0.019069)
(2.795050, 0.019069)
(2.800050, 0.019069)
(2.805050, 0.019069)
(2.810050, 0.019069)
(2.815050, 0.019069)
(2.820050, 0.019069)
(2.825050, 0.019069)
(2.830050, 0.019069)
(2.835050, 0.019069)
(2.840050, 0.019069)
(2.845050, 0.019069)
(2.850050, 0.019069)
(2.855050, 0.019069)
(2.860050, 0.019069)
(2.865050, 0.019069)
(2.870050, 0.019069)
(2.875050, 0.019069)
(2.880050, 0.019069)
(2.885050, 0.019069)
(2.890050, 0.019069)
(2.895050, 0.019069)
(2.900050, 0.019069)
(2.905050, 0.019069)
(2.910050, 0.019069)
(2.915050, 0.019069)
(2.920050, 0.019069)
(2.925050, 0.019069)
(2.930050, 0.019069)
(2.935050, 0.019069)
(2.940050, 0.019069)
(2.945050, 0.019069)
(2.950050, 0.019069)
(2.955050, 0.019069)
(2.960050, 0.019069)
(2.965050, 0.019069)
(2.970050, 0.019069)
(2.975050, 0.019069)
(2.980050, 0.019069)
(2.985050, 0.019069)
(2.990050, 0.019069)
(2.995050, 0.019069)
(3.000050, 0.019069)
(3.005050, 0.019069)
(3.010050, 0.019069)
(3.015050, 0.019069)
(3.020050, 0.019069)
(3.025050, 0.019069)
(3.030050, 0.019069)
(3.035050, 0.019069)
(3.040050, 0.019069)
(3.045050, 0.019069)
(3.050050, 0.019069)
(3.055050, 0.019069)
(3.060050, 0.019069)
(3.065050, 0.019069)
(3.070050, 0.019069)
(3.075050, 0.019069)
(3.080050, 0.019069)
(3.085050, 0.019069)
(3.090050, 0.019069)
(3.095050, 0.019069)
(3.100050, 0.019069)
(3.105050, 0.019069)
(3.110050, 0.019069)
(3.115050, 0.019069)
(3.120050, 0.019069)
(3.125050, 0.019069)
(3.130050, 0.019069)
(3.135050, 0.019069)
(3.140050, 0.019069)
(3.145050, 0.019069)
(3.150050, 0.019069)
(3.155050, 0.019069)
(3.160050, 0.019069)
(3.165050, 0.019069)
(3.170050, 0.019069)
(3.175050, 0.019069)
(3.180050, 0.019069)
(3.185050, 0.019069)
(3.190050, 0.019069)
(3.195050, 0.019069)
(3.200050, 0.019069)
(3.205050, 0.019069)
(3.210050, 0.019069)
(3.215050, 0.019069)
(3.220050, 0.019069)
(3.225050, 0.019069)
(3.230050, 0.019069)
(3.235050, 0.019069)
(3.240050, 0.019069)
(3.245050, 0.019069)
(3.250050, 0.019069)
(3.255050, 0.019069)
(3.260050, 0.019069)
(3.265050, 0.019069)
(3.270050, 0.019069)
(3.275050, 0.019069)
(3.280050, 0.019069)
(3.285050, 0.019069)
(3.290050, 0.019069)
(3.295050, 0.019069)
(3.300050, 0.019069)
(3.305050, 0.019069)
(3.310050, 0.019069)
(3.315050, 0.019069)
(3.320050, 0.019069)
(3.325050, 0.019069)
(3.330050, 0.019069)
(3.335050, 0.019069)
(3.340050, 0.019069)
(3.345050, 0.019069)
(3.350050, 0.019069)
(3.355050, 0.019069)
(3.360050, 0.019069)
(3.365050, 0.019069)
(3.370050, 0.019069)
(3.375050, 0.019069)
(3.380050, 0.019069)
(3.385050, 0.019069)
(3.390050, 0.019069)
(3.395050, 0.019069)
(3.400050, 0.019069)
(3.405050, 0.019069)
(3.410050, 0.019069)
(3.415050, 0.019069)
(3.420050, 0.019069)
(3.425050, 0.019069)
(3.430050, 0.019069)
(3.435050, 0.019069)
(3.440050, 0.019069)
(3.445050, 0.019069)
(3.450050, 0.019069)
(3.455050, 0.019069)
(3.460050, 0.019069)
(3.465050, 0.019069)
(3.470050, 0.019069)
(3.475050, 0.019069)
(3.480050, 0.019069)
(3.485050, 0.019069)
(3.490050, 0.019069)
(3.495050, 0.019069)
(3.500050, 0.019069)
(3.505050, 0.019069)
(3.510050, 0.019069)
(3.515050, 0.019069)
(3.520050, 0.019069)
(3.525050, 0.019069)
(3.530050, 0.019069)
(3.535050, 0.019069)
(3.540050, 0.019069)
(3.545050, 0.019069)
(3.550050, 0.019069)
(3.555050, 0.019069)
(3.560050, 0.019069)
(3.565050, 0.019069)
(3.570050, 0.019069)
(3.575050, 0.019069)
(3.580050, 0.019069)
(3.585050, 0.019069)
(3.590050, 0.019069)
(3.595050, 0.019069)
(3.600050, 0.019069)
(3.605050, 0.019069)
(3.610050, 0.019069)
(3.615050, 0.019069)
(3.620050, 0.019069)
(3.625050, 0.019069)
(3.630050, 0.019069)
(3.635050, 0.019069)
(3.640050, 0.019069)
(3.645050, 0.019069)
(3.650050, 0.019069)
(3.655050, 0.019069)
(3.660050, 0.019069)
(3.665050, 0.019069)
(3.670050, 0.019069)
(3.675050, 0.019069)
(3.680050, 0.019069)
(3.685050, 0.019069)
(3.690050, 0.019069)
(3.695050, 0.019069)
(3.700050, 0.019069)
(3.705050, 0.019069)
(3.710050, 0.019069)
(3.715050, 0.019069)
(3.720050, 0.019069)
(3.725050, 0.019069)
(3.730050, 0.019069)
(3.735050, 0.019069)
(3.740050, 0.019069)
(3.745050, 0.019069)
(3.750050, 0.019069)
(3.755050, 0.019069)
(3.760050, 0.019069)
(3.765050, 0.019069)
(3.770050, 0.019069)
(3.775050, 0.019069)
(3.780050, 0.019069)
(3.785050, 0.019069)
(3.790050, 0.019069)
(3.795050, 0.019069)
(3.800050, 0.019069)
(3.805050, 0.019069)
(3.810050, 0.019069)
(3.815050, 0.019069)
(3.820050, 0.019069)
(3.825050, 0.019069)
(3.830050, 0.019069)
(3.835050, 0.019069)
(3.840050, 0.019069)
(3.845050, 0.019069)
(3.850050, 0.019069)
(3.855050, 0.019069)
(3.860050, 0.019069)
(3.865050, 0.019069)
(3.870050, 0.019069)
(3.875050, 0.019069)
(3.880050, 0.019069)
(3.885050, 0.019069)
(3.890050, 0.019069)
(3.895050, 0.019069)
(3.900050, 0.019069)
(3.905050, 0.019069)
(3.910050, 0.019069)
(3.915050, 0.019069)
(3.920050, 0.019069)
(3.925050, 0.019069)
(3.930050, 0.019069)
(3.935050, 0.019069)
(3.940050, 0.019069)
(3.945050, 0.019069)
(3.950050, 0.019069)
(3.955050, 0.019069)
(3.960050, 0.019069)
(3.965050, 0.019069)
(3.970050, 0.019069)
(3.975050, 0.019069)
(3.980050, 0.019069)
(3.985050, 0.019069)
(3.990050, 0.019069)
(3.995050, 0.019069)
(4.000050, 0.019069)
(4.005050, 0.019069)
(4.010050, 0.019069)
(4.015050, 0.019069)
(4.020050, 0.019069)
(4.025050, 0.019069)
(4.030050, 0.019069)
(4.035050, 0.019069)
(4.040050, 0.019069)
(4.045050, 0.019069)
(4.050050, 0.019069)
(4.055050, 0.019069)
(4.060050, 0.019069)
(4.065050, 0.019069)
(4.070050, 0.019069)
(4.075050, 0.019069)
(4.080050, 0.019069)
(4.085050, 0.019069)
(4.090050, 0.019069)
(4.095050, 0.019069)
(4.100050, 0.019069)
(4.105050, 0.019069)
(4.110050, 0.019069)
(4.115050, 0.019069)
(4.120050, 0.019069)
(4.125050, 0.019069)
(4.130050, 0.019069)
(4.135050, 0.019069)
(4.140050, 0.019069)
(4.145050, 0.019069)
(4.150050, 0.019069)
(4.155050, 0.019069)
(4.160050, 0.019069)
(4.165050, 0.019069)
(4.170050, 0.019069)
(4.175050, 0.019069)
(4.180050, 0.019069)
(4.185050, 0.019069)
(4.190050, 0.019069)
(4.195050, 0.019069)
(4.200050, 0.019069)
(4.205050, 0.019069)
(4.210050, 0.019069)
(4.215050, 0.019069)
(4.220050, 0.019069)
(4.225050, 0.019069)
(4.230050, 0.019069)
(4.235050, 0.019069)
(4.240050, 0.019069)
(4.245050, 0.019069)
(4.250050, 0.019069)
(4.255050, 0.019069)
(4.260050, 0.019069)
(4.265050, 0.019069)
(4.270050, 0.019069)
(4.275050, 0.019069)
(4.280050, 0.019069)
(4.285050, 0.019069)
(4.290050, 0.019069)
(4.295050, 0.019069)
(4.300050, 0.019069)
(4.305050, 0.019069)
(4.310050, 0.019069)
(4.315050, 0.019069)
(4.320050, 0.019069)
(4.325050, 0.019069)
(4.330050, 0.019069)
(4.335050, 0.019069)
(4.340050, 0.019069)
(4.345050, 0.019069)
(4.350050, 0.019069)
(4.355050, 0.019069)
(4.360050, 0.019069)
(4.365050, 0.019069)
(4.370050, 0.019069)
(4.375050, 0.019069)
(4.380050, 0.019069)
(4.385050, 0.019069)
(4.390050, 0.019069)
(4.395050, 0.019069)
(4.400050, 0.019069)
(4.405050, 0.019069)
(4.410050, 0.019069)
(4.415050, 0.019069)
(4.420050, 0.019069)
(4.425050, 0.019069)
(4.430050, 0.019069)
(4.435050, 0.019069)
(4.440050, 0.019069)
(4.445050, 0.019069)
(4.450050, 0.019069)
(4.455050, 0.019069)
(4.460050, 0.019069)
(4.465050, 0.019069)
(4.470050, 0.019069)
(4.475050, 0.019069)
(4.480050, 0.019069)
(4.485050, 0.019069)
(4.490050, 0.019069)
(4.495050, 0.019069)
(4.500050, 0.019069)
(4.505050, 0.019069)
(4.510050, 0.019069)
(4.515050, 0.019069)
(4.520050, 0.019069)
(4.525050, 0.019069)
(4.530050, 0.019069)
(4.535050, 0.019069)
(4.540050, 0.019069)
(4.545050, 0.019069)
(4.550050, 0.019069)
(4.555050, 0.019069)
(4.560050, 0.019069)
(4.565050, 0.019069)
(4.570050, 0.019069)
(4.575050, 0.019069)
(4.580050, 0.019069)
(4.585050, 0.019069)
(4.590050, 0.019069)
(4.595050, 0.019069)
(4.600050, 0.019069)
(4.605050, 0.019069)
(4.610050, 0.019069)
(4.615050, 0.019069)
(4.620050, 0.019069)
(4.625050, 0.019069)
(4.630050, 0.019069)
(4.635050, 0.019069)
(4.640050, 0.019069)
(4.645050, 0.019069)
(4.650050, 0.019069)
(4.655050, 0.019069)
(4.660050, 0.019069)
(4.665050, 0.019069)
(4.670050, 0.019069)
(4.675050, 0.019069)
(4.680050, 0.019069)
(4.685050, 0.019069)
(4.690050, 0.019069)
(4.695050, 0.019069)
(4.700050, 0.019069)
(4.705050, 0.019069)
(4.710050, 0.019069)
(4.715050, 0.019069)
(4.720050, 0.019069)
(4.725050, 0.019069)
(4.730050, 0.019069)
(4.735050, 0.019069)
(4.740050, 0.019069)
(4.745050, 0.019069)
(4.750050, 0.019069)
(4.755050, 0.019069)
(4.760050, 0.019069)
(4.765050, 0.019069)
(4.770050, 0.019069)
(4.775050, 0.019069)
(4.780050, 0.019069)
(4.785050, 0.019069)
(4.790050, 0.019069)
(4.795050, 0.019069)
(4.800050, 0.019069)
(4.805050, 0.019069)
(4.810050, 0.019069)
(4.815050, 0.019069)
(4.820050, 0.019069)
(4.825050, 0.019069)
(4.830050, 0.019069)
(4.835050, 0.019069)
(4.840050, 0.019069)
(4.845050, 0.019069)
(4.850050, 0.019069)
(4.855050, 0.019069)
(4.860050, 0.019069)
(4.865050, 0.019069)
(4.870050, 0.019069)
(4.875050, 0.019069)
(4.880050, 0.019069)
(4.885050, 0.019069)
(4.890050, 0.019069)
(4.895050, 0.019069)
(4.900050, 0.019069)
(4.905050, 0.019069)
(4.910050, 0.019069)
(4.915050, 0.019069)
(4.920050, 0.019069)
(4.925050, 0.019069)
(4.930050, 0.019069)
(4.935050, 0.019069)
(4.940050, 0.019069)
(4.945050, 0.019069)
(4.950050, 0.019069)
(4.955050, 0.019069)
(4.960050, 0.019069)
(4.965050, 0.019069)
(4.970050, 0.019069)
(4.975050, 0.019069)
(4.980050, 0.019069)
(4.985050, 0.019069)
(4.990050, 0.019069)
(4.995050, 0.019069)
};
\label{1e2}
\addplot[
line width=1.0pt, color=blue, solid ]
coordinates {
(0.000050, 0.000050)
(0.005050, 0.005050)
(0.010050, 0.010050)
(0.015050, 0.014893)
(0.020050, 0.017682)
(0.025050, 0.016742)
(0.030050, 0.013793)
(0.035050, 0.010324)
(0.040050, 0.006768)
(0.045050, 0.003212)
(0.050050, -0.000226)
(0.055050, -0.003109)
(0.060050, -0.004722)
(0.065050, -0.004703)
(0.070050, -0.003365)
(0.075050, -0.001294)
(0.080050, 0.001066)
(0.085050, 0.003467)
(0.090050, 0.005705)
(0.095050, 0.007505)
(0.100050, 0.008570)
(0.105050, 0.008758)
(0.110050, 0.008157)
(0.115050, 0.007008)
(0.120050, 0.005571)
(0.125050, 0.004057)
(0.130050, 0.002647)
(0.135050, 0.001514)
(0.140050, 0.000804)
(0.145050, 0.000591)
(0.150050, 0.000845)
(0.155050, 0.001459)
(0.160050, 0.002293)
(0.165050, 0.003207)
(0.170050, 0.004074)
(0.175050, 0.004784)
(0.180050, 0.005253)
(0.185050, 0.005437)
(0.190050, 0.005344)
(0.195050, 0.005024)
(0.200050, 0.004553)
(0.205050, 0.004014)
(0.210050, 0.003489)
(0.215050, 0.003048)
(0.220050, 0.002742)
(0.225050, 0.002600)
(0.230050, 0.002622)
(0.235050, 0.002785)
(0.240050, 0.003047)
(0.245050, 0.003359)
(0.250050, 0.003673)
(0.255050, 0.003944)
(0.260050, 0.004141)
(0.265050, 0.004244)
(0.270050, 0.004251)
(0.275050, 0.004171)
(0.280050, 0.004028)
(0.285050, 0.003849)
(0.290050, 0.003663)
(0.295050, 0.003497)
(0.300050, 0.003372)
(0.305050, 0.003300)
(0.310050, 0.003285)
(0.315050, 0.003321)
(0.320050, 0.003399)
(0.325050, 0.003501)
(0.330050, 0.003610)
(0.335050, 0.003710)
(0.340050, 0.003789)
(0.345050, 0.003838)
(0.350050, 0.003854)
(0.355050, 0.003838)
(0.360050, 0.003797)
(0.365050, 0.003740)
(0.370050, 0.003676)
(0.375050, 0.003615)
(0.380050, 0.003566)
(0.385050, 0.003534)
(0.390050, 0.003521)
(0.395050, 0.003526)
(0.400050, 0.003548)
(0.405050, 0.003580)
(0.410050, 0.003617)
(0.415050, 0.003653)
(0.420050, 0.003683)
(0.425050, 0.003704)
(0.430050, 0.003714)
(0.435050, 0.003713)
(0.440050, 0.003702)
(0.445050, 0.003685)
(0.450050, 0.003664)
(0.455050, 0.003642)
(0.460050, 0.003623)
(0.465050, 0.003610)
(0.470050, 0.003603)
(0.475050, 0.003602)
(0.480050, 0.003607)
(0.485050, 0.003617)
(0.490050, 0.003629)
(0.495050, 0.003642)
(0.500050, 0.003653)
(0.505050, 0.003662)
(0.510050, 0.003667)
(0.515050, 0.003668)
(0.520050, 0.003665)
(0.525050, 0.003660)
(0.530050, 0.003653)
(0.535050, 0.003646)
(0.540050, 0.003639)
(0.545050, 0.003634)
(0.550050, 0.003630)
(0.555050, 0.003629)
(0.560050, 0.003630)
(0.565050, 0.003633)
(0.570050, 0.003637)
(0.575050, 0.003641)
(0.580050, 0.003645)
(0.585050, 0.003648)
(0.590050, 0.003651)
(0.595050, 0.003652)
(0.600050, 0.003651)
(0.605050, 0.003650)
(0.610050, 0.003648)
(0.615050, 0.003645)
(0.620050, 0.003643)
(0.625050, 0.003641)
(0.630050, 0.003639)
(0.635050, 0.003639)
(0.640050, 0.003639)
(0.645050, 0.003639)
(0.650050, 0.003640)
(0.655050, 0.003642)
(0.660050, 0.003643)
(0.665050, 0.003645)
(0.670050, 0.003646)
(0.675050, 0.003646)
(0.680050, 0.003646)
(0.685050, 0.003646)
(0.690050, 0.003645)
(0.695050, 0.003644)
(0.700050, 0.003643)
(0.705050, 0.003643)
(0.710050, 0.003642)
(0.715050, 0.003642)
(0.720050, 0.003642)
(0.725050, 0.003642)
(0.730050, 0.003642)
(0.735050, 0.003643)
(0.740050, 0.003643)
(0.745050, 0.003644)
(0.750050, 0.003644)
(0.755050, 0.003644)
(0.760050, 0.003644)
(0.765050, 0.003644)
(0.770050, 0.003644)
(0.775050, 0.003644)
(0.780050, 0.003644)
(0.785050, 0.003643)
(0.790050, 0.003643)
(0.795050, 0.003643)
(0.800050, 0.003643)
(0.805050, 0.003643)
(0.810050, 0.003643)
(0.815050, 0.003643)
(0.820050, 0.003643)
(0.825050, 0.003643)
(0.830050, 0.003644)
(0.835050, 0.003644)
(0.840050, 0.003644)
(0.845050, 0.003644)
(0.850050, 0.003644)
(0.855050, 0.003644)
(0.860050, 0.003643)
(0.865050, 0.003643)
(0.870050, 0.003643)
(0.875050, 0.003643)
(0.880050, 0.003643)
(0.885050, 0.003643)
(0.890050, 0.003643)
(0.895050, 0.003643)
(0.900050, 0.003643)
(0.905050, 0.003643)
(0.910050, 0.003643)
(0.915050, 0.003643)
(0.920050, 0.003643)
(0.925050, 0.003643)
(0.930050, 0.003643)
(0.935050, 0.003643)
(0.940050, 0.003643)
(0.945050, 0.003643)
(0.950050, 0.003643)
(0.955050, 0.003643)
(0.960050, 0.003643)
(0.965050, 0.003643)
(0.970050, 0.003643)
(0.975050, 0.003643)
(0.980050, 0.003643)
(0.985050, 0.003643)
(0.990050, 0.003643)
(0.995050, 0.003643)
(1.000050, 0.003643)
(1.005050, 0.003643)
(1.010050, 0.003643)
(1.015050, 0.003643)
(1.020050, 0.003643)
(1.025050, 0.003643)
(1.030050, 0.003643)
(1.035050, 0.003643)
(1.040050, 0.003643)
(1.045050, 0.003643)
(1.050050, 0.003643)
(1.055050, 0.003643)
(1.060050, 0.003643)
(1.065050, 0.003643)
(1.070050, 0.003643)
(1.075050, 0.003643)
(1.080050, 0.003643)
(1.085050, 0.003643)
(1.090050, 0.003643)
(1.095050, 0.003643)
(1.100050, 0.003643)
(1.105050, 0.003643)
(1.110050, 0.003643)
(1.115050, 0.003643)
(1.120050, 0.003643)
(1.125050, 0.003643)
(1.130050, 0.003643)
(1.135050, 0.003643)
(1.140050, 0.003643)
(1.145050, 0.003643)
(1.150050, 0.003643)
(1.155050, 0.003643)
(1.160050, 0.003643)
(1.165050, 0.003643)
(1.170050, 0.003643)
(1.175050, 0.003643)
(1.180050, 0.003643)
(1.185050, 0.003643)
(1.190050, 0.003643)
(1.195050, 0.003643)
(1.200050, 0.003643)
(1.205050, 0.003643)
(1.210050, 0.003643)
(1.215050, 0.003643)
(1.220050, 0.003643)
(1.225050, 0.003643)
(1.230050, 0.003643)
(1.235050, 0.003643)
(1.240050, 0.003643)
(1.245050, 0.003643)
(1.250050, 0.003643)
(1.255050, 0.003643)
(1.260050, 0.003643)
(1.265050, 0.003643)
(1.270050, 0.003643)
(1.275050, 0.003643)
(1.280050, 0.003643)
(1.285050, 0.003643)
(1.290050, 0.003643)
(1.295050, 0.003643)
(1.300050, 0.003643)
(1.305050, 0.003643)
(1.310050, 0.003643)
(1.315050, 0.003643)
(1.320050, 0.003643)
(1.325050, 0.003643)
(1.330050, 0.003643)
(1.335050, 0.003643)
(1.340050, 0.003643)
(1.345050, 0.003643)
(1.350050, 0.003643)
(1.355050, 0.003643)
(1.360050, 0.003643)
(1.365050, 0.003643)
(1.370050, 0.003643)
(1.375050, 0.003643)
(1.380050, 0.003643)
(1.385050, 0.003643)
(1.390050, 0.003643)
(1.395050, 0.003643)
(1.400050, 0.003643)
(1.405050, 0.003643)
(1.410050, 0.003643)
(1.415050, 0.003643)
(1.420050, 0.003643)
(1.425050, 0.003643)
(1.430050, 0.003643)
(1.435050, 0.003643)
(1.440050, 0.003643)
(1.445050, 0.003643)
(1.450050, 0.003643)
(1.455050, 0.003643)
(1.460050, 0.003643)
(1.465050, 0.003643)
(1.470050, 0.003643)
(1.475050, 0.003643)
(1.480050, 0.003643)
(1.485050, 0.003643)
(1.490050, 0.003643)
(1.495050, 0.003643)
(1.500050, 0.003643)
(1.505050, 0.003643)
(1.510050, 0.003643)
(1.515050, 0.003643)
(1.520050, 0.003643)
(1.525050, 0.003643)
(1.530050, 0.003643)
(1.535050, 0.003643)
(1.540050, 0.003643)
(1.545050, 0.003643)
(1.550050, 0.003643)
(1.555050, 0.003643)
(1.560050, 0.003643)
(1.565050, 0.003643)
(1.570050, 0.003643)
(1.575050, 0.003643)
(1.580050, 0.003643)
(1.585050, 0.003643)
(1.590050, 0.003643)
(1.595050, 0.003643)
(1.600050, 0.003643)
(1.605050, 0.003643)
(1.610050, 0.003643)
(1.615050, 0.003643)
(1.620050, 0.003643)
(1.625050, 0.003643)
(1.630050, 0.003643)
(1.635050, 0.003643)
(1.640050, 0.003643)
(1.645050, 0.003643)
(1.650050, 0.003643)
(1.655050, 0.003643)
(1.660050, 0.003643)
(1.665050, 0.003643)
(1.670050, 0.003643)
(1.675050, 0.003643)
(1.680050, 0.003643)
(1.685050, 0.003643)
(1.690050, 0.003643)
(1.695050, 0.003643)
(1.700050, 0.003643)
(1.705050, 0.003643)
(1.710050, 0.003643)
(1.715050, 0.003643)
(1.720050, 0.003643)
(1.725050, 0.003643)
(1.730050, 0.003643)
(1.735050, 0.003643)
(1.740050, 0.003643)
(1.745050, 0.003643)
(1.750050, 0.003643)
(1.755050, 0.003643)
(1.760050, 0.003643)
(1.765050, 0.003643)
(1.770050, 0.003643)
(1.775050, 0.003643)
(1.780050, 0.003643)
(1.785050, 0.003643)
(1.790050, 0.003643)
(1.795050, 0.003643)
(1.800050, 0.003643)
(1.805050, 0.003643)
(1.810050, 0.003643)
(1.815050, 0.003643)
(1.820050, 0.003643)
(1.825050, 0.003643)
(1.830050, 0.003643)
(1.835050, 0.003643)
(1.840050, 0.003643)
(1.845050, 0.003643)
(1.850050, 0.003643)
(1.855050, 0.003643)
(1.860050, 0.003643)
(1.865050, 0.003643)
(1.870050, 0.003643)
(1.875050, 0.003643)
(1.880050, 0.003643)
(1.885050, 0.003643)
(1.890050, 0.003643)
(1.895050, 0.003643)
(1.900050, 0.003643)
(1.905050, 0.003643)
(1.910050, 0.003643)
(1.915050, 0.003643)
(1.920050, 0.003643)
(1.925050, 0.003643)
(1.930050, 0.003643)
(1.935050, 0.003643)
(1.940050, 0.003643)
(1.945050, 0.003643)
(1.950050, 0.003643)
(1.955050, 0.003643)
(1.960050, 0.003643)
(1.965050, 0.003643)
(1.970050, 0.003643)
(1.975050, 0.003643)
(1.980050, 0.003643)
(1.985050, 0.003643)
(1.990050, 0.003643)
(1.995050, 0.003643)
};
\label{1e0}
\addplot[
line width=1.0pt, color=red, solid ]
coordinates {
(0.000020, 0.000020)
(0.002020, 0.002020)
(0.004020, 0.004020)
(0.006020, 0.006020)
(0.008020, 0.008020)
(0.010020, 0.010020)
(0.012020, 0.012013)
(0.014020, 0.013944)
(0.016020, 0.015630)
(0.018020, 0.016783)
(0.020020, 0.017181)
(0.022020, 0.016797)
(0.024020, 0.015771)
(0.026020, 0.014304)
(0.028020, 0.012573)
(0.030020, 0.010700)
(0.032020, 0.008757)
(0.034020, 0.006782)
(0.036020, 0.004793)
(0.038020, 0.002798)
(0.040020, 0.000802)
(0.042020, -0.001193)
(0.044020, -0.003181)
(0.046020, -0.005151)
(0.048020, -0.007076)
(0.050020, -0.008909)
(0.052020, -0.010581)
(0.054020, -0.012004)
(0.056020, -0.013092)
(0.058020, -0.013770)
(0.060020, -0.013997)
(0.062020, -0.013770)
(0.064020, -0.013120)
(0.066020, -0.012107)
(0.068020, -0.010799)
(0.070020, -0.009269)
(0.072020, -0.007581)
(0.074020, -0.005785)
(0.076020, -0.003922)
(0.078020, -0.002022)
(0.080020, -0.000109)
(0.082020, 0.001797)
(0.084020, 0.003671)
(0.086020, 0.005483)
(0.088020, 0.007196)
(0.090020, 0.008764)
(0.092020, 0.010138)
(0.094020, 0.011266)
(0.096020, 0.012105)
(0.098020, 0.012621)
(0.100020, 0.012796)
(0.102020, 0.012631)
(0.104020, 0.012141)
(0.106020, 0.011356)
(0.108020, 0.010313)
(0.110020, 0.009058)
(0.112020, 0.007633)
(0.114020, 0.006080)
(0.116020, 0.004438)
(0.118020, 0.002741)
(0.120020, 0.001024)
(0.122020, -0.000684)
(0.124020, -0.002349)
(0.126020, -0.003939)
(0.128020, -0.005416)
(0.130020, -0.006746)
(0.132020, -0.007893)
(0.134020, -0.008823)
(0.136020, -0.009510)
(0.138020, -0.009936)
(0.140020, -0.010090)
(0.142020, -0.009972)
(0.144020, -0.009593)
(0.146020, -0.008971)
(0.148020, -0.008131)
(0.150020, -0.007102)
(0.152020, -0.005916)
(0.154020, -0.004607)
(0.156020, -0.003208)
(0.158020, -0.001752)
(0.160020, -0.000273)
(0.162020, 0.001198)
(0.164020, 0.002628)
(0.166020, 0.003987)
(0.168020, 0.005242)
(0.170020, 0.006364)
(0.172020, 0.007326)
(0.174020, 0.008105)
(0.176020, 0.008681)
(0.178020, 0.009043)
(0.180020, 0.009183)
(0.182020, 0.009102)
(0.184020, 0.008807)
(0.186020, 0.008310)
(0.188020, 0.007629)
(0.190020, 0.006786)
(0.192020, 0.005806)
(0.194020, 0.004716)
(0.196020, 0.003544)
(0.198020, 0.002319)
(0.200020, 0.001071)
(0.202020, -0.000172)
(0.204020, -0.001380)
(0.206020, -0.002525)
(0.208020, -0.003582)
(0.210020, -0.004525)
(0.212020, -0.005334)
(0.214020, -0.005989)
(0.216020, -0.006477)
(0.218020, -0.006787)
(0.220020, -0.006916)
(0.222020, -0.006862)
(0.224020, -0.006631)
(0.226020, -0.006232)
(0.228020, -0.005679)
(0.230020, -0.004989)
(0.232020, -0.004182)
(0.234020, -0.003280)
(0.236020, -0.002305)
(0.238020, -0.001284)
(0.240020, -0.000241)
(0.242020, 0.000799)
(0.244020, 0.001810)
(0.246020, 0.002770)
(0.248020, 0.003656)
(0.250020, 0.004447)
(0.252020, 0.005126)
(0.254020, 0.005678)
(0.256020, 0.006092)
(0.258020, 0.006359)
(0.260020, 0.006476)
(0.262020, 0.006442)
(0.264020, 0.006261)
(0.266020, 0.005940)
(0.268020, 0.005491)
(0.270020, 0.004926)
(0.272020, 0.004262)
(0.274020, 0.003517)
(0.276020, 0.002710)
(0.278020, 0.001862)
(0.280020, 0.000995)
(0.282020, 0.000129)
(0.284020, -0.000715)
(0.286020, -0.001517)
(0.288020, -0.002257)
(0.290020, -0.002920)
(0.292020, -0.003490)
(0.294020, -0.003956)
(0.296020, -0.004306)
(0.298020, -0.004536)
(0.300020, -0.004641)
(0.302020, -0.004621)
(0.304020, -0.004480)
(0.306020, -0.004222)
(0.308020, -0.003856)
(0.310020, -0.003394)
(0.312020, -0.002849)
(0.314020, -0.002234)
(0.316020, -0.001567)
(0.318020, -0.000865)
(0.320020, -0.000145)
(0.322020, 0.000575)
(0.324020, 0.001277)
(0.326020, 0.001946)
(0.328020, 0.002564)
(0.330020, 0.003119)
(0.332020, 0.003597)
(0.334020, 0.003989)
(0.336020, 0.004286)
(0.338020, 0.004483)
(0.340020, 0.004577)
(0.342020, 0.004568)
(0.344020, 0.004457)
(0.346020, 0.004250)
(0.348020, 0.003953)
(0.350020, 0.003574)
(0.352020, 0.003126)
(0.354020, 0.002620)
(0.356020, 0.002069)
(0.358020, 0.001488)
(0.360020, 0.000891)
(0.362020, 0.000293)
(0.364020, -0.000291)
(0.366020, -0.000847)
(0.368020, -0.001364)
(0.370020, -0.001827)
(0.372020, -0.002228)
(0.374020, -0.002558)
(0.376020, -0.002810)
(0.378020, -0.002979)
(0.380020, -0.003062)
(0.382020, -0.003060)
(0.384020, -0.002974)
(0.386020, -0.002807)
(0.388020, -0.002566)
(0.390020, -0.002257)
(0.392020, -0.001889)
(0.394020, -0.001472)
(0.396020, -0.001017)
(0.398020, -0.000536)
(0.400020, -0.000042)
(0.402020, 0.000454)
(0.404020, 0.000940)
(0.406020, 0.001403)
(0.408020, 0.001834)
(0.410020, 0.002221)
(0.412020, 0.002557)
(0.414020, 0.002835)
(0.416020, 0.003048)
(0.418020, 0.003193)
(0.420020, 0.003267)
(0.422020, 0.003270)
(0.424020, 0.003203)
(0.426020, 0.003069)
(0.428020, 0.002872)
(0.430020, 0.002620)
(0.432020, 0.002317)
(0.434020, 0.001974)
(0.436020, 0.001599)
(0.438020, 0.001202)
(0.440020, 0.000792)
(0.442020, 0.000380)
(0.444020, -0.000023)
(0.446020, -0.000409)
(0.448020, -0.000768)
(0.450020, -0.001092)
(0.452020, -0.001373)
(0.454020, -0.001607)
(0.456020, -0.001787)
(0.458020, -0.001911)
(0.460020, -0.001976)
(0.462020, -0.001982)
(0.464020, -0.001931)
(0.466020, -0.001823)
(0.468020, -0.001664)
(0.470020, -0.001457)
(0.472020, -0.001209)
(0.474020, -0.000926)
(0.476020, -0.000617)
(0.478020, -0.000288)
(0.480020, 0.000051)
(0.482020, 0.000392)
(0.484020, 0.000727)
(0.486020, 0.001048)
(0.488020, 0.001348)
(0.490020, 0.001618)
(0.492020, 0.001854)
(0.494020, 0.002050)
(0.496020, 0.002203)
(0.498020, 0.002309)
(0.500020, 0.002366)
(0.502020, 0.002374)
(0.504020, 0.002335)
(0.506020, 0.002248)
(0.508020, 0.002119)
(0.510020, 0.001950)
(0.512020, 0.001746)
(0.514020, 0.001514)
(0.516020, 0.001259)
(0.518020, 0.000987)
(0.520020, 0.000706)
(0.522020, 0.000423)
(0.524020, 0.000145)
(0.526020, -0.000122)
(0.528020, -0.000372)
(0.530020, -0.000598)
(0.532020, -0.000795)
(0.534020, -0.000960)
(0.536020, -0.001089)
(0.538020, -0.001179)
(0.540020, -0.001230)
(0.542020, -0.001239)
(0.544020, -0.001209)
(0.546020, -0.001140)
(0.548020, -0.001035)
(0.550020, -0.000897)
(0.552020, -0.000730)
(0.554020, -0.000538)
(0.556020, -0.000328)
(0.558020, -0.000103)
(0.560020, 0.000129)
(0.562020, 0.000364)
(0.564020, 0.000595)
(0.566020, 0.000817)
(0.568020, 0.001025)
(0.570020, 0.001214)
(0.572020, 0.001379)
(0.574020, 0.001518)
(0.576020, 0.001627)
(0.578020, 0.001704)
(0.580020, 0.001747)
(0.582020, 0.001758)
(0.584020, 0.001734)
(0.586020, 0.001679)
(0.588020, 0.001594)
(0.590020, 0.001481)
(0.592020, 0.001344)
(0.594020, 0.001187)
(0.596020, 0.001013)
(0.598020, 0.000828)
(0.600020, 0.000635)
(0.602020, 0.000441)
(0.604020, 0.000249)
(0.606020, 0.000064)
(0.608020, -0.000109)
(0.610020, -0.000267)
(0.612020, -0.000405)
(0.614020, -0.000522)
(0.616020, -0.000614)
(0.618020, -0.000679)
(0.620020, -0.000717)
(0.622020, -0.000728)
(0.624020, -0.000710)
(0.626020, -0.000666)
(0.628020, -0.000597)
(0.630020, -0.000505)
(0.632020, -0.000393)
(0.634020, -0.000263)
(0.636020, -0.000120)
(0.638020, 0.000033)
(0.640020, 0.000192)
(0.642020, 0.000354)
(0.644020, 0.000513)
(0.646020, 0.000667)
(0.648020, 0.000811)
(0.650020, 0.000943)
(0.652020, 0.001059)
(0.654020, 0.001157)
(0.656020, 0.001234)
(0.658020, 0.001290)
(0.660020, 0.001323)
(0.662020, 0.001333)
(0.664020, 0.001320)
(0.666020, 0.001285)
(0.668020, 0.001229)
(0.670020, 0.001154)
(0.672020, 0.001062)
(0.674020, 0.000955)
(0.676020, 0.000837)
(0.678020, 0.000711)
(0.680020, 0.000579)
(0.682020, 0.000445)
(0.684020, 0.000313)
(0.686020, 0.000185)
(0.688020, 0.000065)
(0.690020, -0.000045)
(0.692020, -0.000142)
(0.694020, -0.000224)
(0.696020, -0.000290)
(0.698020, -0.000337)
(0.700020, -0.000366)
(0.702020, -0.000375)
(0.704020, -0.000366)
(0.706020, -0.000338)
(0.708020, -0.000293)
(0.710020, -0.000231)
(0.712020, -0.000156)
(0.714020, -0.000068)
(0.716020, 0.000029)
(0.718020, 0.000133)
(0.720020, 0.000242)
(0.722020, 0.000353)
(0.724020, 0.000463)
(0.726020, 0.000569)
(0.728020, 0.000670)
(0.730020, 0.000761)
(0.732020, 0.000843)
(0.734020, 0.000911)
(0.736020, 0.000967)
(0.738020, 0.001007)
(0.740020, 0.001032)
(0.742020, 0.001041)
(0.744020, 0.001034)
(0.746020, 0.001012)
(0.748020, 0.000975)
(0.750020, 0.000925)
(0.752020, 0.000863)
(0.754020, 0.000791)
(0.756020, 0.000711)
(0.758020, 0.000625)
(0.760020, 0.000534)
(0.762020, 0.000443)
(0.764020, 0.000351)
(0.766020, 0.000263)
(0.768020, 0.000180)
(0.770020, 0.000103)
(0.772020, 0.000035)
(0.774020, -0.000023)
(0.776020, -0.000069)
(0.778020, -0.000103)
(0.780020, -0.000125)
(0.782020, -0.000133)
(0.784020, -0.000128)
(0.786020, -0.000111)
(0.788020, -0.000081)
(0.790020, -0.000040)
(0.792020, 0.000010)
(0.794020, 0.000069)
(0.796020, 0.000136)
(0.798020, 0.000207)
(0.800020, 0.000281)
(0.802020, 0.000358)
(0.804020, 0.000433)
(0.806020, 0.000507)
(0.808020, 0.000576)
(0.810020, 0.000640)
(0.812020, 0.000697)
(0.814020, 0.000745)
(0.816020, 0.000785)
(0.818020, 0.000814)
(0.820020, 0.000832)
(0.822020, 0.000840)
(0.824020, 0.000836)
(0.826020, 0.000823)
(0.828020, 0.000799)
(0.830020, 0.000765)
(0.832020, 0.000724)
(0.834020, 0.000675)
(0.836020, 0.000621)
(0.838020, 0.000562)
(0.840020, 0.000500)
(0.842020, 0.000437)
(0.844020, 0.000374)
(0.846020, 0.000313)
(0.848020, 0.000255)
(0.850020, 0.000202)
(0.852020, 0.000155)
(0.854020, 0.000114)
(0.856020, 0.000081)
(0.858020, 0.000056)
(0.860020, 0.000040)
(0.862020, 0.000033)
(0.864020, 0.000036)
(0.866020, 0.000047)
(0.868020, 0.000066)
(0.870020, 0.000093)
(0.872020, 0.000127)
(0.874020, 0.000167)
(0.876020, 0.000212)
(0.878020, 0.000260)
(0.880020, 0.000311)
(0.882020, 0.000364)
(0.884020, 0.000416)
(0.886020, 0.000466)
(0.888020, 0.000515)
(0.890020, 0.000559)
(0.892020, 0.000599)
(0.894020, 0.000633)
(0.896020, 0.000661)
(0.898020, 0.000682)
(0.900020, 0.000695)
(0.902020, 0.000702)
(0.904020, 0.000700)
(0.906020, 0.000692)
(0.908020, 0.000676)
(0.910020, 0.000654)
(0.912020, 0.000626)
(0.914020, 0.000593)
(0.916020, 0.000556)
(0.918020, 0.000516)
(0.920020, 0.000474)
(0.922020, 0.000431)
(0.924020, 0.000388)
(0.926020, 0.000345)
(0.928020, 0.000305)
(0.930020, 0.000268)
(0.932020, 0.000235)
(0.934020, 0.000206)
(0.936020, 0.000183)
(0.938020, 0.000165)
(0.940020, 0.000154)
(0.942020, 0.000148)
(0.944020, 0.000149)
(0.946020, 0.000155)
(0.948020, 0.000168)
(0.950020, 0.000186)
(0.952020, 0.000209)
(0.954020, 0.000236)
(0.956020, 0.000266)
(0.958020, 0.000299)
(0.960020, 0.000334)
(0.962020, 0.000370)
};
\label{1e-5}

%% file: figs/data/ha100.tex
\addplot[
line width=1.0pt, color=gray, solid ]
coordinates {
(0.000050, 0.000050)
(0.005050, 0.005050)
(0.010050, 0.009434)
(0.015050, 0.009760)
(0.020050, 0.009523)
(0.025050, 0.009286)
(0.030050, 0.009101)
(0.035050, 0.009070)
(0.040050, 0.009080)
(0.045050, 0.009091)
(0.050050, 0.009099)
(0.055050, 0.009101)
(0.060050, 0.009101)
(0.065050, 0.009100)
(0.070050, 0.009100)
(0.075050, 0.009100)
(0.080050, 0.009100)
(0.085050, 0.009100)
(0.090050, 0.009100)
(0.095050, 0.009100)
(0.100050, 0.009100)
(0.105050, 0.009100)
(0.110050, 0.009100)
(0.115050, 0.009100)
(0.120050, 0.009100)
(0.125050, 0.009100)
(0.130050, 0.009100)
(0.135050, 0.009100)
(0.140050, 0.009100)
(0.145050, 0.009100)
(0.150050, 0.009100)
(0.155050, 0.009100)
(0.160050, 0.009100)
(0.165050, 0.009100)
(0.170050, 0.009100)
(0.175050, 0.009100)
(0.180050, 0.009100)
(0.185050, 0.009100)
(0.190050, 0.009100)
(0.195050, 0.009100)
(0.200050, 0.009100)
(0.205050, 0.009100)
(0.210050, 0.009100)
(0.215050, 0.009100)
(0.220050, 0.009100)
(0.225050, 0.009100)
(0.230050, 0.009100)
(0.235050, 0.009100)
(0.240050, 0.009100)
(0.245050, 0.009100)
(0.250050, 0.009100)
(0.255050, 0.009100)
(0.260050, 0.009100)
(0.265050, 0.009100)
(0.270050, 0.009100)
(0.275050, 0.009100)
(0.280050, 0.009100)
(0.285050, 0.009100)
(0.290050, 0.009100)
(0.295050, 0.009100)
(0.300050, 0.009100)
(0.305050, 0.009100)
(0.310050, 0.009100)
(0.315050, 0.009100)
(0.320050, 0.009100)
(0.325050, 0.009100)
(0.330050, 0.009100)
(0.335050, 0.009100)
(0.340050, 0.009100)
(0.345050, 0.009100)
(0.350050, 0.009100)
(0.355050, 0.009100)
(0.360050, 0.009100)
(0.365050, 0.009100)
(0.370050, 0.009100)
(0.375050, 0.009100)
(0.380050, 0.009100)
(0.385050, 0.009100)
(0.390050, 0.009100)
(0.395050, 0.009100)
(0.400050, 0.009100)
(0.405050, 0.009100)
(0.410050, 0.009100)
(0.415050, 0.009100)
(0.420050, 0.009100)
(0.425050, 0.009100)
(0.430050, 0.009100)
(0.435050, 0.009100)
(0.440050, 0.009100)
(0.445050, 0.009100)
(0.450050, 0.009100)
(0.455050, 0.009100)
(0.460050, 0.009100)
(0.465050, 0.009100)
(0.470050, 0.009100)
(0.475050, 0.009100)
(0.480050, 0.009100)
(0.485050, 0.009100)
(0.490050, 0.009100)
(0.495050, 0.009100)
(0.500050, 0.009100)
(0.505050, 0.009100)
(0.510050, 0.009100)
(0.515050, 0.009100)
(0.520050, 0.009100)
(0.525050, 0.009100)
(0.530050, 0.009100)
(0.535050, 0.009100)
(0.540050, 0.009100)
(0.545050, 0.009100)
(0.550050, 0.009100)
(0.555050, 0.009100)
(0.560050, 0.009100)
(0.565050, 0.009100)
(0.570050, 0.009100)
(0.575050, 0.009100)
(0.580050, 0.009100)
(0.585050, 0.009100)
(0.590050, 0.009100)
(0.595050, 0.009100)
(0.600050, 0.009100)
(0.605050, 0.009100)
(0.610050, 0.009100)
(0.615050, 0.009100)
(0.620050, 0.009100)
(0.625050, 0.009100)
(0.630050, 0.009100)
(0.635050, 0.009100)
(0.640050, 0.009100)
(0.645050, 0.009100)
(0.650050, 0.009100)
(0.655050, 0.009100)
(0.660050, 0.009100)
(0.665050, 0.009100)
(0.670050, 0.009100)
(0.675050, 0.009100)
(0.680050, 0.009100)
(0.685050, 0.009100)
(0.690050, 0.009100)
(0.695050, 0.009100)
(0.700050, 0.009100)
(0.705050, 0.009100)
(0.710050, 0.009100)
(0.715050, 0.009100)
(0.720050, 0.009100)
(0.725050, 0.009100)
(0.730050, 0.009100)
(0.735050, 0.009100)
(0.740050, 0.009100)
(0.745050, 0.009100)
(0.750050, 0.009100)
(0.755050, 0.009100)
(0.760050, 0.009100)
(0.765050, 0.009100)
(0.770050, 0.009100)
(0.775050, 0.009100)
(0.780050, 0.009100)
(0.785050, 0.009100)
(0.790050, 0.009100)
(0.795050, 0.009100)
(0.800050, 0.009100)
(0.805050, 0.009100)
(0.810050, 0.009100)
(0.815050, 0.009100)
(0.820050, 0.009100)
(0.825050, 0.009100)
(0.830050, 0.009100)
(0.835050, 0.009100)
(0.840050, 0.009100)
(0.845050, 0.009100)
(0.850050, 0.009100)
(0.855050, 0.009100)
(0.860050, 0.009100)
(0.865050, 0.009100)
(0.870050, 0.009100)
(0.875050, 0.009100)
(0.880050, 0.009100)
(0.885050, 0.009100)
(0.890050, 0.009100)
(0.895050, 0.009100)
(0.900050, 0.009100)
(0.905050, 0.009100)
(0.910050, 0.009100)
(0.915050, 0.009100)
(0.920050, 0.009100)
(0.925050, 0.009100)
(0.930050, 0.009100)
(0.935050, 0.009100)
(0.940050, 0.009100)
(0.945050, 0.009100)
(0.950050, 0.009100)
(0.955050, 0.009100)
(0.960050, 0.009100)
(0.965050, 0.009100)
(0.970050, 0.009100)
(0.975050, 0.009100)
(0.980050, 0.009100)
(0.985050, 0.009100)
(0.990050, 0.009100)
(0.995050, 0.009100)
(1.000050, 0.009100)
(1.005050, 0.009100)
(1.010050, 0.009100)
(1.015050, 0.009100)
(1.020050, 0.009100)
(1.025050, 0.009100)
(1.030050, 0.009100)
(1.035050, 0.009100)
(1.040050, 0.009100)
(1.045050, 0.009100)
(1.050050, 0.009100)
(1.055050, 0.009100)
(1.060050, 0.009100)
(1.065050, 0.009100)
(1.070050, 0.009100)
(1.075050, 0.009100)
(1.080050, 0.009100)
(1.085050, 0.009100)
(1.090050, 0.009100)
(1.095050, 0.009100)
(1.100050, 0.009100)
(1.105050, 0.009100)
(1.110050, 0.009100)
(1.115050, 0.009100)
(1.120050, 0.009100)
(1.125050, 0.009100)
(1.130050, 0.009100)
(1.135050, 0.009100)
(1.140050, 0.009100)
(1.145050, 0.009100)
(1.150050, 0.009100)
(1.155050, 0.009100)
(1.160050, 0.009100)
(1.165050, 0.009100)
(1.170050, 0.009100)
(1.175050, 0.009100)
(1.180050, 0.009100)
(1.185050, 0.009100)
(1.190050, 0.009100)
(1.195050, 0.009100)
(1.200050, 0.009100)
(1.205050, 0.009100)
(1.210050, 0.009100)
(1.215050, 0.009100)
(1.220050, 0.009100)
(1.225050, 0.009100)
(1.230050, 0.009100)
(1.235050, 0.009100)
(1.240050, 0.009100)
(1.245050, 0.009100)
(1.250050, 0.009100)
(1.255050, 0.009100)
(1.260050, 0.009100)
(1.265050, 0.009100)
(1.270050, 0.009100)
(1.275050, 0.009100)
(1.280050, 0.009100)
(1.285050, 0.009100)
(1.290050, 0.009100)
(1.295050, 0.009100)
(1.300050, 0.009100)
(1.305050, 0.009100)
(1.310050, 0.009100)
(1.315050, 0.009100)
(1.320050, 0.009100)
(1.325050, 0.009100)
(1.330050, 0.009100)
(1.335050, 0.009100)
(1.340050, 0.009100)
(1.345050, 0.009100)
(1.350050, 0.009100)
(1.355050, 0.009100)
(1.360050, 0.009100)
(1.365050, 0.009100)
(1.370050, 0.009100)
(1.375050, 0.009100)
(1.380050, 0.009100)
(1.385050, 0.009100)
(1.390050, 0.009100)
(1.395050, 0.009100)
(1.400050, 0.009100)
(1.405050, 0.009100)
(1.410050, 0.009100)
(1.415050, 0.009100)
(1.420050, 0.009100)
(1.425050, 0.009100)
(1.430050, 0.009100)
(1.435050, 0.009100)
(1.440050, 0.009100)
(1.445050, 0.009100)
(1.450050, 0.009100)
(1.455050, 0.009100)
(1.460050, 0.009100)
(1.465050, 0.009100)
(1.470050, 0.009100)
(1.475050, 0.009100)
(1.480050, 0.009100)
(1.485050, 0.009100)
(1.490050, 0.009100)
(1.495050, 0.009100)
(1.500050, 0.009100)
(1.505050, 0.009100)
(1.510050, 0.009100)
(1.515050, 0.009100)
(1.520050, 0.009100)
(1.525050, 0.009100)
(1.530050, 0.009100)
(1.535050, 0.009100)
(1.540050, 0.009100)
(1.545050, 0.009100)
(1.550050, 0.009100)
(1.555050, 0.009100)
(1.560050, 0.009100)
(1.565050, 0.009100)
(1.570050, 0.009100)
(1.575050, 0.009100)
(1.580050, 0.009100)
(1.585050, 0.009100)
(1.590050, 0.009100)
(1.595050, 0.009100)
(1.600050, 0.009100)
(1.605050, 0.009100)
(1.610050, 0.009100)
(1.615050, 0.009100)
(1.620050, 0.009100)
(1.625050, 0.009100)
(1.630050, 0.009100)
(1.635050, 0.009100)
(1.640050, 0.009100)
(1.645050, 0.009100)
(1.650050, 0.009100)
(1.655050, 0.009100)
(1.660050, 0.009100)
(1.665050, 0.009100)
(1.670050, 0.009100)
(1.675050, 0.009100)
(1.680050, 0.009100)
(1.685050, 0.009100)
(1.690050, 0.009100)
(1.695050, 0.009100)
(1.700050, 0.009100)
(1.705050, 0.009100)
(1.710050, 0.009100)
(1.715050, 0.009100)
(1.720050, 0.009100)
(1.725050, 0.009100)
(1.730050, 0.009100)
(1.735050, 0.009100)
(1.740050, 0.009100)
(1.745050, 0.009100)
(1.750050, 0.009100)
(1.755050, 0.009100)
(1.760050, 0.009100)
(1.765050, 0.009100)
(1.770050, 0.009100)
(1.775050, 0.009100)
(1.780050, 0.009100)
(1.785050, 0.009100)
(1.790050, 0.009100)
(1.795050, 0.009100)
(1.800050, 0.009100)
(1.805050, 0.009100)
(1.810050, 0.009100)
(1.815050, 0.009100)
(1.820050, 0.009100)
(1.825050, 0.009100)
(1.830050, 0.009100)
(1.835050, 0.009100)
(1.840050, 0.009100)
(1.845050, 0.009100)
(1.850050, 0.009100)
(1.855050, 0.009100)
(1.860050, 0.009100)
(1.865050, 0.009100)
(1.870050, 0.009100)
(1.875050, 0.009100)
(1.880050, 0.009100)
(1.885050, 0.009100)
(1.890050, 0.009100)
(1.895050, 0.009100)
(1.900050, 0.009100)
(1.905050, 0.009100)
(1.910050, 0.009100)
(1.915050, 0.009100)
(1.920050, 0.009100)
(1.925050, 0.009100)
(1.930050, 0.009100)
(1.935050, 0.009100)
(1.940050, 0.009100)
(1.945050, 0.009100)
(1.950050, 0.009100)
(1.955050, 0.009100)
(1.960050, 0.009100)
(1.965050, 0.009100)
(1.970050, 0.009100)
(1.975050, 0.009100)
(1.980050, 0.009100)
(1.985050, 0.009100)
(1.990050, 0.009100)
(1.995050, 0.009100)
(2.000050, 0.009100)
(2.005050, 0.009100)
(2.010050, 0.009100)
(2.015050, 0.009100)
(2.020050, 0.009100)
(2.025050, 0.009100)
(2.030050, 0.009100)
(2.035050, 0.009100)
(2.040050, 0.009100)
(2.045050, 0.009100)
(2.050050, 0.009100)
(2.055050, 0.009100)
(2.060050, 0.009100)
(2.065050, 0.009100)
(2.070050, 0.009100)
(2.075050, 0.009100)
(2.080050, 0.009100)
(2.085050, 0.009100)
(2.090050, 0.009100)
(2.095050, 0.009100)
(2.100050, 0.009100)
(2.105050, 0.009100)
(2.110050, 0.009100)
(2.115050, 0.009100)
(2.120050, 0.009100)
(2.125050, 0.009100)
(2.130050, 0.009100)
(2.135050, 0.009100)
(2.140050, 0.009100)
(2.145050, 0.009100)
(2.150050, 0.009100)
(2.155050, 0.009100)
(2.160050, 0.009100)
(2.165050, 0.009100)
(2.170050, 0.009100)
(2.175050, 0.009100)
(2.180050, 0.009100)
(2.185050, 0.009100)
(2.190050, 0.009100)
(2.195050, 0.009100)
(2.200050, 0.009100)
(2.205050, 0.009100)
(2.210050, 0.009100)
(2.215050, 0.009100)
(2.220050, 0.009100)
(2.225050, 0.009100)
(2.230050, 0.009100)
(2.235050, 0.009100)
(2.240050, 0.009100)
(2.245050, 0.009100)
(2.250050, 0.009100)
(2.255050, 0.009100)
(2.260050, 0.009100)
(2.265050, 0.009100)
(2.270050, 0.009100)
(2.275050, 0.009100)
(2.280050, 0.009100)
(2.285050, 0.009100)
(2.290050, 0.009100)
(2.295050, 0.009100)
(2.300050, 0.009100)
(2.305050, 0.009100)
(2.310050, 0.009100)
(2.315050, 0.009100)
(2.320050, 0.009100)
(2.325050, 0.009100)
(2.330050, 0.009100)
(2.335050, 0.009100)
(2.340050, 0.009100)
(2.345050, 0.009100)
(2.350050, 0.009100)
(2.355050, 0.009100)
(2.360050, 0.009100)
(2.365050, 0.009100)
(2.370050, 0.009100)
(2.375050, 0.009100)
(2.380050, 0.009100)
(2.385050, 0.009100)
(2.390050, 0.009100)
(2.395050, 0.009100)
(2.400050, 0.009100)
(2.405050, 0.009100)
(2.410050, 0.009100)
(2.415050, 0.009100)
(2.420050, 0.009100)
(2.425050, 0.009100)
(2.430050, 0.009100)
(2.435050, 0.009100)
(2.440050, 0.009100)
(2.445050, 0.009100)
(2.450050, 0.009100)
(2.455050, 0.009100)
(2.460050, 0.009100)
(2.465050, 0.009100)
(2.470050, 0.009100)
(2.475050, 0.009100)
(2.480050, 0.009100)
(2.485050, 0.009100)
(2.490050, 0.009100)
(2.495050, 0.009100)
(2.500050, 0.009100)
(2.505050, 0.009100)
(2.510050, 0.009100)
(2.515050, 0.009100)
(2.520050, 0.009100)
(2.525050, 0.009100)
(2.530050, 0.009100)
(2.535050, 0.009100)
(2.540050, 0.009100)
(2.545050, 0.009100)
(2.550050, 0.009100)
(2.555050, 0.009100)
(2.560050, 0.009100)
(2.565050, 0.009100)
(2.570050, 0.009100)
(2.575050, 0.009100)
(2.580050, 0.009100)
(2.585050, 0.009100)
(2.590050, 0.009100)
(2.595050, 0.009100)
(2.600050, 0.009100)
(2.605050, 0.009100)
(2.610050, 0.009100)
(2.615050, 0.009100)
(2.620050, 0.009100)
(2.625050, 0.009100)
(2.630050, 0.009100)
(2.635050, 0.009100)
(2.640050, 0.009100)
(2.645050, 0.009100)
(2.650050, 0.009100)
(2.655050, 0.009100)
(2.660050, 0.009100)
(2.665050, 0.009100)
(2.670050, 0.009100)
(2.675050, 0.009100)
(2.680050, 0.009100)
(2.685050, 0.009100)
(2.690050, 0.009100)
(2.695050, 0.009100)
(2.700050, 0.009100)
(2.705050, 0.009100)
(2.710050, 0.009100)
(2.715050, 0.009100)
(2.720050, 0.009100)
(2.725050, 0.009100)
(2.730050, 0.009100)
(2.735050, 0.009100)
(2.740050, 0.009100)
(2.745050, 0.009100)
(2.750050, 0.009100)
(2.755050, 0.009100)
(2.760050, 0.009100)
(2.765050, 0.009100)
(2.770050, 0.009100)
(2.775050, 0.009100)
(2.780050, 0.009100)
(2.785050, 0.009100)
(2.790050, 0.009100)
(2.795050, 0.009100)
(2.800050, 0.009100)
(2.805050, 0.009100)
(2.810050, 0.009100)
(2.815050, 0.009100)
(2.820050, 0.009100)
(2.825050, 0.009100)
(2.830050, 0.009100)
(2.835050, 0.009100)
(2.840050, 0.009100)
(2.845050, 0.009100)
(2.850050, 0.009100)
(2.855050, 0.009100)
(2.860050, 0.009100)
(2.865050, 0.009100)
(2.870050, 0.009100)
(2.875050, 0.009100)
(2.880050, 0.009100)
(2.885050, 0.009100)
(2.890050, 0.009100)
(2.895050, 0.009100)
(2.900050, 0.009100)
(2.905050, 0.009100)
(2.910050, 0.009100)
(2.915050, 0.009100)
(2.920050, 0.009100)
(2.925050, 0.009100)
(2.930050, 0.009100)
(2.935050, 0.009100)
(2.940050, 0.009100)
(2.945050, 0.009100)
(2.950050, 0.009100)
(2.955050, 0.009100)
(2.960050, 0.009100)
(2.965050, 0.009100)
(2.970050, 0.009100)
(2.975050, 0.009100)
(2.980050, 0.009100)
(2.985050, 0.009100)
(2.990050, 0.009100)
(2.995050, 0.009100)
(3.000050, 0.009100)
(3.005050, 0.009100)
(3.010050, 0.009100)
(3.015050, 0.009100)
(3.020050, 0.009100)
(3.025050, 0.009100)
(3.030050, 0.009100)
(3.035050, 0.009100)
(3.040050, 0.009100)
(3.045050, 0.009100)
(3.050050, 0.009100)
(3.055050, 0.009100)
(3.060050, 0.009100)
(3.065050, 0.009100)
(3.070050, 0.009100)
(3.075050, 0.009100)
(3.080050, 0.009100)
(3.085050, 0.009100)
(3.090050, 0.009100)
(3.095050, 0.009100)
(3.100050, 0.009100)
(3.105050, 0.009100)
(3.110050, 0.009100)
(3.115050, 0.009100)
(3.120050, 0.009100)
(3.125050, 0.009100)
(3.130050, 0.009100)
(3.135050, 0.009100)
(3.140050, 0.009100)
(3.145050, 0.009100)
(3.150050, 0.009100)
(3.155050, 0.009100)
(3.160050, 0.009100)
(3.165050, 0.009100)
(3.170050, 0.009100)
(3.175050, 0.009100)
(3.180050, 0.009100)
(3.185050, 0.009100)
(3.190050, 0.009100)
(3.195050, 0.009100)
(3.200050, 0.009100)
(3.205050, 0.009100)
(3.210050, 0.009100)
(3.215050, 0.009100)
(3.220050, 0.009100)
(3.225050, 0.009100)
(3.230050, 0.009100)
(3.235050, 0.009100)
(3.240050, 0.009100)
(3.245050, 0.009100)
(3.250050, 0.009100)
(3.255050, 0.009100)
(3.260050, 0.009100)
(3.265050, 0.009100)
(3.270050, 0.009100)
(3.275050, 0.009100)
(3.280050, 0.009100)
(3.285050, 0.009100)
(3.290050, 0.009100)
(3.295050, 0.009100)
(3.300050, 0.009100)
(3.305050, 0.009100)
(3.310050, 0.009100)
(3.315050, 0.009100)
(3.320050, 0.009100)
(3.325050, 0.009100)
(3.330050, 0.009100)
(3.335050, 0.009100)
(3.340050, 0.009100)
(3.345050, 0.009100)
(3.350050, 0.009100)
(3.355050, 0.009100)
(3.360050, 0.009100)
(3.365050, 0.009100)
(3.370050, 0.009100)
(3.375050, 0.009100)
(3.380050, 0.009100)
(3.385050, 0.009100)
(3.390050, 0.009100)
(3.395050, 0.009100)
(3.400050, 0.009100)
(3.405050, 0.009100)
(3.410050, 0.009100)
(3.415050, 0.009100)
(3.420050, 0.009100)
(3.425050, 0.009100)
(3.430050, 0.009100)
(3.435050, 0.009100)
(3.440050, 0.009100)
(3.445050, 0.009100)
(3.450050, 0.009100)
(3.455050, 0.009100)
(3.460050, 0.009100)
(3.465050, 0.009100)
(3.470050, 0.009100)
(3.475050, 0.009100)
(3.480050, 0.009100)
(3.485050, 0.009100)
(3.490050, 0.009100)
(3.495050, 0.009100)
(3.500050, 0.009100)
(3.505050, 0.009100)
(3.510050, 0.009100)
(3.515050, 0.009100)
(3.520050, 0.009100)
(3.525050, 0.009100)
(3.530050, 0.009100)
(3.535050, 0.009100)
(3.540050, 0.009100)
(3.545050, 0.009100)
(3.550050, 0.009100)
(3.555050, 0.009100)
(3.560050, 0.009100)
(3.565050, 0.009100)
(3.570050, 0.009100)
(3.575050, 0.009100)
(3.580050, 0.009100)
(3.585050, 0.009100)
(3.590050, 0.009100)
(3.595050, 0.009100)
(3.600050, 0.009100)
(3.605050, 0.009100)
(3.610050, 0.009100)
(3.615050, 0.009100)
(3.620050, 0.009100)
(3.625050, 0.009100)
(3.630050, 0.009100)
(3.635050, 0.009100)
(3.640050, 0.009100)
(3.645050, 0.009100)
(3.650050, 0.009100)
(3.655050, 0.009100)
(3.660050, 0.009100)
(3.665050, 0.009100)
(3.670050, 0.009100)
(3.675050, 0.009100)
(3.680050, 0.009100)
(3.685050, 0.009100)
(3.690050, 0.009100)
(3.695050, 0.009100)
(3.700050, 0.009100)
(3.705050, 0.009100)
(3.710050, 0.009100)
(3.715050, 0.009100)
(3.720050, 0.009100)
(3.725050, 0.009100)
(3.730050, 0.009100)
(3.735050, 0.009100)
(3.740050, 0.009100)
(3.745050, 0.009100)
(3.750050, 0.009100)
(3.755050, 0.009100)
(3.760050, 0.009100)
(3.765050, 0.009100)
(3.770050, 0.009100)
(3.775050, 0.009100)
(3.780050, 0.009100)
(3.785050, 0.009100)
(3.790050, 0.009100)
(3.795050, 0.009100)
(3.800050, 0.009100)
(3.805050, 0.009100)
(3.810050, 0.009100)
(3.815050, 0.009100)
(3.820050, 0.009100)
(3.825050, 0.009100)
(3.830050, 0.009100)
(3.835050, 0.009100)
(3.840050, 0.009100)
(3.845050, 0.009100)
(3.850050, 0.009100)
(3.855050, 0.009100)
(3.860050, 0.009100)
(3.865050, 0.009100)
(3.870050, 0.009100)
(3.875050, 0.009100)
(3.880050, 0.009100)
(3.885050, 0.009100)
(3.890050, 0.009100)
(3.895050, 0.009100)
(3.900050, 0.009100)
(3.905050, 0.009100)
(3.910050, 0.009100)
(3.915050, 0.009100)
(3.920050, 0.009100)
(3.925050, 0.009100)
(3.930050, 0.009100)
(3.935050, 0.009100)
(3.940050, 0.009100)
(3.945050, 0.009100)
(3.950050, 0.009100)
(3.955050, 0.009100)
(3.960050, 0.009100)
(3.965050, 0.009100)
(3.970050, 0.009100)
(3.975050, 0.009100)
(3.980050, 0.009100)
(3.985050, 0.009100)
(3.990050, 0.009100)
(3.995050, 0.009100)
(4.000050, 0.009100)
(4.005050, 0.009100)
(4.010050, 0.009100)
(4.015050, 0.009100)
(4.020050, 0.009100)
(4.025050, 0.009100)
(4.030050, 0.009100)
(4.035050, 0.009100)
(4.040050, 0.009100)
(4.045050, 0.009100)
(4.050050, 0.009100)
(4.055050, 0.009100)
(4.060050, 0.009100)
(4.065050, 0.009100)
(4.070050, 0.009100)
(4.075050, 0.009100)
(4.080050, 0.009100)
(4.085050, 0.009100)
(4.090050, 0.009100)
(4.095050, 0.009100)
(4.100050, 0.009100)
(4.105050, 0.009100)
(4.110050, 0.009100)
(4.115050, 0.009100)
(4.120050, 0.009100)
(4.125050, 0.009100)
(4.130050, 0.009100)
(4.135050, 0.009100)
(4.140050, 0.009100)
(4.145050, 0.009100)
(4.150050, 0.009100)
(4.155050, 0.009100)
(4.160050, 0.009100)
(4.165050, 0.009100)
(4.170050, 0.009100)
(4.175050, 0.009100)
(4.180050, 0.009100)
(4.185050, 0.009100)
(4.190050, 0.009100)
(4.195050, 0.009100)
(4.200050, 0.009100)
(4.205050, 0.009100)
(4.210050, 0.009100)
(4.215050, 0.009100)
(4.220050, 0.009100)
(4.225050, 0.009100)
(4.230050, 0.009100)
(4.235050, 0.009100)
(4.240050, 0.009100)
(4.245050, 0.009100)
(4.250050, 0.009100)
(4.255050, 0.009100)
(4.260050, 0.009100)
(4.265050, 0.009100)
(4.270050, 0.009100)
(4.275050, 0.009100)
(4.280050, 0.009100)
(4.285050, 0.009100)
(4.290050, 0.009100)
(4.295050, 0.009100)
(4.300050, 0.009100)
(4.305050, 0.009100)
(4.310050, 0.009100)
(4.315050, 0.009100)
(4.320050, 0.009100)
(4.325050, 0.009100)
(4.330050, 0.009100)
(4.335050, 0.009100)
(4.340050, 0.009100)
(4.345050, 0.009100)
(4.350050, 0.009100)
(4.355050, 0.009100)
(4.360050, 0.009100)
(4.365050, 0.009100)
(4.370050, 0.009100)
(4.375050, 0.009100)
(4.380050, 0.009100)
(4.385050, 0.009100)
(4.390050, 0.009100)
(4.395050, 0.009100)
(4.400050, 0.009100)
(4.405050, 0.009100)
(4.410050, 0.009100)
(4.415050, 0.009100)
(4.420050, 0.009100)
(4.425050, 0.009100)
(4.430050, 0.009100)
(4.435050, 0.009100)
(4.440050, 0.009100)
(4.445050, 0.009100)
(4.450050, 0.009100)
(4.455050, 0.009100)
(4.460050, 0.009100)
(4.465050, 0.009100)
(4.470050, 0.009100)
(4.475050, 0.009100)
(4.480050, 0.009100)
(4.485050, 0.009100)
(4.490050, 0.009100)
(4.495050, 0.009100)
(4.500050, 0.009100)
(4.505050, 0.009100)
(4.510050, 0.009100)
(4.515050, 0.009100)
(4.520050, 0.009100)
(4.525050, 0.009100)
(4.530050, 0.009100)
(4.535050, 0.009100)
(4.540050, 0.009100)
(4.545050, 0.009100)
(4.550050, 0.009100)
(4.555050, 0.009100)
(4.560050, 0.009100)
(4.565050, 0.009100)
(4.570050, 0.009100)
(4.575050, 0.009100)
(4.580050, 0.009100)
(4.585050, 0.009100)
(4.590050, 0.009100)
(4.595050, 0.009100)
(4.600050, 0.009100)
(4.605050, 0.009100)
(4.610050, 0.009100)
(4.615050, 0.009100)
(4.620050, 0.009100)
(4.625050, 0.009100)
(4.630050, 0.009100)
(4.635050, 0.009100)
(4.640050, 0.009100)
(4.645050, 0.009100)
(4.650050, 0.009100)
(4.655050, 0.009100)
(4.660050, 0.009100)
(4.665050, 0.009100)
(4.670050, 0.009100)
(4.675050, 0.009100)
(4.680050, 0.009100)
(4.685050, 0.009100)
(4.690050, 0.009100)
(4.695050, 0.009100)
(4.700050, 0.009100)
(4.705050, 0.009100)
(4.710050, 0.009100)
(4.715050, 0.009100)
(4.720050, 0.009100)
(4.725050, 0.009100)
(4.730050, 0.009100)
(4.735050, 0.009100)
(4.740050, 0.009100)
(4.745050, 0.009100)
(4.750050, 0.009100)
(4.755050, 0.009100)
(4.760050, 0.009100)
(4.765050, 0.009100)
(4.770050, 0.009100)
(4.775050, 0.009100)
(4.780050, 0.009100)
(4.785050, 0.009100)
(4.790050, 0.009100)
(4.795050, 0.009100)
(4.800050, 0.009100)
(4.805050, 0.009100)
(4.810050, 0.009100)
(4.815050, 0.009100)
(4.820050, 0.009100)
(4.825050, 0.009100)
(4.830050, 0.009100)
(4.835050, 0.009100)
(4.840050, 0.009100)
(4.845050, 0.009100)
(4.850050, 0.009100)
(4.855050, 0.009100)
(4.860050, 0.009100)
(4.865050, 0.009100)
(4.870050, 0.009100)
(4.875050, 0.009100)
(4.880050, 0.009100)
(4.885050, 0.009100)
(4.890050, 0.009100)
(4.895050, 0.009100)
(4.900050, 0.009100)
(4.905050, 0.009100)
(4.910050, 0.009100)
(4.915050, 0.009100)
(4.920050, 0.009100)
(4.925050, 0.009100)
(4.930050, 0.009100)
(4.935050, 0.009100)
(4.940050, 0.009100)
(4.945050, 0.009100)
(4.950050, 0.009100)
(4.955050, 0.009100)
(4.960050, 0.009100)
(4.965050, 0.009100)
(4.970050, 0.009100)
(4.975050, 0.009100)
(4.980050, 0.009100)
(4.985050, 0.009100)
(4.990050, 0.009100)
(4.995050, 0.009100)
};
\label{1e2}
\addplot[
line width=1.0pt, color=blue, solid ]
coordinates {
(0.000050, 0.000050)
(0.015050, 0.006365)
(0.030050, -0.004989)
(0.045050, 0.003041)
(0.060050, 0.002426)
(0.075050, -0.002174)
(0.090050, 0.003438)
(0.105050, 0.000551)
(0.120050, -0.000099)
(0.135050, 0.002514)
(0.150050, 0.000122)
(0.165050, 0.000946)
(0.180050, 0.001636)
(0.195050, 0.000356)
(0.210050, 0.001272)
(0.225050, 0.001134)
(0.240050, 0.000682)
(0.255050, 0.001255)
(0.270050, 0.000939)
(0.285050, 0.000901)
(0.300050, 0.001148)
(0.315050, 0.000910)
(0.330050, 0.001003)
(0.345050, 0.001059)
(0.360050, 0.000938)
(0.375050, 0.001030)
(0.390050, 0.001011)
(0.405050, 0.000971)
(0.420050, 0.001026)
(0.435050, 0.000994)
(0.450050, 0.000993)
(0.465050, 0.001015)
(0.480050, 0.000992)
(0.495050, 0.001002)
(0.510050, 0.001006)
(0.525050, 0.000995)
(0.540050, 0.001004)
(0.555050, 0.001002)
(0.570050, 0.000998)
(0.585050, 0.001004)
(0.600050, 0.001000)
(0.615050, 0.001000)
(0.630050, 0.001002)
(0.645050, 0.001000)
(0.660050, 0.001001)
(0.675050, 0.001002)
(0.690050, 0.001000)
(0.705050, 0.001001)
(0.720050, 0.001001)
(0.735050, 0.001001)
(0.750050, 0.001001)
(0.765050, 0.001001)
(0.780050, 0.001001)
(0.795050, 0.001001)
(0.810050, 0.001001)
(0.825050, 0.001001)
(0.840050, 0.001001)
(0.855050, 0.001001)
(0.870050, 0.001001)
(0.885050, 0.001001)
(0.900050, 0.001001)
(0.915050, 0.001001)
(0.930050, 0.001001)
(0.945050, 0.001001)
(0.960050, 0.001001)
(0.975050, 0.001001)
(0.990050, 0.001001)
(1.005050, 0.001001)
(1.020050, 0.001001)
(1.035050, 0.001001)
(1.050050, 0.001001)
(1.065050, 0.001001)
(1.080050, 0.001001)
(1.095050, 0.001001)
(1.110050, 0.001001)
(1.125050, 0.001001)
(1.140050, 0.001001)
(1.155050, 0.001001)
(1.170050, 0.001001)
(1.185050, 0.001001)
(1.200050, 0.001001)
(1.215050, 0.001001)
(1.230050, 0.001001)
(1.245050, 0.001001)
(1.260050, 0.001001)
(1.275050, 0.001001)
(1.290050, 0.001001)
(1.305050, 0.001001)
(1.320050, 0.001001)
(1.335050, 0.001001)
(1.350050, 0.001001)
(1.365050, 0.001001)
(1.380050, 0.001001)
(1.395050, 0.001001)
(1.410050, 0.001001)
(1.425050, 0.001001)
(1.440050, 0.001001)
(1.455050, 0.001001)
(1.470050, 0.001001)
(1.485050, 0.001001)
(1.500050, 0.001001)
(1.515050, 0.001001)
(1.530050, 0.001001)
(1.545050, 0.001001)
(1.560050, 0.001001)
(1.575050, 0.001001)
(1.590050, 0.001001)
(1.605050, 0.001001)
(1.620050, 0.001001)
(1.635050, 0.001001)
(1.650050, 0.001001)
(1.665050, 0.001001)
(1.680050, 0.001001)
(1.695050, 0.001001)
(1.710050, 0.001001)
(1.725050, 0.001001)
(1.740050, 0.001001)
(1.755050, 0.001001)
(1.770050, 0.001001)
(1.785050, 0.001001)
(1.800050, 0.001001)
(1.815050, 0.001001)
(1.830050, 0.001001)
(1.845050, 0.001001)
(1.860050, 0.001001)
(1.875050, 0.001001)
(1.890050, 0.001001)
(1.905050, 0.001001)
(1.920050, 0.001001)
(1.935050, 0.001001)
(1.950050, 0.001001)
(1.965050, 0.001001)
(1.980050, 0.001001)
(1.995050, 0.001001)
(2.010050, 0.001001)
(2.025050, 0.001001)
(2.040050, 0.001001)
(2.055050, 0.001001)
(2.070050, 0.001001)
(2.085050, 0.001001)
(2.100050, 0.001001)
(2.115050, 0.001001)
(2.130050, 0.001001)
(2.145050, 0.001001)
(2.160050, 0.001001)
(2.175050, 0.001001)
(2.190050, 0.001001)
(2.205050, 0.001001)
(2.220050, 0.001001)
(2.235050, 0.001001)
(2.250050, 0.001001)
(2.265050, 0.001001)
(2.280050, 0.001001)
(2.295050, 0.001001)
(2.310050, 0.001001)
(2.325050, 0.001001)
(2.340050, 0.001001)
(2.355050, 0.001001)
(2.370050, 0.001001)
(2.385050, 0.001001)
(2.400050, 0.001001)
(2.415050, 0.001001)
(2.430050, 0.001001)
(2.445050, 0.001001)
(2.460050, 0.001001)
(2.475050, 0.001001)
(2.490050, 0.001001)
};
\label{1e0}
\addplot[
line width=1.0pt, color=red, solid ]
coordinates {
(0.000020, 0.000020)
(0.006020, 0.006020)
(0.012020, 0.008024)
(0.018020, 0.002187)
(0.024020, -0.003808)
(0.030020, -0.007940)
(0.036020, -0.003962)
(0.042020, 0.002000)
(0.048020, 0.007070)
(0.054020, 0.005806)
(0.060020, 0.000213)
(0.066020, -0.005269)
(0.072020, -0.006428)
(0.078020, -0.001940)
(0.084020, 0.003631)
(0.090020, 0.006525)
(0.096020, 0.003767)
(0.102020, -0.001529)
(0.108020, -0.005490)
(0.114020, -0.004649)
(0.120020, -0.000089)
(0.126020, 0.004400)
(0.132020, 0.005275)
(0.138020, 0.001876)
(0.144020, -0.002683)
(0.150020, -0.004876)
(0.156020, -0.002884)
(0.162020, 0.001328)
(0.168020, 0.004388)
(0.174020, 0.003825)
(0.180020, 0.000321)
(0.186020, -0.003172)
(0.192020, -0.003872)
(0.198020, -0.001335)
(0.204020, 0.002193)
(0.210020, 0.003878)
(0.216020, 0.002430)
(0.222020, -0.000789)
(0.228020, -0.003132)
(0.234020, -0.002755)
(0.240020, -0.000116)
(0.246020, 0.002552)
(0.252020, 0.003115)
(0.258020, 0.001231)
(0.264020, -0.001449)
(0.270020, -0.002747)
(0.276020, -0.001691)
(0.282020, 0.000737)
(0.288020, 0.002524)
(0.294020, 0.002272)
(0.300020, 0.000297)
(0.306020, -0.001726)
(0.312020, -0.002177)
(0.318020, -0.000780)
(0.324020, 0.001243)
(0.330020, 0.002242)
(0.336020, 0.001471)
(0.342020, -0.000353)
(0.348020, -0.001714)
(0.354020, -0.001548)
(0.360020, -0.000072)
(0.366020, 0.001459)
(0.372020, 0.001819)
(0.378020, 0.000784)
(0.384020, -0.000741)
(0.390020, -0.001508)
(0.396020, -0.000946)
(0.402020, 0.000422)
(0.408020, 0.001457)
(0.414020, 0.001349)
(0.420020, 0.000248)
(0.426020, -0.000910)
(0.432020, -0.001196)
(0.438020, -0.000429)
(0.444020, 0.000719)
(0.450020, 0.001308)
(0.456020, 0.000898)
(0.462020, -0.000127)
(0.468020, -0.000914)
(0.474020, -0.000846)
(0.480020, -0.000024)
(0.486020, 0.000851)
(0.492020, 0.001077)
(0.498020, 0.000509)
(0.504020, -0.000354)
(0.510020, -0.000807)
(0.516020, -0.000508)
(0.522020, 0.000261)
(0.528020, 0.000858)
(0.534020, 0.000817)
(0.540020, 0.000204)
(0.546020, -0.000457)
(0.552020, -0.000636)
(0.558020, -0.000216)
(0.564020, 0.000434)
(0.570020, 0.000781)
(0.576020, 0.000564)
(0.582020, -0.000012)
(0.588020, -0.000466)
(0.594020, -0.000442)
(0.600020, 0.000015)
(0.606020, 0.000515)
(0.612020, 0.000655)
(0.618020, 0.000344)
(0.624020, -0.000145)
(0.630020, -0.000411)
(0.636020, -0.000253)
(0.642020, 0.000179)
(0.648020, 0.000523)
(0.654020, 0.000511)
(0.660020, 0.000170)
(0.666020, -0.000207)
(0.672020, -0.000318)
(0.678020, -0.000088)
(0.684020, 0.000280)
(0.690020, 0.000484)
(0.696020, 0.000369)
(0.702020, 0.000046)
(0.708020, -0.000216)
(0.714020, -0.000210)
(0.720020, 0.000043)
(0.726020, 0.000328)
(0.732020, 0.000415)
(0.738020, 0.000245)
(0.744020, -0.000031)
(0.750020, -0.000187)
(0.756020, -0.000104)
(0.762020, 0.000138)
(0.768020, 0.000336)
(0.774020, 0.000335)
(0.780020, 0.000146)
(0.786020, -0.000069)
(0.792020, -0.000137)
(0.798020, -0.000011)
(0.804020, 0.000197)
(0.810020, 0.000316)
(0.816020, 0.000256)
(0.822020, 0.000075)
(0.828020, -0.000076)
(0.834020, -0.000077)
(0.840020, 0.000063)
(0.846020, 0.000226)
(0.852020, 0.000279)
(0.858020, 0.000186)
(0.864020, 0.000030)
(0.870020, -0.000062)
(0.876020, -0.000018)
(0.882020, 0.000117)
(0.888020, 0.000232)
(0.894020, 0.000235)
(0.900020, 0.000130)
(0.906020, 0.000007)
(0.912020, -0.000034)
(0.918020, 0.000034)
(0.924020, 0.000152)
(0.930020, 0.000222)
(0.936020, 0.000190)
(0.942020, 0.000089)
(0.948020, 0.000002)
(0.954020, -0.000001)
(0.960020, 0.000077)
(0.966020, 0.000169)
(0.972020, 0.000201)
(0.978020, 0.000151)
(0.984020, 0.000063)
(0.990020, 0.000009)
(0.996020, 0.000032)
(1.002020, 0.000108)
(1.008020, 0.000173)
(1.014020, 0.000177)
(1.020020, 0.000119)
(1.026020, 0.000049)
(1.032020, 0.000024)
(1.038020, 0.000061)
(1.044020, 0.000128)
(1.050020, 0.000168)
(1.056020, 0.000152)
(1.062020, 0.000095)
(1.068020, 0.000046)
(1.074020, 0.000042)
(1.080020, 0.000085)
(1.086020, 0.000138)
(1.092020, 0.000158)
(1.098020, 0.000130)
(1.104020, 0.000080)
(1.110020, 0.000049)
(1.116020, 0.000061)
(1.122020, 0.000103)
(1.128020, 0.000141)
(1.134020, 0.000144)
(1.140020, 0.000112)
(1.146020, 0.000072)
(1.152020, 0.000057)
(1.158020, 0.000077)
(1.164020, 0.000115)
(1.170020, 0.000138)
(1.176020, 0.000130)
(1.182020, 0.000098)
(1.188020, 0.000070)
(1.194020, 0.000067)
(1.200020, 0.000091)
(1.206020, 0.000121)
(1.212020, 0.000133)
(1.218020, 0.000118)
(1.224020, 0.000090)
(1.230020, 0.000072)
(1.236020, 0.000077)
(1.242020, 0.000101)
(1.248020, 0.000123)
(1.254020, 0.000125)
(1.260020, 0.000107)
(1.266020, 0.000085)
(1.272020, 0.000076)
(1.278020, 0.000087)
(1.284020, 0.000108)
(1.290020, 0.000122)
(1.296020, 0.000118)
(1.302020, 0.000100)
(1.308020, 0.000083)
(1.314020, 0.000081)
(1.320020, 0.000095)
(1.326020, 0.000112)
(1.332020, 0.000119)
(1.338020, 0.000111)
(1.344020, 0.000095)
(1.350020, 0.000084)
(1.356020, 0.000087)
(1.362020, 0.000100)
(1.368020, 0.000113)
(1.374020, 0.000114)
(1.380020, 0.000105)
(1.386020, 0.000092)
(1.392020, 0.000086)
(1.398020, 0.000092)
(1.404020, 0.000104)
(1.410020, 0.000112)
(1.416020, 0.000110)
(1.422020, 0.000100)
(1.428020, 0.000091)
(1.434020, 0.000089)
(1.440020, 0.000097)
(1.446020, 0.000106)
(1.452020, 0.000111)
(1.458020, 0.000106)
(1.464020, 0.000097)
(1.470020, 0.000091)
(1.476020, 0.000093)
(1.482020, 0.000100)
(1.488020, 0.000107)
(1.494020, 0.000108)
(1.500020, 0.000103)
(1.506020, 0.000096)
(1.512020, 0.000092)
(1.518020, 0.000096)
(1.524020, 0.000102)
(1.530020, 0.000107)
(1.536020, 0.000106)
(1.542020, 0.000100)
(1.548020, 0.000095)
(1.554020, 0.000094)
(1.560020, 0.000098)
(1.566020, 0.000104)
(1.572020, 0.000106)
(1.578020, 0.000104)
(1.584020, 0.000099)
(1.590020, 0.000095)
(1.596020, 0.000096)
(1.602020, 0.000100)
(1.608020, 0.000104)
(1.614020, 0.000105)
(1.620020, 0.000102)
(1.626020, 0.000098)
(1.632020, 0.000096)
(1.638020, 0.000097)
(1.644020, 0.000101)
(1.650020, 0.000104)
(1.656020, 0.000103)
(1.662020, 0.000100)
(1.668020, 0.000097)
(1.674020, 0.000097)
(1.680020, 0.000099)
(1.686020, 0.000102)
(1.692020, 0.000103)
(1.698020, 0.000102)
(1.704020, 0.000099)
(1.710020, 0.000097)
(1.716020, 0.000098)
(1.722020, 0.000100)
(1.728020, 0.000102)
(1.734020, 0.000103)
(1.740020, 0.000101)
(1.746020, 0.000099)
(1.752020, 0.000098)
(1.758020, 0.000099)
(1.764020, 0.000101)
(1.770020, 0.000102)
(1.776020, 0.000102)
(1.782020, 0.000100)
(1.788020, 0.000099)
(1.794020, 0.000098)
(1.800020, 0.000099)
(1.806020, 0.000101)
(1.812020, 0.000102)
(1.818020, 0.000101)
(1.824020, 0.000100)
(1.830020, 0.000099)
(1.836020, 0.000099)
(1.842020, 0.000100)
(1.848020, 0.000101)
(1.854020, 0.000102)
(1.860020, 0.000101)
(1.866020, 0.000099)
(1.872020, 0.000099)
(1.878020, 0.000099)
(1.884020, 0.000100)
(1.890020, 0.000101)
(1.896020, 0.000101)
(1.902020, 0.000100)
(1.908020, 0.000099)
(1.914020, 0.000099)
(1.920020, 0.000100)
(1.926020, 0.000101)
(1.932020, 0.000101)
(1.938020, 0.000101)
(1.944020, 0.000100)
(1.950020, 0.000099)
(1.956020, 0.000099)
(1.962020, 0.000100)
(1.968020, 0.000101)
(1.974020, 0.000101)
(1.980020, 0.000101)
(1.986020, 0.000100)
(1.992020, 0.000099)
(1.998020, 0.000100)
(2.004020, 0.000100)
(2.010020, 0.000101)
(2.016020, 0.000101)
(2.022020, 0.000100)
(2.028020, 0.000100)
(2.034020, 0.000100)
(2.040020, 0.000100)
(2.046020, 0.000100)
(2.052020, 0.000101)
(2.058020, 0.000101)
(2.064020, 0.000100)
(2.070020, 0.000100)
(2.076020, 0.000100)
(2.082020, 0.000100)
(2.088020, 0.000101)
(2.094020, 0.000101)
(2.100020, 0.000100)
(2.106020, 0.000100)
(2.112020, 0.000100)
(2.118020, 0.000100)
(2.124020, 0.000100)
(2.130020, 0.000101)
(2.136020, 0.000101)
(2.142020, 0.000100)
(2.148020, 0.000100)
(2.154020, 0.000100)
(2.160020, 0.000100)
(2.166020, 0.000100)
(2.172020, 0.000101)
(2.178020, 0.000100)
(2.184020, 0.000100)
(2.190020, 0.000100)
(2.196020, 0.000100)
(2.202020, 0.000100)
(2.208020, 0.000100)
(2.214020, 0.000100)
(2.220020, 0.000100)
(2.226020, 0.000100)
(2.232020, 0.000100)
(2.238020, 0.000100)
(2.244020, 0.000100)
(2.250020, 0.000100)
(2.256020, 0.000100)
(2.262020, 0.000100)
(2.268020, 0.000100)
(2.274020, 0.000100)
(2.280020, 0.000100)
(2.286020, 0.000100)
(2.292020, 0.000100)
(2.298020, 0.000100)
(2.304020, 0.000100)
(2.310020, 0.000100)
(2.316020, 0.000100)
(2.322020, 0.000100)
(2.328020, 0.000100)
(2.334020, 0.000100)
(2.340020, 0.000100)
(2.346020, 0.000100)
(2.352020, 0.000100)
(2.358020, 0.000100)
(2.364020, 0.000100)
(2.370020, 0.000100)
(2.376020, 0.000100)
(2.382020, 0.000100)
(2.388020, 0.000100)
(2.394020, 0.000100)
(2.400020, 0.000100)
};
\label{1e-5}

%% file: tex/fig_his2.tex
\begin{figure}[t]
  \centering
  \begin{tikzpicture}
    \begin{axis}[
        xlabel={Time},
        ylabel={$u_z(0,0,t)$},
        width=4.8cm,
        height=6cm,
        xmax=2,
        legend pos= north east,
        ymajorgrids=true,
        grid style=dashed,
        y tick label style={/pgf/number format/fixed, /pgf/number format/precision=2},
        scaled y ticks=false
    ] \input{figs/data/nonlinearfitha010.tex}
    \end{axis}
    \end{tikzpicture}
    \hfill
  \begin{tikzpicture}
  \begin{axis}[
      xlabel={Time},
      xmax =1,
      width=4.8cm,
      height=6cm,
      legend pos= north east,
      ymajorgrids=true,
      grid style=dashed,
      y tick label style={/pgf/number format/fixed, /pgf/number format/precision=2},
      scaled y ticks=false
  ] \input{figs/data/nonlinearfitha050.tex}
  \end{axis}
  \end{tikzpicture}
  \hfill
  \begin{tikzpicture}
  \begin{axis}[
      xlabel={Time},
      xmax =1,
      width=4.8cm,
      height=6cm,
      legend pos= north east,
      ymajorgrids=true,
      grid style=dashed,
      y tick label style={/pgf/number format/fixed, /pgf/number format/precision=3},
      scaled y ticks=false
  ] \input{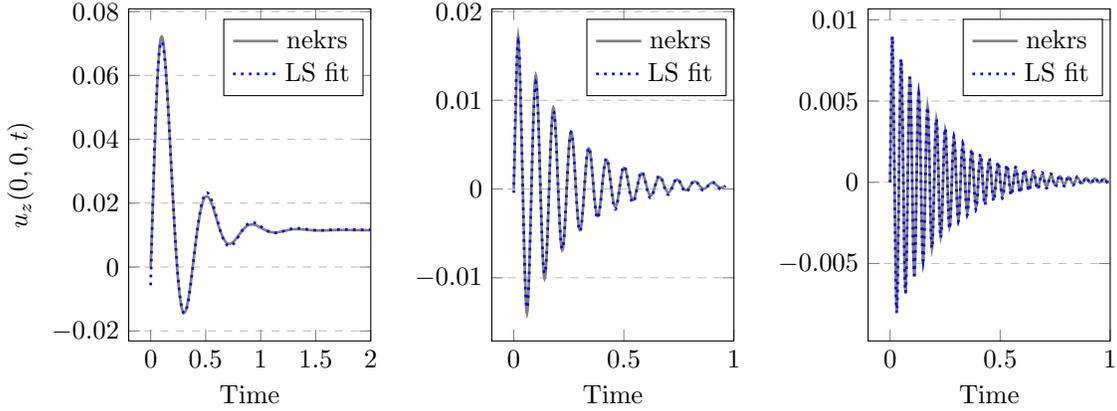}
  \end{axis}
  \end{tikzpicture}
  \caption{
Model fit of the center-point velocity to the NekRS results for $\ha = 10, 50,
100$ (left, middle, and right) with conducting walls ($r_w=10^{-5}$) for $Re =
Rm = 1$.  A nonlinear fit is used to approximate $\hat{u}_{0}, s, \omega, \phi$
and $s_0$ in $\hat{u}=\hat{u}_{0} e^{-s t} \sin (\omega t+\phi)+ s_0$.}
\label{fig:time_his2}    
\end{figure}

%% file: figs/data/nonlinearfitha010.tex
\addplot[
line width=1.0pt, color=gray, solid ]
coordinates {
(0.000010, 0.000010)
(0.005010, 0.005010)
(0.010010, 0.010010)
(0.015010, 0.015010)
(0.020010, 0.020010)
(0.025010, 0.025008)
(0.030010, 0.029997)
(0.035010, 0.034954)
(0.040010, 0.039836)
(0.045010, 0.044585)
(0.050010, 0.049128)
(0.055010, 0.053394)
(0.060010, 0.057315)
(0.065010, 0.060835)
(0.070010, 0.063909)
(0.075010, 0.066507)
(0.080010, 0.068614)
(0.085010, 0.070224)
(0.090010, 0.071344)
(0.095010, 0.071988)
(0.100010, 0.072176)
(0.105010, 0.071932)
(0.110010, 0.071284)
(0.115010, 0.070263)
(0.120010, 0.068898)
(0.125010, 0.067221)
(0.130010, 0.065260)
(0.135010, 0.063046)
(0.140010, 0.060606)
(0.145010, 0.057968)
(0.150010, 0.055158)
(0.155010, 0.052201)
(0.160010, 0.049121)
(0.165010, 0.045941)
(0.170010, 0.042684)
(0.175010, 0.039371)
(0.180010, 0.036024)
(0.185010, 0.032663)
(0.190010, 0.029309)
(0.195010, 0.025980)
(0.200010, 0.022696)
(0.205010, 0.019475)
(0.210010, 0.016333)
(0.215010, 0.013286)
(0.220010, 0.010350)
(0.225010, 0.007539)
(0.230010, 0.004865)
(0.235010, 0.002340)
(0.240010, -0.000026)
(0.245010, -0.002225)
(0.250010, -0.004249)
(0.255010, -0.006093)
(0.260010, -0.007753)
(0.265010, -0.009225)
(0.270010, -0.010507)
(0.275010, -0.011601)
(0.280010, -0.012506)
(0.285010, -0.013224)
(0.290010, -0.013760)
(0.295010, -0.014118)
(0.300010, -0.014302)
(0.305010, -0.014320)
(0.310010, -0.014177)
(0.315010, -0.013882)
(0.320010, -0.013443)
(0.325010, -0.012869)
(0.330010, -0.012169)
(0.335010, -0.011353)
(0.340010, -0.010430)
(0.345010, -0.009412)
(0.350010, -0.008307)
(0.355010, -0.007128)
(0.360010, -0.005883)
(0.365010, -0.004584)
(0.370010, -0.003241)
(0.375010, -0.001864)
(0.380010, -0.000462)
(0.385010, 0.000954)
(0.390010, 0.002375)
(0.395010, 0.003792)
(0.400010, 0.005197)
(0.405010, 0.006581)
(0.410010, 0.007938)
(0.415010, 0.009259)
(0.420010, 0.010538)
(0.425010, 0.011769)
(0.430010, 0.012947)
(0.435010, 0.014066)
(0.440010, 0.015122)
(0.445010, 0.016111)
(0.450010, 0.017030)
(0.455010, 0.017877)
(0.460010, 0.018649)
(0.465010, 0.019344)
(0.470010, 0.019962)
(0.475010, 0.020502)
(0.480010, 0.020963)
(0.485010, 0.021348)
(0.490010, 0.021655)
(0.495010, 0.021888)
(0.500010, 0.022047)
(0.505010, 0.022135)
(0.510010, 0.022154)
(0.515010, 0.022107)
(0.520010, 0.021997)
(0.525010, 0.021828)
(0.530010, 0.021602)
(0.535010, 0.021325)
(0.540010, 0.020999)
(0.545010, 0.020629)
(0.550010, 0.020219)
(0.555010, 0.019773)
(0.560010, 0.019295)
(0.565010, 0.018791)
(0.570010, 0.018263)
(0.575010, 0.017717)
(0.580010, 0.017156)
(0.585010, 0.016584)
(0.590010, 0.016006)
(0.595010, 0.015426)
(0.600010, 0.014846)
(0.605010, 0.014271)
(0.610010, 0.013704)
(0.615010, 0.013147)
(0.620010, 0.012605)
(0.625010, 0.012080)
(0.630010, 0.011573)
(0.635010, 0.011089)
(0.640010, 0.010628)
(0.645010, 0.010192)
(0.650010, 0.009784)
(0.655010, 0.009403)
(0.660010, 0.009053)
(0.665010, 0.008733)
(0.670010, 0.008444)
(0.675010, 0.008187)
(0.680010, 0.007961)
(0.685010, 0.007767)
(0.690010, 0.007605)
(0.695010, 0.007475)
(0.700010, 0.007375)
(0.705010, 0.007305)
(0.710010, 0.007265)
(0.715010, 0.007253)
(0.720010, 0.007268)
(0.725010, 0.007308)
(0.730010, 0.007374)
(0.735010, 0.007462)
(0.740010, 0.007572)
(0.745010, 0.007702)
(0.750010, 0.007851)
(0.755010, 0.008015)
(0.760010, 0.008195)
(0.765010, 0.008388)
(0.770010, 0.008592)
(0.775010, 0.008806)
(0.780010, 0.009028)
(0.785010, 0.009256)
(0.790010, 0.009489)
(0.795010, 0.009724)
(0.800010, 0.009961)
(0.805010, 0.010198)
(0.810010, 0.010433)
(0.815010, 0.010665)
(0.820010, 0.010893)
(0.825010, 0.011115)
(0.830010, 0.011330)
(0.835010, 0.011538)
(0.840010, 0.011737)
(0.845010, 0.011927)
(0.850010, 0.012106)
(0.855010, 0.012275)
(0.860010, 0.012432)
(0.865010, 0.012577)
(0.870010, 0.012709)
(0.875010, 0.012829)
(0.880010, 0.012936)
(0.885010, 0.013030)
(0.890010, 0.013111)
(0.895010, 0.013179)
(0.900010, 0.013234)
(0.905010, 0.013276)
(0.910010, 0.013307)
(0.915010, 0.013325)
(0.920010, 0.013331)
(0.925010, 0.013327)
(0.930010, 0.013312)
(0.935010, 0.013286)
(0.940010, 0.013252)
(0.945010, 0.013208)
(0.950010, 0.013156)
(0.955010, 0.013096)
(0.960010, 0.013030)
(0.965010, 0.012957)
(0.970010, 0.012879)
(0.975010, 0.012796)
(0.980010, 0.012710)
(0.985010, 0.012619)
(0.990010, 0.012527)
(0.995010, 0.012432)
(1.000010, 0.012336)
(1.005010, 0.012239)
(1.010010, 0.012142)
(1.015010, 0.012046)
(1.020010, 0.011951)
(1.025010, 0.011858)
(1.030010, 0.011766)
(1.035010, 0.011678)
(1.040010, 0.011592)
(1.045010, 0.011510)
(1.050010, 0.011432)
(1.055010, 0.011358)
(1.060010, 0.011289)
(1.065010, 0.011224)
(1.070010, 0.011164)
(1.075010, 0.011109)
(1.080010, 0.011059)
(1.085010, 0.011015)
(1.090010, 0.010976)
(1.095010, 0.010942)
(1.100010, 0.010913)
(1.105010, 0.010890)
(1.110010, 0.010871)
(1.115010, 0.010858)
(1.120010, 0.010850)
(1.125010, 0.010847)
(1.130010, 0.010848)
(1.135010, 0.010853)
(1.140010, 0.010863)
(1.145010, 0.010876)
(1.150010, 0.010894)
(1.155010, 0.010914)
(1.160010, 0.010938)
(1.165010, 0.010965)
(1.170010, 0.010994)
(1.175010, 0.011025)
(1.180010, 0.011059)
(1.185010, 0.011094)
(1.190010, 0.011131)
(1.195010, 0.011168)
(1.200010, 0.011207)
(1.205010, 0.011246)
(1.210010, 0.011285)
(1.215010, 0.011325)
(1.220010, 0.011364)
(1.225010, 0.011403)
(1.230010, 0.011441)
(1.235010, 0.011478)
(1.240010, 0.011515)
(1.245010, 0.011550)
(1.250010, 0.011583)
(1.255010, 0.011615)
(1.260010, 0.011646)
(1.265010, 0.011674)
(1.270010, 0.011701)
(1.275010, 0.011726)
(1.280010, 0.011748)
(1.285010, 0.011769)
(1.290010, 0.011788)
(1.295010, 0.011804)
(1.300010, 0.011818)
(1.305010, 0.011830)
(1.310010, 0.011840)
(1.315010, 0.011847)
(1.320010, 0.011853)
(1.325010, 0.011857)
(1.330010, 0.011858)
(1.335010, 0.011858)
(1.340010, 0.011856)
(1.345010, 0.011852)
(1.350010, 0.011847)
(1.355010, 0.011840)
(1.360010, 0.011832)
(1.365010, 0.011822)
(1.370010, 0.011811)
(1.375010, 0.011800)
(1.380010, 0.011787)
(1.385010, 0.011773)
(1.390010, 0.011759)
(1.395010, 0.011744)
(1.400010, 0.011729)
(1.405010, 0.011713)
(1.410010, 0.011697)
(1.415010, 0.011681)
(1.420010, 0.011665)
(1.425010, 0.011649)
(1.430010, 0.011633)
(1.435010, 0.011617)
(1.440010, 0.011602)
(1.445010, 0.011587)
(1.450010, 0.011572)
(1.455010, 0.011558)
(1.460010, 0.011545)
(1.465010, 0.011533)
(1.470010, 0.011521)
(1.475010, 0.011510)
(1.480010, 0.011500)
(1.485010, 0.011490)
(1.490010, 0.011481)
(1.495010, 0.011474)
(1.500010, 0.011467)
(1.505010, 0.011461)
(1.510010, 0.011456)
(1.515010, 0.011452)
(1.520010, 0.011448)
(1.525010, 0.011446)
(1.530010, 0.011444)
(1.535010, 0.011444)
(1.540010, 0.011443)
(1.545010, 0.011444)
(1.550010, 0.011445)
(1.555010, 0.011448)
(1.560010, 0.011450)
(1.565010, 0.011453)
(1.570010, 0.011457)
(1.575010, 0.011462)
(1.580010, 0.011466)
(1.585010, 0.011471)
(1.590010, 0.011477)
(1.595010, 0.011483)
(1.600010, 0.011489)
(1.605010, 0.011495)
(1.610010, 0.011501)
(1.615010, 0.011508)
(1.620010, 0.011514)
(1.625010, 0.011521)
(1.630010, 0.011527)
(1.635010, 0.011534)
(1.640010, 0.011540)
(1.645010, 0.011546)
(1.650010, 0.011552)
(1.655010, 0.011558)
(1.660010, 0.011564)
(1.665010, 0.011569)
(1.670010, 0.011575)
(1.675010, 0.011579)
(1.680010, 0.011584)
(1.685010, 0.011588)
(1.690010, 0.011592)
(1.695010, 0.011596)
(1.700010, 0.011599)
(1.705010, 0.011602)
(1.710010, 0.011604)
(1.715010, 0.011606)
(1.720010, 0.011608)
(1.725010, 0.011609)
(1.730010, 0.011610)
(1.735010, 0.011611)
(1.740010, 0.011611)
(1.745010, 0.011611)
(1.750010, 0.011611)
(1.755010, 0.011610)
(1.760010, 0.011610)
(1.765010, 0.011609)
(1.770010, 0.011607)
(1.775010, 0.011606)
(1.780010, 0.011604)
(1.785010, 0.011602)
(1.790010, 0.011600)
(1.795010, 0.011598)
(1.800010, 0.011595)
(1.805010, 0.011593)
(1.810010, 0.011590)
(1.815010, 0.011588)
(1.820010, 0.011585)
(1.825010, 0.011582)
(1.830010, 0.011580)
(1.835010, 0.011577)
(1.840010, 0.011574)
(1.845010, 0.011571)
(1.850010, 0.011569)
(1.855010, 0.011566)
(1.860010, 0.011564)
(1.865010, 0.011562)
(1.870010, 0.011559)
(1.875010, 0.011557)
(1.880010, 0.011555)
(1.885010, 0.011553)
(1.890010, 0.011551)
(1.895010, 0.011550)
(1.900010, 0.011548)
(1.905010, 0.011547)
(1.910010, 0.011546)
(1.915010, 0.011545)
(1.920010, 0.011544)
(1.925010, 0.011543)
(1.930010, 0.011543)
(1.935010, 0.011542)
(1.940010, 0.011542)
(1.945010, 0.011542)
(1.950010, 0.011541)
(1.955010, 0.011542)
(1.960010, 0.011542)
(1.965010, 0.011542)
(1.970010, 0.011542)
(1.975010, 0.011543)
(1.980010, 0.011543)
(1.985010, 0.011544)
(1.990010, 0.011545)
(1.995010, 0.011546)
(2.000010, 0.011547)
(2.005010, 0.011548)
(2.010010, 0.011549)
(2.015010, 0.011550)
(2.020010, 0.011551)
(2.025010, 0.011552)
(2.030010, 0.011553)
(2.035010, 0.011554)
(2.040010, 0.011555)
(2.045010, 0.011556)
(2.050010, 0.011557)
(2.055010, 0.011558)
(2.060010, 0.011559)
(2.065010, 0.011560)
(2.070010, 0.011561)
(2.075010, 0.011562)
(2.080010, 0.011563)
(2.085010, 0.011564)
(2.090010, 0.011564)
(2.095010, 0.011565)
(2.100010, 0.011566)
(2.105010, 0.011566)
(2.110010, 0.011567)
(2.115010, 0.011567)
(2.120010, 0.011568)
(2.125010, 0.011568)
(2.130010, 0.011568)
(2.135010, 0.011569)
(2.140010, 0.011569)
(2.145010, 0.011569)
(2.150010, 0.011569)
(2.155010, 0.011569)
(2.160010, 0.011569)
(2.165010, 0.011569)
(2.170010, 0.011569)
(2.175010, 0.011568)
(2.180010, 0.011568)
(2.185010, 0.011568)
(2.190010, 0.011568)
(2.195010, 0.011567)
(2.200010, 0.011567)
(2.205010, 0.011567)
(2.210010, 0.011566)
(2.215010, 0.011566)
(2.220010, 0.011565)
(2.225010, 0.011565)
(2.230010, 0.011564)
(2.235010, 0.011564)
(2.240010, 0.011564)
(2.245010, 0.011563)
(2.250010, 0.011563)
(2.255010, 0.011562)
(2.260010, 0.011562)
(2.265010, 0.011561)
(2.270010, 0.011561)
(2.275010, 0.011560)
(2.280010, 0.011560)
(2.285010, 0.011560)
(2.290010, 0.011559)
(2.295010, 0.011559)
(2.300010, 0.011559)
(2.305010, 0.011558)
(2.310010, 0.011558)
(2.315010, 0.011558)
(2.320010, 0.011558)
(2.325010, 0.011557)
(2.330010, 0.011557)
(2.335010, 0.011557)
(2.340010, 0.011557)
(2.345010, 0.011557)
(2.350010, 0.011557)
(2.355010, 0.011557)
(2.360010, 0.011557)
(2.365010, 0.011557)
(2.370010, 0.011557)
(2.375010, 0.011557)
(2.380010, 0.011557)
(2.385010, 0.011557)
(2.390010, 0.011557)
(2.395010, 0.011557)
(2.400010, 0.011557)
(2.405010, 0.011557)
(2.410010, 0.011557)
(2.415010, 0.011558)
(2.420010, 0.011558)
(2.425010, 0.011558)
(2.430010, 0.011558)
(2.435010, 0.011558)
(2.440010, 0.011558)
(2.445010, 0.011559)
(2.450010, 0.011559)
(2.455010, 0.011559)
(2.460010, 0.011559)
(2.465010, 0.011559)
(2.470010, 0.011559)
(2.475010, 0.011560)
(2.480010, 0.011560)
(2.485010, 0.011560)
(2.490010, 0.011560)
(2.495010, 0.011560)
(2.500010, 0.011560)
(2.505010, 0.011560)
(2.510010, 0.011560)
(2.515010, 0.011561)
(2.520010, 0.011561)
(2.525010, 0.011561)
(2.530010, 0.011561)
(2.535010, 0.011561)
(2.540010, 0.011561)
(2.545010, 0.011561)
(2.550010, 0.011561)
(2.555010, 0.011561)
(2.560010, 0.011561)
(2.565010, 0.011561)
(2.570010, 0.011561)
(2.575010, 0.011561)
(2.580010, 0.011561)
(2.585010, 0.011561)
(2.590010, 0.011561)
(2.595010, 0.011561)
(2.600010, 0.011561)
(2.605010, 0.011561)
(2.610010, 0.011560)
(2.615010, 0.011560)
(2.620010, 0.011560)
(2.625010, 0.011560)
(2.630010, 0.011560)
(2.635010, 0.011560)
(2.640010, 0.011560)
(2.645010, 0.011560)
(2.650010, 0.011560)
(2.655010, 0.011560)
(2.660010, 0.011560)
(2.665010, 0.011560)
(2.670010, 0.011560)
(2.675010, 0.011559)
(2.680010, 0.011559)
(2.685010, 0.011559)
(2.690010, 0.011559)
(2.695010, 0.011559)
(2.700010, 0.011559)
(2.705010, 0.011559)
(2.710010, 0.011559)
(2.715010, 0.011559)
(2.720010, 0.011559)
(2.725010, 0.011559)
(2.730010, 0.011559)
(2.735010, 0.011559)
(2.740010, 0.011559)
(2.745010, 0.011559)
(2.750010, 0.011559)
(2.755010, 0.011559)
(2.760010, 0.011559)
(2.765010, 0.011558)
(2.770010, 0.011558)
(2.775010, 0.011558)
(2.780010, 0.011558)
(2.785010, 0.011558)
(2.790010, 0.011558)
(2.795010, 0.011558)
(2.800010, 0.011558)
(2.805010, 0.011558)
(2.810010, 0.011558)
(2.815010, 0.011558)
(2.820010, 0.011559)
(2.825010, 0.011559)
(2.830010, 0.011559)
(2.835010, 0.011559)
(2.840010, 0.011559)
(2.845010, 0.011559)
(2.850010, 0.011559)
(2.855010, 0.011559)
(2.860010, 0.011559)
(2.865010, 0.011559)
(2.870010, 0.011559)
(2.875010, 0.011559)
(2.880010, 0.011559)
(2.885010, 0.011559)
(2.890010, 0.011559)
(2.895010, 0.011559)
(2.900010, 0.011559)
(2.905010, 0.011559)
(2.910010, 0.011559)
(2.915010, 0.011559)
(2.920010, 0.011559)
(2.925010, 0.011559)
(2.930010, 0.011559)
(2.935010, 0.011559)
(2.940010, 0.011559)
(2.945010, 0.011559)
(2.950010, 0.011559)
(2.955010, 0.011559)
(2.960010, 0.011559)
(2.965010, 0.011559)
(2.970010, 0.011559)
(2.975010, 0.011559)
(2.980010, 0.011559)
(2.985010, 0.011559)
(2.990010, 0.011559)
(2.995010, 0.011559)
(3.000010, 0.011559)
(3.005010, 0.011559)
(3.010010, 0.011559)
(3.015010, 0.011559)
(3.020010, 0.011559)
(3.025010, 0.011559)
(3.030010, 0.011559)
(3.035010, 0.011559)
(3.040010, 0.011559)
(3.045010, 0.011559)
(3.050010, 0.011559)
(3.055010, 0.011559)
(3.060010, 0.011559)
(3.065010, 0.011559)
(3.070010, 0.011558)
(3.075010, 0.011558)
(3.080010, 0.011558)
(3.085010, 0.011558)
(3.090010, 0.011558)
(3.095010, 0.011558)
(3.100010, 0.011558)
(3.105010, 0.011558)
(3.110010, 0.011558)
(3.115010, 0.011558)
(3.120010, 0.011558)
(3.125010, 0.011558)
(3.130010, 0.011558)
(3.135010, 0.011558)
(3.140010, 0.011558)
(3.145010, 0.011558)
(3.150010, 0.011558)
(3.155010, 0.011558)
(3.160010, 0.011558)
(3.165010, 0.011558)
(3.170010, 0.011558)
(3.175010, 0.011558)
(3.180010, 0.011558)
(3.185010, 0.011558)
(3.190010, 0.011558)
(3.195010, 0.011558)
(3.200010, 0.011558)
(3.205010, 0.011558)
(3.210010, 0.011558)
(3.215010, 0.011558)
(3.220010, 0.011558)
(3.225010, 0.011558)
(3.230010, 0.011558)
(3.235010, 0.011558)
(3.240010, 0.011558)
(3.245010, 0.011558)
(3.250010, 0.011558)
(3.255010, 0.011558)
(3.260010, 0.011558)
(3.265010, 0.011558)
(3.270010, 0.011558)
(3.275010, 0.011558)
(3.280010, 0.011558)
(3.285010, 0.011558)
(3.290010, 0.011558)
(3.295010, 0.011558)
(3.300010, 0.011558)
(3.305010, 0.011558)
(3.310010, 0.011558)
(3.315010, 0.011558)
(3.320010, 0.011558)
(3.325010, 0.011558)
(3.330010, 0.011558)
(3.335010, 0.011558)
(3.340010, 0.011558)
(3.345010, 0.011558)
(3.350010, 0.011558)
(3.355010, 0.011558)
(3.360010, 0.011558)
(3.365010, 0.011558)
(3.370010, 0.011558)
(3.375010, 0.011558)
(3.380010, 0.011558)
(3.385010, 0.011558)
(3.390010, 0.011558)
(3.395010, 0.011558)
(3.400010, 0.011558)
(3.405010, 0.011558)
(3.410010, 0.011558)
(3.415010, 0.011558)
(3.420010, 0.011558)
(3.425010, 0.011558)
(3.430010, 0.011558)
(3.435010, 0.011558)
(3.440010, 0.011558)
(3.445010, 0.011558)
(3.450010, 0.011558)
(3.455010, 0.011558)
(3.460010, 0.011558)
(3.465010, 0.011558)
(3.470010, 0.011558)
(3.475010, 0.011558)
(3.480010, 0.011558)
(3.485010, 0.011558)
(3.490010, 0.011558)
(3.495010, 0.011558)
(3.500010, 0.011558)
(3.505010, 0.011558)
(3.510010, 0.011558)
(3.515010, 0.011558)
(3.520010, 0.011558)
(3.525010, 0.011558)
(3.530010, 0.011558)
(3.535010, 0.011558)
(3.540010, 0.011558)
(3.545010, 0.011558)
(3.550010, 0.011558)
(3.555010, 0.011558)
(3.560010, 0.011558)
(3.565010, 0.011558)
(3.570010, 0.011558)
(3.575010, 0.011558)
(3.580010, 0.011558)
(3.585010, 0.011558)
(3.590010, 0.011558)
(3.595010, 0.011558)
(3.600010, 0.011557)
(3.605010, 0.011557)
(3.610010, 0.011557)
(3.615010, 0.011557)
(3.620010, 0.011557)
(3.625010, 0.011557)
(3.630010, 0.011557)
(3.635010, 0.011557)
(3.640010, 0.011557)
(3.645010, 0.011557)
(3.650010, 0.011557)
(3.655010, 0.011557)
(3.660010, 0.011557)
(3.665010, 0.011557)
(3.670010, 0.011557)
(3.675010, 0.011557)
(3.680010, 0.011557)
(3.685010, 0.011557)
(3.690010, 0.011557)
(3.695010, 0.011557)
(3.700010, 0.011557)
(3.705010, 0.011557)
(3.710010, 0.011557)
(3.715010, 0.011557)
(3.720010, 0.011557)
(3.725010, 0.011557)
(3.730010, 0.011557)
(3.735010, 0.011557)
(3.740010, 0.011557)
(3.745010, 0.011557)
(3.750010, 0.011557)
(3.755010, 0.011557)
(3.760010, 0.011557)
(3.765010, 0.011557)
(3.770010, 0.011557)
(3.775010, 0.011557)
(3.780010, 0.011557)
(3.785010, 0.011557)
(3.790010, 0.011557)
(3.795010, 0.011557)
(3.800010, 0.011557)
(3.805010, 0.011557)
(3.810010, 0.011557)
(3.815010, 0.011557)
(3.820010, 0.011557)
(3.825010, 0.011557)
(3.830010, 0.011557)
(3.835010, 0.011557)
(3.840010, 0.011557)
(3.845010, 0.011557)
(3.850010, 0.011557)
(3.855010, 0.011557)
(3.860010, 0.011557)
(3.865010, 0.011557)
(3.870010, 0.011557)
(3.875010, 0.011557)
(3.880010, 0.011557)
(3.885010, 0.011557)
(3.890010, 0.011557)
(3.895010, 0.011557)
(3.900010, 0.011557)
(3.905010, 0.011557)
(3.910010, 0.011557)
(3.915010, 0.011557)
(3.920010, 0.011557)
(3.925010, 0.011557)
(3.930010, 0.011557)
(3.935010, 0.011557)
(3.940010, 0.011557)
(3.945010, 0.011557)
(3.950010, 0.011557)
(3.955010, 0.011557)
(3.960010, 0.011557)
(3.965010, 0.011557)
(3.970010, 0.011557)
(3.975010, 0.011557)
(3.980010, 0.011557)
(3.985010, 0.011557)
(3.990010, 0.011557)
(3.995010, 0.011557)
(4.000010, 0.011557)
(4.005010, 0.011557)
(4.010010, 0.011557)
(4.015010, 0.011557)
(4.020010, 0.011557)
(4.025010, 0.011557)
(4.030010, 0.011557)
(4.035010, 0.011557)
(4.040010, 0.011557)
(4.045010, 0.011557)
(4.050010, 0.011557)
(4.055010, 0.011557)
(4.060010, 0.011557)
(4.065010, 0.011557)
(4.070010, 0.011557)
(4.075010, 0.011557)
(4.080010, 0.011557)
(4.085010, 0.011557)
(4.090010, 0.011557)
(4.095010, 0.011557)
(4.100010, 0.011557)
(4.105010, 0.011557)
(4.110010, 0.011557)
(4.115010, 0.011557)
(4.120010, 0.011557)
(4.125010, 0.011557)
(4.130010, 0.011557)
(4.135010, 0.011557)
(4.140010, 0.011557)
(4.145010, 0.011557)
(4.150010, 0.011557)
(4.155010, 0.011557)
(4.160010, 0.011557)
(4.165010, 0.011557)
(4.170010, 0.011557)
(4.175010, 0.011557)
(4.180010, 0.011557)
(4.185010, 0.011557)
(4.190010, 0.011557)
(4.195010, 0.011557)
(4.200010, 0.011557)
(4.205010, 0.011557)
(4.210010, 0.011557)
(4.215010, 0.011557)
(4.220010, 0.011557)
(4.225010, 0.011557)
(4.230010, 0.011557)
(4.235010, 0.011557)
(4.240010, 0.011557)
(4.245010, 0.011557)
(4.250010, 0.011557)
(4.255010, 0.011557)
(4.260010, 0.011557)
(4.265010, 0.011557)
(4.270010, 0.011557)
(4.275010, 0.011557)
(4.280010, 0.011557)
(4.285010, 0.011557)
(4.290010, 0.011557)
(4.295010, 0.011557)
(4.300010, 0.011557)
(4.305010, 0.011557)
(4.310010, 0.011557)
(4.315010, 0.011557)
(4.320010, 0.011557)
(4.325010, 0.011557)
(4.330010, 0.011557)
(4.335010, 0.011557)
(4.340010, 0.011557)
(4.345010, 0.011557)
(4.350010, 0.011557)
(4.355010, 0.011557)
(4.360010, 0.011557)
(4.365010, 0.011557)
(4.370010, 0.011557)
(4.375010, 0.011557)
(4.380010, 0.011556)
(4.385010, 0.011556)
(4.390010, 0.011556)
(4.395010, 0.011556)
(4.400010, 0.011556)
(4.405010, 0.011556)
(4.410010, 0.011556)
(4.415010, 0.011556)
(4.420010, 0.011556)
(4.425010, 0.011556)
(4.430010, 0.011556)
(4.435010, 0.011556)
(4.440010, 0.011556)
(4.445010, 0.011556)
(4.450010, 0.011556)
(4.455010, 0.011556)
(4.460010, 0.011556)
(4.465010, 0.011556)
(4.470010, 0.011556)
(4.475010, 0.011556)
(4.480010, 0.011556)
(4.485010, 0.011556)
(4.490010, 0.011556)
(4.495010, 0.011556)
(4.500010, 0.011556)
(4.505010, 0.011556)
(4.510010, 0.011556)
(4.515010, 0.011556)
(4.520010, 0.011556)
(4.525010, 0.011556)
(4.530010, 0.011556)
(4.535010, 0.011556)
(4.540010, 0.011556)
(4.545010, 0.011556)
(4.550010, 0.011556)
(4.555010, 0.011556)
(4.560010, 0.011556)
(4.565010, 0.011556)
(4.570010, 0.011556)
(4.575010, 0.011556)
(4.580010, 0.011556)
(4.585010, 0.011556)
(4.590010, 0.011556)
(4.595010, 0.011556)
(4.600010, 0.011556)
(4.605010, 0.011556)
(4.610010, 0.011556)
(4.615010, 0.011556)
(4.620010, 0.011556)
(4.625010, 0.011556)
(4.630010, 0.011556)
(4.635010, 0.011556)
(4.640010, 0.011556)
(4.645010, 0.011556)
(4.650010, 0.011556)
(4.655010, 0.011556)
(4.660010, 0.011556)
(4.665010, 0.011556)
(4.670010, 0.011556)
(4.675010, 0.011556)
(4.680010, 0.011556)
(4.685010, 0.011556)
(4.690010, 0.011556)
(4.695010, 0.011556)
(4.700010, 0.011556)
(4.705010, 0.011556)
(4.710010, 0.011556)
(4.715010, 0.011556)
(4.720010, 0.011556)
(4.725010, 0.011556)
(4.730010, 0.011556)
(4.735010, 0.011556)
(4.740010, 0.011556)
(4.745010, 0.011556)
(4.750010, 0.011556)
(4.755010, 0.011556)
(4.760010, 0.011556)
(4.765010, 0.011556)
(4.770010, 0.011556)
(4.775010, 0.011556)
(4.780010, 0.011556)
(4.785010, 0.011556)
(4.790010, 0.011556)
(4.795010, 0.011556)
(4.800010, 0.011556)
(4.805010, 0.011556)
(4.810010, 0.011556)
(4.815010, 0.011556)
(4.820010, 0.011556)
(4.825010, 0.011556)
(4.830010, 0.011556)
(4.835010, 0.011556)
(4.840010, 0.011556)
(4.845010, 0.011556)
(4.850010, 0.011556)
(4.855010, 0.011556)
(4.860010, 0.011556)
(4.865010, 0.011556)
(4.870010, 0.011556)
(4.875010, 0.011556)
(4.880010, 0.011556)
(4.885010, 0.011556)
(4.890010, 0.011556)
(4.895010, 0.011556)
(4.900010, 0.011556)
(4.905010, 0.011556)
(4.910010, 0.011556)
(4.915010, 0.011556)
(4.920010, 0.011556)
(4.925010, 0.011556)
(4.930010, 0.011556)
(4.935010, 0.011556)
(4.940010, 0.011556)
(4.945010, 0.011556)
(4.950010, 0.011556)
(4.955010, 0.011556)
(4.960010, 0.011556)
(4.965010, 0.011556)
(4.970010, 0.011556)
(4.975010, 0.011556)
(4.980010, 0.011556)
(4.985010, 0.011556)
(4.990010, 0.011556)
(4.995010, 0.011556)
};
\addlegendentry{nekrs}
\addplot[
line width=1.0pt, color=blue, dotted ]
coordinates {
(0.000010, -0.005696)
(0.005010, 0.001295)
(0.010010, 0.008080)
(0.015010, 0.014623)
(0.020010, 0.020896)
(0.025010, 0.026869)
(0.030010, 0.032519)
(0.035010, 0.037823)
(0.040010, 0.042762)
(0.045010, 0.047322)
(0.050010, 0.051487)
(0.055010, 0.055249)
(0.060010, 0.058599)
(0.065010, 0.061535)
(0.070010, 0.064053)
(0.075010, 0.066154)
(0.080010, 0.067843)
(0.085010, 0.069123)
(0.090010, 0.070005)
(0.095010, 0.070497)
(0.100010, 0.070612)
(0.105010, 0.070363)
(0.110010, 0.069767)
(0.115010, 0.068840)
(0.120010, 0.067600)
(0.125010, 0.066068)
(0.130010, 0.064264)
(0.135010, 0.062210)
(0.140010, 0.059927)
(0.145010, 0.057439)
(0.150010, 0.054769)
(0.155010, 0.051939)
(0.160010, 0.048974)
(0.165010, 0.045896)
(0.170010, 0.042730)
(0.175010, 0.039496)
(0.180010, 0.036219)
(0.185010, 0.032919)
(0.190010, 0.029617)
(0.195010, 0.026334)
(0.200010, 0.023089)
(0.205010, 0.019900)
(0.210010, 0.016784)
(0.215010, 0.013758)
(0.220010, 0.010836)
(0.225010, 0.008033)
(0.230010, 0.005360)
(0.235010, 0.002830)
(0.240010, 0.000451)
(0.245010, -0.001767)
(0.250010, -0.003817)
(0.255010, -0.005694)
(0.260010, -0.007392)
(0.265010, -0.008908)
(0.270010, -0.010240)
(0.275010, -0.011387)
(0.280010, -0.012349)
(0.285010, -0.013127)
(0.290010, -0.013724)
(0.295010, -0.014143)
(0.300010, -0.014389)
(0.305010, -0.014466)
(0.310010, -0.014381)
(0.315010, -0.014140)
(0.320010, -0.013752)
(0.325010, -0.013224)
(0.330010, -0.012564)
(0.335010, -0.011782)
(0.340010, -0.010888)
(0.345010, -0.009890)
(0.350010, -0.008800)
(0.355010, -0.007627)
(0.360010, -0.006382)
(0.365010, -0.005075)
(0.370010, -0.003717)
(0.375010, -0.002317)
(0.380010, -0.000887)
(0.385010, 0.000565)
(0.390010, 0.002028)
(0.395010, 0.003494)
(0.400010, 0.004953)
(0.405010, 0.006396)
(0.410010, 0.007816)
(0.415010, 0.009204)
(0.420010, 0.010553)
(0.425010, 0.011858)
(0.430010, 0.013110)
(0.435010, 0.014306)
(0.440010, 0.015439)
(0.445010, 0.016505)
(0.450010, 0.017501)
(0.455010, 0.018423)
(0.460010, 0.019268)
(0.465010, 0.020035)
(0.470010, 0.020720)
(0.475010, 0.021324)
(0.480010, 0.021846)
(0.485010, 0.022286)
(0.490010, 0.022645)
(0.495010, 0.022922)
(0.500010, 0.023121)
(0.505010, 0.023242)
(0.510010, 0.023288)
(0.515010, 0.023261)
(0.520010, 0.023165)
(0.525010, 0.023003)
(0.530010, 0.022778)
(0.535010, 0.022495)
(0.540010, 0.022156)
(0.545010, 0.021767)
(0.550010, 0.021331)
(0.555010, 0.020854)
(0.560010, 0.020339)
(0.565010, 0.019791)
(0.570010, 0.019216)
(0.575010, 0.018616)
(0.580010, 0.017998)
(0.585010, 0.017365)
(0.590010, 0.016722)
(0.595010, 0.016073)
(0.600010, 0.015423)
(0.605010, 0.014775)
(0.610010, 0.014133)
(0.615010, 0.013501)
(0.620010, 0.012883)
(0.625010, 0.012281)
(0.630010, 0.011699)
(0.635010, 0.011139)
(0.640010, 0.010604)
(0.645010, 0.010097)
(0.650010, 0.009619)
(0.655010, 0.009172)
(0.660010, 0.008757)
(0.665010, 0.008377)
(0.670010, 0.008031)
(0.675010, 0.007721)
(0.680010, 0.007447)
(0.685010, 0.007210)
(0.690010, 0.007009)
(0.695010, 0.006844)
(0.700010, 0.006715)
(0.705010, 0.006622)
(0.710010, 0.006563)
(0.715010, 0.006537)
(0.720010, 0.006544)
(0.725010, 0.006581)
(0.730010, 0.006649)
(0.735010, 0.006744)
(0.740010, 0.006866)
(0.745010, 0.007012)
(0.750010, 0.007182)
(0.755010, 0.007372)
(0.760010, 0.007581)
(0.765010, 0.007807)
(0.770010, 0.008048)
(0.775010, 0.008301)
(0.780010, 0.008566)
(0.785010, 0.008839)
(0.790010, 0.009119)
(0.795010, 0.009404)
(0.800010, 0.009691)
(0.805010, 0.009980)
(0.810010, 0.010268)
(0.815010, 0.010553)
(0.820010, 0.010834)
(0.825010, 0.011110)
(0.830010, 0.011378)
(0.835010, 0.011638)
(0.840010, 0.011888)
(0.845010, 0.012127)
(0.850010, 0.012354)
(0.855010, 0.012568)
(0.860010, 0.012769)
(0.865010, 0.012955)
(0.870010, 0.013126)
(0.875010, 0.013282)
(0.880010, 0.013422)
(0.885010, 0.013546)
(0.890010, 0.013654)
(0.895010, 0.013746)
(0.900010, 0.013822)
(0.905010, 0.013881)
(0.910010, 0.013925)
(0.915010, 0.013954)
(0.920010, 0.013968)
(0.925010, 0.013967)
(0.930010, 0.013953)
(0.935010, 0.013925)
(0.940010, 0.013884)
(0.945010, 0.013832)
(0.950010, 0.013769)
(0.955010, 0.013695)
(0.960010, 0.013612)
(0.965010, 0.013521)
(0.970010, 0.013422)
(0.975010, 0.013316)
(0.980010, 0.013204)
(0.985010, 0.013088)
(0.990010, 0.012967)
(0.995010, 0.012843)
(1.000010, 0.012717)
(1.005010, 0.012589)
(1.010010, 0.012461)
(1.015010, 0.012334)
(1.020010, 0.012207)
(1.025010, 0.012082)
(1.030010, 0.011959)
(1.035010, 0.011839)
(1.040010, 0.011723)
(1.045010, 0.011612)
(1.050010, 0.011505)
(1.055010, 0.011403)
(1.060010, 0.011307)
(1.065010, 0.011217)
(1.070010, 0.011134)
(1.075010, 0.011056)
(1.080010, 0.010986)
(1.085010, 0.010923)
(1.090010, 0.010867)
(1.095010, 0.010818)
(1.100010, 0.010776)
(1.105010, 0.010741)
(1.110010, 0.010714)
(1.115010, 0.010693)
(1.120010, 0.010679)
(1.125010, 0.010672)
(1.130010, 0.010671)
(1.135010, 0.010677)
(1.140010, 0.010688)
(1.145010, 0.010705)
(1.150010, 0.010728)
(1.155010, 0.010755)
(1.160010, 0.010787)
(1.165010, 0.010823)
(1.170010, 0.010863)
(1.175010, 0.010907)
(1.180010, 0.010953)
(1.185010, 0.011002)
(1.190010, 0.011054)
(1.195010, 0.011107)
(1.200010, 0.011162)
(1.205010, 0.011218)
(1.210010, 0.011274)
(1.215010, 0.011331)
(1.220010, 0.011388)
(1.225010, 0.011444)
(1.230010, 0.011500)
(1.235010, 0.011554)
(1.240010, 0.011608)
(1.245010, 0.011659)
(1.250010, 0.011709)
(1.255010, 0.011757)
(1.260010, 0.011802)
(1.265010, 0.011845)
(1.270010, 0.011886)
(1.275010, 0.011923)
(1.280010, 0.011958)
(1.285010, 0.011990)
(1.290010, 0.012018)
(1.295010, 0.012044)
(1.300010, 0.012066)
(1.305010, 0.012085)
(1.310010, 0.012101)
(1.315010, 0.012114)
(1.320010, 0.012123)
(1.325010, 0.012130)
(1.330010, 0.012134)
(1.335010, 0.012134)
(1.340010, 0.012132)
(1.345010, 0.012128)
(1.350010, 0.012121)
(1.355010, 0.012111)
(1.360010, 0.012099)
(1.365010, 0.012085)
(1.370010, 0.012070)
(1.375010, 0.012052)
(1.380010, 0.012033)
(1.385010, 0.012013)
(1.390010, 0.011991)
(1.395010, 0.011968)
(1.400010, 0.011945)
(1.405010, 0.011921)
(1.410010, 0.011896)
(1.415010, 0.011871)
(1.420010, 0.011846)
(1.425010, 0.011820)
(1.430010, 0.011795)
(1.435010, 0.011771)
(1.440010, 0.011746)
(1.445010, 0.011723)
(1.450010, 0.011699)
(1.455010, 0.011677)
(1.460010, 0.011656)
(1.465010, 0.011635)
(1.470010, 0.011616)
(1.475010, 0.011598)
(1.480010, 0.011581)
(1.485010, 0.011566)
(1.490010, 0.011551)
(1.495010, 0.011538)
(1.500010, 0.011527)
(1.505010, 0.011517)
(1.510010, 0.011508)
(1.515010, 0.011501)
(1.520010, 0.011495)
(1.525010, 0.011490)
(1.530010, 0.011487)
(1.535010, 0.011486)
(1.540010, 0.011485)
(1.545010, 0.011486)
(1.550010, 0.011488)
(1.555010, 0.011491)
(1.560010, 0.011495)
(1.565010, 0.011500)
(1.570010, 0.011506)
(1.575010, 0.011513)
(1.580010, 0.011520)
(1.585010, 0.011529)
(1.590010, 0.011538)
(1.595010, 0.011547)
(1.600010, 0.011557)
(1.605010, 0.011567)
(1.610010, 0.011578)
(1.615010, 0.011589)
(1.620010, 0.011600)
(1.625010, 0.011611)
(1.630010, 0.011623)
(1.635010, 0.011634)
(1.640010, 0.011645)
(1.645010, 0.011656)
(1.650010, 0.011666)
(1.655010, 0.011676)
(1.660010, 0.011686)
(1.665010, 0.011696)
(1.670010, 0.011705)
(1.675010, 0.011714)
(1.680010, 0.011722)
(1.685010, 0.011729)
(1.690010, 0.011736)
(1.695010, 0.011743)
(1.700010, 0.011749)
(1.705010, 0.011754)
(1.710010, 0.011758)
(1.715010, 0.011762)
(1.720010, 0.011766)
(1.725010, 0.011768)
(1.730010, 0.011770)
(1.735010, 0.011772)
(1.740010, 0.011773)
(1.745010, 0.011773)
(1.750010, 0.011773)
(1.755010, 0.011772)
(1.760010, 0.011771)
(1.765010, 0.011769)
(1.770010, 0.011767)
(1.775010, 0.011764)
(1.780010, 0.011761)
(1.785010, 0.011758)
(1.790010, 0.011754)
(1.795010, 0.011750)
(1.800010, 0.011746)
(1.805010, 0.011742)
(1.810010, 0.011737)
(1.815010, 0.011733)
(1.820010, 0.011728)
(1.825010, 0.011723)
(1.830010, 0.011718)
(1.835010, 0.011713)
(1.840010, 0.011708)
(1.845010, 0.011703)
(1.850010, 0.011698)
(1.855010, 0.011694)
(1.860010, 0.011689)
(1.865010, 0.011685)
(1.870010, 0.011680)
(1.875010, 0.011676)
(1.880010, 0.011672)
(1.885010, 0.011669)
(1.890010, 0.011665)
(1.895010, 0.011662)
(1.900010, 0.011659)
(1.905010, 0.011657)
(1.910010, 0.011654)
(1.915010, 0.011652)
(1.920010, 0.011650)
(1.925010, 0.011649)
(1.930010, 0.011648)
(1.935010, 0.011647)
(1.940010, 0.011646)
(1.945010, 0.011645)
(1.950010, 0.011645)
(1.955010, 0.011645)
(1.960010, 0.011646)
(1.965010, 0.011646)
(1.970010, 0.011647)
(1.975010, 0.011648)
(1.980010, 0.011649)
(1.985010, 0.011650)
(1.990010, 0.011652)
(1.995010, 0.011653)
(2.000010, 0.011655)
(2.005010, 0.011657)
(2.010010, 0.011659)
(2.015010, 0.011661)
(2.020010, 0.011663)
(2.025010, 0.011665)
(2.030010, 0.011667)
(2.035010, 0.011669)
(2.040010, 0.011672)
(2.045010, 0.011674)
(2.050010, 0.011676)
(2.055010, 0.011678)
(2.060010, 0.011680)
(2.065010, 0.011682)
(2.070010, 0.011684)
(2.075010, 0.011686)
(2.080010, 0.011688)
(2.085010, 0.011690)
(2.090010, 0.011691)
(2.095010, 0.011693)
(2.100010, 0.011694)
(2.105010, 0.011696)
(2.110010, 0.011697)
(2.115010, 0.011698)
(2.120010, 0.011699)
(2.125010, 0.011700)
(2.130010, 0.011700)
(2.135010, 0.011701)
(2.140010, 0.011701)
(2.145010, 0.011702)
(2.150010, 0.011702)
(2.155010, 0.011702)
(2.160010, 0.011702)
(2.165010, 0.011702)
(2.170010, 0.011702)
(2.175010, 0.011701)
(2.180010, 0.011701)
(2.185010, 0.011700)
(2.190010, 0.011700)
(2.195010, 0.011699)
(2.200010, 0.011699)
(2.205010, 0.011698)
(2.210010, 0.011697)
(2.215010, 0.011696)
(2.220010, 0.011695)
(2.225010, 0.011694)
(2.230010, 0.011693)
(2.235010, 0.011692)
(2.240010, 0.011691)
(2.245010, 0.011690)
(2.250010, 0.011689)
(2.255010, 0.011689)
(2.260010, 0.011688)
(2.265010, 0.011687)
(2.270010, 0.011686)
(2.275010, 0.011685)
(2.280010, 0.011684)
(2.285010, 0.011683)
(2.290010, 0.011682)
(2.295010, 0.011682)
(2.300010, 0.011681)
(2.305010, 0.011680)
(2.310010, 0.011680)
(2.315010, 0.011679)
(2.320010, 0.011679)
(2.325010, 0.011678)
(2.330010, 0.011678)
(2.335010, 0.011678)
(2.340010, 0.011677)
(2.345010, 0.011677)
(2.350010, 0.011677)
(2.355010, 0.011677)
(2.360010, 0.011677)
(2.365010, 0.011677)
(2.370010, 0.011677)
(2.375010, 0.011677)
(2.380010, 0.011677)
(2.385010, 0.011677)
(2.390010, 0.011677)
(2.395010, 0.011678)
(2.400010, 0.011678)
(2.405010, 0.011678)
(2.410010, 0.011679)
(2.415010, 0.011679)
(2.420010, 0.011679)
(2.425010, 0.011680)
(2.430010, 0.011680)
(2.435010, 0.011681)
(2.440010, 0.011681)
(2.445010, 0.011681)
(2.450010, 0.011682)
(2.455010, 0.011682)
(2.460010, 0.011683)
(2.465010, 0.011683)
(2.470010, 0.011684)
(2.475010, 0.011684)
(2.480010, 0.011684)
(2.485010, 0.011685)
(2.490010, 0.011685)
(2.495010, 0.011685)
(2.500010, 0.011686)
(2.505010, 0.011686)
(2.510010, 0.011686)
(2.515010, 0.011687)
(2.520010, 0.011687)
(2.525010, 0.011687)
(2.530010, 0.011687)
(2.535010, 0.011687)
(2.540010, 0.011688)
(2.545010, 0.011688)
(2.550010, 0.011688)
(2.555010, 0.011688)
(2.560010, 0.011688)
(2.565010, 0.011688)
(2.570010, 0.011688)
(2.575010, 0.011688)
(2.580010, 0.011688)
(2.585010, 0.011688)
(2.590010, 0.011688)
(2.595010, 0.011688)
(2.600010, 0.011688)
(2.605010, 0.011688)
(2.610010, 0.011687)
(2.615010, 0.011687)
(2.620010, 0.011687)
(2.625010, 0.011687)
(2.630010, 0.011687)
(2.635010, 0.011687)
(2.640010, 0.011686)
(2.645010, 0.011686)
(2.650010, 0.011686)
(2.655010, 0.011686)
(2.660010, 0.011686)
(2.665010, 0.011685)
(2.670010, 0.011685)
(2.675010, 0.011685)
(2.680010, 0.011685)
(2.685010, 0.011685)
(2.690010, 0.011685)
(2.695010, 0.011684)
(2.700010, 0.011684)
(2.705010, 0.011684)
(2.710010, 0.011684)
(2.715010, 0.011684)
(2.720010, 0.011684)
(2.725010, 0.011684)
(2.730010, 0.011683)
(2.735010, 0.011683)
(2.740010, 0.011683)
(2.745010, 0.011683)
(2.750010, 0.011683)
(2.755010, 0.011683)
(2.760010, 0.011683)
(2.765010, 0.011683)
(2.770010, 0.011683)
(2.775010, 0.011683)
(2.780010, 0.011683)
(2.785010, 0.011683)
(2.790010, 0.011683)
(2.795010, 0.011683)
(2.800010, 0.011683)
(2.805010, 0.011683)
(2.810010, 0.011683)
(2.815010, 0.011683)
(2.820010, 0.011683)
(2.825010, 0.011683)
(2.830010, 0.011684)
(2.835010, 0.011684)
(2.840010, 0.011684)
(2.845010, 0.011684)
(2.850010, 0.011684)
(2.855010, 0.011684)
(2.860010, 0.011684)
(2.865010, 0.011684)
(2.870010, 0.011684)
(2.875010, 0.011684)
(2.880010, 0.011684)
(2.885010, 0.011684)
(2.890010, 0.011685)
(2.895010, 0.011685)
(2.900010, 0.011685)
(2.905010, 0.011685)
(2.910010, 0.011685)
(2.915010, 0.011685)
(2.920010, 0.011685)
(2.925010, 0.011685)
(2.930010, 0.011685)
(2.935010, 0.011685)
(2.940010, 0.011685)
(2.945010, 0.011685)
(2.950010, 0.011685)
(2.955010, 0.011685)
(2.960010, 0.011685)
(2.965010, 0.011685)
(2.970010, 0.011685)
(2.975010, 0.011685)
(2.980010, 0.011685)
(2.985010, 0.011685)
(2.990010, 0.011685)
(2.995010, 0.011685)
(3.000010, 0.011685)
(3.005010, 0.011685)
(3.010010, 0.011685)
(3.015010, 0.011685)
(3.020010, 0.011685)
(3.025010, 0.011685)
(3.030010, 0.011685)
(3.035010, 0.011685)
(3.040010, 0.011685)
(3.045010, 0.011685)
(3.050010, 0.011685)
(3.055010, 0.011685)
(3.060010, 0.011685)
(3.065010, 0.011685)
(3.070010, 0.011685)
(3.075010, 0.011685)
(3.080010, 0.011685)
(3.085010, 0.011685)
(3.090010, 0.011685)
(3.095010, 0.011685)
(3.100010, 0.011685)
(3.105010, 0.011685)
(3.110010, 0.011685)
(3.115010, 0.011684)
(3.120010, 0.011684)
(3.125010, 0.011684)
(3.130010, 0.011684)
(3.135010, 0.011684)
(3.140010, 0.011684)
(3.145010, 0.011684)
(3.150010, 0.011684)
(3.155010, 0.011684)
(3.160010, 0.011684)
(3.165010, 0.011684)
(3.170010, 0.011684)
(3.175010, 0.011684)
(3.180010, 0.011684)
(3.185010, 0.011684)
(3.190010, 0.011684)
(3.195010, 0.011684)
(3.200010, 0.011684)
(3.205010, 0.011684)
(3.210010, 0.011684)
(3.215010, 0.011684)
(3.220010, 0.011684)
(3.225010, 0.011684)
(3.230010, 0.011684)
(3.235010, 0.011684)
(3.240010, 0.011684)
(3.245010, 0.011684)
(3.250010, 0.011684)
(3.255010, 0.011684)
(3.260010, 0.011684)
(3.265010, 0.011684)
(3.270010, 0.011684)
(3.275010, 0.011684)
(3.280010, 0.011684)
(3.285010, 0.011685)
(3.290010, 0.011685)
(3.295010, 0.011685)
(3.300010, 0.011685)
(3.305010, 0.011685)
(3.310010, 0.011685)
(3.315010, 0.011685)
(3.320010, 0.011685)
(3.325010, 0.011685)
(3.330010, 0.011685)
(3.335010, 0.011685)
(3.340010, 0.011685)
(3.345010, 0.011685)
(3.350010, 0.011685)
(3.355010, 0.011685)
(3.360010, 0.011685)
(3.365010, 0.011685)
(3.370010, 0.011685)
(3.375010, 0.011685)
(3.380010, 0.011685)
(3.385010, 0.011685)
(3.390010, 0.011685)
(3.395010, 0.011685)
(3.400010, 0.011685)
(3.405010, 0.011685)
(3.410010, 0.011685)
(3.415010, 0.011685)
(3.420010, 0.011685)
(3.425010, 0.011685)
(3.430010, 0.011685)
(3.435010, 0.011685)
(3.440010, 0.011685)
(3.445010, 0.011685)
(3.450010, 0.011685)
(3.455010, 0.011685)
(3.460010, 0.011685)
(3.465010, 0.011685)
(3.470010, 0.011685)
(3.475010, 0.011685)
(3.480010, 0.011685)
(3.485010, 0.011685)
(3.490010, 0.011685)
(3.495010, 0.011685)
(3.500010, 0.011685)
(3.505010, 0.011685)
(3.510010, 0.011685)
(3.515010, 0.011685)
(3.520010, 0.011685)
(3.525010, 0.011685)
(3.530010, 0.011685)
(3.535010, 0.011685)
(3.540010, 0.011685)
(3.545010, 0.011685)
(3.550010, 0.011685)
(3.555010, 0.011685)
(3.560010, 0.011685)
(3.565010, 0.011685)
(3.570010, 0.011685)
(3.575010, 0.011685)
(3.580010, 0.011685)
(3.585010, 0.011685)
(3.590010, 0.011685)
(3.595010, 0.011685)
(3.600010, 0.011685)
(3.605010, 0.011685)
(3.610010, 0.011685)
(3.615010, 0.011685)
(3.620010, 0.011685)
(3.625010, 0.011685)
(3.630010, 0.011685)
(3.635010, 0.011685)
(3.640010, 0.011685)
(3.645010, 0.011685)
(3.650010, 0.011685)
(3.655010, 0.011685)
(3.660010, 0.011685)
(3.665010, 0.011685)
(3.670010, 0.011685)
(3.675010, 0.011685)
(3.680010, 0.011685)
(3.685010, 0.011685)
(3.690010, 0.011685)
(3.695010, 0.011685)
(3.700010, 0.011685)
(3.705010, 0.011685)
(3.710010, 0.011685)
(3.715010, 0.011685)
(3.720010, 0.011685)
(3.725010, 0.011685)
(3.730010, 0.011685)
(3.735010, 0.011685)
(3.740010, 0.011685)
(3.745010, 0.011685)
(3.750010, 0.011685)
(3.755010, 0.011685)
(3.760010, 0.011685)
(3.765010, 0.011685)
(3.770010, 0.011685)
(3.775010, 0.011685)
(3.780010, 0.011685)
(3.785010, 0.011685)
(3.790010, 0.011685)
(3.795010, 0.011685)
(3.800010, 0.011685)
(3.805010, 0.011685)
(3.810010, 0.011685)
(3.815010, 0.011685)
(3.820010, 0.011685)
(3.825010, 0.011685)
(3.830010, 0.011685)
(3.835010, 0.011685)
(3.840010, 0.011685)
(3.845010, 0.011685)
(3.850010, 0.011685)
(3.855010, 0.011685)
(3.860010, 0.011685)
(3.865010, 0.011685)
(3.870010, 0.011685)
(3.875010, 0.011685)
(3.880010, 0.011685)
(3.885010, 0.011685)
(3.890010, 0.011685)
(3.895010, 0.011685)
(3.900010, 0.011685)
(3.905010, 0.011685)
(3.910010, 0.011685)
(3.915010, 0.011685)
(3.920010, 0.011685)
(3.925010, 0.011685)
(3.930010, 0.011685)
(3.935010, 0.011685)
(3.940010, 0.011685)
(3.945010, 0.011685)
(3.950010, 0.011685)
(3.955010, 0.011685)
(3.960010, 0.011685)
(3.965010, 0.011685)
(3.970010, 0.011685)
(3.975010, 0.011685)
(3.980010, 0.011685)
(3.985010, 0.011685)
(3.990010, 0.011685)
(3.995010, 0.011685)
(4.000010, 0.011685)
(4.005010, 0.011685)
(4.010010, 0.011685)
(4.015010, 0.011685)
(4.020010, 0.011685)
(4.025010, 0.011685)
(4.030010, 0.011685)
(4.035010, 0.011685)
(4.040010, 0.011685)
(4.045010, 0.011685)
(4.050010, 0.011685)
(4.055010, 0.011685)
(4.060010, 0.011685)
(4.065010, 0.011685)
(4.070010, 0.011685)
(4.075010, 0.011685)
(4.080010, 0.011685)
(4.085010, 0.011685)
(4.090010, 0.011685)
(4.095010, 0.011685)
(4.100010, 0.011685)
(4.105010, 0.011685)
(4.110010, 0.011685)
(4.115010, 0.011685)
(4.120010, 0.011685)
(4.125010, 0.011685)
(4.130010, 0.011685)
(4.135010, 0.011685)
(4.140010, 0.011685)
(4.145010, 0.011685)
(4.150010, 0.011685)
(4.155010, 0.011685)
(4.160010, 0.011685)
(4.165010, 0.011685)
(4.170010, 0.011685)
(4.175010, 0.011685)
(4.180010, 0.011685)
(4.185010, 0.011685)
(4.190010, 0.011685)
(4.195010, 0.011685)
(4.200010, 0.011685)
(4.205010, 0.011685)
(4.210010, 0.011685)
(4.215010, 0.011685)
(4.220010, 0.011685)
(4.225010, 0.011685)
(4.230010, 0.011685)
(4.235010, 0.011685)
(4.240010, 0.011685)
(4.245010, 0.011685)
(4.250010, 0.011685)
(4.255010, 0.011685)
(4.260010, 0.011685)
(4.265010, 0.011685)
(4.270010, 0.011685)
(4.275010, 0.011685)
(4.280010, 0.011685)
(4.285010, 0.011685)
(4.290010, 0.011685)
(4.295010, 0.011685)
(4.300010, 0.011685)
(4.305010, 0.011685)
(4.310010, 0.011685)
(4.315010, 0.011685)
(4.320010, 0.011685)
(4.325010, 0.011685)
(4.330010, 0.011685)
(4.335010, 0.011685)
(4.340010, 0.011685)
(4.345010, 0.011685)
(4.350010, 0.011685)
(4.355010, 0.011685)
(4.360010, 0.011685)
(4.365010, 0.011685)
(4.370010, 0.011685)
(4.375010, 0.011685)
(4.380010, 0.011685)
(4.385010, 0.011685)
(4.390010, 0.011685)
(4.395010, 0.011685)
(4.400010, 0.011685)
(4.405010, 0.011685)
(4.410010, 0.011685)
(4.415010, 0.011685)
(4.420010, 0.011685)
(4.425010, 0.011685)
(4.430010, 0.011685)
(4.435010, 0.011685)
(4.440010, 0.011685)
(4.445010, 0.011685)
(4.450010, 0.011685)
(4.455010, 0.011685)
(4.460010, 0.011685)
(4.465010, 0.011685)
(4.470010, 0.011685)
(4.475010, 0.011685)
(4.480010, 0.011685)
(4.485010, 0.011685)
(4.490010, 0.011685)
(4.495010, 0.011685)
(4.500010, 0.011685)
(4.505010, 0.011685)
(4.510010, 0.011685)
(4.515010, 0.011685)
(4.520010, 0.011685)
(4.525010, 0.011685)
(4.530010, 0.011685)
(4.535010, 0.011685)
(4.540010, 0.011685)
(4.545010, 0.011685)
(4.550010, 0.011685)
(4.555010, 0.011685)
(4.560010, 0.011685)
(4.565010, 0.011685)
(4.570010, 0.011685)
(4.575010, 0.011685)
(4.580010, 0.011685)
(4.585010, 0.011685)
(4.590010, 0.011685)
(4.595010, 0.011685)
(4.600010, 0.011685)
(4.605010, 0.011685)
(4.610010, 0.011685)
(4.615010, 0.011685)
(4.620010, 0.011685)
(4.625010, 0.011685)
(4.630010, 0.011685)
(4.635010, 0.011685)
(4.640010, 0.011685)
(4.645010, 0.011685)
(4.650010, 0.011685)
(4.655010, 0.011685)
(4.660010, 0.011685)
(4.665010, 0.011685)
(4.670010, 0.011685)
(4.675010, 0.011685)
(4.680010, 0.011685)
(4.685010, 0.011685)
(4.690010, 0.011685)
(4.695010, 0.011685)
(4.700010, 0.011685)
(4.705010, 0.011685)
(4.710010, 0.011685)
(4.715010, 0.011685)
(4.720010, 0.011685)
(4.725010, 0.011685)
(4.730010, 0.011685)
(4.735010, 0.011685)
(4.740010, 0.011685)
(4.745010, 0.011685)
(4.750010, 0.011685)
(4.755010, 0.011685)
(4.760010, 0.011685)
(4.765010, 0.011685)
(4.770010, 0.011685)
(4.775010, 0.011685)
(4.780010, 0.011685)
(4.785010, 0.011685)
(4.790010, 0.011685)
(4.795010, 0.011685)
(4.800010, 0.011685)
(4.805010, 0.011685)
(4.810010, 0.011685)
(4.815010, 0.011685)
(4.820010, 0.011685)
(4.825010, 0.011685)
(4.830010, 0.011685)
(4.835010, 0.011685)
(4.840010, 0.011685)
(4.845010, 0.011685)
(4.850010, 0.011685)
(4.855010, 0.011685)
(4.860010, 0.011685)
(4.865010, 0.011685)
(4.870010, 0.011685)
(4.875010, 0.011685)
(4.880010, 0.011685)
(4.885010, 0.011685)
(4.890010, 0.011685)
(4.895010, 0.011685)
(4.900010, 0.011685)
(4.905010, 0.011685)
(4.910010, 0.011685)
(4.915010, 0.011685)
(4.920010, 0.011685)
(4.925010, 0.011685)
(4.930010, 0.011685)
(4.935010, 0.011685)
(4.940010, 0.011685)
(4.945010, 0.011685)
(4.950010, 0.011685)
(4.955010, 0.011685)
(4.960010, 0.011685)
(4.965010, 0.011685)
(4.970010, 0.011685)
(4.975010, 0.011685)
(4.980010, 0.011685)
(4.985010, 0.011685)
(4.990010, 0.011685)
(4.995010, 0.011685)
};
\addlegendentry{LS fit}

%% file: figs/data/nonlinearfitha050.tex
\addplot[
line width=1.0pt, color=gray, solid ]
coordinates {
(0.000020, 0.000020)
(0.002020, 0.002020)
(0.004020, 0.004020)
(0.006020, 0.006020)
(0.008020, 0.008020)
(0.010020, 0.010020)
(0.012020, 0.012013)
(0.014020, 0.013944)
(0.016020, 0.015630)
(0.018020, 0.016783)
(0.020020, 0.017181)
(0.022020, 0.016797)
(0.024020, 0.015771)
(0.026020, 0.014304)
(0.028020, 0.012573)
(0.030020, 0.010700)
(0.032020, 0.008757)
(0.034020, 0.006782)
(0.036020, 0.004793)
(0.038020, 0.002798)
(0.040020, 0.000802)
(0.042020, -0.001193)
(0.044020, -0.003181)
(0.046020, -0.005151)
(0.048020, -0.007076)
(0.050020, -0.008909)
(0.052020, -0.010581)
(0.054020, -0.012004)
(0.056020, -0.013092)
(0.058020, -0.013770)
(0.060020, -0.013997)
(0.062020, -0.013770)
(0.064020, -0.013120)
(0.066020, -0.012107)
(0.068020, -0.010799)
(0.070020, -0.009269)
(0.072020, -0.007581)
(0.074020, -0.005785)
(0.076020, -0.003922)
(0.078020, -0.002022)
(0.080020, -0.000109)
(0.082020, 0.001797)
(0.084020, 0.003671)
(0.086020, 0.005483)
(0.088020, 0.007196)
(0.090020, 0.008764)
(0.092020, 0.010138)
(0.094020, 0.011266)
(0.096020, 0.012105)
(0.098020, 0.012621)
(0.100020, 0.012796)
(0.102020, 0.012631)
(0.104020, 0.012141)
(0.106020, 0.011356)
(0.108020, 0.010313)
(0.110020, 0.009058)
(0.112020, 0.007633)
(0.114020, 0.006080)
(0.116020, 0.004438)
(0.118020, 0.002741)
(0.120020, 0.001024)
(0.122020, -0.000684)
(0.124020, -0.002349)
(0.126020, -0.003939)
(0.128020, -0.005416)
(0.130020, -0.006746)
(0.132020, -0.007893)
(0.134020, -0.008823)
(0.136020, -0.009510)
(0.138020, -0.009936)
(0.140020, -0.010090)
(0.142020, -0.009972)
(0.144020, -0.009593)
(0.146020, -0.008971)
(0.148020, -0.008131)
(0.150020, -0.007102)
(0.152020, -0.005916)
(0.154020, -0.004607)
(0.156020, -0.003208)
(0.158020, -0.001752)
(0.160020, -0.000273)
(0.162020, 0.001198)
(0.164020, 0.002628)
(0.166020, 0.003987)
(0.168020, 0.005242)
(0.170020, 0.006364)
(0.172020, 0.007326)
(0.174020, 0.008105)
(0.176020, 0.008681)
(0.178020, 0.009043)
(0.180020, 0.009183)
(0.182020, 0.009102)
(0.184020, 0.008807)
(0.186020, 0.008310)
(0.188020, 0.007629)
(0.190020, 0.006786)
(0.192020, 0.005806)
(0.194020, 0.004716)
(0.196020, 0.003544)
(0.198020, 0.002319)
(0.200020, 0.001071)
(0.202020, -0.000172)
(0.204020, -0.001380)
(0.206020, -0.002525)
(0.208020, -0.003582)
(0.210020, -0.004525)
(0.212020, -0.005334)
(0.214020, -0.005989)
(0.216020, -0.006477)
(0.218020, -0.006787)
(0.220020, -0.006916)
(0.222020, -0.006862)
(0.224020, -0.006631)
(0.226020, -0.006232)
(0.228020, -0.005679)
(0.230020, -0.004989)
(0.232020, -0.004182)
(0.234020, -0.003280)
(0.236020, -0.002305)
(0.238020, -0.001284)
(0.240020, -0.000241)
(0.242020, 0.000799)
(0.244020, 0.001810)
(0.246020, 0.002770)
(0.248020, 0.003656)
(0.250020, 0.004447)
(0.252020, 0.005126)
(0.254020, 0.005678)
(0.256020, 0.006092)
(0.258020, 0.006359)
(0.260020, 0.006476)
(0.262020, 0.006442)
(0.264020, 0.006261)
(0.266020, 0.005940)
(0.268020, 0.005491)
(0.270020, 0.004926)
(0.272020, 0.004262)
(0.274020, 0.003517)
(0.276020, 0.002710)
(0.278020, 0.001862)
(0.280020, 0.000995)
(0.282020, 0.000129)
(0.284020, -0.000715)
(0.286020, -0.001517)
(0.288020, -0.002257)
(0.290020, -0.002920)
(0.292020, -0.003490)
(0.294020, -0.003956)
(0.296020, -0.004306)
(0.298020, -0.004536)
(0.300020, -0.004641)
(0.302020, -0.004621)
(0.304020, -0.004480)
(0.306020, -0.004222)
(0.308020, -0.003856)
(0.310020, -0.003394)
(0.312020, -0.002849)
(0.314020, -0.002234)
(0.316020, -0.001567)
(0.318020, -0.000865)
(0.320020, -0.000145)
(0.322020, 0.000575)
(0.324020, 0.001277)
(0.326020, 0.001946)
(0.328020, 0.002564)
(0.330020, 0.003119)
(0.332020, 0.003597)
(0.334020, 0.003989)
(0.336020, 0.004286)
(0.338020, 0.004483)
(0.340020, 0.004577)
(0.342020, 0.004568)
(0.344020, 0.004457)
(0.346020, 0.004250)
(0.348020, 0.003953)
(0.350020, 0.003574)
(0.352020, 0.003126)
(0.354020, 0.002620)
(0.356020, 0.002069)
(0.358020, 0.001488)
(0.360020, 0.000891)
(0.362020, 0.000293)
(0.364020, -0.000291)
(0.366020, -0.000847)
(0.368020, -0.001364)
(0.370020, -0.001827)
(0.372020, -0.002228)
(0.374020, -0.002558)
(0.376020, -0.002810)
(0.378020, -0.002979)
(0.380020, -0.003062)
(0.382020, -0.003060)
(0.384020, -0.002974)
(0.386020, -0.002807)
(0.388020, -0.002566)
(0.390020, -0.002257)
(0.392020, -0.001889)
(0.394020, -0.001472)
(0.396020, -0.001017)
(0.398020, -0.000536)
(0.400020, -0.000042)
(0.402020, 0.000454)
(0.404020, 0.000940)
(0.406020, 0.001403)
(0.408020, 0.001834)
(0.410020, 0.002221)
(0.412020, 0.002557)
(0.414020, 0.002835)
(0.416020, 0.003048)
(0.418020, 0.003193)
(0.420020, 0.003267)
(0.422020, 0.003270)
(0.424020, 0.003203)
(0.426020, 0.003069)
(0.428020, 0.002872)
(0.430020, 0.002620)
(0.432020, 0.002317)
(0.434020, 0.001974)
(0.436020, 0.001599)
(0.438020, 0.001202)
(0.440020, 0.000792)
(0.442020, 0.000380)
(0.444020, -0.000023)
(0.446020, -0.000409)
(0.448020, -0.000768)
(0.450020, -0.001092)
(0.452020, -0.001373)
(0.454020, -0.001607)
(0.456020, -0.001787)
(0.458020, -0.001911)
(0.460020, -0.001976)
(0.462020, -0.001982)
(0.464020, -0.001931)
(0.466020, -0.001823)
(0.468020, -0.001664)
(0.470020, -0.001457)
(0.472020, -0.001209)
(0.474020, -0.000926)
(0.476020, -0.000617)
(0.478020, -0.000288)
(0.480020, 0.000051)
(0.482020, 0.000392)
(0.484020, 0.000727)
(0.486020, 0.001048)
(0.488020, 0.001348)
(0.490020, 0.001618)
(0.492020, 0.001854)
(0.494020, 0.002050)
(0.496020, 0.002203)
(0.498020, 0.002309)
(0.500020, 0.002366)
(0.502020, 0.002374)
(0.504020, 0.002335)
(0.506020, 0.002248)
(0.508020, 0.002119)
(0.510020, 0.001950)
(0.512020, 0.001746)
(0.514020, 0.001514)
(0.516020, 0.001259)
(0.518020, 0.000987)
(0.520020, 0.000706)
(0.522020, 0.000423)
(0.524020, 0.000145)
(0.526020, -0.000122)
(0.528020, -0.000372)
(0.530020, -0.000598)
(0.532020, -0.000795)
(0.534020, -0.000960)
(0.536020, -0.001089)
(0.538020, -0.001179)
(0.540020, -0.001230)
(0.542020, -0.001239)
(0.544020, -0.001209)
(0.546020, -0.001140)
(0.548020, -0.001035)
(0.550020, -0.000897)
(0.552020, -0.000730)
(0.554020, -0.000538)
(0.556020, -0.000328)
(0.558020, -0.000103)
(0.560020, 0.000129)
(0.562020, 0.000364)
(0.564020, 0.000595)
(0.566020, 0.000817)
(0.568020, 0.001025)
(0.570020, 0.001214)
(0.572020, 0.001379)
(0.574020, 0.001518)
(0.576020, 0.001627)
(0.578020, 0.001704)
(0.580020, 0.001747)
(0.582020, 0.001758)
(0.584020, 0.001734)
(0.586020, 0.001679)
(0.588020, 0.001594)
(0.590020, 0.001481)
(0.592020, 0.001344)
(0.594020, 0.001187)
(0.596020, 0.001013)
(0.598020, 0.000828)
(0.600020, 0.000635)
(0.602020, 0.000441)
(0.604020, 0.000249)
(0.606020, 0.000064)
(0.608020, -0.000109)
(0.610020, -0.000267)
(0.612020, -0.000405)
(0.614020, -0.000522)
(0.616020, -0.000614)
(0.618020, -0.000679)
(0.620020, -0.000717)
(0.622020, -0.000728)
(0.624020, -0.000710)
(0.626020, -0.000666)
(0.628020, -0.000597)
(0.630020, -0.000505)
(0.632020, -0.000393)
(0.634020, -0.000263)
(0.636020, -0.000120)
(0.638020, 0.000033)
(0.640020, 0.000192)
(0.642020, 0.000354)
(0.644020, 0.000513)
(0.646020, 0.000667)
(0.648020, 0.000811)
(0.650020, 0.000943)
(0.652020, 0.001059)
(0.654020, 0.001157)
(0.656020, 0.001234)
(0.658020, 0.001290)
(0.660020, 0.001323)
(0.662020, 0.001333)
(0.664020, 0.001320)
(0.666020, 0.001285)
(0.668020, 0.001229)
(0.670020, 0.001154)
(0.672020, 0.001062)
(0.674020, 0.000955)
(0.676020, 0.000837)
(0.678020, 0.000711)
(0.680020, 0.000579)
(0.682020, 0.000445)
(0.684020, 0.000313)
(0.686020, 0.000185)
(0.688020, 0.000065)
(0.690020, -0.000045)
(0.692020, -0.000142)
(0.694020, -0.000224)
(0.696020, -0.000290)
(0.698020, -0.000337)
(0.700020, -0.000366)
(0.702020, -0.000375)
(0.704020, -0.000366)
(0.706020, -0.000338)
(0.708020, -0.000293)
(0.710020, -0.000231)
(0.712020, -0.000156)
(0.714020, -0.000068)
(0.716020, 0.000029)
(0.718020, 0.000133)
(0.720020, 0.000242)
(0.722020, 0.000353)
(0.724020, 0.000463)
(0.726020, 0.000569)
(0.728020, 0.000670)
(0.730020, 0.000761)
(0.732020, 0.000843)
(0.734020, 0.000911)
(0.736020, 0.000967)
(0.738020, 0.001007)
(0.740020, 0.001032)
(0.742020, 0.001041)
(0.744020, 0.001034)
(0.746020, 0.001012)
(0.748020, 0.000975)
(0.750020, 0.000925)
(0.752020, 0.000863)
(0.754020, 0.000791)
(0.756020, 0.000711)
(0.758020, 0.000625)
(0.760020, 0.000534)
(0.762020, 0.000443)
(0.764020, 0.000351)
(0.766020, 0.000263)
(0.768020, 0.000180)
(0.770020, 0.000103)
(0.772020, 0.000035)
(0.774020, -0.000023)
(0.776020, -0.000069)
(0.778020, -0.000103)
(0.780020, -0.000125)
(0.782020, -0.000133)
(0.784020, -0.000128)
(0.786020, -0.000111)
(0.788020, -0.000081)
(0.790020, -0.000040)
(0.792020, 0.000010)
(0.794020, 0.000069)
(0.796020, 0.000136)
(0.798020, 0.000207)
(0.800020, 0.000281)
(0.802020, 0.000358)
(0.804020, 0.000433)
(0.806020, 0.000507)
(0.808020, 0.000576)
(0.810020, 0.000640)
(0.812020, 0.000697)
(0.814020, 0.000745)
(0.816020, 0.000785)
(0.818020, 0.000814)
(0.820020, 0.000832)
(0.822020, 0.000840)
(0.824020, 0.000836)
(0.826020, 0.000823)
(0.828020, 0.000799)
(0.830020, 0.000765)
(0.832020, 0.000724)
(0.834020, 0.000675)
(0.836020, 0.000621)
(0.838020, 0.000562)
(0.840020, 0.000500)
(0.842020, 0.000437)
(0.844020, 0.000374)
(0.846020, 0.000313)
(0.848020, 0.000255)
(0.850020, 0.000202)
(0.852020, 0.000155)
(0.854020, 0.000114)
(0.856020, 0.000081)
(0.858020, 0.000056)
(0.860020, 0.000040)
(0.862020, 0.000033)
(0.864020, 0.000036)
(0.866020, 0.000047)
(0.868020, 0.000066)
(0.870020, 0.000093)
(0.872020, 0.000127)
(0.874020, 0.000167)
(0.876020, 0.000212)
(0.878020, 0.000260)
(0.880020, 0.000311)
(0.882020, 0.000364)
(0.884020, 0.000416)
(0.886020, 0.000466)
(0.888020, 0.000515)
(0.890020, 0.000559)
(0.892020, 0.000599)
(0.894020, 0.000633)
(0.896020, 0.000661)
(0.898020, 0.000682)
(0.900020, 0.000695)
(0.902020, 0.000702)
(0.904020, 0.000700)
(0.906020, 0.000692)
(0.908020, 0.000676)
(0.910020, 0.000654)
(0.912020, 0.000626)
(0.914020, 0.000593)
(0.916020, 0.000556)
(0.918020, 0.000516)
(0.920020, 0.000474)
(0.922020, 0.000431)
(0.924020, 0.000388)
(0.926020, 0.000345)
(0.928020, 0.000305)
(0.930020, 0.000268)
(0.932020, 0.000235)
(0.934020, 0.000206)
(0.936020, 0.000183)
(0.938020, 0.000165)
(0.940020, 0.000154)
(0.942020, 0.000148)
(0.944020, 0.000149)
(0.946020, 0.000155)
(0.948020, 0.000168)
(0.950020, 0.000186)
(0.952020, 0.000209)
(0.954020, 0.000236)
(0.956020, 0.000266)
(0.958020, 0.000299)
(0.960020, 0.000334)
(0.962020, 0.000370)
};
\addlegendentry{nekrs} 
\addplot[
line width=1.0pt, color=blue, dotted ]
coordinates {
(0.000020, -0.000425)
(0.002020, 0.002319)
(0.004020, 0.004970)
(0.006020, 0.007466)
(0.008020, 0.009747)
(0.010020, 0.011763)
(0.012020, 0.013467)
(0.014020, 0.014824)
(0.016020, 0.015807)
(0.018020, 0.016397)
(0.020020, 0.016586)
(0.022020, 0.016376)
(0.024020, 0.015780)
(0.026020, 0.014817)
(0.028020, 0.013518)
(0.030020, 0.011919)
(0.032020, 0.010064)
(0.034020, 0.008002)
(0.036020, 0.005788)
(0.038020, 0.003478)
(0.040020, 0.001129)
(0.042020, -0.001201)
(0.044020, -0.003455)
(0.046020, -0.005579)
(0.048020, -0.007524)
(0.050020, -0.009245)
(0.052020, -0.010704)
(0.054020, -0.011871)
(0.056020, -0.012720)
(0.058020, -0.013236)
(0.060020, -0.013414)
(0.062020, -0.013252)
(0.064020, -0.012762)
(0.066020, -0.011960)
(0.068020, -0.010871)
(0.070020, -0.009527)
(0.072020, -0.007963)
(0.074020, -0.006223)
(0.076020, -0.004350)
(0.078020, -0.002393)
(0.080020, -0.000401)
(0.082020, 0.001576)
(0.084020, 0.003492)
(0.086020, 0.005300)
(0.088020, 0.006958)
(0.090020, 0.008428)
(0.092020, 0.009677)
(0.094020, 0.010679)
(0.096020, 0.011413)
(0.098020, 0.011865)
(0.100020, 0.012029)
(0.102020, 0.011906)
(0.104020, 0.011504)
(0.106020, 0.010836)
(0.108020, 0.009924)
(0.110020, 0.008794)
(0.112020, 0.007476)
(0.114020, 0.006006)
(0.116020, 0.004422)
(0.118020, 0.002765)
(0.120020, 0.001076)
(0.122020, -0.000602)
(0.124020, -0.002231)
(0.126020, -0.003769)
(0.128020, -0.005183)
(0.130020, -0.006438)
(0.132020, -0.007507)
(0.134020, -0.008367)
(0.136020, -0.009001)
(0.138020, -0.009397)
(0.140020, -0.009548)
(0.142020, -0.009456)
(0.144020, -0.009126)
(0.146020, -0.008570)
(0.148020, -0.007806)
(0.150020, -0.006856)
(0.152020, -0.005745)
(0.154020, -0.004504)
(0.156020, -0.003165)
(0.158020, -0.001762)
(0.160020, -0.000330)
(0.162020, 0.001095)
(0.164020, 0.002479)
(0.166020, 0.003789)
(0.168020, 0.004993)
(0.170020, 0.006065)
(0.172020, 0.006980)
(0.174020, 0.007719)
(0.176020, 0.008266)
(0.178020, 0.008611)
(0.180020, 0.008750)
(0.182020, 0.008682)
(0.184020, 0.008411)
(0.186020, 0.007949)
(0.188020, 0.007309)
(0.190020, 0.006511)
(0.192020, 0.005575)
(0.194020, 0.004527)
(0.196020, 0.003395)
(0.198020, 0.002206)
(0.200020, 0.000993)
(0.202020, -0.000217)
(0.204020, -0.001393)
(0.206020, -0.002508)
(0.208020, -0.003534)
(0.210020, -0.004449)
(0.212020, -0.005232)
(0.214020, -0.005866)
(0.216020, -0.006338)
(0.218020, -0.006640)
(0.220020, -0.006766)
(0.222020, -0.006717)
(0.224020, -0.006496)
(0.226020, -0.006111)
(0.228020, -0.005576)
(0.230020, -0.004904)
(0.232020, -0.004116)
(0.234020, -0.003231)
(0.236020, -0.002274)
(0.238020, -0.001268)
(0.240020, -0.000239)
(0.242020, 0.000788)
(0.244020, 0.001787)
(0.246020, 0.002736)
(0.248020, 0.003610)
(0.250020, 0.004391)
(0.252020, 0.005061)
(0.254020, 0.005606)
(0.256020, 0.006013)
(0.258020, 0.006276)
(0.260020, 0.006390)
(0.262020, 0.006355)
(0.264020, 0.006175)
(0.266020, 0.005856)
(0.268020, 0.005407)
(0.270020, 0.004843)
(0.272020, 0.004179)
(0.274020, 0.003432)
(0.276020, 0.002623)
(0.278020, 0.001771)
(0.280020, 0.000899)
(0.282020, 0.000027)
(0.284020, -0.000822)
(0.286020, -0.001629)
(0.288020, -0.002374)
(0.290020, -0.003041)
(0.292020, -0.003614)
(0.294020, -0.004081)
(0.296020, -0.004432)
(0.298020, -0.004661)
(0.300020, -0.004764)
(0.302020, -0.004741)
(0.304020, -0.004594)
(0.306020, -0.004329)
(0.308020, -0.003954)
(0.310020, -0.003480)
(0.312020, -0.002920)
(0.314020, -0.002290)
(0.316020, -0.001606)
(0.318020, -0.000885)
(0.320020, -0.000145)
(0.322020, 0.000594)
(0.324020, 0.001316)
(0.326020, 0.002002)
(0.328020, 0.002637)
(0.330020, 0.003206)
(0.332020, 0.003697)
(0.334020, 0.004097)
(0.336020, 0.004400)
(0.338020, 0.004600)
(0.340020, 0.004693)
(0.342020, 0.004678)
(0.344020, 0.004558)
(0.346020, 0.004338)
(0.348020, 0.004024)
(0.350020, 0.003626)
(0.352020, 0.003155)
(0.354020, 0.002623)
(0.356020, 0.002044)
(0.358020, 0.001434)
(0.360020, 0.000807)
(0.362020, 0.000180)
(0.364020, -0.000433)
(0.366020, -0.001017)
(0.368020, -0.001558)
(0.370020, -0.002044)
(0.372020, -0.002463)
(0.374020, -0.002807)
(0.376020, -0.003068)
(0.378020, -0.003242)
(0.380020, -0.003325)
(0.382020, -0.003317)
(0.384020, -0.003220)
(0.386020, -0.003037)
(0.388020, -0.002774)
(0.390020, -0.002440)
(0.392020, -0.002043)
(0.394020, -0.001594)
(0.396020, -0.001105)
(0.398020, -0.000589)
(0.400020, -0.000057)
(0.402020, 0.000475)
(0.404020, 0.000996)
(0.406020, 0.001493)
(0.408020, 0.001954)
(0.410020, 0.002368)
(0.412020, 0.002727)
(0.414020, 0.003021)
(0.416020, 0.003246)
(0.418020, 0.003397)
(0.420020, 0.003472)
(0.422020, 0.003469)
(0.424020, 0.003390)
(0.426020, 0.003238)
(0.428020, 0.003019)
(0.430020, 0.002738)
(0.432020, 0.002404)
(0.434020, 0.002025)
(0.436020, 0.001612)
(0.438020, 0.001174)
(0.440020, 0.000724)
(0.442020, 0.000272)
(0.444020, -0.000170)
(0.446020, -0.000593)
(0.448020, -0.000985)
(0.450020, -0.001339)
(0.452020, -0.001645)
(0.454020, -0.001898)
(0.456020, -0.002092)
(0.458020, -0.002223)
(0.460020, -0.002289)
(0.462020, -0.002290)
(0.464020, -0.002226)
(0.466020, -0.002101)
(0.468020, -0.001917)
(0.470020, -0.001682)
(0.472020, -0.001400)
(0.474020, -0.001081)
(0.476020, -0.000731)
(0.478020, -0.000361)
(0.480020, 0.000020)
(0.482020, 0.000404)
(0.484020, 0.000780)
(0.486020, 0.001139)
(0.488020, 0.001474)
(0.490020, 0.001775)
(0.492020, 0.002037)
(0.494020, 0.002254)
(0.496020, 0.002421)
(0.498020, 0.002535)
(0.500020, 0.002594)
(0.502020, 0.002597)
(0.504020, 0.002545)
(0.506020, 0.002441)
(0.508020, 0.002288)
(0.510020, 0.002090)
(0.512020, 0.001853)
(0.514020, 0.001584)
(0.516020, 0.001289)
(0.518020, 0.000975)
(0.520020, 0.000652)
(0.522020, 0.000327)
(0.524020, 0.000007)
(0.526020, -0.000298)
(0.528020, -0.000583)
(0.530020, -0.000840)
(0.532020, -0.001064)
(0.534020, -0.001250)
(0.536020, -0.001394)
(0.538020, -0.001493)
(0.540020, -0.001545)
(0.542020, -0.001550)
(0.544020, -0.001509)
(0.546020, -0.001423)
(0.548020, -0.001295)
(0.550020, -0.001129)
(0.552020, -0.000929)
(0.554020, -0.000702)
(0.556020, -0.000452)
(0.558020, -0.000187)
(0.560020, 0.000087)
(0.562020, 0.000363)
(0.564020, 0.000634)
(0.566020, 0.000894)
(0.568020, 0.001137)
(0.570020, 0.001356)
(0.572020, 0.001548)
(0.574020, 0.001707)
(0.576020, 0.001831)
(0.578020, 0.001916)
(0.580020, 0.001963)
(0.582020, 0.001969)
(0.584020, 0.001936)
(0.586020, 0.001864)
(0.588020, 0.001757)
(0.590020, 0.001618)
(0.592020, 0.001450)
(0.594020, 0.001258)
(0.596020, 0.001048)
(0.598020, 0.000823)
(0.600020, 0.000591)
(0.602020, 0.000357)
(0.604020, 0.000127)
(0.606020, -0.000094)
(0.608020, -0.000301)
(0.610020, -0.000488)
(0.612020, -0.000652)
(0.614020, -0.000788)
(0.616020, -0.000895)
(0.618020, -0.000969)
(0.620020, -0.001010)
(0.622020, -0.001017)
(0.624020, -0.000990)
(0.626020, -0.000931)
(0.628020, -0.000842)
(0.630020, -0.000725)
(0.632020, -0.000584)
(0.634020, -0.000422)
(0.636020, -0.000244)
(0.638020, -0.000054)
(0.640020, 0.000142)
(0.642020, 0.000341)
(0.644020, 0.000537)
(0.646020, 0.000725)
(0.648020, 0.000900)
(0.650020, 0.001060)
(0.652020, 0.001200)
(0.654020, 0.001317)
(0.656020, 0.001409)
(0.658020, 0.001473)
(0.660020, 0.001509)
(0.662020, 0.001516)
(0.664020, 0.001495)
(0.666020, 0.001447)
(0.668020, 0.001372)
(0.670020, 0.001274)
(0.672020, 0.001155)
(0.674020, 0.001019)
(0.676020, 0.000868)
(0.678020, 0.000708)
(0.680020, 0.000541)
(0.682020, 0.000372)
(0.684020, 0.000206)
(0.686020, 0.000047)
(0.688020, -0.000103)
(0.690020, -0.000239)
(0.692020, -0.000359)
(0.694020, -0.000459)
(0.696020, -0.000538)
(0.698020, -0.000594)
(0.700020, -0.000625)
(0.702020, -0.000633)
(0.704020, -0.000616)
(0.706020, -0.000576)
(0.708020, -0.000514)
(0.710020, -0.000432)
(0.712020, -0.000332)
(0.714020, -0.000216)
(0.716020, -0.000089)
(0.718020, 0.000047)
(0.720020, 0.000188)
(0.722020, 0.000331)
(0.724020, 0.000472)
(0.726020, 0.000608)
(0.728020, 0.000735)
(0.730020, 0.000851)
(0.732020, 0.000953)
(0.734020, 0.001039)
(0.736020, 0.001107)
(0.738020, 0.001155)
(0.740020, 0.001183)
(0.742020, 0.001190)
(0.744020, 0.001177)
(0.746020, 0.001144)
(0.748020, 0.001092)
(0.750020, 0.001023)
(0.752020, 0.000939)
(0.754020, 0.000842)
(0.756020, 0.000735)
(0.758020, 0.000620)
(0.760020, 0.000500)
(0.762020, 0.000379)
(0.764020, 0.000259)
(0.766020, 0.000144)
(0.768020, 0.000035)
(0.770020, -0.000064)
(0.772020, -0.000151)
(0.774020, -0.000225)
(0.776020, -0.000283)
(0.778020, -0.000325)
(0.780020, -0.000349)
(0.782020, -0.000356)
(0.784020, -0.000346)
(0.786020, -0.000319)
(0.788020, -0.000275)
(0.790020, -0.000218)
(0.792020, -0.000147)
(0.794020, -0.000065)
(0.796020, 0.000026)
(0.798020, 0.000123)
(0.800020, 0.000224)
(0.802020, 0.000327)
(0.804020, 0.000429)
(0.806020, 0.000527)
(0.808020, 0.000619)
(0.810020, 0.000704)
(0.812020, 0.000778)
(0.814020, 0.000841)
(0.816020, 0.000891)
(0.818020, 0.000927)
(0.820020, 0.000949)
(0.822020, 0.000956)
(0.824020, 0.000948)
(0.826020, 0.000925)
(0.828020, 0.000889)
(0.830020, 0.000841)
(0.832020, 0.000781)
(0.834020, 0.000712)
(0.836020, 0.000635)
(0.838020, 0.000553)
(0.840020, 0.000468)
(0.842020, 0.000380)
(0.844020, 0.000294)
(0.846020, 0.000210)
(0.848020, 0.000132)
(0.850020, 0.000060)
(0.852020, -0.000004)
(0.854020, -0.000058)
(0.856020, -0.000101)
(0.858020, -0.000132)
(0.860020, -0.000151)
(0.862020, -0.000157)
(0.864020, -0.000151)
(0.866020, -0.000132)
(0.868020, -0.000102)
(0.870020, -0.000062)
(0.872020, -0.000012)
(0.874020, 0.000046)
(0.876020, 0.000111)
(0.878020, 0.000181)
(0.880020, 0.000253)
(0.882020, 0.000327)
(0.884020, 0.000401)
(0.886020, 0.000472)
(0.888020, 0.000538)
(0.890020, 0.000600)
(0.892020, 0.000654)
(0.894020, 0.000700)
(0.896020, 0.000737)
(0.898020, 0.000764)
(0.900020, 0.000781)
(0.902020, 0.000787)
(0.904020, 0.000782)
(0.906020, 0.000767)
(0.908020, 0.000742)
(0.910020, 0.000708)
(0.912020, 0.000666)
(0.914020, 0.000617)
(0.916020, 0.000562)
(0.918020, 0.000503)
(0.920020, 0.000442)
(0.922020, 0.000379)
(0.924020, 0.000317)
(0.926020, 0.000256)
(0.928020, 0.000199)
(0.930020, 0.000147)
(0.932020, 0.000101)
(0.934020, 0.000061)
(0.936020, 0.000029)
(0.938020, 0.000006)
(0.940020, -0.000008)
(0.942020, -0.000014)
(0.944020, -0.000010)
(0.946020, 0.000002)
(0.948020, 0.000023)
(0.950020, 0.000052)
(0.952020, 0.000087)
(0.954020, 0.000128)
(0.956020, 0.000174)
(0.958020, 0.000224)
(0.960020, 0.000276)
(0.962020, 0.000329)
};
\addlegendentry{LS fit} 

%% file: tex/fig_his3.tex

\begin{figure}[h]
      \centering
      \begin{tikzpicture}
        \begin{axis}[
            xlabel={Time},
            ylabel={$u_z(0,0,t)$},
            width=4.8cm,
            height=6cm,
            xmax=2,
            legend pos= north east,
            ymajorgrids=true,
            grid style=dashed,
            y tick label style={/pgf/number format/fixed, /pgf/number format/precision=2},
            scaled y ticks=false
        ] \input{figs/data/lsfitha010.tex}
        \end{axis}
        \end{tikzpicture}
        \hfill
      \begin{tikzpicture}
      \begin{axis}[
          xlabel={Time},
          xmax =1,
          width=4.8cm,
          height=6cm,
          legend pos= north east,
          ymajorgrids=true,
          grid style=dashed,
          y tick label style={/pgf/number format/fixed, /pgf/number format/precision=2},
          scaled y ticks=false
      ] \input{figs/data/lsfitha050.tex}
      \end{axis}
      \end{tikzpicture}
      \hfill
      \begin{tikzpicture}
      \begin{axis}[
          xlabel={Time},
          xmax =1,
          width=4.8cm,
          height=6cm,
          legend pos= north east,
          ymajorgrids=true,
          grid style=dashed,
          y tick label style={/pgf/number format/fixed, /pgf/number format/precision=3},
          scaled y ticks=false
      ] \input{figs/data/lsfitha100.tex}
      \end{axis}
      \end{tikzpicture}
\caption{Time history of center-point velocity for $\ha = 10, 50, 100$ with
conducting walls ($r_w=10^{-5}$) for $Re = Rm = 1$.  A nonlinear fit is used to
approximte $\hat{u}_{0}$ and $\phi$ in the model $\hat{u}=\hat{u}_{0} e^{-s t}
\sin (\omega t+\phi)+\frac{1}{H a^{2}}$.  Parameters $s$ and $\omega$ are
calculated from (\ref{eq:eig1}).} \label{fig:time_his3}    
\end{figure}
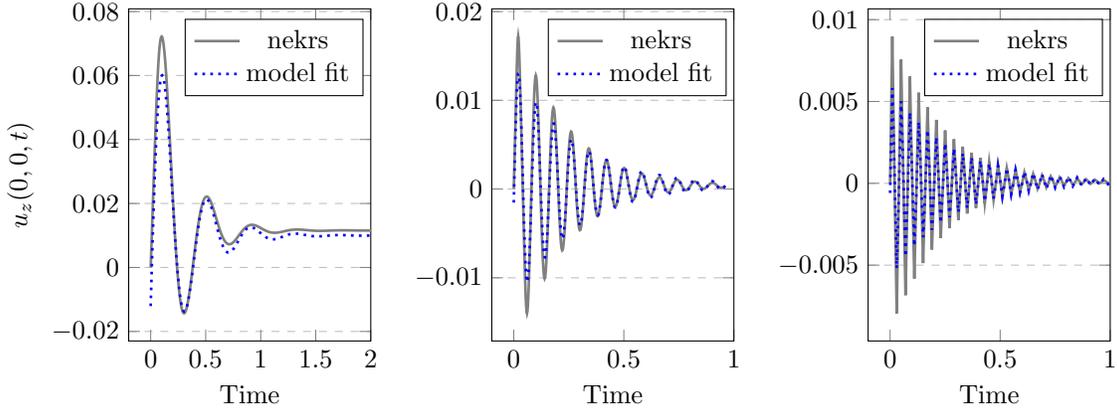

%% file: figs/data/lsfitha010.tex
\addplot[
line width=1.0pt, color=gray, solid ]
coordinates {
(0.000010, 0.000010)
(0.005010, 0.005010)
(0.010010, 0.010010)
(0.015010, 0.015010)
(0.020010, 0.020010)
(0.025010, 0.025008)
(0.030010, 0.029997)
(0.035010, 0.034954)
(0.040010, 0.039836)
(0.045010, 0.044585)
(0.050010, 0.049128)
(0.055010, 0.053394)
(0.060010, 0.057315)
(0.065010, 0.060835)
(0.070010, 0.063909)
(0.075010, 0.066507)
(0.080010, 0.068614)
(0.085010, 0.070224)
(0.090010, 0.071344)
(0.095010, 0.071988)
(0.100010, 0.072176)
(0.105010, 0.071932)
(0.110010, 0.071284)
(0.115010, 0.070263)
(0.120010, 0.068898)
(0.125010, 0.067221)
(0.130010, 0.065260)
(0.135010, 0.063046)
(0.140010, 0.060606)
(0.145010, 0.057968)
(0.150010, 0.055158)
(0.155010, 0.052201)
(0.160010, 0.049121)
(0.165010, 0.045941)
(0.170010, 0.042684)
(0.175010, 0.039371)
(0.180010, 0.036024)
(0.185010, 0.032663)
(0.190010, 0.029309)
(0.195010, 0.025980)
(0.200010, 0.022696)
(0.205010, 0.019475)
(0.210010, 0.016333)
(0.215010, 0.013286)
(0.220010, 0.010350)
(0.225010, 0.007539)
(0.230010, 0.004865)
(0.235010, 0.002340)
(0.240010, -0.000026)
(0.245010, -0.002225)
(0.250010, -0.004249)
(0.255010, -0.006093)
(0.260010, -0.007753)
(0.265010, -0.009225)
(0.270010, -0.010507)
(0.275010, -0.011601)
(0.280010, -0.012506)
(0.285010, -0.013224)
(0.290010, -0.013760)
(0.295010, -0.014118)
(0.300010, -0.014302)
(0.305010, -0.014320)
(0.310010, -0.014177)
(0.315010, -0.013882)
(0.320010, -0.013443)
(0.325010, -0.012869)
(0.330010, -0.012169)
(0.335010, -0.011353)
(0.340010, -0.010430)
(0.345010, -0.009412)
(0.350010, -0.008307)
(0.355010, -0.007128)
(0.360010, -0.005883)
(0.365010, -0.004584)
(0.370010, -0.003241)
(0.375010, -0.001864)
(0.380010, -0.000462)
(0.385010, 0.000954)
(0.390010, 0.002375)
(0.395010, 0.003792)
(0.400010, 0.005197)
(0.405010, 0.006581)
(0.410010, 0.007938)
(0.415010, 0.009259)
(0.420010, 0.010538)
(0.425010, 0.011769)
(0.430010, 0.012947)
(0.435010, 0.014066)
(0.440010, 0.015122)
(0.445010, 0.016111)
(0.450010, 0.017030)
(0.455010, 0.017877)
(0.460010, 0.018649)
(0.465010, 0.019344)
(0.470010, 0.019962)
(0.475010, 0.020502)
(0.480010, 0.020963)
(0.485010, 0.021348)
(0.490010, 0.021655)
(0.495010, 0.021888)
(0.500010, 0.022047)
(0.505010, 0.022135)
(0.510010, 0.022154)
(0.515010, 0.022107)
(0.520010, 0.021997)
(0.525010, 0.021828)
(0.530010, 0.021602)
(0.535010, 0.021325)
(0.540010, 0.020999)
(0.545010, 0.020629)
(0.550010, 0.020219)
(0.555010, 0.019773)
(0.560010, 0.019295)
(0.565010, 0.018791)
(0.570010, 0.018263)
(0.575010, 0.017717)
(0.580010, 0.017156)
(0.585010, 0.016584)
(0.590010, 0.016006)
(0.595010, 0.015426)
(0.600010, 0.014846)
(0.605010, 0.014271)
(0.610010, 0.013704)
(0.615010, 0.013147)
(0.620010, 0.012605)
(0.625010, 0.012080)
(0.630010, 0.011573)
(0.635010, 0.011089)
(0.640010, 0.010628)
(0.645010, 0.010192)
(0.650010, 0.009784)
(0.655010, 0.009403)
(0.660010, 0.009053)
(0.665010, 0.008733)
(0.670010, 0.008444)
(0.675010, 0.008187)
(0.680010, 0.007961)
(0.685010, 0.007767)
(0.690010, 0.007605)
(0.695010, 0.007475)
(0.700010, 0.007375)
(0.705010, 0.007305)
(0.710010, 0.007265)
(0.715010, 0.007253)
(0.720010, 0.007268)
(0.725010, 0.007308)
(0.730010, 0.007374)
(0.735010, 0.007462)
(0.740010, 0.007572)
(0.745010, 0.007702)
(0.750010, 0.007851)
(0.755010, 0.008015)
(0.760010, 0.008195)
(0.765010, 0.008388)
(0.770010, 0.008592)
(0.775010, 0.008806)
(0.780010, 0.009028)
(0.785010, 0.009256)
(0.790010, 0.009489)
(0.795010, 0.009724)
(0.800010, 0.009961)
(0.805010, 0.010198)
(0.810010, 0.010433)
(0.815010, 0.010665)
(0.820010, 0.010893)
(0.825010, 0.011115)
(0.830010, 0.011330)
(0.835010, 0.011538)
(0.840010, 0.011737)
(0.845010, 0.011927)
(0.850010, 0.012106)
(0.855010, 0.012275)
(0.860010, 0.012432)
(0.865010, 0.012577)
(0.870010, 0.012709)
(0.875010, 0.012829)
(0.880010, 0.012936)
(0.885010, 0.013030)
(0.890010, 0.013111)
(0.895010, 0.013179)
(0.900010, 0.013234)
(0.905010, 0.013276)
(0.910010, 0.013307)
(0.915010, 0.013325)
(0.920010, 0.013331)
(0.925010, 0.013327)
(0.930010, 0.013312)
(0.935010, 0.013286)
(0.940010, 0.013252)
(0.945010, 0.013208)
(0.950010, 0.013156)
(0.955010, 0.013096)
(0.960010, 0.013030)
(0.965010, 0.012957)
(0.970010, 0.012879)
(0.975010, 0.012796)
(0.980010, 0.012710)
(0.985010, 0.012619)
(0.990010, 0.012527)
(0.995010, 0.012432)
(1.000010, 0.012336)
(1.005010, 0.012239)
(1.010010, 0.012142)
(1.015010, 0.012046)
(1.020010, 0.011951)
(1.025010, 0.011858)
(1.030010, 0.011766)
(1.035010, 0.011678)
(1.040010, 0.011592)
(1.045010, 0.011510)
(1.050010, 0.011432)
(1.055010, 0.011358)
(1.060010, 0.011289)
(1.065010, 0.011224)
(1.070010, 0.011164)
(1.075010, 0.011109)
(1.080010, 0.011059)
(1.085010, 0.011015)
(1.090010, 0.010976)
(1.095010, 0.010942)
(1.100010, 0.010913)
(1.105010, 0.010890)
(1.110010, 0.010871)
(1.115010, 0.010858)
(1.120010, 0.010850)
(1.125010, 0.010847)
(1.130010, 0.010848)
(1.135010, 0.010853)
(1.140010, 0.010863)
(1.145010, 0.010876)
(1.150010, 0.010894)
(1.155010, 0.010914)
(1.160010, 0.010938)
(1.165010, 0.010965)
(1.170010, 0.010994)
(1.175010, 0.011025)
(1.180010, 0.011059)
(1.185010, 0.011094)
(1.190010, 0.011131)
(1.195010, 0.011168)
(1.200010, 0.011207)
(1.205010, 0.011246)
(1.210010, 0.011285)
(1.215010, 0.011325)
(1.220010, 0.011364)
(1.225010, 0.011403)
(1.230010, 0.011441)
(1.235010, 0.011478)
(1.240010, 0.011515)
(1.245010, 0.011550)
(1.250010, 0.011583)
(1.255010, 0.011615)
(1.260010, 0.011646)
(1.265010, 0.011674)
(1.270010, 0.011701)
(1.275010, 0.011726)
(1.280010, 0.011748)
(1.285010, 0.011769)
(1.290010, 0.011788)
(1.295010, 0.011804)
(1.300010, 0.011818)
(1.305010, 0.011830)
(1.310010, 0.011840)
(1.315010, 0.011847)
(1.320010, 0.011853)
(1.325010, 0.011857)
(1.330010, 0.011858)
(1.335010, 0.011858)
(1.340010, 0.011856)
(1.345010, 0.011852)
(1.350010, 0.011847)
(1.355010, 0.011840)
(1.360010, 0.011832)
(1.365010, 0.011822)
(1.370010, 0.011811)
(1.375010, 0.011800)
(1.380010, 0.011787)
(1.385010, 0.011773)
(1.390010, 0.011759)
(1.395010, 0.011744)
(1.400010, 0.011729)
(1.405010, 0.011713)
(1.410010, 0.011697)
(1.415010, 0.011681)
(1.420010, 0.011665)
(1.425010, 0.011649)
(1.430010, 0.011633)
(1.435010, 0.011617)
(1.440010, 0.011602)
(1.445010, 0.011587)
(1.450010, 0.011572)
(1.455010, 0.011558)
(1.460010, 0.011545)
(1.465010, 0.011533)
(1.470010, 0.011521)
(1.475010, 0.011510)
(1.480010, 0.011500)
(1.485010, 0.011490)
(1.490010, 0.011481)
(1.495010, 0.011474)
(1.500010, 0.011467)
(1.505010, 0.011461)
(1.510010, 0.011456)
(1.515010, 0.011452)
(1.520010, 0.011448)
(1.525010, 0.011446)
(1.530010, 0.011444)
(1.535010, 0.011444)
(1.540010, 0.011443)
(1.545010, 0.011444)
(1.550010, 0.011445)
(1.555010, 0.011448)
(1.560010, 0.011450)
(1.565010, 0.011453)
(1.570010, 0.011457)
(1.575010, 0.011462)
(1.580010, 0.011466)
(1.585010, 0.011471)
(1.590010, 0.011477)
(1.595010, 0.011483)
(1.600010, 0.011489)
(1.605010, 0.011495)
(1.610010, 0.011501)
(1.615010, 0.011508)
(1.620010, 0.011514)
(1.625010, 0.011521)
(1.630010, 0.011527)
(1.635010, 0.011534)
(1.640010, 0.011540)
(1.645010, 0.011546)
(1.650010, 0.011552)
(1.655010, 0.011558)
(1.660010, 0.011564)
(1.665010, 0.011569)
(1.670010, 0.011575)
(1.675010, 0.011579)
(1.680010, 0.011584)
(1.685010, 0.011588)
(1.690010, 0.011592)
(1.695010, 0.011596)
(1.700010, 0.011599)
(1.705010, 0.011602)
(1.710010, 0.011604)
(1.715010, 0.011606)
(1.720010, 0.011608)
(1.725010, 0.011609)
(1.730010, 0.011610)
(1.735010, 0.011611)
(1.740010, 0.011611)
(1.745010, 0.011611)
(1.750010, 0.011611)
(1.755010, 0.011610)
(1.760010, 0.011610)
(1.765010, 0.011609)
(1.770010, 0.011607)
(1.775010, 0.011606)
(1.780010, 0.011604)
(1.785010, 0.011602)
(1.790010, 0.011600)
(1.795010, 0.011598)
(1.800010, 0.011595)
(1.805010, 0.011593)
(1.810010, 0.011590)
(1.815010, 0.011588)
(1.820010, 0.011585)
(1.825010, 0.011582)
(1.830010, 0.011580)
(1.835010, 0.011577)
(1.840010, 0.011574)
(1.845010, 0.011571)
(1.850010, 0.011569)
(1.855010, 0.011566)
(1.860010, 0.011564)
(1.865010, 0.011562)
(1.870010, 0.011559)
(1.875010, 0.011557)
(1.880010, 0.011555)
(1.885010, 0.011553)
(1.890010, 0.011551)
(1.895010, 0.011550)
(1.900010, 0.011548)
(1.905010, 0.011547)
(1.910010, 0.011546)
(1.915010, 0.011545)
(1.920010, 0.011544)
(1.925010, 0.011543)
(1.930010, 0.011543)
(1.935010, 0.011542)
(1.940010, 0.011542)
(1.945010, 0.011542)
(1.950010, 0.011541)
(1.955010, 0.011542)
(1.960010, 0.011542)
(1.965010, 0.011542)
(1.970010, 0.011542)
(1.975010, 0.011543)
(1.980010, 0.011543)
(1.985010, 0.011544)
(1.990010, 0.011545)
(1.995010, 0.011546)
(2.000010, 0.011547)
(2.005010, 0.011548)
(2.010010, 0.011549)
(2.015010, 0.011550)
(2.020010, 0.011551)
(2.025010, 0.011552)
(2.030010, 0.011553)
(2.035010, 0.011554)
(2.040010, 0.011555)
(2.045010, 0.011556)
(2.050010, 0.011557)
(2.055010, 0.011558)
(2.060010, 0.011559)
(2.065010, 0.011560)
(2.070010, 0.011561)
(2.075010, 0.011562)
(2.080010, 0.011563)
(2.085010, 0.011564)
(2.090010, 0.011564)
(2.095010, 0.011565)
(2.100010, 0.011566)
(2.105010, 0.011566)
(2.110010, 0.011567)
(2.115010, 0.011567)
(2.120010, 0.011568)
(2.125010, 0.011568)
(2.130010, 0.011568)
(2.135010, 0.011569)
(2.140010, 0.011569)
(2.145010, 0.011569)
(2.150010, 0.011569)
(2.155010, 0.011569)
(2.160010, 0.011569)
(2.165010, 0.011569)
(2.170010, 0.011569)
(2.175010, 0.011568)
(2.180010, 0.011568)
(2.185010, 0.011568)
(2.190010, 0.011568)
(2.195010, 0.011567)
(2.200010, 0.011567)
(2.205010, 0.011567)
(2.210010, 0.011566)
(2.215010, 0.011566)
(2.220010, 0.011565)
(2.225010, 0.011565)
(2.230010, 0.011564)
(2.235010, 0.011564)
(2.240010, 0.011564)
(2.245010, 0.011563)
(2.250010, 0.011563)
(2.255010, 0.011562)
(2.260010, 0.011562)
(2.265010, 0.011561)
(2.270010, 0.011561)
(2.275010, 0.011560)
(2.280010, 0.011560)
(2.285010, 0.011560)
(2.290010, 0.011559)
(2.295010, 0.011559)
(2.300010, 0.011559)
(2.305010, 0.011558)
(2.310010, 0.011558)
(2.315010, 0.011558)
(2.320010, 0.011558)
(2.325010, 0.011557)
(2.330010, 0.011557)
(2.335010, 0.011557)
(2.340010, 0.011557)
(2.345010, 0.011557)
(2.350010, 0.011557)
(2.355010, 0.011557)
(2.360010, 0.011557)
(2.365010, 0.011557)
(2.370010, 0.011557)
(2.375010, 0.011557)
(2.380010, 0.011557)
(2.385010, 0.011557)
(2.390010, 0.011557)
(2.395010, 0.011557)
(2.400010, 0.011557)
(2.405010, 0.011557)
(2.410010, 0.011557)
(2.415010, 0.011558)
(2.420010, 0.011558)
(2.425010, 0.011558)
(2.430010, 0.011558)
(2.435010, 0.011558)
(2.440010, 0.011558)
(2.445010, 0.011559)
(2.450010, 0.011559)
(2.455010, 0.011559)
(2.460010, 0.011559)
(2.465010, 0.011559)
(2.470010, 0.011559)
(2.475010, 0.011560)
(2.480010, 0.011560)
(2.485010, 0.011560)
(2.490010, 0.011560)
(2.495010, 0.011560)
(2.500010, 0.011560)
(2.505010, 0.011560)
(2.510010, 0.011560)
(2.515010, 0.011561)
(2.520010, 0.011561)
(2.525010, 0.011561)
(2.530010, 0.011561)
(2.535010, 0.011561)
(2.540010, 0.011561)
(2.545010, 0.011561)
(2.550010, 0.011561)
(2.555010, 0.011561)
(2.560010, 0.011561)
(2.565010, 0.011561)
(2.570010, 0.011561)
(2.575010, 0.011561)
(2.580010, 0.011561)
(2.585010, 0.011561)
(2.590010, 0.011561)
(2.595010, 0.011561)
(2.600010, 0.011561)
(2.605010, 0.011561)
(2.610010, 0.011560)
(2.615010, 0.011560)
(2.620010, 0.011560)
(2.625010, 0.011560)
(2.630010, 0.011560)
(2.635010, 0.011560)
(2.640010, 0.011560)
(2.645010, 0.011560)
(2.650010, 0.011560)
(2.655010, 0.011560)
(2.660010, 0.011560)
(2.665010, 0.011560)
(2.670010, 0.011560)
(2.675010, 0.011559)
(2.680010, 0.011559)
(2.685010, 0.011559)
(2.690010, 0.011559)
(2.695010, 0.011559)
(2.700010, 0.011559)
(2.705010, 0.011559)
(2.710010, 0.011559)
(2.715010, 0.011559)
(2.720010, 0.011559)
(2.725010, 0.011559)
(2.730010, 0.011559)
(2.735010, 0.011559)
(2.740010, 0.011559)
(2.745010, 0.011559)
(2.750010, 0.011559)
(2.755010, 0.011559)
(2.760010, 0.011559)
(2.765010, 0.011558)
(2.770010, 0.011558)
(2.775010, 0.011558)
(2.780010, 0.011558)
(2.785010, 0.011558)
(2.790010, 0.011558)
(2.795010, 0.011558)
(2.800010, 0.011558)
(2.805010, 0.011558)
(2.810010, 0.011558)
(2.815010, 0.011558)
(2.820010, 0.011559)
(2.825010, 0.011559)
(2.830010, 0.011559)
(2.835010, 0.011559)
(2.840010, 0.011559)
(2.845010, 0.011559)
(2.850010, 0.011559)
(2.855010, 0.011559)
(2.860010, 0.011559)
(2.865010, 0.011559)
(2.870010, 0.011559)
(2.875010, 0.011559)
(2.880010, 0.011559)
(2.885010, 0.011559)
(2.890010, 0.011559)
(2.895010, 0.011559)
(2.900010, 0.011559)
(2.905010, 0.011559)
(2.910010, 0.011559)
(2.915010, 0.011559)
(2.920010, 0.011559)
(2.925010, 0.011559)
(2.930010, 0.011559)
(2.935010, 0.011559)
(2.940010, 0.011559)
(2.945010, 0.011559)
(2.950010, 0.011559)
(2.955010, 0.011559)
(2.960010, 0.011559)
(2.965010, 0.011559)
(2.970010, 0.011559)
(2.975010, 0.011559)
(2.980010, 0.011559)
(2.985010, 0.011559)
(2.990010, 0.011559)
(2.995010, 0.011559)
(3.000010, 0.011559)
(3.005010, 0.011559)
(3.010010, 0.011559)
(3.015010, 0.011559)
(3.020010, 0.011559)
(3.025010, 0.011559)
(3.030010, 0.011559)
(3.035010, 0.011559)
(3.040010, 0.011559)
(3.045010, 0.011559)
(3.050010, 0.011559)
(3.055010, 0.011559)
(3.060010, 0.011559)
(3.065010, 0.011559)
(3.070010, 0.011558)
(3.075010, 0.011558)
(3.080010, 0.011558)
(3.085010, 0.011558)
(3.090010, 0.011558)
(3.095010, 0.011558)
(3.100010, 0.011558)
(3.105010, 0.011558)
(3.110010, 0.011558)
(3.115010, 0.011558)
(3.120010, 0.011558)
(3.125010, 0.011558)
(3.130010, 0.011558)
(3.135010, 0.011558)
(3.140010, 0.011558)
(3.145010, 0.011558)
(3.150010, 0.011558)
(3.155010, 0.011558)
(3.160010, 0.011558)
(3.165010, 0.011558)
(3.170010, 0.011558)
(3.175010, 0.011558)
(3.180010, 0.011558)
(3.185010, 0.011558)
(3.190010, 0.011558)
(3.195010, 0.011558)
(3.200010, 0.011558)
(3.205010, 0.011558)
(3.210010, 0.011558)
(3.215010, 0.011558)
(3.220010, 0.011558)
(3.225010, 0.011558)
(3.230010, 0.011558)
(3.235010, 0.011558)
(3.240010, 0.011558)
(3.245010, 0.011558)
(3.250010, 0.011558)
(3.255010, 0.011558)
(3.260010, 0.011558)
(3.265010, 0.011558)
(3.270010, 0.011558)
(3.275010, 0.011558)
(3.280010, 0.011558)
(3.285010, 0.011558)
(3.290010, 0.011558)
(3.295010, 0.011558)
(3.300010, 0.011558)
(3.305010, 0.011558)
(3.310010, 0.011558)
(3.315010, 0.011558)
(3.320010, 0.011558)
(3.325010, 0.011558)
(3.330010, 0.011558)
(3.335010, 0.011558)
(3.340010, 0.011558)
(3.345010, 0.011558)
(3.350010, 0.011558)
(3.355010, 0.011558)
(3.360010, 0.011558)
(3.365010, 0.011558)
(3.370010, 0.011558)
(3.375010, 0.011558)
(3.380010, 0.011558)
(3.385010, 0.011558)
(3.390010, 0.011558)
(3.395010, 0.011558)
(3.400010, 0.011558)
(3.405010, 0.011558)
(3.410010, 0.011558)
(3.415010, 0.011558)
(3.420010, 0.011558)
(3.425010, 0.011558)
(3.430010, 0.011558)
(3.435010, 0.011558)
(3.440010, 0.011558)
(3.445010, 0.011558)
(3.450010, 0.011558)
(3.455010, 0.011558)
(3.460010, 0.011558)
(3.465010, 0.011558)
(3.470010, 0.011558)
(3.475010, 0.011558)
(3.480010, 0.011558)
(3.485010, 0.011558)
(3.490010, 0.011558)
(3.495010, 0.011558)
(3.500010, 0.011558)
(3.505010, 0.011558)
(3.510010, 0.011558)
(3.515010, 0.011558)
(3.520010, 0.011558)
(3.525010, 0.011558)
(3.530010, 0.011558)
(3.535010, 0.011558)
(3.540010, 0.011558)
(3.545010, 0.011558)
(3.550010, 0.011558)
(3.555010, 0.011558)
(3.560010, 0.011558)
(3.565010, 0.011558)
(3.570010, 0.011558)
(3.575010, 0.011558)
(3.580010, 0.011558)
(3.585010, 0.011558)
(3.590010, 0.011558)
(3.595010, 0.011558)
(3.600010, 0.011557)
(3.605010, 0.011557)
(3.610010, 0.011557)
(3.615010, 0.011557)
(3.620010, 0.011557)
(3.625010, 0.011557)
(3.630010, 0.011557)
(3.635010, 0.011557)
(3.640010, 0.011557)
(3.645010, 0.011557)
(3.650010, 0.011557)
(3.655010, 0.011557)
(3.660010, 0.011557)
(3.665010, 0.011557)
(3.670010, 0.011557)
(3.675010, 0.011557)
(3.680010, 0.011557)
(3.685010, 0.011557)
(3.690010, 0.011557)
(3.695010, 0.011557)
(3.700010, 0.011557)
(3.705010, 0.011557)
(3.710010, 0.011557)
(3.715010, 0.011557)
(3.720010, 0.011557)
(3.725010, 0.011557)
(3.730010, 0.011557)
(3.735010, 0.011557)
(3.740010, 0.011557)
(3.745010, 0.011557)
(3.750010, 0.011557)
(3.755010, 0.011557)
(3.760010, 0.011557)
(3.765010, 0.011557)
(3.770010, 0.011557)
(3.775010, 0.011557)
(3.780010, 0.011557)
(3.785010, 0.011557)
(3.790010, 0.011557)
(3.795010, 0.011557)
(3.800010, 0.011557)
(3.805010, 0.011557)
(3.810010, 0.011557)
(3.815010, 0.011557)
(3.820010, 0.011557)
(3.825010, 0.011557)
(3.830010, 0.011557)
(3.835010, 0.011557)
(3.840010, 0.011557)
(3.845010, 0.011557)
(3.850010, 0.011557)
(3.855010, 0.011557)
(3.860010, 0.011557)
(3.865010, 0.011557)
(3.870010, 0.011557)
(3.875010, 0.011557)
(3.880010, 0.011557)
(3.885010, 0.011557)
(3.890010, 0.011557)
(3.895010, 0.011557)
(3.900010, 0.011557)
(3.905010, 0.011557)
(3.910010, 0.011557)
(3.915010, 0.011557)
(3.920010, 0.011557)
(3.925010, 0.011557)
(3.930010, 0.011557)
(3.935010, 0.011557)
(3.940010, 0.011557)
(3.945010, 0.011557)
(3.950010, 0.011557)
(3.955010, 0.011557)
(3.960010, 0.011557)
(3.965010, 0.011557)
(3.970010, 0.011557)
(3.975010, 0.011557)
(3.980010, 0.011557)
(3.985010, 0.011557)
(3.990010, 0.011557)
(3.995010, 0.011557)
(4.000010, 0.011557)
(4.005010, 0.011557)
(4.010010, 0.011557)
(4.015010, 0.011557)
(4.020010, 0.011557)
(4.025010, 0.011557)
(4.030010, 0.011557)
(4.035010, 0.011557)
(4.040010, 0.011557)
(4.045010, 0.011557)
(4.050010, 0.011557)
(4.055010, 0.011557)
(4.060010, 0.011557)
(4.065010, 0.011557)
(4.070010, 0.011557)
(4.075010, 0.011557)
(4.080010, 0.011557)
(4.085010, 0.011557)
(4.090010, 0.011557)
(4.095010, 0.011557)
(4.100010, 0.011557)
(4.105010, 0.011557)
(4.110010, 0.011557)
(4.115010, 0.011557)
(4.120010, 0.011557)
(4.125010, 0.011557)
(4.130010, 0.011557)
(4.135010, 0.011557)
(4.140010, 0.011557)
(4.145010, 0.011557)
(4.150010, 0.011557)
(4.155010, 0.011557)
(4.160010, 0.011557)
(4.165010, 0.011557)
(4.170010, 0.011557)
(4.175010, 0.011557)
(4.180010, 0.011557)
(4.185010, 0.011557)
(4.190010, 0.011557)
(4.195010, 0.011557)
(4.200010, 0.011557)
(4.205010, 0.011557)
(4.210010, 0.011557)
(4.215010, 0.011557)
(4.220010, 0.011557)
(4.225010, 0.011557)
(4.230010, 0.011557)
(4.235010, 0.011557)
(4.240010, 0.011557)
(4.245010, 0.011557)
(4.250010, 0.011557)
(4.255010, 0.011557)
(4.260010, 0.011557)
(4.265010, 0.011557)
(4.270010, 0.011557)
(4.275010, 0.011557)
(4.280010, 0.011557)
(4.285010, 0.011557)
(4.290010, 0.011557)
(4.295010, 0.011557)
(4.300010, 0.011557)
(4.305010, 0.011557)
(4.310010, 0.011557)
(4.315010, 0.011557)
(4.320010, 0.011557)
(4.325010, 0.011557)
(4.330010, 0.011557)
(4.335010, 0.011557)
(4.340010, 0.011557)
(4.345010, 0.011557)
(4.350010, 0.011557)
(4.355010, 0.011557)
(4.360010, 0.011557)
(4.365010, 0.011557)
(4.370010, 0.011557)
(4.375010, 0.011557)
(4.380010, 0.011556)
(4.385010, 0.011556)
(4.390010, 0.011556)
(4.395010, 0.011556)
(4.400010, 0.011556)
(4.405010, 0.011556)
(4.410010, 0.011556)
(4.415010, 0.011556)
(4.420010, 0.011556)
(4.425010, 0.011556)
(4.430010, 0.011556)
(4.435010, 0.011556)
(4.440010, 0.011556)
(4.445010, 0.011556)
(4.450010, 0.011556)
(4.455010, 0.011556)
(4.460010, 0.011556)
(4.465010, 0.011556)
(4.470010, 0.011556)
(4.475010, 0.011556)
(4.480010, 0.011556)
(4.485010, 0.011556)
(4.490010, 0.011556)
(4.495010, 0.011556)
(4.500010, 0.011556)
(4.505010, 0.011556)
(4.510010, 0.011556)
(4.515010, 0.011556)
(4.520010, 0.011556)
(4.525010, 0.011556)
(4.530010, 0.011556)
(4.535010, 0.011556)
(4.540010, 0.011556)
(4.545010, 0.011556)
(4.550010, 0.011556)
(4.555010, 0.011556)
(4.560010, 0.011556)
(4.565010, 0.011556)
(4.570010, 0.011556)
(4.575010, 0.011556)
(4.580010, 0.011556)
(4.585010, 0.011556)
(4.590010, 0.011556)
(4.595010, 0.011556)
(4.600010, 0.011556)
(4.605010, 0.011556)
(4.610010, 0.011556)
(4.615010, 0.011556)
(4.620010, 0.011556)
(4.625010, 0.011556)
(4.630010, 0.011556)
(4.635010, 0.011556)
(4.640010, 0.011556)
(4.645010, 0.011556)
(4.650010, 0.011556)
(4.655010, 0.011556)
(4.660010, 0.011556)
(4.665010, 0.011556)
(4.670010, 0.011556)
(4.675010, 0.011556)
(4.680010, 0.011556)
(4.685010, 0.011556)
(4.690010, 0.011556)
(4.695010, 0.011556)
(4.700010, 0.011556)
(4.705010, 0.011556)
(4.710010, 0.011556)
(4.715010, 0.011556)
(4.720010, 0.011556)
(4.725010, 0.011556)
(4.730010, 0.011556)
(4.735010, 0.011556)
(4.740010, 0.011556)
(4.745010, 0.011556)
(4.750010, 0.011556)
(4.755010, 0.011556)
(4.760010, 0.011556)
(4.765010, 0.011556)
(4.770010, 0.011556)
(4.775010, 0.011556)
(4.780010, 0.011556)
(4.785010, 0.011556)
(4.790010, 0.011556)
(4.795010, 0.011556)
(4.800010, 0.011556)
(4.805010, 0.011556)
(4.810010, 0.011556)
(4.815010, 0.011556)
(4.820010, 0.011556)
(4.825010, 0.011556)
(4.830010, 0.011556)
(4.835010, 0.011556)
(4.840010, 0.011556)
(4.845010, 0.011556)
(4.850010, 0.011556)
(4.855010, 0.011556)
(4.860010, 0.011556)
(4.865010, 0.011556)
(4.870010, 0.011556)
(4.875010, 0.011556)
(4.880010, 0.011556)
(4.885010, 0.011556)
(4.890010, 0.011556)
(4.895010, 0.011556)
(4.900010, 0.011556)
(4.905010, 0.011556)
(4.910010, 0.011556)
(4.915010, 0.011556)
(4.920010, 0.011556)
(4.925010, 0.011556)
(4.930010, 0.011556)
(4.935010, 0.011556)
(4.940010, 0.011556)
(4.945010, 0.011556)
(4.950010, 0.011556)
(4.955010, 0.011556)
(4.960010, 0.011556)
(4.965010, 0.011556)
(4.970010, 0.011556)
(4.975010, 0.011556)
(4.980010, 0.011556)
(4.985010, 0.011556)
(4.990010, 0.011556)
(4.995010, 0.011556)
};
\addlegendentry{nekrs} 
\addplot[
line width=1.0pt, color=blue, dotted ]
coordinates {
(0.000010, -0.012069)
(0.005010, -0.006048)
(0.010010, -0.000144)
(0.015010, 0.005609)
(0.020010, 0.011181)
(0.025010, 0.016544)
(0.030010, 0.021670)
(0.035010, 0.026536)
(0.040010, 0.031121)
(0.045010, 0.035406)
(0.050010, 0.039375)
(0.055010, 0.043014)
(0.060010, 0.046312)
(0.065010, 0.049262)
(0.070010, 0.051856)
(0.075010, 0.054092)
(0.080010, 0.055968)
(0.085010, 0.057486)
(0.090010, 0.058649)
(0.095010, 0.059462)
(0.100010, 0.059934)
(0.105010, 0.060073)
(0.110010, 0.059891)
(0.115010, 0.059400)
(0.120010, 0.058616)
(0.125010, 0.057553)
(0.130010, 0.056228)
(0.135010, 0.054660)
(0.140010, 0.052866)
(0.145010, 0.050868)
(0.150010, 0.048683)
(0.155010, 0.046334)
(0.160010, 0.043841)
(0.165010, 0.041225)
(0.170010, 0.038507)
(0.175010, 0.035707)
(0.180010, 0.032847)
(0.185010, 0.029946)
(0.190010, 0.027025)
(0.195010, 0.024103)
(0.200010, 0.021198)
(0.205010, 0.018327)
(0.210010, 0.015509)
(0.215010, 0.012758)
(0.220010, 0.010089)
(0.225010, 0.007517)
(0.230010, 0.005054)
(0.235010, 0.002712)
(0.240010, 0.000501)
(0.245010, -0.001570)
(0.250010, -0.003493)
(0.255010, -0.005260)
(0.260010, -0.006867)
(0.265010, -0.008309)
(0.270010, -0.009583)
(0.275010, -0.010687)
(0.280010, -0.011620)
(0.285010, -0.012383)
(0.290010, -0.012977)
(0.295010, -0.013404)
(0.300010, -0.013668)
(0.305010, -0.013773)
(0.310010, -0.013724)
(0.315010, -0.013527)
(0.320010, -0.013189)
(0.325010, -0.012716)
(0.330010, -0.012118)
(0.335010, -0.011401)
(0.340010, -0.010575)
(0.345010, -0.009650)
(0.350010, -0.008633)
(0.355010, -0.007537)
(0.360010, -0.006369)
(0.365010, -0.005140)
(0.370010, -0.003861)
(0.375010, -0.002540)
(0.380010, -0.001188)
(0.385010, 0.000185)
(0.390010, 0.001570)
(0.395010, 0.002959)
(0.400010, 0.004341)
(0.405010, 0.005709)
(0.410010, 0.007054)
(0.415010, 0.008369)
(0.420010, 0.009646)
(0.425010, 0.010879)
(0.430010, 0.012062)
(0.435010, 0.013189)
(0.440010, 0.014255)
(0.445010, 0.015255)
(0.450010, 0.016186)
(0.455010, 0.017044)
(0.460010, 0.017826)
(0.465010, 0.018530)
(0.470010, 0.019155)
(0.475010, 0.019700)
(0.480010, 0.020163)
(0.485010, 0.020545)
(0.490010, 0.020847)
(0.495010, 0.021070)
(0.500010, 0.021214)
(0.505010, 0.021282)
(0.510010, 0.021277)
(0.515010, 0.021201)
(0.520010, 0.021057)
(0.525010, 0.020848)
(0.530010, 0.020578)
(0.535010, 0.020251)
(0.540010, 0.019872)
(0.545010, 0.019444)
(0.550010, 0.018971)
(0.555010, 0.018460)
(0.560010, 0.017913)
(0.565010, 0.017337)
(0.570010, 0.016735)
(0.575010, 0.016112)
(0.580010, 0.015473)
(0.585010, 0.014823)
(0.590010, 0.014167)
(0.595010, 0.013507)
(0.600010, 0.012850)
(0.605010, 0.012198)
(0.610010, 0.011557)
(0.615010, 0.010929)
(0.620010, 0.010317)
(0.625010, 0.009726)
(0.630010, 0.009158)
(0.635010, 0.008616)
(0.640010, 0.008103)
(0.645010, 0.007620)
(0.650010, 0.007170)
(0.655010, 0.006754)
(0.660010, 0.006373)
(0.665010, 0.006029)
(0.670010, 0.005723)
(0.675010, 0.005455)
(0.680010, 0.005226)
(0.685010, 0.005035)
(0.690010, 0.004882)
(0.695010, 0.004767)
(0.700010, 0.004689)
(0.705010, 0.004648)
(0.710010, 0.004642)
(0.715010, 0.004670)
(0.720010, 0.004730)
(0.725010, 0.004822)
(0.730010, 0.004943)
(0.735010, 0.005091)
(0.740010, 0.005266)
(0.745010, 0.005463)
(0.750010, 0.005683)
(0.755010, 0.005921)
(0.760010, 0.006177)
(0.765010, 0.006447)
(0.770010, 0.006730)
(0.775010, 0.007024)
(0.780010, 0.007325)
(0.785010, 0.007633)
(0.790010, 0.007944)
(0.795010, 0.008257)
(0.800010, 0.008569)
(0.805010, 0.008879)
(0.810010, 0.009185)
(0.815010, 0.009486)
(0.820010, 0.009778)
(0.825010, 0.010061)
(0.830010, 0.010334)
(0.835010, 0.010594)
(0.840010, 0.010841)
(0.845010, 0.011074)
(0.850010, 0.011292)
(0.855010, 0.011494)
(0.860010, 0.011679)
(0.865010, 0.011846)
(0.870010, 0.011996)
(0.875010, 0.012128)
(0.880010, 0.012241)
(0.885010, 0.012337)
(0.890010, 0.012414)
(0.895010, 0.012473)
(0.900010, 0.012514)
(0.905010, 0.012538)
(0.910010, 0.012545)
(0.915010, 0.012536)
(0.920010, 0.012511)
(0.925010, 0.012471)
(0.930010, 0.012417)
(0.935010, 0.012349)
(0.940010, 0.012270)
(0.945010, 0.012178)
(0.950010, 0.012077)
(0.955010, 0.011966)
(0.960010, 0.011846)
(0.965010, 0.011720)
(0.970010, 0.011587)
(0.975010, 0.011448)
(0.980010, 0.011306)
(0.985010, 0.011161)
(0.990010, 0.011013)
(0.995010, 0.010865)
(1.000010, 0.010716)
(1.005010, 0.010569)
(1.010010, 0.010423)
(1.015010, 0.010279)
(1.020010, 0.010140)
(1.025010, 0.010004)
(1.030010, 0.009873)
(1.035010, 0.009748)
(1.040010, 0.009629)
(1.045010, 0.009517)
(1.050010, 0.009412)
(1.055010, 0.009314)
(1.060010, 0.009224)
(1.065010, 0.009142)
(1.070010, 0.009069)
(1.075010, 0.009004)
(1.080010, 0.008948)
(1.085010, 0.008901)
(1.090010, 0.008862)
(1.095010, 0.008832)
(1.100010, 0.008810)
(1.105010, 0.008797)
(1.110010, 0.008792)
(1.115010, 0.008794)
(1.120010, 0.008804)
(1.125010, 0.008821)
(1.130010, 0.008845)
(1.135010, 0.008876)
(1.140010, 0.008912)
(1.145010, 0.008954)
(1.150010, 0.009001)
(1.155010, 0.009053)
(1.160010, 0.009109)
(1.165010, 0.009168)
(1.170010, 0.009231)
(1.175010, 0.009296)
(1.180010, 0.009363)
(1.185010, 0.009432)
(1.190010, 0.009501)
(1.195010, 0.009572)
(1.200010, 0.009642)
(1.205010, 0.009713)
(1.210010, 0.009782)
(1.215010, 0.009851)
(1.220010, 0.009917)
(1.225010, 0.009982)
(1.230010, 0.010045)
(1.235010, 0.010105)
(1.240010, 0.010162)
(1.245010, 0.010216)
(1.250010, 0.010267)
(1.255010, 0.010315)
(1.260010, 0.010358)
(1.265010, 0.010398)
(1.270010, 0.010434)
(1.275010, 0.010465)
(1.280010, 0.010493)
(1.285010, 0.010517)
(1.290010, 0.010536)
(1.295010, 0.010551)
(1.300010, 0.010563)
(1.305010, 0.010570)
(1.310010, 0.010573)
(1.315010, 0.010573)
(1.320010, 0.010569)
(1.325010, 0.010562)
(1.330010, 0.010551)
(1.335010, 0.010538)
(1.340010, 0.010521)
(1.345010, 0.010502)
(1.350010, 0.010480)
(1.355010, 0.010456)
(1.360010, 0.010430)
(1.365010, 0.010402)
(1.370010, 0.010373)
(1.375010, 0.010342)
(1.380010, 0.010311)
(1.385010, 0.010278)
(1.390010, 0.010245)
(1.395010, 0.010212)
(1.400010, 0.010178)
(1.405010, 0.010145)
(1.410010, 0.010112)
(1.415010, 0.010079)
(1.420010, 0.010047)
(1.425010, 0.010016)
(1.430010, 0.009986)
(1.435010, 0.009957)
(1.440010, 0.009930)
(1.445010, 0.009903)
(1.450010, 0.009879)
(1.455010, 0.009856)
(1.460010, 0.009835)
(1.465010, 0.009816)
(1.470010, 0.009798)
(1.475010, 0.009783)
(1.480010, 0.009769)
(1.485010, 0.009757)
(1.490010, 0.009748)
(1.495010, 0.009740)
(1.500010, 0.009734)
(1.505010, 0.009730)
(1.510010, 0.009728)
(1.515010, 0.009728)
(1.520010, 0.009729)
(1.525010, 0.009732)
(1.530010, 0.009737)
(1.535010, 0.009743)
(1.540010, 0.009751)
(1.545010, 0.009759)
(1.550010, 0.009769)
(1.555010, 0.009781)
(1.560010, 0.009793)
(1.565010, 0.009806)
(1.570010, 0.009819)
(1.575010, 0.009834)
(1.580010, 0.009849)
(1.585010, 0.009864)
(1.590010, 0.009880)
(1.595010, 0.009896)
(1.600010, 0.009911)
(1.605010, 0.009927)
(1.610010, 0.009943)
(1.615010, 0.009959)
(1.620010, 0.009974)
(1.625010, 0.009989)
(1.630010, 0.010003)
(1.635010, 0.010017)
(1.640010, 0.010030)
(1.645010, 0.010043)
(1.650010, 0.010055)
(1.655010, 0.010066)
(1.660010, 0.010076)
(1.665010, 0.010085)
(1.670010, 0.010094)
(1.675010, 0.010101)
(1.680010, 0.010108)
(1.685010, 0.010114)
(1.690010, 0.010119)
(1.695010, 0.010123)
(1.700010, 0.010126)
(1.705010, 0.010128)
(1.710010, 0.010129)
(1.715010, 0.010129)
(1.720010, 0.010129)
(1.725010, 0.010128)
(1.730010, 0.010126)
(1.735010, 0.010123)
(1.740010, 0.010119)
(1.745010, 0.010115)
(1.750010, 0.010111)
(1.755010, 0.010106)
(1.760010, 0.010100)
(1.765010, 0.010094)
(1.770010, 0.010087)
(1.775010, 0.010081)
(1.780010, 0.010074)
(1.785010, 0.010066)
(1.790010, 0.010059)
(1.795010, 0.010051)
(1.800010, 0.010044)
(1.805010, 0.010036)
(1.810010, 0.010029)
(1.815010, 0.010021)
(1.820010, 0.010014)
(1.825010, 0.010007)
(1.830010, 0.010000)
(1.835010, 0.009994)
(1.840010, 0.009987)
(1.845010, 0.009981)
(1.850010, 0.009975)
(1.855010, 0.009970)
(1.860010, 0.009965)
(1.865010, 0.009961)
(1.870010, 0.009956)
(1.875010, 0.009953)
(1.880010, 0.009949)
(1.885010, 0.009947)
(1.890010, 0.009944)
(1.895010, 0.009942)
(1.900010, 0.009941)
(1.905010, 0.009940)
(1.910010, 0.009939)
(1.915010, 0.009939)
(1.920010, 0.009939)
(1.925010, 0.009939)
(1.930010, 0.009940)
(1.935010, 0.009941)
(1.940010, 0.009943)
(1.945010, 0.009945)
(1.950010, 0.009947)
(1.955010, 0.009949)
(1.960010, 0.009952)
(1.965010, 0.009955)
(1.970010, 0.009958)
(1.975010, 0.009961)
(1.980010, 0.009964)
(1.985010, 0.009968)
(1.990010, 0.009971)
(1.995010, 0.009975)
(2.000010, 0.009978)
(2.005010, 0.009982)
(2.010010, 0.009985)
(2.015010, 0.009989)
(2.020010, 0.009992)
(2.025010, 0.009996)
(2.030010, 0.009999)
(2.035010, 0.010002)
(2.040010, 0.010005)
(2.045010, 0.010008)
(2.050010, 0.010011)
(2.055010, 0.010014)
(2.060010, 0.010016)
(2.065010, 0.010018)
(2.070010, 0.010020)
(2.075010, 0.010022)
(2.080010, 0.010024)
(2.085010, 0.010025)
(2.090010, 0.010026)
(2.095010, 0.010027)
(2.100010, 0.010028)
(2.105010, 0.010029)
(2.110010, 0.010029)
(2.115010, 0.010029)
(2.120010, 0.010029)
(2.125010, 0.010029)
(2.130010, 0.010029)
(2.135010, 0.010028)
(2.140010, 0.010027)
(2.145010, 0.010026)
(2.150010, 0.010025)
(2.155010, 0.010024)
(2.160010, 0.010023)
(2.165010, 0.010022)
(2.170010, 0.010020)
(2.175010, 0.010019)
(2.180010, 0.010017)
(2.185010, 0.010016)
(2.190010, 0.010014)
(2.195010, 0.010012)
(2.200010, 0.010011)
(2.205010, 0.010009)
(2.210010, 0.010007)
(2.215010, 0.010006)
(2.220010, 0.010004)
(2.225010, 0.010002)
(2.230010, 0.010001)
(2.235010, 0.009999)
(2.240010, 0.009998)
(2.245010, 0.009996)
(2.250010, 0.009995)
(2.255010, 0.009994)
(2.260010, 0.009993)
(2.265010, 0.009992)
(2.270010, 0.009991)
(2.275010, 0.009990)
(2.280010, 0.009989)
(2.285010, 0.009988)
(2.290010, 0.009988)
(2.295010, 0.009987)
(2.300010, 0.009987)
(2.305010, 0.009986)
(2.310010, 0.009986)
(2.315010, 0.009986)
(2.320010, 0.009986)
(2.325010, 0.009986)
(2.330010, 0.009986)
(2.335010, 0.009987)
(2.340010, 0.009987)
(2.345010, 0.009987)
(2.350010, 0.009988)
(2.355010, 0.009988)
(2.360010, 0.009989)
(2.365010, 0.009989)
(2.370010, 0.009990)
(2.375010, 0.009991)
(2.380010, 0.009992)
(2.385010, 0.009992)
(2.390010, 0.009993)
(2.395010, 0.009994)
(2.400010, 0.009995)
(2.405010, 0.009996)
(2.410010, 0.009996)
(2.415010, 0.009997)
(2.420010, 0.009998)
(2.425010, 0.009999)
(2.430010, 0.009999)
(2.435010, 0.010000)
(2.440010, 0.010001)
(2.445010, 0.010002)
(2.450010, 0.010002)
(2.455010, 0.010003)
(2.460010, 0.010003)
(2.465010, 0.010004)
(2.470010, 0.010004)
(2.475010, 0.010005)
(2.480010, 0.010005)
(2.485010, 0.010005)
(2.490010, 0.010006)
(2.495010, 0.010006)
(2.500010, 0.010006)
(2.505010, 0.010006)
(2.510010, 0.010006)
(2.515010, 0.010007)
(2.520010, 0.010007)
(2.525010, 0.010007)
(2.530010, 0.010006)
(2.535010, 0.010006)
(2.540010, 0.010006)
(2.545010, 0.010006)
(2.550010, 0.010006)
(2.555010, 0.010006)
(2.560010, 0.010005)
(2.565010, 0.010005)
(2.570010, 0.010005)
(2.575010, 0.010004)
(2.580010, 0.010004)
(2.585010, 0.010004)
(2.590010, 0.010003)
(2.595010, 0.010003)
(2.600010, 0.010003)
(2.605010, 0.010002)
(2.610010, 0.010002)
(2.615010, 0.010001)
(2.620010, 0.010001)
(2.625010, 0.010001)
(2.630010, 0.010000)
(2.635010, 0.010000)
(2.640010, 0.010000)
(2.645010, 0.009999)
(2.650010, 0.009999)
(2.655010, 0.009999)
(2.660010, 0.009998)
(2.665010, 0.009998)
(2.670010, 0.009998)
(2.675010, 0.009998)
(2.680010, 0.009998)
(2.685010, 0.009997)
(2.690010, 0.009997)
(2.695010, 0.009997)
(2.700010, 0.009997)
(2.705010, 0.009997)
(2.710010, 0.009997)
(2.715010, 0.009997)
(2.720010, 0.009997)
(2.725010, 0.009997)
(2.730010, 0.009997)
(2.735010, 0.009997)
(2.740010, 0.009997)
(2.745010, 0.009997)
(2.750010, 0.009997)
(2.755010, 0.009997)
(2.760010, 0.009997)
(2.765010, 0.009998)
(2.770010, 0.009998)
(2.775010, 0.009998)
(2.780010, 0.009998)
(2.785010, 0.009998)
(2.790010, 0.009998)
(2.795010, 0.009999)
(2.800010, 0.009999)
(2.805010, 0.009999)
(2.810010, 0.009999)
(2.815010, 0.009999)
(2.820010, 0.009999)
(2.825010, 0.010000)
(2.830010, 0.010000)
(2.835010, 0.010000)
(2.840010, 0.010000)
(2.845010, 0.010000)
(2.850010, 0.010000)
(2.855010, 0.010001)
(2.860010, 0.010001)
(2.865010, 0.010001)
(2.870010, 0.010001)
(2.875010, 0.010001)
(2.880010, 0.010001)
(2.885010, 0.010001)
(2.890010, 0.010001)
(2.895010, 0.010001)
(2.900010, 0.010001)
(2.905010, 0.010001)
(2.910010, 0.010001)
(2.915010, 0.010001)
(2.920010, 0.010001)
(2.925010, 0.010001)
(2.930010, 0.010001)
(2.935010, 0.010001)
(2.940010, 0.010001)
(2.945010, 0.010001)
(2.950010, 0.010001)
(2.955010, 0.010001)
(2.960010, 0.010001)
(2.965010, 0.010001)
(2.970010, 0.010001)
(2.975010, 0.010001)
(2.980010, 0.010001)
(2.985010, 0.010001)
(2.990010, 0.010001)
(2.995010, 0.010001)
(3.000010, 0.010001)
(3.005010, 0.010001)
(3.010010, 0.010000)
(3.015010, 0.010000)
(3.020010, 0.010000)
(3.025010, 0.010000)
(3.030010, 0.010000)
(3.035010, 0.010000)
(3.040010, 0.010000)
(3.045010, 0.010000)
(3.050010, 0.010000)
(3.055010, 0.010000)
(3.060010, 0.010000)
(3.065010, 0.010000)
(3.070010, 0.010000)
(3.075010, 0.010000)
(3.080010, 0.009999)
(3.085010, 0.009999)
(3.090010, 0.009999)
(3.095010, 0.009999)
(3.100010, 0.009999)
(3.105010, 0.009999)
(3.110010, 0.009999)
(3.115010, 0.009999)
(3.120010, 0.009999)
(3.125010, 0.009999)
(3.130010, 0.009999)
(3.135010, 0.009999)
(3.140010, 0.009999)
(3.145010, 0.009999)
(3.150010, 0.009999)
(3.155010, 0.009999)
(3.160010, 0.009999)
(3.165010, 0.009999)
(3.170010, 0.009999)
(3.175010, 0.009999)
(3.180010, 0.010000)
(3.185010, 0.010000)
(3.190010, 0.010000)
(3.195010, 0.010000)
(3.200010, 0.010000)
(3.205010, 0.010000)
(3.210010, 0.010000)
(3.215010, 0.010000)
(3.220010, 0.010000)
(3.225010, 0.010000)
(3.230010, 0.010000)
(3.235010, 0.010000)
(3.240010, 0.010000)
(3.245010, 0.010000)
(3.250010, 0.010000)
(3.255010, 0.010000)
(3.260010, 0.010000)
(3.265010, 0.010000)
(3.270010, 0.010000)
(3.275010, 0.010000)
(3.280010, 0.010000)
(3.285010, 0.010000)
(3.290010, 0.010000)
(3.295010, 0.010000)
(3.300010, 0.010000)
(3.305010, 0.010000)
(3.310010, 0.010000)
(3.315010, 0.010000)
(3.320010, 0.010000)
(3.325010, 0.010000)
(3.330010, 0.010000)
(3.335010, 0.010000)
(3.340010, 0.010000)
(3.345010, 0.010000)
(3.350010, 0.010000)
(3.355010, 0.010000)
(3.360010, 0.010000)
(3.365010, 0.010000)
(3.370010, 0.010000)
(3.375010, 0.010000)
(3.380010, 0.010000)
(3.385010, 0.010000)
(3.390010, 0.010000)
(3.395010, 0.010000)
(3.400010, 0.010000)
(3.405010, 0.010000)
(3.410010, 0.010000)
(3.415010, 0.010000)
(3.420010, 0.010000)
(3.425010, 0.010000)
(3.430010, 0.010000)
(3.435010, 0.010000)
(3.440010, 0.010000)
(3.445010, 0.010000)
(3.450010, 0.010000)
(3.455010, 0.010000)
(3.460010, 0.010000)
(3.465010, 0.010000)
(3.470010, 0.010000)
(3.475010, 0.010000)
(3.480010, 0.010000)
(3.485010, 0.010000)
(3.490010, 0.010000)
(3.495010, 0.010000)
(3.500010, 0.010000)
(3.505010, 0.010000)
(3.510010, 0.010000)
(3.515010, 0.010000)
(3.520010, 0.010000)
(3.525010, 0.010000)
(3.530010, 0.010000)
(3.535010, 0.010000)
(3.540010, 0.010000)
(3.545010, 0.010000)
(3.550010, 0.010000)
(3.555010, 0.010000)
(3.560010, 0.010000)
(3.565010, 0.010000)
(3.570010, 0.010000)
(3.575010, 0.010000)
(3.580010, 0.010000)
(3.585010, 0.010000)
(3.590010, 0.010000)
(3.595010, 0.010000)
(3.600010, 0.010000)
(3.605010, 0.010000)
(3.610010, 0.010000)
(3.615010, 0.010000)
(3.620010, 0.010000)
(3.625010, 0.010000)
(3.630010, 0.010000)
(3.635010, 0.010000)
(3.640010, 0.010000)
(3.645010, 0.010000)
(3.650010, 0.010000)
(3.655010, 0.010000)
(3.660010, 0.010000)
(3.665010, 0.010000)
(3.670010, 0.010000)
(3.675010, 0.010000)
(3.680010, 0.010000)
(3.685010, 0.010000)
(3.690010, 0.010000)
(3.695010, 0.010000)
(3.700010, 0.010000)
(3.705010, 0.010000)
(3.710010, 0.010000)
(3.715010, 0.010000)
(3.720010, 0.010000)
(3.725010, 0.010000)
(3.730010, 0.010000)
(3.735010, 0.010000)
(3.740010, 0.010000)
(3.745010, 0.010000)
(3.750010, 0.010000)
(3.755010, 0.010000)
(3.760010, 0.010000)
(3.765010, 0.010000)
(3.770010, 0.010000)
(3.775010, 0.010000)
(3.780010, 0.010000)
(3.785010, 0.010000)
(3.790010, 0.010000)
(3.795010, 0.010000)
(3.800010, 0.010000)
(3.805010, 0.010000)
(3.810010, 0.010000)
(3.815010, 0.010000)
(3.820010, 0.010000)
(3.825010, 0.010000)
(3.830010, 0.010000)
(3.835010, 0.010000)
(3.840010, 0.010000)
(3.845010, 0.010000)
(3.850010, 0.010000)
(3.855010, 0.010000)
(3.860010, 0.010000)
(3.865010, 0.010000)
(3.870010, 0.010000)
(3.875010, 0.010000)
(3.880010, 0.010000)
(3.885010, 0.010000)
(3.890010, 0.010000)
(3.895010, 0.010000)
(3.900010, 0.010000)
(3.905010, 0.010000)
(3.910010, 0.010000)
(3.915010, 0.010000)
(3.920010, 0.010000)
(3.925010, 0.010000)
(3.930010, 0.010000)
(3.935010, 0.010000)
(3.940010, 0.010000)
(3.945010, 0.010000)
(3.950010, 0.010000)
(3.955010, 0.010000)
(3.960010, 0.010000)
(3.965010, 0.010000)
(3.970010, 0.010000)
(3.975010, 0.010000)
(3.980010, 0.010000)
(3.985010, 0.010000)
(3.990010, 0.010000)
(3.995010, 0.010000)
(4.000010, 0.010000)
(4.005010, 0.010000)
(4.010010, 0.010000)
(4.015010, 0.010000)
(4.020010, 0.010000)
(4.025010, 0.010000)
(4.030010, 0.010000)
(4.035010, 0.010000)
(4.040010, 0.010000)
(4.045010, 0.010000)
(4.050010, 0.010000)
(4.055010, 0.010000)
(4.060010, 0.010000)
(4.065010, 0.010000)
(4.070010, 0.010000)
(4.075010, 0.010000)
(4.080010, 0.010000)
(4.085010, 0.010000)
(4.090010, 0.010000)
(4.095010, 0.010000)
(4.100010, 0.010000)
(4.105010, 0.010000)
(4.110010, 0.010000)
(4.115010, 0.010000)
(4.120010, 0.010000)
(4.125010, 0.010000)
(4.130010, 0.010000)
(4.135010, 0.010000)
(4.140010, 0.010000)
(4.145010, 0.010000)
(4.150010, 0.010000)
(4.155010, 0.010000)
(4.160010, 0.010000)
(4.165010, 0.010000)
(4.170010, 0.010000)
(4.175010, 0.010000)
(4.180010, 0.010000)
(4.185010, 0.010000)
(4.190010, 0.010000)
(4.195010, 0.010000)
(4.200010, 0.010000)
(4.205010, 0.010000)
(4.210010, 0.010000)
(4.215010, 0.010000)
(4.220010, 0.010000)
(4.225010, 0.010000)
(4.230010, 0.010000)
(4.235010, 0.010000)
(4.240010, 0.010000)
(4.245010, 0.010000)
(4.250010, 0.010000)
(4.255010, 0.010000)
(4.260010, 0.010000)
(4.265010, 0.010000)
(4.270010, 0.010000)
(4.275010, 0.010000)
(4.280010, 0.010000)
(4.285010, 0.010000)
(4.290010, 0.010000)
(4.295010, 0.010000)
(4.300010, 0.010000)
(4.305010, 0.010000)
(4.310010, 0.010000)
(4.315010, 0.010000)
(4.320010, 0.010000)
(4.325010, 0.010000)
(4.330010, 0.010000)
(4.335010, 0.010000)
(4.340010, 0.010000)
(4.345010, 0.010000)
(4.350010, 0.010000)
(4.355010, 0.010000)
(4.360010, 0.010000)
(4.365010, 0.010000)
(4.370010, 0.010000)
(4.375010, 0.010000)
(4.380010, 0.010000)
(4.385010, 0.010000)
(4.390010, 0.010000)
(4.395010, 0.010000)
(4.400010, 0.010000)
(4.405010, 0.010000)
(4.410010, 0.010000)
(4.415010, 0.010000)
(4.420010, 0.010000)
(4.425010, 0.010000)
(4.430010, 0.010000)
(4.435010, 0.010000)
(4.440010, 0.010000)
(4.445010, 0.010000)
(4.450010, 0.010000)
(4.455010, 0.010000)
(4.460010, 0.010000)
(4.465010, 0.010000)
(4.470010, 0.010000)
(4.475010, 0.010000)
(4.480010, 0.010000)
(4.485010, 0.010000)
(4.490010, 0.010000)
(4.495010, 0.010000)
(4.500010, 0.010000)
(4.505010, 0.010000)
(4.510010, 0.010000)
(4.515010, 0.010000)
(4.520010, 0.010000)
(4.525010, 0.010000)
(4.530010, 0.010000)
(4.535010, 0.010000)
(4.540010, 0.010000)
(4.545010, 0.010000)
(4.550010, 0.010000)
(4.555010, 0.010000)
(4.560010, 0.010000)
(4.565010, 0.010000)
(4.570010, 0.010000)
(4.575010, 0.010000)
(4.580010, 0.010000)
(4.585010, 0.010000)
(4.590010, 0.010000)
(4.595010, 0.010000)
(4.600010, 0.010000)
(4.605010, 0.010000)
(4.610010, 0.010000)
(4.615010, 0.010000)
(4.620010, 0.010000)
(4.625010, 0.010000)
(4.630010, 0.010000)
(4.635010, 0.010000)
(4.640010, 0.010000)
(4.645010, 0.010000)
(4.650010, 0.010000)
(4.655010, 0.010000)
(4.660010, 0.010000)
(4.665010, 0.010000)
(4.670010, 0.010000)
(4.675010, 0.010000)
(4.680010, 0.010000)
(4.685010, 0.010000)
(4.690010, 0.010000)
(4.695010, 0.010000)
(4.700010, 0.010000)
(4.705010, 0.010000)
(4.710010, 0.010000)
(4.715010, 0.010000)
(4.720010, 0.010000)
(4.725010, 0.010000)
(4.730010, 0.010000)
(4.735010, 0.010000)
(4.740010, 0.010000)
(4.745010, 0.010000)
(4.750010, 0.010000)
(4.755010, 0.010000)
(4.760010, 0.010000)
(4.765010, 0.010000)
(4.770010, 0.010000)
(4.775010, 0.010000)
(4.780010, 0.010000)
(4.785010, 0.010000)
(4.790010, 0.010000)
(4.795010, 0.010000)
(4.800010, 0.010000)
(4.805010, 0.010000)
(4.810010, 0.010000)
(4.815010, 0.010000)
(4.820010, 0.010000)
(4.825010, 0.010000)
(4.830010, 0.010000)
(4.835010, 0.010000)
(4.840010, 0.010000)
(4.845010, 0.010000)
(4.850010, 0.010000)
(4.855010, 0.010000)
(4.860010, 0.010000)
(4.865010, 0.010000)
(4.870010, 0.010000)
(4.875010, 0.010000)
(4.880010, 0.010000)
(4.885010, 0.010000)
(4.890010, 0.010000)
(4.895010, 0.010000)
(4.900010, 0.010000)
(4.905010, 0.010000)
(4.910010, 0.010000)
(4.915010, 0.010000)
(4.920010, 0.010000)
(4.925010, 0.010000)
(4.930010, 0.010000)
(4.935010, 0.010000)
(4.940010, 0.010000)
(4.945010, 0.010000)
(4.950010, 0.010000)
(4.955010, 0.010000)
(4.960010, 0.010000)
(4.965010, 0.010000)
(4.970010, 0.010000)
(4.975010, 0.010000)
(4.980010, 0.010000)
(4.985010, 0.010000)
(4.990010, 0.010000)
(4.995010, 0.010000)
};
\addlegendentry{model fit} 

%% file: figs/data/lsfitha050.tex
\addplot[
line width=1.0pt, color=gray, solid ]
coordinates {
(0.000020, 0.000020)
(0.002020, 0.002020)
(0.004020, 0.004020)
(0.006020, 0.006020)
(0.008020, 0.008020)
(0.010020, 0.010020)
(0.012020, 0.012013)
(0.014020, 0.013944)
(0.016020, 0.015630)
(0.018020, 0.016783)
(0.020020, 0.017181)
(0.022020, 0.016797)
(0.024020, 0.015771)
(0.026020, 0.014304)
(0.028020, 0.012573)
(0.030020, 0.010700)
(0.032020, 0.008757)
(0.034020, 0.006782)
(0.036020, 0.004793)
(0.038020, 0.002798)
(0.040020, 0.000802)
(0.042020, -0.001193)
(0.044020, -0.003181)
(0.046020, -0.005151)
(0.048020, -0.007076)
(0.050020, -0.008909)
(0.052020, -0.010581)
(0.054020, -0.012004)
(0.056020, -0.013092)
(0.058020, -0.013770)
(0.060020, -0.013997)
(0.062020, -0.013770)
(0.064020, -0.013120)
(0.066020, -0.012107)
(0.068020, -0.010799)
(0.070020, -0.009269)
(0.072020, -0.007581)
(0.074020, -0.005785)
(0.076020, -0.003922)
(0.078020, -0.002022)
(0.080020, -0.000109)
(0.082020, 0.001797)
(0.084020, 0.003671)
(0.086020, 0.005483)
(0.088020, 0.007196)
(0.090020, 0.008764)
(0.092020, 0.010138)
(0.094020, 0.011266)
(0.096020, 0.012105)
(0.098020, 0.012621)
(0.100020, 0.012796)
(0.102020, 0.012631)
(0.104020, 0.012141)
(0.106020, 0.011356)
(0.108020, 0.010313)
(0.110020, 0.009058)
(0.112020, 0.007633)
(0.114020, 0.006080)
(0.116020, 0.004438)
(0.118020, 0.002741)
(0.120020, 0.001024)
(0.122020, -0.000684)
(0.124020, -0.002349)
(0.126020, -0.003939)
(0.128020, -0.005416)
(0.130020, -0.006746)
(0.132020, -0.007893)
(0.134020, -0.008823)
(0.136020, -0.009510)
(0.138020, -0.009936)
(0.140020, -0.010090)
(0.142020, -0.009972)
(0.144020, -0.009593)
(0.146020, -0.008971)
(0.148020, -0.008131)
(0.150020, -0.007102)
(0.152020, -0.005916)
(0.154020, -0.004607)
(0.156020, -0.003208)
(0.158020, -0.001752)
(0.160020, -0.000273)
(0.162020, 0.001198)
(0.164020, 0.002628)
(0.166020, 0.003987)
(0.168020, 0.005242)
(0.170020, 0.006364)
(0.172020, 0.007326)
(0.174020, 0.008105)
(0.176020, 0.008681)
(0.178020, 0.009043)
(0.180020, 0.009183)
(0.182020, 0.009102)
(0.184020, 0.008807)
(0.186020, 0.008310)
(0.188020, 0.007629)
(0.190020, 0.006786)
(0.192020, 0.005806)
(0.194020, 0.004716)
(0.196020, 0.003544)
(0.198020, 0.002319)
(0.200020, 0.001071)
(0.202020, -0.000172)
(0.204020, -0.001380)
(0.206020, -0.002525)
(0.208020, -0.003582)
(0.210020, -0.004525)
(0.212020, -0.005334)
(0.214020, -0.005989)
(0.216020, -0.006477)
(0.218020, -0.006787)
(0.220020, -0.006916)
(0.222020, -0.006862)
(0.224020, -0.006631)
(0.226020, -0.006232)
(0.228020, -0.005679)
(0.230020, -0.004989)
(0.232020, -0.004182)
(0.234020, -0.003280)
(0.236020, -0.002305)
(0.238020, -0.001284)
(0.240020, -0.000241)
(0.242020, 0.000799)
(0.244020, 0.001810)
(0.246020, 0.002770)
(0.248020, 0.003656)
(0.250020, 0.004447)
(0.252020, 0.005126)
(0.254020, 0.005678)
(0.256020, 0.006092)
(0.258020, 0.006359)
(0.260020, 0.006476)
(0.262020, 0.006442)
(0.264020, 0.006261)
(0.266020, 0.005940)
(0.268020, 0.005491)
(0.270020, 0.004926)
(0.272020, 0.004262)
(0.274020, 0.003517)
(0.276020, 0.002710)
(0.278020, 0.001862)
(0.280020, 0.000995)
(0.282020, 0.000129)
(0.284020, -0.000715)
(0.286020, -0.001517)
(0.288020, -0.002257)
(0.290020, -0.002920)
(0.292020, -0.003490)
(0.294020, -0.003956)
(0.296020, -0.004306)
(0.298020, -0.004536)
(0.300020, -0.004641)
(0.302020, -0.004621)
(0.304020, -0.004480)
(0.306020, -0.004222)
(0.308020, -0.003856)
(0.310020, -0.003394)
(0.312020, -0.002849)
(0.314020, -0.002234)
(0.316020, -0.001567)
(0.318020, -0.000865)
(0.320020, -0.000145)
(0.322020, 0.000575)
(0.324020, 0.001277)
(0.326020, 0.001946)
(0.328020, 0.002564)
(0.330020, 0.003119)
(0.332020, 0.003597)
(0.334020, 0.003989)
(0.336020, 0.004286)
(0.338020, 0.004483)
(0.340020, 0.004577)
(0.342020, 0.004568)
(0.344020, 0.004457)
(0.346020, 0.004250)
(0.348020, 0.003953)
(0.350020, 0.003574)
(0.352020, 0.003126)
(0.354020, 0.002620)
(0.356020, 0.002069)
(0.358020, 0.001488)
(0.360020, 0.000891)
(0.362020, 0.000293)
(0.364020, -0.000291)
(0.366020, -0.000847)
(0.368020, -0.001364)
(0.370020, -0.001827)
(0.372020, -0.002228)
(0.374020, -0.002558)
(0.376020, -0.002810)
(0.378020, -0.002979)
(0.380020, -0.003062)
(0.382020, -0.003060)
(0.384020, -0.002974)
(0.386020, -0.002807)
(0.388020, -0.002566)
(0.390020, -0.002257)
(0.392020, -0.001889)
(0.394020, -0.001472)
(0.396020, -0.001017)
(0.398020, -0.000536)
(0.400020, -0.000042)
(0.402020, 0.000454)
(0.404020, 0.000940)
(0.406020, 0.001403)
(0.408020, 0.001834)
(0.410020, 0.002221)
(0.412020, 0.002557)
(0.414020, 0.002835)
(0.416020, 0.003048)
(0.418020, 0.003193)
(0.420020, 0.003267)
(0.422020, 0.003270)
(0.424020, 0.003203)
(0.426020, 0.003069)
(0.428020, 0.002872)
(0.430020, 0.002620)
(0.432020, 0.002317)
(0.434020, 0.001974)
(0.436020, 0.001599)
(0.438020, 0.001202)
(0.440020, 0.000792)
(0.442020, 0.000380)
(0.444020, -0.000023)
(0.446020, -0.000409)
(0.448020, -0.000768)
(0.450020, -0.001092)
(0.452020, -0.001373)
(0.454020, -0.001607)
(0.456020, -0.001787)
(0.458020, -0.001911)
(0.460020, -0.001976)
(0.462020, -0.001982)
(0.464020, -0.001931)
(0.466020, -0.001823)
(0.468020, -0.001664)
(0.470020, -0.001457)
(0.472020, -0.001209)
(0.474020, -0.000926)
(0.476020, -0.000617)
(0.478020, -0.000288)
(0.480020, 0.000051)
(0.482020, 0.000392)
(0.484020, 0.000727)
(0.486020, 0.001048)
(0.488020, 0.001348)
(0.490020, 0.001618)
(0.492020, 0.001854)
(0.494020, 0.002050)
(0.496020, 0.002203)
(0.498020, 0.002309)
(0.500020, 0.002366)
(0.502020, 0.002374)
(0.504020, 0.002335)
(0.506020, 0.002248)
(0.508020, 0.002119)
(0.510020, 0.001950)
(0.512020, 0.001746)
(0.514020, 0.001514)
(0.516020, 0.001259)
(0.518020, 0.000987)
(0.520020, 0.000706)
(0.522020, 0.000423)
(0.524020, 0.000145)
(0.526020, -0.000122)
(0.528020, -0.000372)
(0.530020, -0.000598)
(0.532020, -0.000795)
(0.534020, -0.000960)
(0.536020, -0.001089)
(0.538020, -0.001179)
(0.540020, -0.001230)
(0.542020, -0.001239)
(0.544020, -0.001209)
(0.546020, -0.001140)
(0.548020, -0.001035)
(0.550020, -0.000897)
(0.552020, -0.000730)
(0.554020, -0.000538)
(0.556020, -0.000328)
(0.558020, -0.000103)
(0.560020, 0.000129)
(0.562020, 0.000364)
(0.564020, 0.000595)
(0.566020, 0.000817)
(0.568020, 0.001025)
(0.570020, 0.001214)
(0.572020, 0.001379)
(0.574020, 0.001518)
(0.576020, 0.001627)
(0.578020, 0.001704)
(0.580020, 0.001747)
(0.582020, 0.001758)
(0.584020, 0.001734)
(0.586020, 0.001679)
(0.588020, 0.001594)
(0.590020, 0.001481)
(0.592020, 0.001344)
(0.594020, 0.001187)
(0.596020, 0.001013)
(0.598020, 0.000828)
(0.600020, 0.000635)
(0.602020, 0.000441)
(0.604020, 0.000249)
(0.606020, 0.000064)
(0.608020, -0.000109)
(0.610020, -0.000267)
(0.612020, -0.000405)
(0.614020, -0.000522)
(0.616020, -0.000614)
(0.618020, -0.000679)
(0.620020, -0.000717)
(0.622020, -0.000728)
(0.624020, -0.000710)
(0.626020, -0.000666)
(0.628020, -0.000597)
(0.630020, -0.000505)
(0.632020, -0.000393)
(0.634020, -0.000263)
(0.636020, -0.000120)
(0.638020, 0.000033)
(0.640020, 0.000192)
(0.642020, 0.000354)
(0.644020, 0.000513)
(0.646020, 0.000667)
(0.648020, 0.000811)
(0.650020, 0.000943)
(0.652020, 0.001059)
(0.654020, 0.001157)
(0.656020, 0.001234)
(0.658020, 0.001290)
(0.660020, 0.001323)
(0.662020, 0.001333)
(0.664020, 0.001320)
(0.666020, 0.001285)
(0.668020, 0.001229)
(0.670020, 0.001154)
(0.672020, 0.001062)
(0.674020, 0.000955)
(0.676020, 0.000837)
(0.678020, 0.000711)
(0.680020, 0.000579)
(0.682020, 0.000445)
(0.684020, 0.000313)
(0.686020, 0.000185)
(0.688020, 0.000065)
(0.690020, -0.000045)
(0.692020, -0.000142)
(0.694020, -0.000224)
(0.696020, -0.000290)
(0.698020, -0.000337)
(0.700020, -0.000366)
(0.702020, -0.000375)
(0.704020, -0.000366)
(0.706020, -0.000338)
(0.708020, -0.000293)
(0.710020, -0.000231)
(0.712020, -0.000156)
(0.714020, -0.000068)
(0.716020, 0.000029)
(0.718020, 0.000133)
(0.720020, 0.000242)
(0.722020, 0.000353)
(0.724020, 0.000463)
(0.726020, 0.000569)
(0.728020, 0.000670)
(0.730020, 0.000761)
(0.732020, 0.000843)
(0.734020, 0.000911)
(0.736020, 0.000967)
(0.738020, 0.001007)
(0.740020, 0.001032)
(0.742020, 0.001041)
(0.744020, 0.001034)
(0.746020, 0.001012)
(0.748020, 0.000975)
(0.750020, 0.000925)
(0.752020, 0.000863)
(0.754020, 0.000791)
(0.756020, 0.000711)
(0.758020, 0.000625)
(0.760020, 0.000534)
(0.762020, 0.000443)
(0.764020, 0.000351)
(0.766020, 0.000263)
(0.768020, 0.000180)
(0.770020, 0.000103)
(0.772020, 0.000035)
(0.774020, -0.000023)
(0.776020, -0.000069)
(0.778020, -0.000103)
(0.780020, -0.000125)
(0.782020, -0.000133)
(0.784020, -0.000128)
(0.786020, -0.000111)
(0.788020, -0.000081)
(0.790020, -0.000040)
(0.792020, 0.000010)
(0.794020, 0.000069)
(0.796020, 0.000136)
(0.798020, 0.000207)
(0.800020, 0.000281)
(0.802020, 0.000358)
(0.804020, 0.000433)
(0.806020, 0.000507)
(0.808020, 0.000576)
(0.810020, 0.000640)
(0.812020, 0.000697)
(0.814020, 0.000745)
(0.816020, 0.000785)
(0.818020, 0.000814)
(0.820020, 0.000832)
(0.822020, 0.000840)
(0.824020, 0.000836)
(0.826020, 0.000823)
(0.828020, 0.000799)
(0.830020, 0.000765)
(0.832020, 0.000724)
(0.834020, 0.000675)
(0.836020, 0.000621)
(0.838020, 0.000562)
(0.840020, 0.000500)
(0.842020, 0.000437)
(0.844020, 0.000374)
(0.846020, 0.000313)
(0.848020, 0.000255)
(0.850020, 0.000202)
(0.852020, 0.000155)
(0.854020, 0.000114)
(0.856020, 0.000081)
(0.858020, 0.000056)
(0.860020, 0.000040)
(0.862020, 0.000033)
(0.864020, 0.000036)
(0.866020, 0.000047)
(0.868020, 0.000066)
(0.870020, 0.000093)
(0.872020, 0.000127)
(0.874020, 0.000167)
(0.876020, 0.000212)
(0.878020, 0.000260)
(0.880020, 0.000311)
(0.882020, 0.000364)
(0.884020, 0.000416)
(0.886020, 0.000466)
(0.888020, 0.000515)
(0.890020, 0.000559)
(0.892020, 0.000599)
(0.894020, 0.000633)
(0.896020, 0.000661)
(0.898020, 0.000682)
(0.900020, 0.000695)
(0.902020, 0.000702)
(0.904020, 0.000700)
(0.906020, 0.000692)
(0.908020, 0.000676)
(0.910020, 0.000654)
(0.912020, 0.000626)
(0.914020, 0.000593)
(0.916020, 0.000556)
(0.918020, 0.000516)
(0.920020, 0.000474)
(0.922020, 0.000431)
(0.924020, 0.000388)
(0.926020, 0.000345)
(0.928020, 0.000305)
(0.930020, 0.000268)
(0.932020, 0.000235)
(0.934020, 0.000206)
(0.936020, 0.000183)
(0.938020, 0.000165)
(0.940020, 0.000154)
(0.942020, 0.000148)
(0.944020, 0.000149)
(0.946020, 0.000155)
(0.948020, 0.000168)
(0.950020, 0.000186)
(0.952020, 0.000209)
(0.954020, 0.000236)
(0.956020, 0.000266)
(0.958020, 0.000299)
(0.960020, 0.000334)
(0.962020, 0.000370)
};
\addlegendentry{nekrs} 
\addplot[
line width=1.0pt, color=blue, dotted ]
coordinates {
(0.000020, -0.001522)
(0.002020, 0.000621)
(0.004020, 0.002726)
(0.006020, 0.004744)
(0.008020, 0.006626)
(0.010020, 0.008328)
(0.012020, 0.009811)
(0.014020, 0.011041)
(0.016020, 0.011993)
(0.018020, 0.012647)
(0.020020, 0.012992)
(0.022020, 0.013023)
(0.024020, 0.012745)
(0.026020, 0.012168)
(0.028020, 0.011312)
(0.030020, 0.010201)
(0.032020, 0.008867)
(0.034020, 0.007345)
(0.036020, 0.005675)
(0.038020, 0.003901)
(0.040020, 0.002067)
(0.042020, 0.000219)
(0.044020, -0.001597)
(0.046020, -0.003338)
(0.048020, -0.004962)
(0.050020, -0.006430)
(0.052020, -0.007710)
(0.054020, -0.008772)
(0.056020, -0.009594)
(0.058020, -0.010159)
(0.060020, -0.010458)
(0.062020, -0.010486)
(0.064020, -0.010247)
(0.066020, -0.009751)
(0.068020, -0.009014)
(0.070020, -0.008057)
(0.072020, -0.006908)
(0.074020, -0.005596)
(0.076020, -0.004156)
(0.078020, -0.002627)
(0.080020, -0.001045)
(0.082020, 0.000548)
(0.084020, 0.002115)
(0.086020, 0.003616)
(0.088020, 0.005017)
(0.090020, 0.006284)
(0.092020, 0.007389)
(0.094020, 0.008306)
(0.096020, 0.009016)
(0.098020, 0.009504)
(0.100020, 0.009763)
(0.102020, 0.009788)
(0.104020, 0.009583)
(0.106020, 0.009157)
(0.108020, 0.008522)
(0.110020, 0.007698)
(0.112020, 0.006707)
(0.114020, 0.005576)
(0.116020, 0.004336)
(0.118020, 0.003017)
(0.120020, 0.001653)
(0.122020, 0.000279)
(0.124020, -0.001072)
(0.126020, -0.002367)
(0.128020, -0.003576)
(0.130020, -0.004669)
(0.132020, -0.005622)
(0.134020, -0.006414)
(0.136020, -0.007027)
(0.138020, -0.007449)
(0.140020, -0.007673)
(0.142020, -0.007696)
(0.144020, -0.007521)
(0.146020, -0.007154)
(0.148020, -0.006607)
(0.150020, -0.005897)
(0.152020, -0.005044)
(0.154020, -0.004069)
(0.156020, -0.002999)
(0.158020, -0.001862)
(0.160020, -0.000686)
(0.162020, 0.000499)
(0.164020, 0.001664)
(0.166020, 0.002781)
(0.168020, 0.003824)
(0.170020, 0.004767)
(0.172020, 0.005590)
(0.174020, 0.006273)
(0.176020, 0.006803)
(0.178020, 0.007168)
(0.180020, 0.007362)
(0.182020, 0.007382)
(0.184020, 0.007232)
(0.186020, 0.006916)
(0.188020, 0.006446)
(0.190020, 0.005834)
(0.192020, 0.005098)
(0.194020, 0.004258)
(0.196020, 0.003336)
(0.198020, 0.002356)
(0.200020, 0.001342)
(0.202020, 0.000320)
(0.204020, -0.000685)
(0.206020, -0.001649)
(0.208020, -0.002548)
(0.210020, -0.003362)
(0.212020, -0.004072)
(0.214020, -0.004662)
(0.216020, -0.005120)
(0.218020, -0.005435)
(0.220020, -0.005603)
(0.222020, -0.005622)
(0.224020, -0.005492)
(0.226020, -0.005221)
(0.228020, -0.004816)
(0.230020, -0.004289)
(0.232020, -0.003655)
(0.234020, -0.002931)
(0.236020, -0.002136)
(0.238020, -0.001291)
(0.240020, -0.000417)
(0.242020, 0.000465)
(0.244020, 0.001332)
(0.246020, 0.002163)
(0.248020, 0.002939)
(0.250020, 0.003641)
(0.252020, 0.004254)
(0.254020, 0.004763)
(0.256020, 0.005158)
(0.258020, 0.005431)
(0.260020, 0.005576)
(0.262020, 0.005593)
(0.264020, 0.005482)
(0.266020, 0.005249)
(0.268020, 0.004900)
(0.270020, 0.004446)
(0.272020, 0.003900)
(0.274020, 0.003276)
(0.276020, 0.002591)
(0.278020, 0.001862)
(0.280020, 0.001108)
(0.282020, 0.000348)
(0.284020, -0.000400)
(0.286020, -0.001117)
(0.288020, -0.001786)
(0.290020, -0.002392)
(0.292020, -0.002921)
(0.294020, -0.003361)
(0.296020, -0.003702)
(0.298020, -0.003938)
(0.300020, -0.004064)
(0.302020, -0.004078)
(0.304020, -0.003983)
(0.306020, -0.003783)
(0.308020, -0.003482)
(0.310020, -0.003091)
(0.312020, -0.002621)
(0.314020, -0.002083)
(0.316020, -0.001492)
(0.318020, -0.000864)
(0.320020, -0.000214)
(0.322020, 0.000442)
(0.324020, 0.001087)
(0.326020, 0.001705)
(0.328020, 0.002283)
(0.330020, 0.002805)
(0.332020, 0.003262)
(0.334020, 0.003641)
(0.336020, 0.003936)
(0.338020, 0.004140)
(0.340020, 0.004249)
(0.342020, 0.004262)
(0.344020, 0.004181)
(0.346020, 0.004008)
(0.348020, 0.003749)
(0.350020, 0.003413)
(0.352020, 0.003007)
(0.354020, 0.002543)
(0.356020, 0.002034)
(0.358020, 0.001493)
(0.360020, 0.000932)
(0.362020, 0.000367)
(0.364020, -0.000189)
(0.366020, -0.000723)
(0.368020, -0.001221)
(0.370020, -0.001672)
(0.372020, -0.002066)
(0.374020, -0.002394)
(0.376020, -0.002648)
(0.378020, -0.002824)
(0.380020, -0.002919)
(0.382020, -0.002931)
(0.384020, -0.002861)
(0.386020, -0.002712)
(0.388020, -0.002490)
(0.390020, -0.002199)
(0.392020, -0.001850)
(0.394020, -0.001450)
(0.396020, -0.001012)
(0.398020, -0.000545)
(0.400020, -0.000061)
(0.402020, 0.000426)
(0.404020, 0.000906)
(0.406020, 0.001366)
(0.408020, 0.001796)
(0.410020, 0.002185)
(0.412020, 0.002525)
(0.414020, 0.002808)
(0.416020, 0.003028)
(0.418020, 0.003180)
(0.420020, 0.003262)
(0.422020, 0.003272)
(0.424020, 0.003213)
(0.426020, 0.003085)
(0.428020, 0.002893)
(0.430020, 0.002643)
(0.432020, 0.002342)
(0.434020, 0.001998)
(0.436020, 0.001619)
(0.438020, 0.001217)
(0.440020, 0.000800)
(0.442020, 0.000379)
(0.444020, -0.000034)
(0.446020, -0.000431)
(0.448020, -0.000802)
(0.450020, -0.001138)
(0.452020, -0.001431)
(0.454020, -0.001676)
(0.456020, -0.001865)
(0.458020, -0.001997)
(0.460020, -0.002068)
(0.462020, -0.002077)
(0.464020, -0.002026)
(0.466020, -0.001916)
(0.468020, -0.001751)
(0.470020, -0.001536)
(0.472020, -0.001276)
(0.474020, -0.000979)
(0.476020, -0.000653)
(0.478020, -0.000306)
(0.480020, 0.000053)
(0.482020, 0.000416)
(0.484020, 0.000773)
(0.486020, 0.001115)
(0.488020, 0.001435)
(0.490020, 0.001725)
(0.492020, 0.001978)
(0.494020, 0.002189)
(0.496020, 0.002353)
(0.498020, 0.002467)
(0.500020, 0.002528)
(0.502020, 0.002536)
(0.504020, 0.002492)
(0.506020, 0.002398)
(0.508020, 0.002256)
(0.510020, 0.002070)
(0.512020, 0.001847)
(0.514020, 0.001591)
(0.516020, 0.001310)
(0.518020, 0.001010)
(0.520020, 0.000700)
(0.522020, 0.000388)
(0.524020, 0.000080)
(0.526020, -0.000215)
(0.528020, -0.000491)
(0.530020, -0.000741)
(0.532020, -0.000960)
(0.534020, -0.001142)
(0.536020, -0.001283)
(0.538020, -0.001382)
(0.540020, -0.001435)
(0.542020, -0.001442)
(0.544020, -0.001405)
(0.546020, -0.001323)
(0.548020, -0.001201)
(0.550020, -0.001041)
(0.552020, -0.000848)
(0.554020, -0.000628)
(0.556020, -0.000386)
(0.558020, -0.000128)
(0.560020, 0.000140)
(0.562020, 0.000409)
(0.564020, 0.000675)
(0.566020, 0.000929)
(0.568020, 0.001168)
(0.570020, 0.001383)
(0.572020, 0.001572)
(0.574020, 0.001729)
(0.576020, 0.001851)
(0.578020, 0.001936)
(0.580020, 0.001982)
(0.582020, 0.001989)
(0.584020, 0.001957)
(0.586020, 0.001887)
(0.588020, 0.001781)
(0.590020, 0.001644)
(0.592020, 0.001477)
(0.594020, 0.001287)
(0.596020, 0.001079)
(0.598020, 0.000856)
(0.600020, 0.000626)
(0.602020, 0.000393)
(0.604020, 0.000164)
(0.606020, -0.000056)
(0.608020, -0.000261)
(0.610020, -0.000447)
(0.612020, -0.000610)
(0.614020, -0.000745)
(0.616020, -0.000851)
(0.618020, -0.000924)
(0.620020, -0.000964)
(0.622020, -0.000970)
(0.624020, -0.000942)
(0.626020, -0.000882)
(0.628020, -0.000792)
(0.630020, -0.000673)
(0.632020, -0.000530)
(0.634020, -0.000366)
(0.636020, -0.000186)
(0.638020, 0.000006)
(0.640020, 0.000204)
(0.642020, 0.000405)
(0.644020, 0.000602)
(0.646020, 0.000792)
(0.648020, 0.000969)
(0.650020, 0.001130)
(0.652020, 0.001270)
(0.654020, 0.001387)
(0.656020, 0.001478)
(0.658020, 0.001542)
(0.660020, 0.001576)
(0.662020, 0.001582)
(0.664020, 0.001558)
(0.666020, 0.001506)
(0.668020, 0.001428)
(0.670020, 0.001326)
(0.672020, 0.001203)
(0.674020, 0.001061)
(0.676020, 0.000906)
(0.678020, 0.000741)
(0.680020, 0.000570)
(0.682020, 0.000397)
(0.684020, 0.000226)
(0.686020, 0.000063)
(0.688020, -0.000090)
(0.690020, -0.000229)
(0.692020, -0.000350)
(0.694020, -0.000451)
(0.696020, -0.000530)
(0.698020, -0.000585)
(0.700020, -0.000614)
(0.702020, -0.000619)
(0.704020, -0.000599)
(0.706020, -0.000554)
(0.708020, -0.000487)
(0.710020, -0.000399)
(0.712020, -0.000293)
(0.714020, -0.000171)
(0.716020, -0.000037)
(0.718020, 0.000105)
(0.720020, 0.000253)
(0.722020, 0.000402)
(0.724020, 0.000549)
(0.726020, 0.000690)
(0.728020, 0.000822)
(0.730020, 0.000942)
(0.732020, 0.001046)
(0.734020, 0.001133)
(0.736020, 0.001201)
(0.738020, 0.001249)
(0.740020, 0.001275)
(0.742020, 0.001279)
(0.744020, 0.001261)
(0.746020, 0.001223)
(0.748020, 0.001165)
(0.750020, 0.001089)
(0.752020, 0.000998)
(0.754020, 0.000893)
(0.756020, 0.000778)
(0.758020, 0.000655)
(0.760020, 0.000527)
(0.762020, 0.000399)
(0.764020, 0.000272)
(0.766020, 0.000150)
(0.768020, 0.000037)
(0.770020, -0.000067)
(0.772020, -0.000157)
(0.774020, -0.000232)
(0.776020, -0.000291)
(0.778020, -0.000332)
(0.780020, -0.000354)
(0.782020, -0.000358)
(0.784020, -0.000343)
(0.786020, -0.000310)
(0.788020, -0.000260)
(0.790020, -0.000195)
(0.792020, -0.000116)
(0.794020, -0.000026)
(0.796020, 0.000074)
(0.798020, 0.000180)
(0.800020, 0.000290)
(0.802020, 0.000401)
(0.804020, 0.000510)
(0.806020, 0.000615)
(0.808020, 0.000713)
(0.810020, 0.000802)
(0.812020, 0.000880)
(0.814020, 0.000945)
(0.816020, 0.000996)
(0.818020, 0.001031)
(0.820020, 0.001050)
(0.822020, 0.001054)
(0.824020, 0.001041)
(0.826020, 0.001012)
(0.828020, 0.000970)
(0.830020, 0.000913)
(0.832020, 0.000845)
(0.834020, 0.000767)
(0.836020, 0.000682)
(0.838020, 0.000590)
(0.840020, 0.000496)
(0.842020, 0.000400)
(0.844020, 0.000306)
(0.846020, 0.000215)
(0.848020, 0.000131)
(0.850020, 0.000054)
(0.852020, -0.000014)
(0.854020, -0.000070)
(0.856020, -0.000113)
(0.858020, -0.000144)
(0.860020, -0.000161)
(0.862020, -0.000164)
(0.864020, -0.000153)
(0.866020, -0.000128)
(0.868020, -0.000091)
(0.870020, -0.000043)
(0.872020, 0.000016)
(0.874020, 0.000083)
(0.876020, 0.000157)
(0.878020, 0.000235)
(0.880020, 0.000317)
(0.882020, 0.000400)
(0.884020, 0.000481)
(0.886020, 0.000559)
(0.888020, 0.000632)
(0.890020, 0.000698)
(0.892020, 0.000756)
(0.894020, 0.000805)
(0.896020, 0.000843)
(0.898020, 0.000869)
(0.900020, 0.000884)
(0.902020, 0.000886)
(0.904020, 0.000877)
(0.906020, 0.000856)
(0.908020, 0.000824)
(0.910020, 0.000782)
(0.912020, 0.000732)
(0.914020, 0.000674)
(0.916020, 0.000610)
(0.918020, 0.000542)
(0.920020, 0.000472)
(0.922020, 0.000401)
(0.924020, 0.000331)
(0.926020, 0.000263)
(0.928020, 0.000200)
(0.930020, 0.000143)
(0.932020, 0.000093)
(0.934020, 0.000051)
(0.936020, 0.000018)
(0.938020, -0.000004)
(0.940020, -0.000017)
(0.942020, -0.000019)
(0.944020, -0.000011)
(0.946020, 0.000007)
(0.948020, 0.000034)
(0.950020, 0.000070)
(0.952020, 0.000114)
(0.954020, 0.000164)
(0.956020, 0.000218)
(0.958020, 0.000277)
(0.960020, 0.000338)
(0.962020, 0.000399)
};
\addlegendentry{model fit} 

%% file: figs/data/lsfitha100.tex
\addplot[
line width=1.0pt, color=gray, solid ]
coordinates {
(0.000020, 0.000020)
(0.010020, 0.008967)
(0.020020, 0.000187)
(0.030020, -0.007940)
(0.040020, 0.000008)
(0.050020, 0.007559)
(0.060020, 0.000213)
(0.070020, -0.006835)
(0.080020, -0.000040)
(0.090020, 0.006525)
(0.100020, 0.000268)
(0.110020, -0.005821)
(0.120020, -0.000089)
(0.130020, 0.005537)
(0.140020, 0.000306)
(0.150020, -0.004876)
(0.160020, -0.000115)
(0.170020, 0.004646)
(0.180020, 0.000321)
(0.190020, -0.004046)
(0.200020, -0.000121)
(0.210020, 0.003878)
(0.220020, 0.000320)
(0.230020, -0.003339)
(0.240020, -0.000116)
(0.250020, 0.003231)
(0.260020, 0.000311)
(0.270020, -0.002747)
(0.280020, -0.000104)
(0.290020, 0.002690)
(0.300020, 0.000297)
(0.310020, -0.002255)
(0.320020, -0.000089)
(0.330020, 0.002242)
(0.340020, 0.000281)
(0.350020, -0.001846)
(0.360020, -0.000072)
(0.370020, 0.001870)
(0.380020, 0.000265)
(0.390020, -0.001508)
(0.400020, -0.000056)
(0.410020, 0.001562)
(0.420020, 0.000248)
(0.430020, -0.001229)
(0.440020, -0.000040)
(0.450020, 0.001308)
(0.460020, 0.000232)
(0.470020, -0.000998)
(0.480020, -0.000024)
(0.490020, 0.001098)
(0.500020, 0.000218)
(0.510020, -0.000807)
(0.520020, -0.000010)
(0.530020, 0.000924)
(0.540020, 0.000204)
(0.550020, -0.000649)
(0.560020, 0.000003)
(0.570020, 0.000781)
(0.580020, 0.000192)
(0.590020, -0.000518)
(0.600020, 0.000015)
(0.610020, 0.000662)
(0.620020, 0.000180)
(0.630020, -0.000411)
(0.640020, 0.000025)
(0.650020, 0.000564)
(0.660020, 0.000170)
(0.670020, -0.000322)
(0.680020, 0.000035)
(0.690020, 0.000484)
(0.700020, 0.000161)
(0.710020, -0.000248)
(0.720020, 0.000043)
(0.730020, 0.000417)
(0.740020, 0.000153)
(0.750020, -0.000187)
(0.760020, 0.000051)
(0.770020, 0.000362)
(0.780020, 0.000146)
(0.790020, -0.000137)
(0.800020, 0.000058)
(0.810020, 0.000316)
(0.820020, 0.000140)
(0.830020, -0.000096)
(0.840020, 0.000063)
(0.850020, 0.000278)
(0.860020, 0.000135)
(0.870020, -0.000062)
(0.880020, 0.000068)
(0.890020, 0.000247)
(0.900020, 0.000130)
(0.910020, -0.000033)
(0.920020, 0.000073)
(0.930020, 0.000222)
(0.940020, 0.000126)
(0.950020, -0.000010)
(0.960020, 0.000077)
(0.970020, 0.000200)
(0.980020, 0.000122)
(0.990020, 0.000009)
(1.000020, 0.000080)
(1.010020, 0.000183)
(1.020020, 0.000119)
(1.030020, 0.000025)
(1.040020, 0.000083)
(1.050020, 0.000168)
(1.060020, 0.000116)
(1.070020, 0.000038)
(1.080020, 0.000085)
(1.090020, 0.000156)
(1.100020, 0.000114)
(1.110020, 0.000049)
(1.120020, 0.000088)
(1.130020, 0.000147)
(1.140020, 0.000112)
(1.150020, 0.000058)
(1.160020, 0.000089)
(1.170020, 0.000138)
(1.180020, 0.000110)
(1.190020, 0.000065)
(1.200020, 0.000091)
(1.210020, 0.000132)
(1.220020, 0.000109)
(1.230020, 0.000072)
(1.240020, 0.000092)
(1.250020, 0.000126)
(1.260020, 0.000107)
(1.270020, 0.000077)
(1.280020, 0.000094)
(1.290020, 0.000122)
(1.300020, 0.000106)
(1.310020, 0.000081)
(1.320020, 0.000095)
(1.330020, 0.000118)
(1.340020, 0.000105)
(1.350020, 0.000084)
(1.360020, 0.000095)
(1.370020, 0.000115)
(1.380020, 0.000105)
(1.390020, 0.000087)
(1.400020, 0.000096)
(1.410020, 0.000112)
(1.420020, 0.000104)
(1.430020, 0.000089)
(1.440020, 0.000097)
(1.450020, 0.000110)
(1.460020, 0.000103)
(1.470020, 0.000091)
(1.480020, 0.000097)
(1.490020, 0.000108)
(1.500020, 0.000103)
(1.510020, 0.000093)
(1.520020, 0.000098)
(1.530020, 0.000107)
(1.540020, 0.000102)
(1.550020, 0.000094)
(1.560020, 0.000098)
(1.570020, 0.000106)
(1.580020, 0.000102)
(1.590020, 0.000095)
(1.600020, 0.000098)
(1.610020, 0.000105)
(1.620020, 0.000102)
(1.630020, 0.000096)
(1.640020, 0.000099)
(1.650020, 0.000104)
(1.660020, 0.000102)
(1.670020, 0.000097)
(1.680020, 0.000099)
(1.690020, 0.000103)
(1.700020, 0.000101)
(1.710020, 0.000097)
(1.720020, 0.000099)
(1.730020, 0.000103)
(1.740020, 0.000101)
(1.750020, 0.000098)
(1.760020, 0.000099)
(1.770020, 0.000102)
(1.780020, 0.000101)
(1.790020, 0.000098)
(1.800020, 0.000099)
(1.810020, 0.000102)
(1.820020, 0.000101)
(1.830020, 0.000099)
(1.840020, 0.000100)
(1.850020, 0.000102)
(1.860020, 0.000101)
(1.870020, 0.000099)
(1.880020, 0.000100)
(1.890020, 0.000101)
(1.900020, 0.000101)
(1.910020, 0.000099)
(1.920020, 0.000100)
(1.930020, 0.000101)
(1.940020, 0.000101)
(1.950020, 0.000099)
(1.960020, 0.000100)
(1.970020, 0.000101)
(1.980020, 0.000101)
(1.990020, 0.000099)
(2.000020, 0.000100)
(2.010020, 0.000101)
(2.020020, 0.000100)
(2.030020, 0.000100)
(2.040020, 0.000100)
(2.050020, 0.000101)
(2.060020, 0.000100)
(2.070020, 0.000100)
(2.080020, 0.000100)
(2.090020, 0.000101)
(2.100020, 0.000100)
(2.110020, 0.000100)
(2.120020, 0.000100)
(2.130020, 0.000101)
(2.140020, 0.000100)
(2.150020, 0.000100)
(2.160020, 0.000100)
(2.170020, 0.000100)
(2.180020, 0.000100)
(2.190020, 0.000100)
(2.200020, 0.000100)
(2.210020, 0.000100)
(2.220020, 0.000100)
(2.230020, 0.000100)
(2.240020, 0.000100)
(2.250020, 0.000100)
(2.260020, 0.000100)
(2.270020, 0.000100)
(2.280020, 0.000100)
(2.290020, 0.000100)
(2.300020, 0.000100)
(2.310020, 0.000100)
(2.320020, 0.000100)
(2.330020, 0.000100)
(2.340020, 0.000100)
(2.350020, 0.000100)
(2.360020, 0.000100)
(2.370020, 0.000100)
(2.380020, 0.000100)
(2.390020, 0.000100)
(2.400020, 0.000100)
};
\addlegendentry{nekrs} 
\addplot[
line width=1.0pt, color=blue, dotted ]
coordinates {
(0.000020, -0.000591)
(0.010020, 0.005804)
(0.020020, 0.000743)
(0.030020, -0.005197)
(0.040020, -0.000498)
(0.050020, 0.005019)
(0.060020, 0.000656)
(0.070020, -0.004467)
(0.080020, -0.000418)
(0.090020, 0.004342)
(0.100020, 0.000581)
(0.110020, -0.003839)
(0.120020, -0.000348)
(0.130020, 0.003758)
(0.140020, 0.000516)
(0.150020, -0.003297)
(0.160020, -0.000287)
(0.170020, 0.003254)
(0.180020, 0.000460)
(0.190020, -0.002829)
(0.200020, -0.000235)
(0.210020, 0.002820)
(0.220020, 0.000412)
(0.230020, -0.002426)
(0.240020, -0.000190)
(0.250020, 0.002446)
(0.260020, 0.000370)
(0.270020, -0.002078)
(0.280020, -0.000151)
(0.290020, 0.002123)
(0.300020, 0.000333)
(0.310020, -0.001778)
(0.320020, -0.000117)
(0.330020, 0.001844)
(0.340020, 0.000302)
(0.350020, -0.001520)
(0.360020, -0.000088)
(0.370020, 0.001604)
(0.380020, 0.000275)
(0.390020, -0.001297)
(0.400020, -0.000063)
(0.410020, 0.001397)
(0.420020, 0.000251)
(0.430020, -0.001105)
(0.440020, -0.000041)
(0.450020, 0.001219)
(0.460020, 0.000231)
(0.470020, -0.000939)
(0.480020, -0.000022)
(0.490020, 0.001065)
(0.500020, 0.000213)
(0.510020, -0.000796)
(0.520020, -0.000005)
(0.530020, 0.000932)
(0.540020, 0.000198)
(0.550020, -0.000673)
(0.560020, 0.000009)
(0.570020, 0.000817)
(0.580020, 0.000185)
(0.590020, -0.000566)
(0.600020, 0.000021)
(0.610020, 0.000719)
(0.620020, 0.000173)
(0.630020, -0.000474)
(0.640020, 0.000032)
(0.650020, 0.000633)
(0.660020, 0.000163)
(0.670020, -0.000395)
(0.680020, 0.000041)
(0.690020, 0.000560)
(0.700020, 0.000155)
(0.710020, -0.000327)
(0.720020, 0.000049)
(0.730020, 0.000497)
(0.740020, 0.000147)
(0.750020, -0.000268)
(0.760020, 0.000056)
(0.770020, 0.000442)
(0.780020, 0.000141)
(0.790020, -0.000218)
(0.800020, 0.000062)
(0.810020, 0.000395)
(0.820020, 0.000135)
(0.830020, -0.000174)
(0.840020, 0.000067)
(0.850020, 0.000354)
(0.860020, 0.000131)
(0.870020, -0.000136)
(0.880020, 0.000071)
(0.890020, 0.000319)
(0.900020, 0.000127)
(0.910020, -0.000104)
(0.920020, 0.000075)
(0.930020, 0.000289)
(0.940020, 0.000123)
(0.950020, -0.000076)
(0.960020, 0.000079)
(0.970020, 0.000263)
(0.980020, 0.000120)
(0.990020, -0.000052)
(1.000020, 0.000082)
(1.010020, 0.000241)
(1.020020, 0.000117)
(1.030020, -0.000031)
(1.040020, 0.000084)
(1.050020, 0.000221)
(1.060020, 0.000115)
(1.070020, -0.000013)
(1.080020, 0.000086)
(1.090020, 0.000205)
(1.100020, 0.000113)
(1.110020, 0.000003)
(1.120020, 0.000088)
(1.130020, 0.000190)
(1.140020, 0.000111)
(1.150020, 0.000016)
(1.160020, 0.000090)
(1.170020, 0.000178)
(1.180020, 0.000110)
(1.190020, 0.000028)
(1.200020, 0.000091)
(1.210020, 0.000167)
(1.220020, 0.000108)
(1.230020, 0.000038)
(1.240020, 0.000092)
(1.250020, 0.000158)
(1.260020, 0.000107)
(1.270020, 0.000046)
(1.280020, 0.000093)
(1.290020, 0.000150)
(1.300020, 0.000106)
(1.310020, 0.000054)
(1.320020, 0.000094)
(1.330020, 0.000143)
(1.340020, 0.000105)
(1.350020, 0.000060)
(1.360020, 0.000095)
(1.370020, 0.000137)
(1.380020, 0.000105)
(1.390020, 0.000066)
(1.400020, 0.000096)
(1.410020, 0.000132)
(1.420020, 0.000104)
(1.430020, 0.000070)
(1.440020, 0.000096)
(1.450020, 0.000128)
(1.460020, 0.000103)
(1.470020, 0.000074)
(1.480020, 0.000097)
(1.490020, 0.000124)
(1.500020, 0.000103)
(1.510020, 0.000078)
(1.520020, 0.000097)
(1.530020, 0.000121)
(1.540020, 0.000103)
(1.550020, 0.000081)
(1.560020, 0.000098)
(1.570020, 0.000118)
(1.580020, 0.000102)
(1.590020, 0.000084)
(1.600020, 0.000098)
(1.610020, 0.000115)
(1.620020, 0.000102)
(1.630020, 0.000086)
(1.640020, 0.000098)
(1.650020, 0.000113)
(1.660020, 0.000102)
(1.670020, 0.000088)
(1.680020, 0.000098)
(1.690020, 0.000111)
(1.700020, 0.000101)
(1.710020, 0.000089)
(1.720020, 0.000099)
(1.730020, 0.000110)
(1.740020, 0.000101)
(1.750020, 0.000091)
(1.760020, 0.000099)
(1.770020, 0.000108)
(1.780020, 0.000101)
(1.790020, 0.000092)
(1.800020, 0.000099)
(1.810020, 0.000107)
(1.820020, 0.000101)
(1.830020, 0.000093)
(1.840020, 0.000099)
(1.850020, 0.000106)
(1.860020, 0.000101)
(1.870020, 0.000094)
(1.880020, 0.000099)
(1.890020, 0.000105)
(1.900020, 0.000101)
(1.910020, 0.000095)
(1.920020, 0.000099)
(1.930020, 0.000105)
(1.940020, 0.000101)
(1.950020, 0.000096)
(1.960020, 0.000099)
(1.970020, 0.000104)
(1.980020, 0.000101)
(1.990020, 0.000096)
(2.000020, 0.000100)
(2.010020, 0.000103)
(2.020020, 0.000100)
(2.030020, 0.000097)
(2.040020, 0.000100)
(2.050020, 0.000103)
(2.060020, 0.000100)
(2.070020, 0.000097)
(2.080020, 0.000100)
(2.090020, 0.000103)
(2.100020, 0.000100)
(2.110020, 0.000098)
(2.120020, 0.000100)
(2.130020, 0.000102)
(2.140020, 0.000100)
(2.150020, 0.000098)
(2.160020, 0.000100)
(2.170020, 0.000102)
(2.180020, 0.000100)
(2.190020, 0.000098)
(2.200020, 0.000100)
(2.210020, 0.000102)
(2.220020, 0.000100)
(2.230020, 0.000098)
(2.240020, 0.000100)
(2.250020, 0.000101)
(2.260020, 0.000100)
(2.270020, 0.000099)
(2.280020, 0.000100)
(2.290020, 0.000101)
(2.300020, 0.000100)
(2.310020, 0.000099)
(2.320020, 0.000100)
(2.330020, 0.000101)
(2.340020, 0.000100)
(2.350020, 0.000099)
(2.360020, 0.000100)
(2.370020, 0.000101)
(2.380020, 0.000100)
(2.390020, 0.000099)
(2.400020, 0.000100)
};
\addlegendentry{model fit} 

%% file: tex/table1.tex
\begin{table}[]
  \centering
  \footnotesize
  \begin{tabular}{c|cc|cc|cc}
  \hline
 Hartmann number & \multicolumn{2}{c|}{$\ha=10$} & \multicolumn{2}{c|}{$\ha=50$} & \multicolumn{2}{c}{$\ha=100$} \\ \hline
 Data & approx. & analytic & approx. & analytic & approx. & analytic \\\hline
  decay rate $s$ & 3.95 & 3.70 &  4.19  &  3.70 &   4.11 &  3.70 \\
  frequency  $\omega$
   &  15.27& 15.61 & 156.87 & 157.07 & 78.35 &  78.52  \\
  steady state $s_0$ &  0.012& 0.01 & 0.97$\times 10^{-4}$ & $ 1\times 10^{-4} $ & $3.5 \times 10^{-4}$ &  $4\times 10^{-4}$ \\
   \hline
  \end{tabular}
  \normalsize
  \caption{Comparison of decay rate, frequency, and steady state of center-point velocity fitted by Levenberg-Marquardt method.}
  \label{table:compare}
\end{table}

%% file: tex/table2.tex
\begin{table}[t]
    \centering
    \footnotesize
    \begin{tabular}{c|c|c|c}
    \hline
    \multirow{1}{*}{} &    
    \multirow{1}{*}{${r_w}$} &  \multirow{1}{*}{$\gg 1$} & \multirow{1}{*}{$\ll 1$} \\ 
    \cline{2-4}
    \multirow{2}{*}{} &    
    \multirow{2}{*}{Solid Domain} &  \multirow{ 2}{*}{Insulating}  & \multirow{ 2}{*}{Conducting}  \\
    \multirow{2}{*}{$\bB$} &    
    \multirow{2}{*}{} &  \multirow{ 2}{*}{}  & \multirow{ 2}{*}{}  \\   
    \cline{2-4}
    \multirow{2}{*}{} &    
    \multirow{2}{*}{Boundary Behaviors} & \multirow{2}{*}{Dirichlet} & \multirow{2}{*}{Neumann} \\
    \multirow{2}{*}{} &    
    \multirow{2}{*}{} & \multirow{2}{*}{} &  \multirow{2}{*}{} \\ 
    \cline{2-4}
    \hline
    \multirow{1}{*}{} &    
    Decay Rate & --  & $\frac{\pi^2}{4}\left(\frac{1}{R e}+\frac{1}{R m}\right) $  \\
    \cline{2-4}
    \multirow{1}{*}{$\bu$} &    
    Frequency & -- & $\frac{\pi}{2} \frac{\ha}{\sqrt{R e R m}}\left(1-\frac{\pi^2}{4 \ha^2} 
                     \frac{(R m-R e)^2}{R m R e}\right)^{\frac{1}{2}}$ \\
    \cline{2-4}
    \multirow{1}{*}{} &    
    Steady Solution  $u(0,0)$ & $\frac{1}{\ha}$ & $\frac{1}{\ha^2}$ \\ \hline
    \end{tabular}
    \normalsize 
    \caption{Description of the behaviors of $\bB$ on the solid domain $\Omega_S$ 
             and on the boundary $\d\Omega_{MD}$,
             and the decay rate, frequency and steady state solution of $\bu$
             for the Hartmann flows,
             depending on the electric conductivity ratio $r_w$.
             Dirichlet boundary conditions are used for both $\bB$ and $\bu$ for 
             the Hartmann flow simulations.}
    \label{tab:my_label}
\end{table}

%% file: tex/yichen.tex
\section{Test Cases}

\begin{figure}[htbp]
  \centering
  \includegraphics[width=0.7\textwidth]{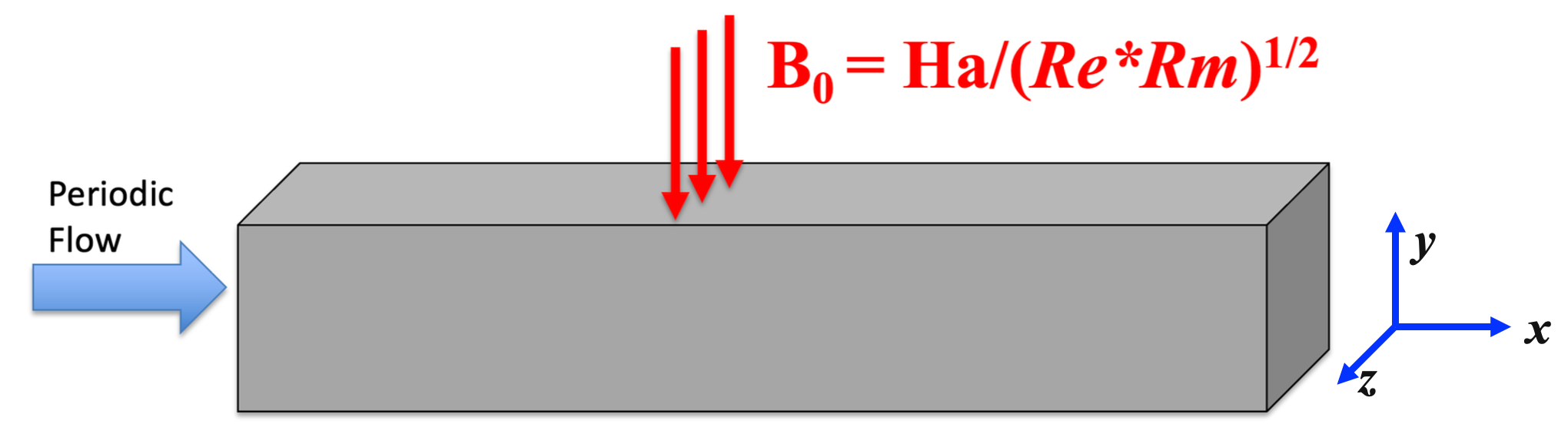}
  \caption{Laminar flow in a square duct with an applied magnetic field.}
  \label{fig:duct3d}
  \end{figure} 

In this section, we explore several Hartmann flow cases using the axially-periodic duct
configuration of Fig. \ref{fig:duct3d}.   Here, the flow is forced in the $x$ direction
and an external magnetic field of strength $B_0$ is applied in the $-y$ direction.
The first two cases, Shercliff flow and Hunt flow, have closed-form steady-state solutions
that can be used to establish spatial convergence.
Subsequently, we explore novel asymptotic approximations to the time-transient
response of impulsively-started Hartmann flow and compare these with the observed
response of NekRS in order to check correctness of the MHD temporal discretization.

As an initial verification case, we consider laminar  flow in a square duct
with a fixed mean pressure gradient $\frac{\partial p}{\partial x} =
-\frac{1}{Re}$ and an applied magnetic field, $\bB_0:=(0,B_0,0)$, as
illustrated in Fig.~\ref{fig:duct_3d}.  We impulsively start the flow in the
$x$-direction.  The simulations are carried out with periodic boundary
conditions in $x$.  We define the fluid domain as $\Omega_F = [0,
L]\times[-1,1]^2$ with Dirichlet boundary conditions $\bu|_{\dO_{FD}} =
(0,0,0)$.  The solid domain and magnetic boundary conditions are defined later
based on the problem.  The initial conditions of velocity and magnetic field
are zero.  With the applied pressure gradient, the flow evolves to a steady
state for the relatively low Reynolds numbers considered in these examples.

\subsection{Shercliff flow}

\begin{figure}
  \centering
  \begin{subfigure}[t]{0.34\textwidth} 
    \centering
      \makebox[7pt]{\raisebox{60pt}{\rotatebox[origin=c]{0}{
          $\bu$
      }}}%
      \includegraphics[width=6cm]{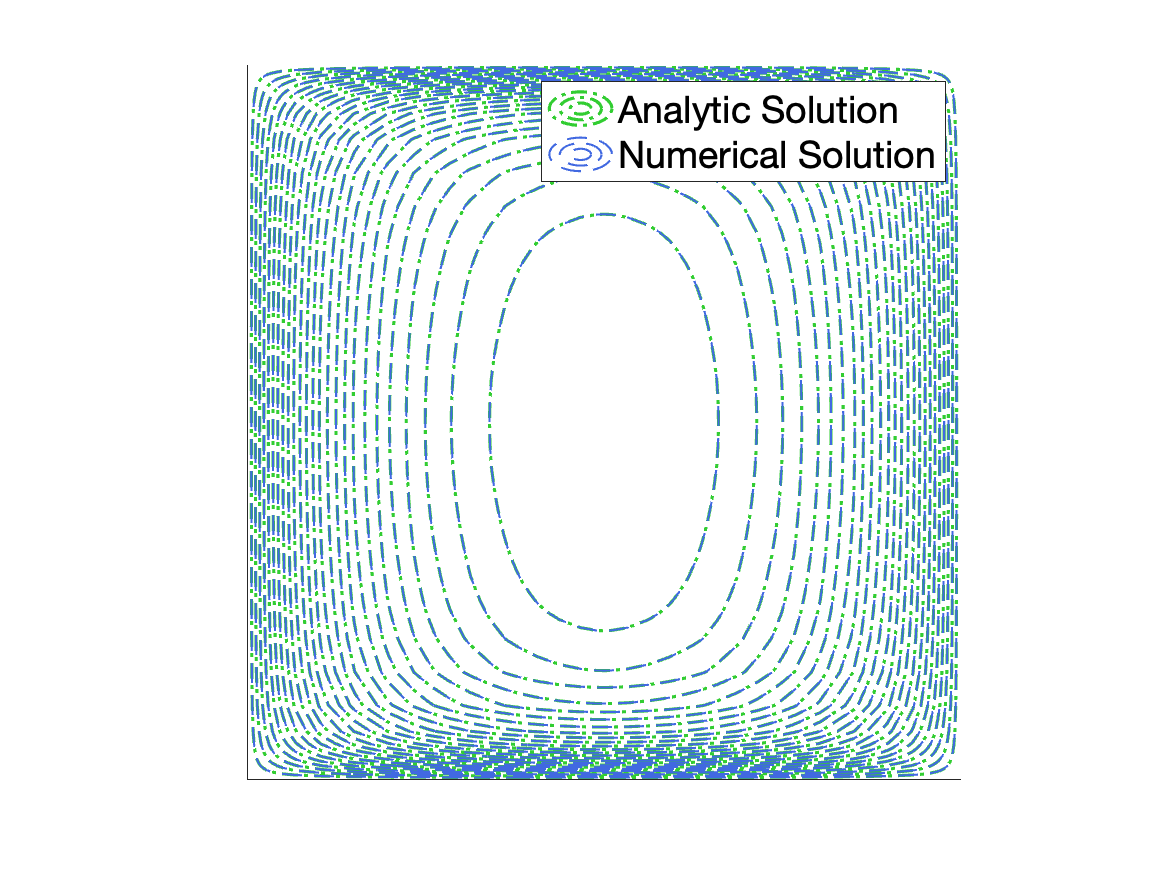}\\[5pt]
      \makebox[7pt]{\raisebox{60pt}{\rotatebox[origin=c]{0}{
          $\bB$
      }}}%
      \includegraphics[width=6cm]{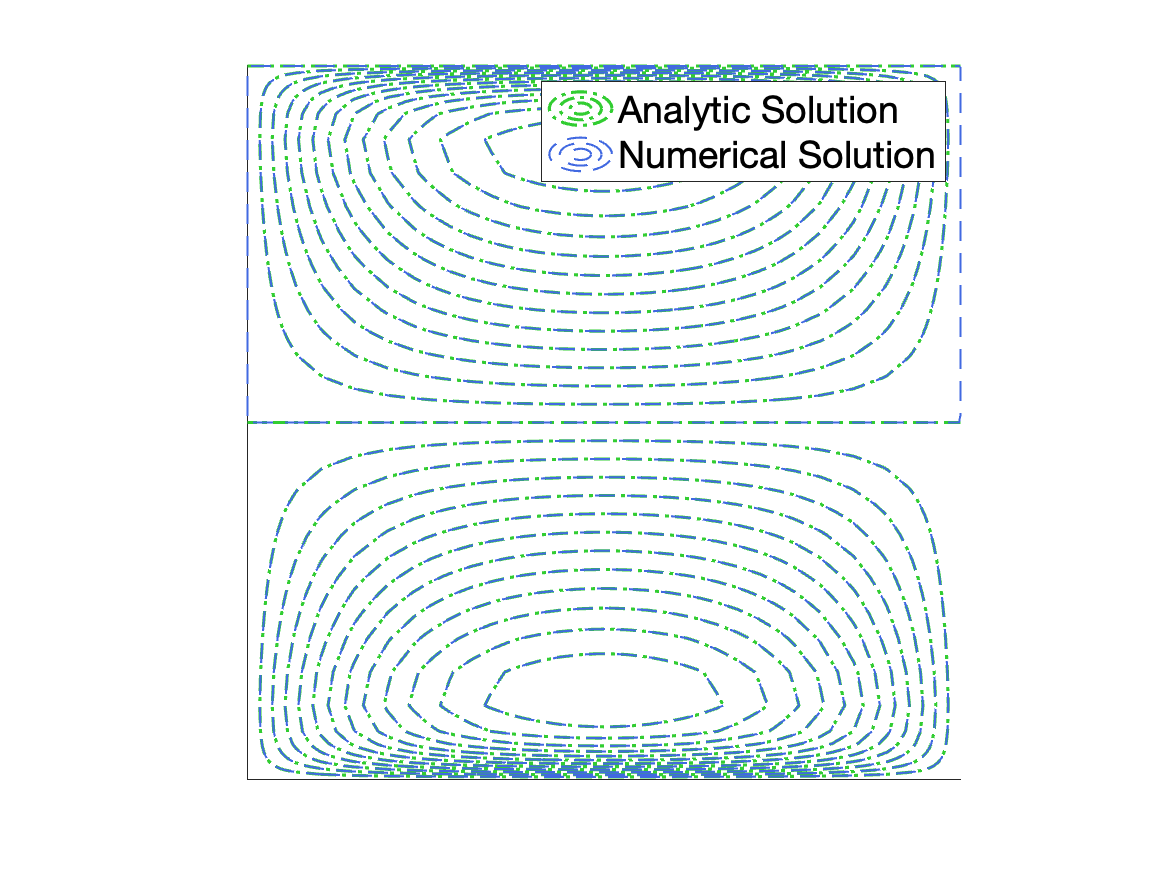}
      \caption{$\ha = 10$} 
  \end{subfigure}
  \hfill
  \begin{subfigure}[t]{0.33\textwidth} 
    \centering
      \includegraphics[width=6cm]{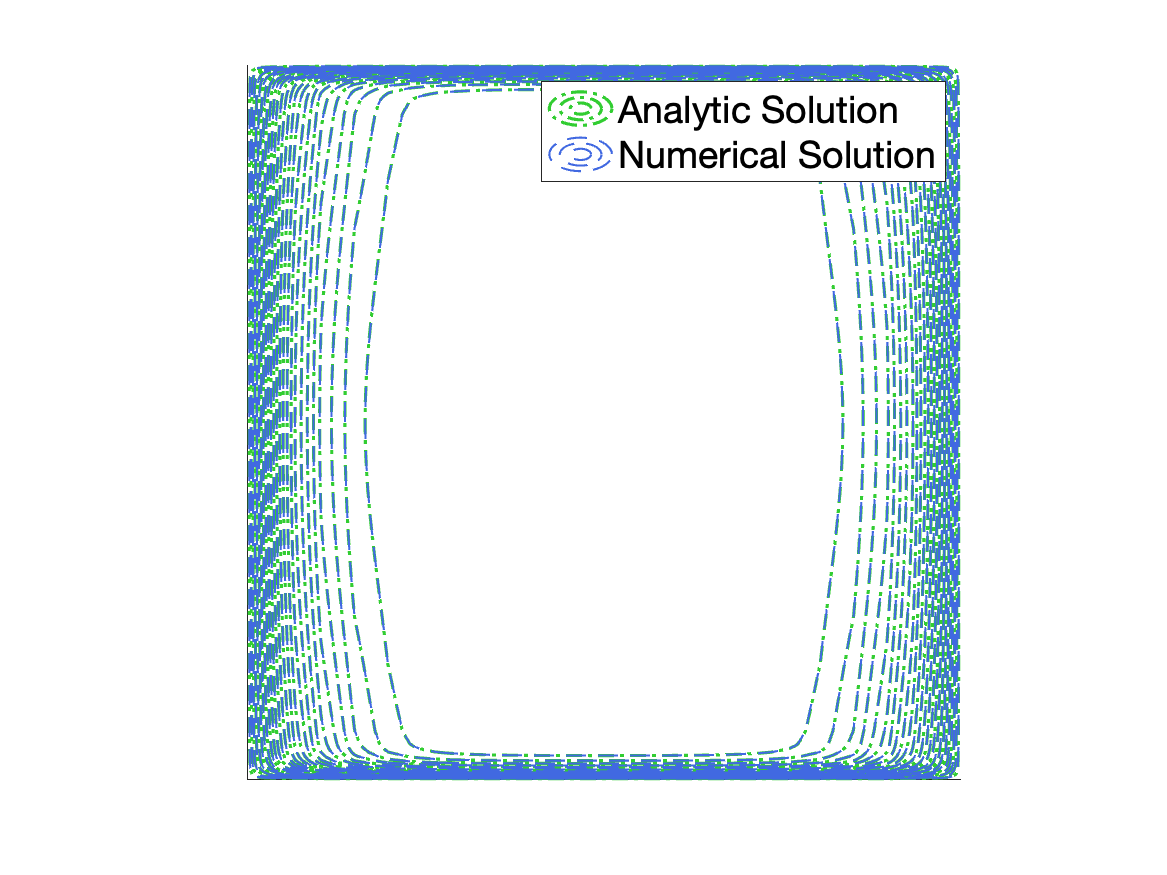}\\[5pt]
      \includegraphics[width=6cm]{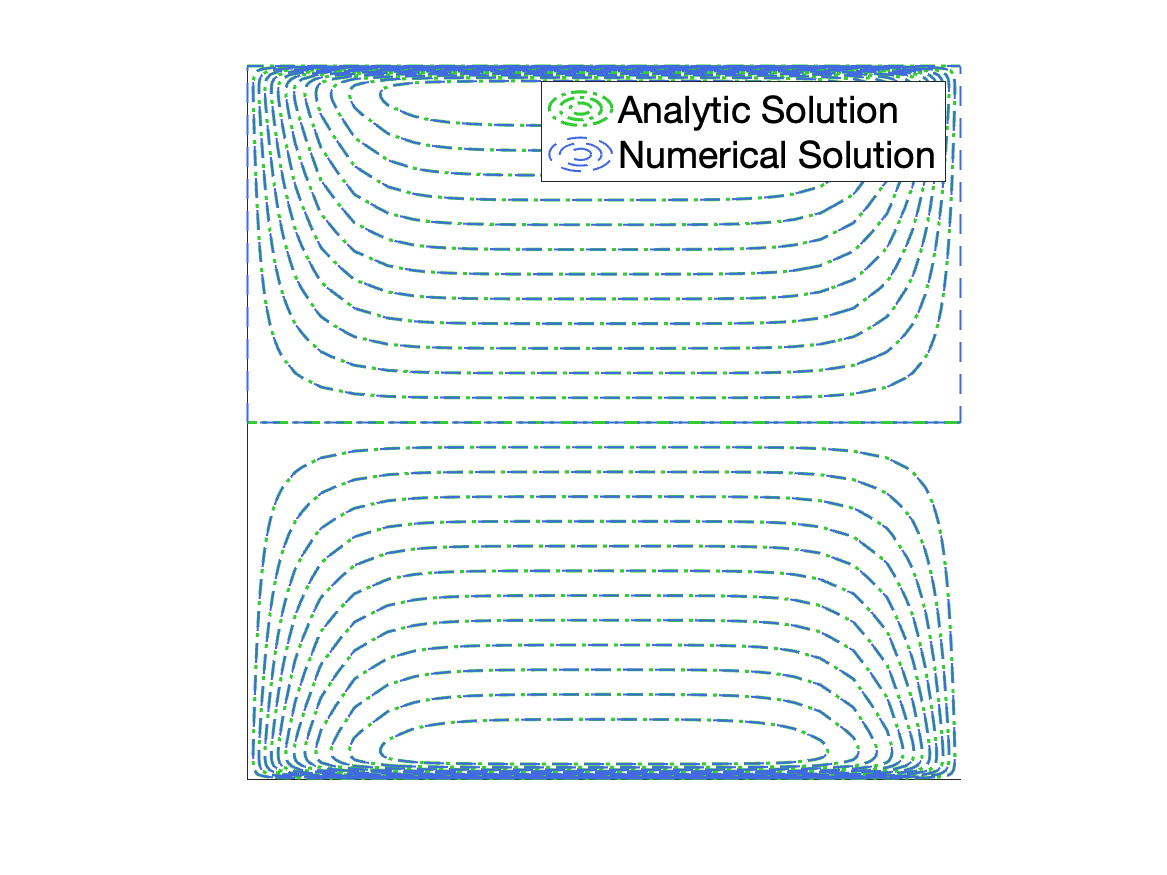}
      \caption{$\ha = 50$}
  \end{subfigure}
  \hfill
  \begin{subfigure}[t]{0.3\textwidth} 
    \centering
      \includegraphics[width=6cm]{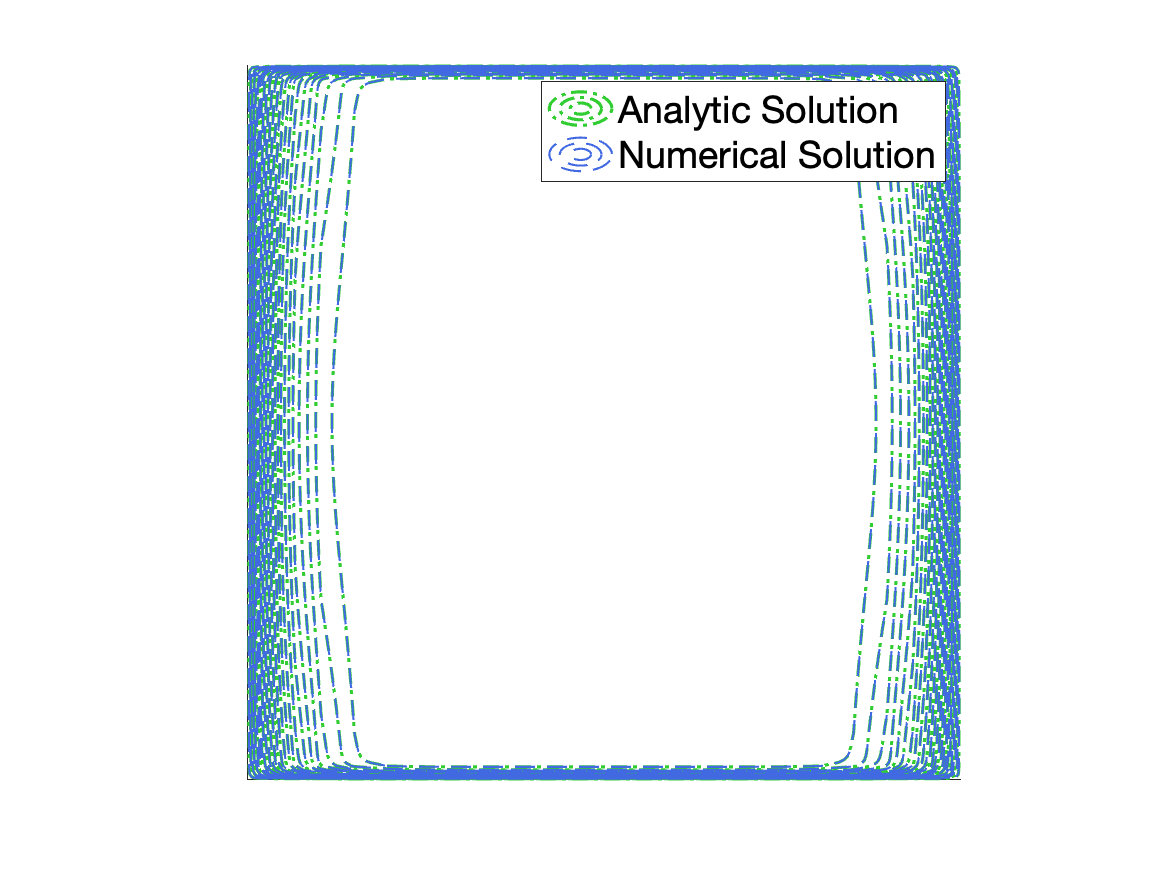}\\[5pt]
      \includegraphics[width=6cm]{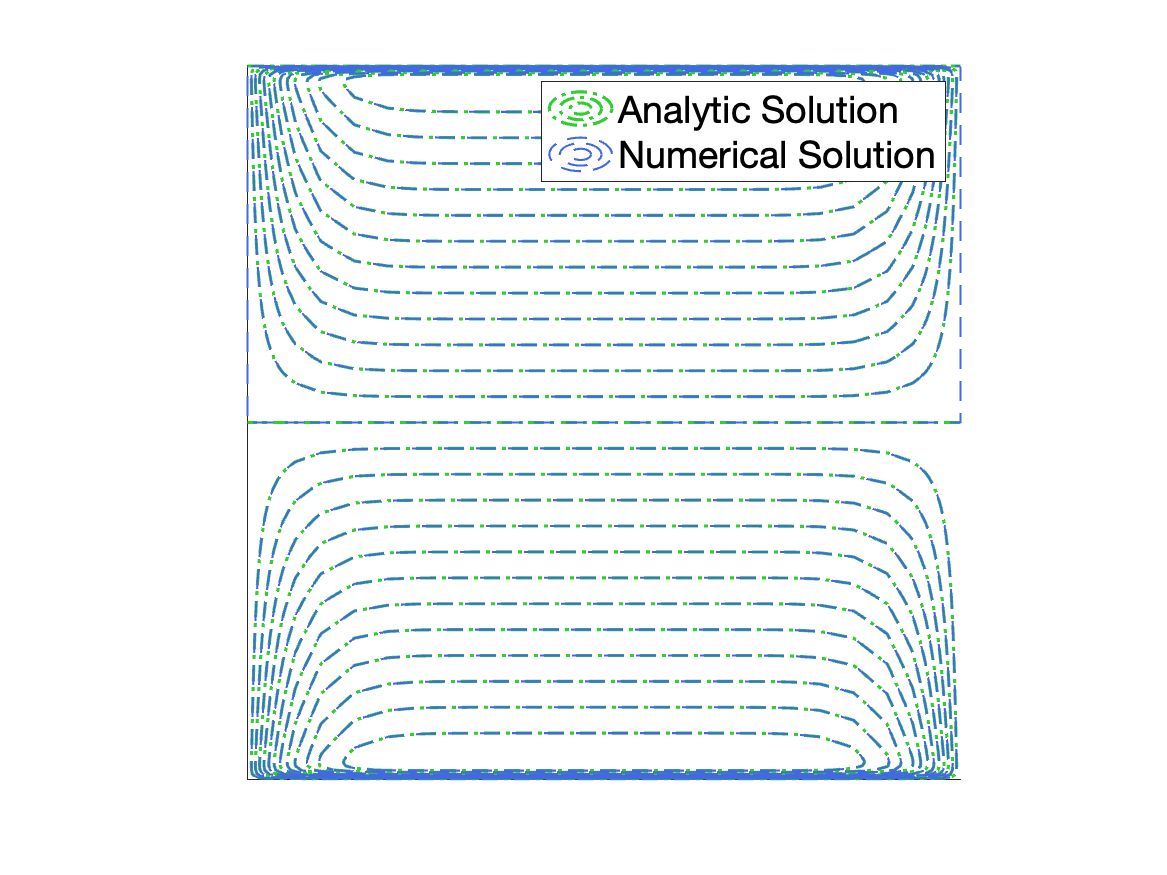}
      \caption{$\ha = 100$} 
  \end{subfigure}
  \caption{Velocity (top) and magnetic (bottom) distribution contour of the numerical solution 
  and the analytic solution for $\ha=10,50,100$ and polynomial order $N=6$.} 
  \label{compareMono}
  \end{figure}

\begin{figure}
  \centering
  \begin{tikzpicture}
    \begin{axis}[
      width=7cm, 
      height=6cm, 
      xlabel={Polynomial degree $N$}, 
      ylabel={$L^{\infty}$ error of velocity}, 
      grid=both, 
      legend style={at={(1.0,1)}, anchor=north east}, 
      xtick={2,4,6,8}, 
      ymode=log, 
      ]
      \addplot[color=blue, mark=*] coordinates {
        ( 2, 1.996360260688301e-02 ) 
        ( 4, 6.123753448923111e-04 ) 
        ( 6, 9.576270083684248e-06 ) 
        ( 8, 2.964274466838751e-07 )  
      };
    \addlegendentry{$\ha = 10$}
    \addplot[color=orange, mark=triangle*] coordinates {
      ( 2, 1.051879585061456e-01 ) 
      ( 4, 4.654242552134231e-03 ) 
      ( 6, 1.080659049710304e-04 ) 
      ( 8, 3.223775644510393e-06 ) 
      };
    \addlegendentry{$\ha = 100$}
    \end{axis}
    \end{tikzpicture}
    \hfill
    \begin{tikzpicture}
      \begin{axis}[
        width=7cm, 
        height=6cm, 
        xlabel={Polynomial degree $N$}, 
        ylabel={$L^{\infty}$ error of magnetic}, 
        grid=both, 
        legend style={at={(1.0,1)}, anchor=north east}, 
        xtick={2,4,6,8}, 
        ymode=log, 
        ]
        \addplot[color=blue, mark=*] coordinates {
          ( 2, 3.222371873071808e-02 ) 
          ( 4, 9.964802102786741e-04 ) 
          ( 6, 1.516709747882900e-05 ) 
          ( 8, 3.504790010170631e-07 ) 
        };
      \addlegendentry{$\ha = 10$}
      \addplot[color=orange, mark=triangle*] coordinates {
        ( 2, 1.113124071894065e-01 ) 
        ( 4, 4.940203228480270e-03 ) 
        ( 6, 1.126279402214886e-04 ) 
        ( 8, 3.540397437505826e-06 ) 
        };
      \addlegendentry{$\ha = 100$}
      \end{axis}
      \end{tikzpicture}
\caption{Convergence for Shercliff flow: the relative $L^{\infty}$ error of
velocity (left) and magnetic (right) fields with the polynomial degree $N$
varying from 2 to 8 and using $4\times 10 \times 10 $ elements.  The problem
considered is flow in rectangular channels on the domain $ [0,4]\times [-1,1]
\times [-1,1]$ with insulating walls and $Re=Rm=1$.  An analytic solution
\cite[eq. (4.37)]{muller2001magnetofluiddynamics, eardley2023scalable} is used to compute the error.}
 \label{fig:convergence}
  \end{figure}
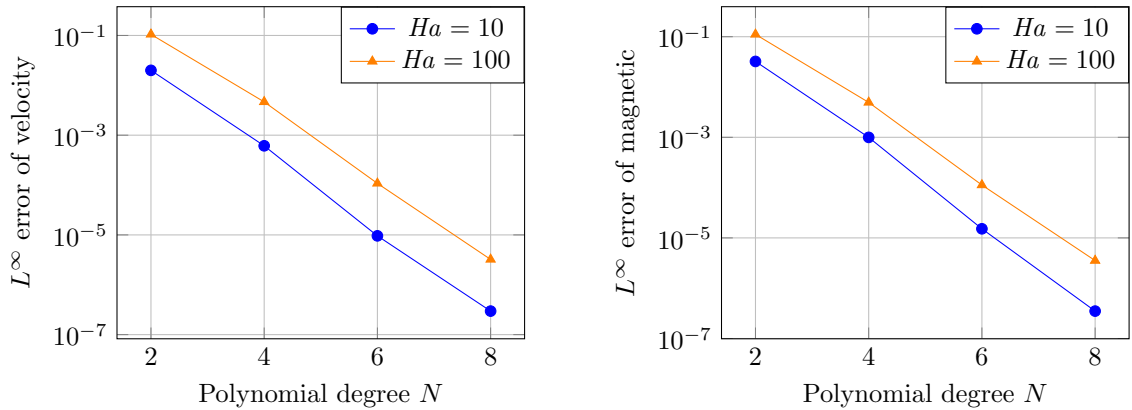

In this subsection, we examine a square channel flow with insulating walls
using the analytic solutions provided in \cite{muller2001magnetofluiddynamics}
for $\ha=10,50,100$.  For the first case, we consider $\Omega_M \equiv
\Omega_F$, applying homogeneous Dirichlet boundary conditions for both the
velocity and magnetic fields on $\partial \Omega_F$.  Figure~\ref{compareMono}
shows the numerical and analytic solutions for velocity and magnetic fields at
$\ha = 10, 50, 100$, using a polynomial order of $N = 6$.  The strong agreement
between the two demonstrates correctness of the spatial operators.
We establish the convergence rate by varying the polynomial degree from $N=2$ to $8$.
Exponential convergence is clear in the semi-log plot of Fig.~\ref{fig:convergence}, 
which shows the relative $L^{\infty}$ error for the velocity and magnetic fields
versus $N$.


  \subsection{Hunt flow}
\begin{figure}
  \centering
  \begin{subfigure}[t]{0.4\textwidth} 
      \makebox[25pt]{\raisebox{40pt}{\rotatebox[origin=c]{0}{
        $\bu$
      }}}%
      \includegraphics[width=\textwidth]{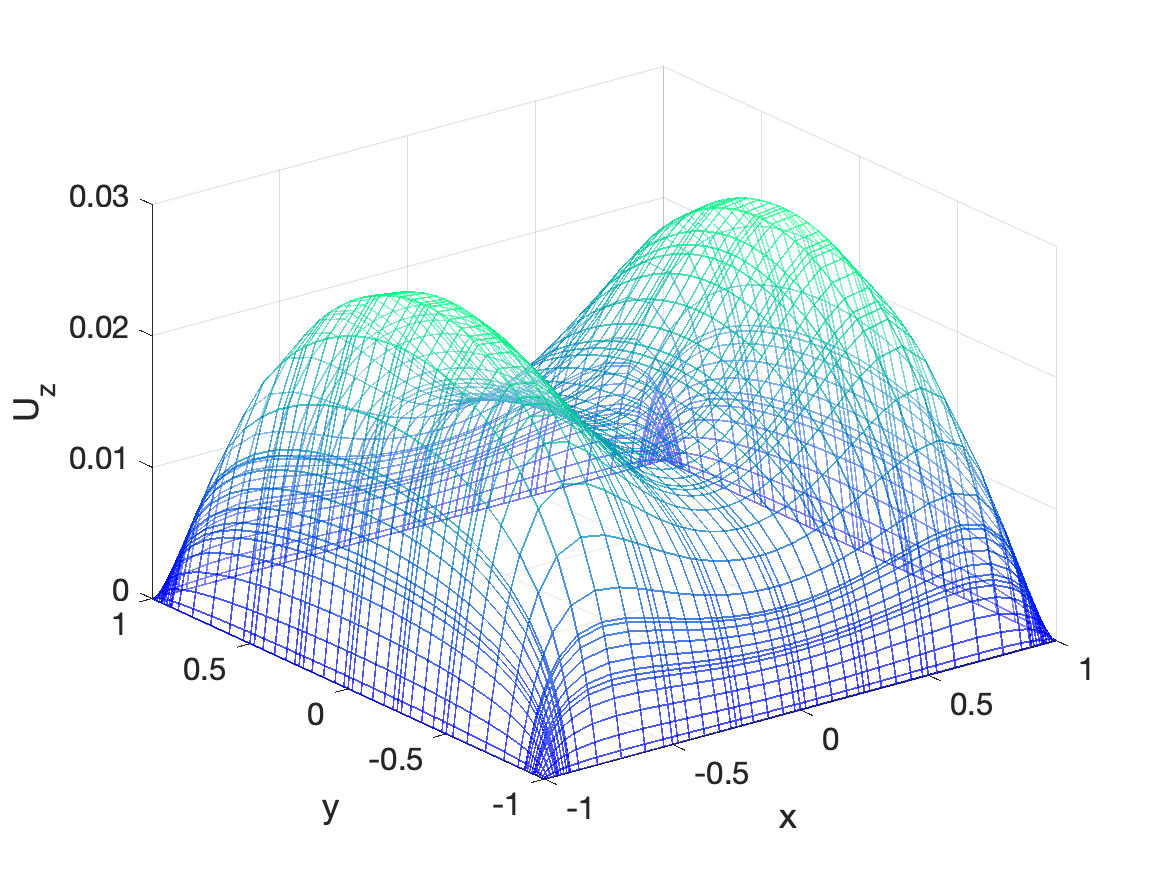}\\[5pt]
      \makebox[25pt]{\raisebox{40pt}{\rotatebox[origin=c]{0}{
        $\bB$
      }}}%
      \includegraphics[width=\textwidth]{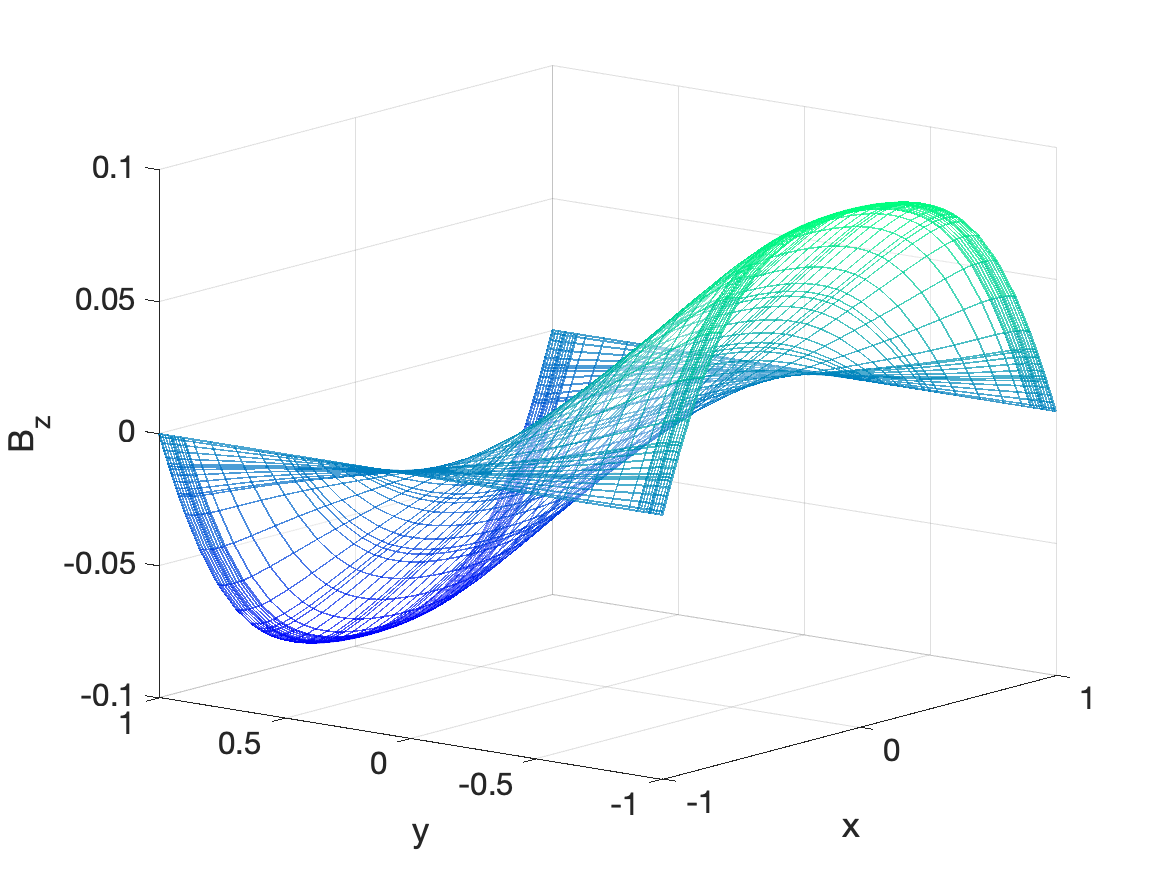}
      \caption{$\ha = 10$} 
  \end{subfigure}
  \hfill
  \begin{subfigure}[t]{0.4\textwidth} 
      \includegraphics[width=\textwidth]{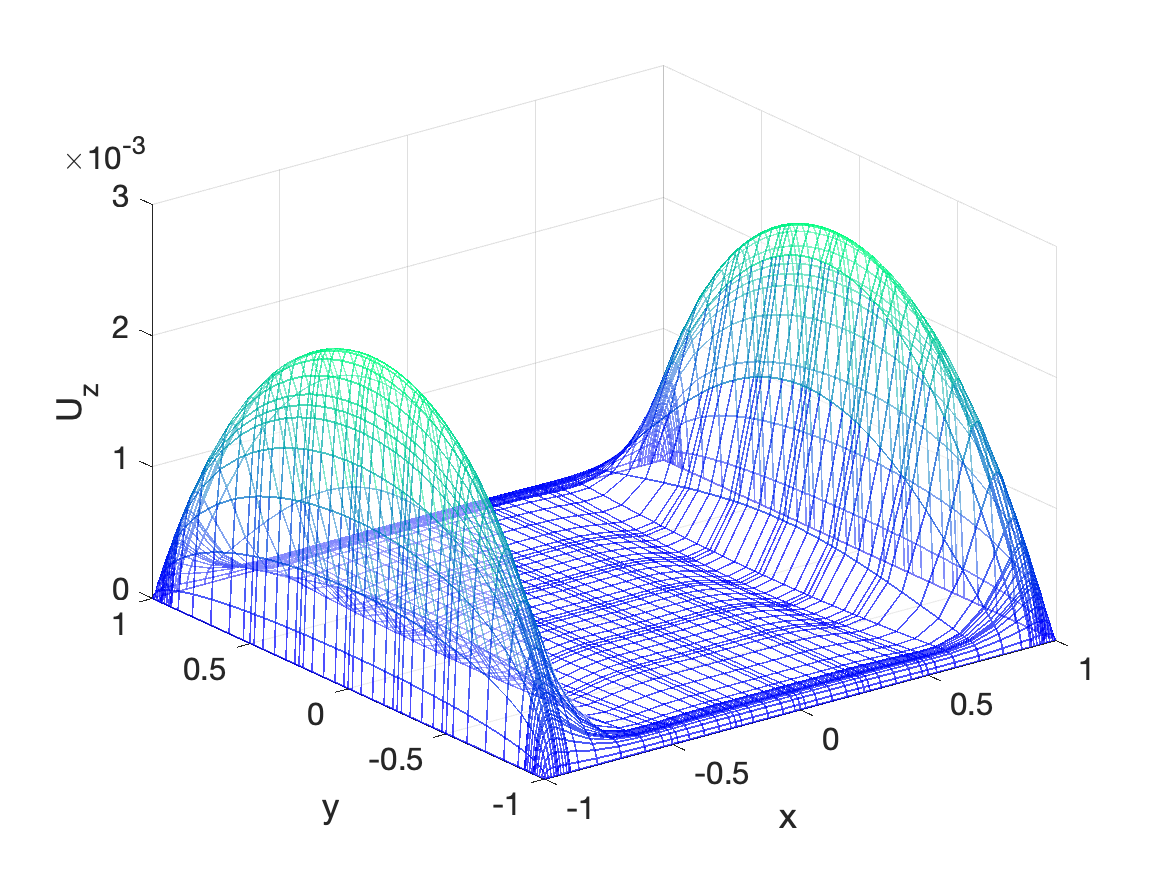}\\[5pt]
      \includegraphics[width=\textwidth]{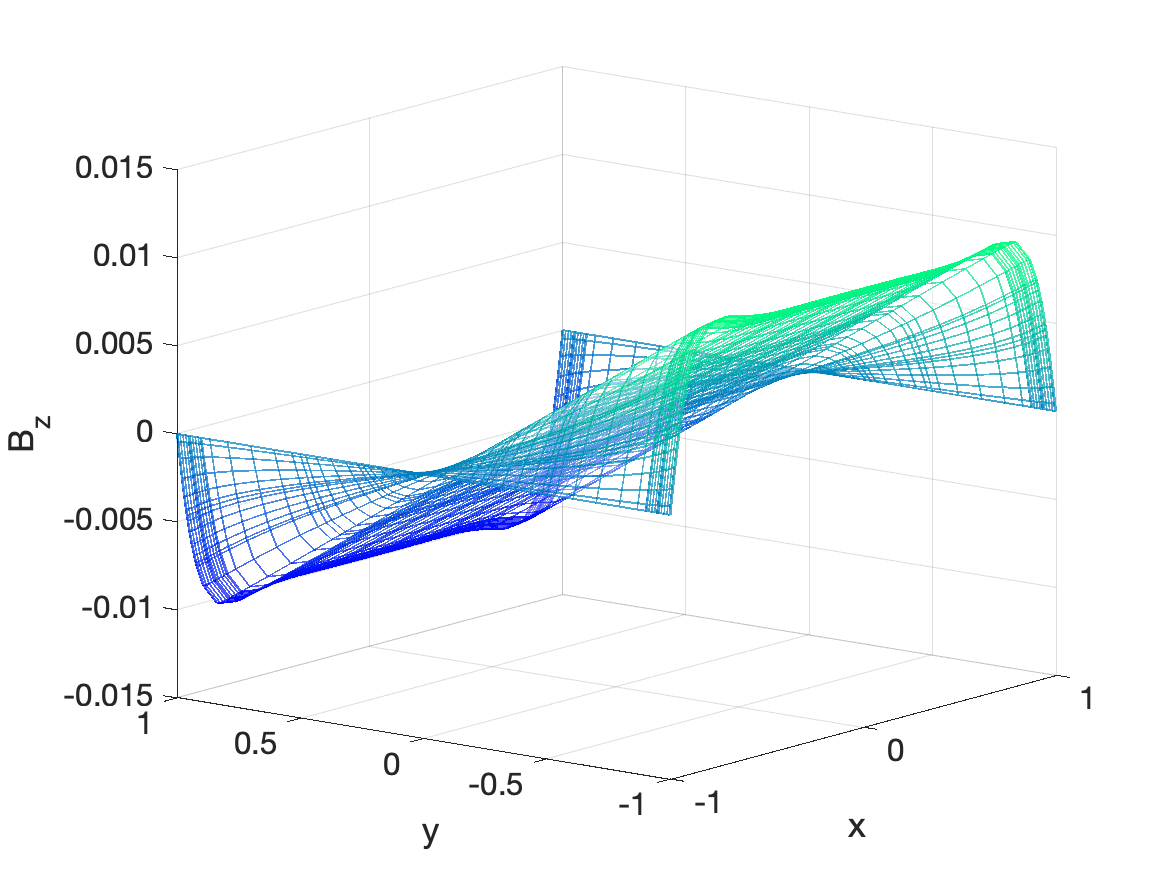}
      \caption{$\ha = 100$}
  \end{subfigure}
  \caption{Velocity(above) and magnetic(below) distribution of the numerical solution for Hunt flow 
 with $\ha=10$ and $\ha=100$, perfectly conducting and insulating walls, and polynomial order $N=6$.} 

  \label{compareTwoDomHunt}
  \end{figure}

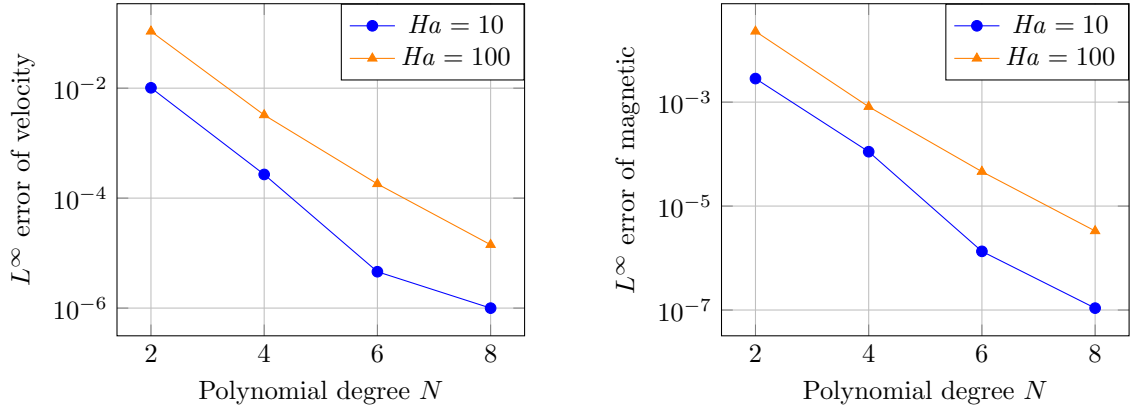
\begin{figure}
    \centering
    \begin{tikzpicture}
      \begin{axis}[
        width=7cm, 
        height=6cm, 
        xlabel={Polynomial degree $N$}, 
        ylabel={$L^{\infty}$ error of velocity}, 
        grid=both, 
        legend style={at={(1.0,1)}, anchor=north east}, 
        xtick={2,4,6, 8}, 
        ymode=log, 
        ]
        \addplot[color=blue, mark=*] coordinates {
        ( 2, 0.010145715798186 ) 
        ( 4, 0.000269431095501 ) 
        ( 6, 0.000004587058459 ) 
        ( 8, 0.000000998076168)
        };
      \addlegendentry{$\ha = 10$}
      \addplot[color=orange, mark=triangle*] coordinates {
        ( 2, 0.107797624438995 ) 
        ( 4, 0.003247270756889 ) 
        ( 6, 0.000181263926369 ) 
        ( 8, 0.000014189370395 )
        };
      \addlegendentry{$\ha = 100$}
      \end{axis}
      \end{tikzpicture}
      \hfill
      \begin{tikzpicture}
        \begin{axis}[
          width=7cm, 
          height=6cm, 
          xlabel={Polynomial degree $N$}, 
          ylabel={$L^{\infty}$ error of magnetic}, 
          grid=both, 
          legend style={at={(1.0,1)}, anchor=north east}, 
          xtick={2,4,6, 8}, 
          ymode=log, 
          ]
          \addplot[color=blue, mark=*] coordinates {
            ( 2, 0.002836823551710  ) 
            ( 4, 0.000111132418367  ) 
            ( 6, 0.000001345196084  ) 
            ( 8, 0.000000108742823  )
          };
        \addlegendentry{$\ha = 10$}
        \addplot[color=orange, mark=triangle*] coordinates {
          ( 2,  0.022949622909536 ) 
          ( 4,  0.000815480179298 ) 
          ( 6,  0.000045965057025 ) 
          ( 8,  0.000003315988255 )
          };
        \addlegendentry{$\ha = 100$}
        \end{axis}
        \end{tikzpicture}
        \caption{Convergence of the relative $L^{\infty}$ error of velocity (left) and magnetic (right) fields 
        with the polynomial degree $N$ varying from 2 to 8 and using $ 4 \times 10 \times 10 $ elements. 
        The problem considered is Hunt flow in rectangular channels on the domain $ [0,4]\times [-1,1] \times [-1,1] $ 
        with perfectly conducting and insulating walls and $Re=Rm=1$. 
        An analytic solution \cite{hunt1965magnetohydrodynamic, eardley2023scalable} is used to compute the error.}
    \label{fig:convergenceHunt}
    \end{figure}

  In \cite{hunt1965magnetohydrodynamic}, an analytic solution is provided for the Hunt flow problem. 
  The duct has perfectly conducting walls perpendicular to the applied magnetic field 
  and insulating walls parallel to it.
  Velocity has homogeneous Dirichlet boundary conditions on the walls, 
  while the magnetic field has homogeneous Neumann boundary conditions on the conducting walls 
  and homogeneous Dirichlet boundary conditions on the insulating walls.
  Figure~\ref{compareTwoDomHunt} shows the numerical solution of velocity and magnetic fields
  for $\ha=10$ and $\ha=100$ with $N=6$ and $4 \times 10 \times 10 $ elements. Comparing with the analytic
  solution, the convergence of the relative $L^{\infty}$ error of velocity 
  and magnetic fields is shown in Fig.~\ref{fig:convergenceHunt} with varying polynomial degrees from 2 to 8.

%% file: tex/conclusion.tex
\section{Conclusion}

This article introduces a new MHD capability into the highly-scalable
thermal-fluid simulation code, Nek5000/RS, and demonstrates basic verification
results including exponential convergence to known analytical steady-state
solutions and consistent results with novel time-transient responses based on
asymptotic analysis.   
The code has been recently applied to turbulent MHD Hartmann flows and
validation of these results is ongoing.

%% file: tex/acknowledgments.tex
\section*{Acknowledgments}
    This material is based upon work supported by the U.S. Department of Energy,
    Office of Science, under contract DE-AC02-06CH11357  and by the Exascale
    Computing Project (17-SC-20-SC), a collaborative effort of two U.S.
    Department of Energy organizations (Office of Science and the National
    Nuclear Security Administration).

%% file: fst24-fusion.bbl
\begin{thebibliography}{10}
\newcommand{\enquote}[1]{``#1''}
\providecommand{\url}[1]{\texttt{#1}}
\providecommand{\urlprefix}{URL }
\expandafter\ifx\csname urlstyle\endcsname\relax
  \providecommand{\doi}[1]{doi:\discretionary{}{}{}#1}\else
  \providecommand{\doi}{doi:\discretionary{}{}{}\begingroup
  \urlstyle{rm}\Url}\fi

\bibitem{fischer15}
\textsc{P.~Fischer}, \textsc{K.~Heisey}, and \textsc{M.~Min}, \enquote{Scaling
  Limits for {PDE}-Based Simulation (Invited),} \emph{22nd AIAA Computational
  Fluid Dynamics Conference, AIAA Aviation}, AIAA 2015-3049 (2015).

\bibitem{gb23}
\textsc{E.~Merzari}, \textsc{S.~Hamilton}, \textsc{T.~Evans},
  \textsc{P.~Romano}, \textsc{P.~Fischer}, \textsc{M.~Min},
  \textsc{S.~Kerkemeier}, \textsc{Y.~Lan}, \textsc{J.~Fang},
  \textsc{M.~Phillips}, \textsc{T.~Rathnayake}, \textsc{E.~Biondo},
  \textsc{K.~Royston}, \textsc{N.~Chalmers}, and \textsc{T.~Warburton},
  \enquote{Exascale Multiphysics Nuclear Reactor Simulations for Advanced
  Designs ({G}ordon {B}ell {P}rize {F}inalist paper),} \emph{Proc. of SC23:
  Int. Conf. for High Performance Computing, Networking, Storage and Analysis},
  IEEE (2023).

\bibitem{patel18}
\textsc{S.~Patel}, \textsc{P.~Fischer}, \textsc{M.~Min}, and
  \textsc{A.~Tomboulides}, \enquote{A Characteristic-Based, Spectral Element
  Method for Moving-Domain Problems,} \emph{J. Sci. Comp.}, \textbf{79}, 564
  (2019).

\bibitem{fischer17}
\textsc{P.~Fischer}, \textsc{M.~Schmitt}, and \textsc{A.~Tomboulides},
  \emph{Recent developments in spectral element simulations of moving-domain
  problems}, vol.~79, 213--244, Fields Institute Communications, Springer.

\bibitem{pat84}
\textsc{A.~Patera}, \enquote{A spectral element method for fluid dynamics :
  laminar flow in a channel expansion,} \emph{J. Comput. Phys.}, \textbf{54},
  468 (1984).

\bibitem{dfm02}
\textsc{M.~Deville}, \textsc{P.~Fischer}, and \textsc{E.~Mund},
  \emph{High-order methods for incompressible fluid flow}, Cambridge University
  Press, Cambridge (2002).

\bibitem{tomboulides89}
\textsc{A.~Tomboulides}, \textsc{M.~Israeli}, and \textsc{G.~Karniadakis},
  \enquote{Efficient Removal of Boundary-Divergence Errors in Time-Splitting
  Methods,} \emph{J. Sci. Comput.}, \textbf{4}, 291 (1989).

\bibitem{sao80}
\textsc{S.~Orszag}, \enquote{Spectral methods for problems in complex
  geometry,} \emph{J. Comput. Phys.}, \textbf{37}, 70 (1980).

\bibitem{johan13}
\textsc{J.~Malm}, \textsc{P.~Schlatter}, \textsc{P.~Fischer}, and
  \textsc{D.~Henningson}, \enquote{Stabilization of the Spectral-Element Method
  in convection dominated flows by recovery of skew symmetry,} \emph{J. Sci.
  Comp.}, \textbf{57}, 254 (2013).

\bibitem{muller2001magnetofluiddynamics}
\textsc{U.~M{\"u}ller} and \textsc{L.~B{\"u}hler}, \emph{Magnetofluiddynamics
  in channels and containers}, Springer Science \& Business Media (2001).

\bibitem{hunt1965magnetohydrodynamic}
\textsc{J.~Hunt}, \enquote{Magnetohydrodynamic flow in rectangular ducts,}
  \emph{Journal of fluid mechanics}, \textbf{21}, \emph{4}, 577 (1965).

\end{thebibliography}
